\documentclass[12pt]{article}
\usepackage{a4wide}
\usepackage{amssymb}
\usepackage{mathrsfs}
\usepackage{comment}
\usepackage{tensor}
\usepackage{amsmath}

\usepackage[dvipsnames]{xcolor}

\begin{document}
{\renewcommand{\thefootnote}{\fnsymbol{footnote}}
\hfill  \\
\medskip
\begin{center}
{\LARGE Perturbative emergent modified gravity\\[2mm] on cosmological backgrounds: Kinematics}\\
\vspace{1.5em}

Martin Bojowald,$^1$\footnote{e-mail address: {\tt bojowald@psu.edu}}
Manuel D\'{\i}az,$^2$\footnote{e-mail
address: {\tt manueldiaz@umass.edu}}
and
Erick I. Duque$^1$\footnote{e-mail address: {\tt eqd5272@psu.edu}}

\vspace{1em}
$^1$ Institute for Gravitation and the Cosmos,
The Pennsylvania State
University,\\
104 Davey Lab, University Park, PA 16802, USA\\
\vspace{0.5em}

$^2$  Amherst Center for Fundamental Interactions,\\ Department of Physics,
University of Massachusetts Amherst,\\
426 Lederle Graduate Research Tower,
Amherst, MA 01003 USA\\

\vspace{1.5em}
\end{center}
}

\setcounter{footnote}{0}

\begin{abstract}
  Emergent modified gravity has shown that the canonical formulation of
  general relativity gives rise to a larger class of covariant modifications
  than action-based approaches, so far in symmetry-reduced models. This
  outcome is made possible by distinguishing between the space-time metric on
  a given solution, and the basic field degrees of freedom in which equations
  of motion are formulated. In this general treatment, the metric is no longer
  fundamental but emerges after field equations and covariance conditions are
  solved. Here, the results are extended to perturbative inhomogeneity on a
  spatially flat cosmological background, showing that new modifications are
  possible while maintaining the classical derivative order and setting the
  stage for dynamical equations suitable for detailed studies of
  early-universe models.
\end{abstract}

\tableofcontents

\section{Introduction}

Having a sufficiently large class of modified theories of gravity is important
for detailed observational tests of general relativity. At the same time,
suitable theories must obey important conditions for their physical viability,
such as stability \cite{OstrogradskiProblem} and at most a tiny difference
between the propagation speeds for gravitational and electromagnetic waves
\cite{MultiMess1,MultiMess2}. While the standard approach based on
higher-curvature actions or other covariant modifications can provide a large
class of modified gravity theories, the general physical conditions are
fulfilled only in a limited number of cases.

Compared with action-based modification of general relativity, the canonical
formulation provides a different approach in which the
crucial dynamical objects, the Hamiltonian and diffeomorphism constraints, are
spatial integrals within a foliation of space-time. Unlike the action, these
expressions do not require an independent space-time integration measure and
can potentially be generalized or modified more broadly. This expectation has
been realized so far in spherically symmetric \cite{Higher,HigherCov,SphSymmMinCoup} and
Gowdy models \cite{EmergentGowdy}, where new modifications are possible even
at the classical order of derivatives in the field equations. Starting with
generic expressions for the Hamiltonian constraint that contains all terms up
to a given derivative order, gauge consistency and covariance conditions can
be imposed and solved, resulting in an effective theory of covariant
modifications of general relativity. With a successful solution of covariance
conditions, every solution of the field equations can be equipped with a
geometrical interpretation by deriving a compatible emergent space-time
metric. In contrast to action-based modifications, the emergent space-time
metric is a function of the fundamental fields but not necessarily equal to
one of them. This property relaxes implicit assumptions in previous theories
of modified gravity and leads to new consistent theories. Examples for new
effects are non-singular black-hole models \cite{SphSymmEff,EmergentMubar} and
MOND-like properties in a fully relativistic formulation without extra fields
\cite{HigherMOND}.

Here, we initiate a formulation of emergent modified gravity for perturbative
inhomogeneity on a spatially flat cosmological background. The resulting
dynamical equations will be suitable for a generalized effective field theory
of cosmology in which various effects can be modeled, even at the classical
order of derivatives. These equations are also suitable for effective theories
of canonical quantum gravity or other approaches to quantum gravity in which a
continuum space-time metric is expected to emerge but not to be
fundamental. An application to loop quantum cosmology is discussed in
\cite{EMGCosmo}, where previously proposed equations are shown to be either
non-covariant or singular. This shortcoming can be addressed by using the
general results found here within emergent modified gravity.

In a cosmological context, perturbative inhomogeneity was first formulated
canonically in \cite{HamGaugePert}. However, several ingredients such as
gauge-invariant combinations of the perturbations were not derived within the
canonical theory but rather taken directly from the 4-dimensional coordinate
picture. In modified theories, gauge transformations are expected to change
along with the dynamics. It is therefore important to have a complete
procedure for the derivation of potentially observable quantities within the
canonical formulation. This step was completed in
\cite{ScalarGaugeInv,ScalarGaugeInvErratum}, having in mind applications in
loop quantum cosmology \cite{LivRev}. We will follow the same methods here in
the broader context of emergent modified gravity, including a complete set of
modification functions suitable for an effective theory at the classical order
of derivatives.

Following this work, and also for convenience, we use Schwinger-type canonical
variables \cite{Schwinger} given by a densitized triad $E^a_i$, which provides
the spatial metric via $\delta^{ij} E^a_iE^b_j=\det (q) q^{ab}$ or
$q^{a b} = \delta^{ij} E^a_iE^b_j / |\det E|$, and its conjugate counterpart $K_a^i$,
classically related to the extrinsic curvature $K_{ab}$ by
$K_a^i=\delta^{ij}E^b_jK_{ab}/\sqrt{|\det (E)|}$. We use Greek letters for space-time
indices and Latin letters for spatial indices. Among the latter, we reserve
$a,b,c,\ldots$ for tangent-space indices, and $i,j,k,\ldots$ for internal
indices that label different components of the triad. In the emergent context,
the spatial functions $E^a_i$ and $K_b^j$ still present phase-space degrees of
freedom with canonical Poisson brackets, but they lose their geometrical roles
as components of a densitized triad and extrinsic curvature. These geometrical
concepts are emergent and can be derived from the emergent space-time metric after
equations of motion have been solved.

Section~\ref{s:CosmoPert} provides a brief introduction into the underlying classical
formalism of cosmological perturbation theory for scalar modes, which is generalized to
emergent modified gravity in Section~\ref{s:Vacuum} for vacuum space-times, in
Section~\ref{s:Fluid} to gravity coupled to perfect fluids, and in
Section~\ref{s:Scalar} to gravity coupled to scalar matter. The consistency and
covariance conditions become successively more complicated, but we will show
that the final and most complete version does have non-trivial solutions for
modification functions not equal to their classical versions. We therefore
demonstrate the existence of new covariant theories of emergent modified
gravity in the setting of cosmological perturbation theory. Conceptual
implications are presented in our discussion Section~\ref{s:Discussion}. The main
text presents only technical details that are necessary for following the
underlying procedures. Intermediate derivations are collected in a set of appendices.

%%%%%%%%

%%%%%%%%%

\section{Cosmological perturbation theory in canonical form}
\label{s:CosmoPert}

The Friedmann--Lemaitre--Robertson--Walker (FLRW) line element
\begin{equation} \label{FRW}
  {\rm d}s^2=-N^2{\rm d}t^2+a^2 \delta_{ab}{\rm d}x^a{\rm d}x^b
\end{equation}
(assuming flat spatial slices) has a preferred foliation of constant-$t$
hypersurfaces that can be used for a canonical formulation. The
spatial metric $q_{ab}$ of the general canonical line element
\begin{equation}
    {\rm d} s^2 = - N^2 {\rm d} t^2 + q_{a b} ( {\rm d} x^a + N^a {\rm d} t ) ( {\rm d} x^b + N^b {\rm d} t )
    \label{eq:ADM line element}
\end{equation}
is then uniquely determined by the scale factor $a(t)$, while the lapse
function $N(t)$ determines the specific form of the time coordinate. The
spatial shift vector $N^a$ always vanishes in a strictly isotropic slicing.

\subsection{Canonical fields}

For formal reasons related to the derivative structure of compatible
Hamiltonians, it is convenient to develop the canonical formulation using
$\bar{p}=a^2$ as the canonical configuration variable, turning the line
element into
\begin{equation}
    {\rm d} s^2 = - \bar{N}^2 {\rm d} t^2 + \bar{p} \delta_{a b} {\rm d} x^a {\rm d} x^b
    \,.
    \label{eq:ADM line element - background}
\end{equation}
(We use bars in order to emphasize the spatially homogeneous nature of the
variables.) The spatial tensor determined by $\bar{p}$ can be identified with
the spatial metric $\bar{q}_{ab}$, or with the densitized triad $\bar{E}^a_i$
such that
\begin{equation}
  \bar{q}^{ab}\det(\bar{q})=\bar{E}^a_i\bar{E}^b_j\delta^{ij}\,.
\end{equation}
Just like $\delta_{ab}$, $\delta_{ij}$ denotes the flat spatial metric and is
used to raise and lower spatial and internal indices, respectively. We use
$\delta^i_a$ to convert from spatial to internal indices and vice versa.

The canonical momentum conjugate to $\bar{E}^a_i$ is given by
\begin{equation}
  \bar{K}_a^i=\frac{\bar{K}_{ab}\bar{E}^b_j\delta^{ij}}{\sqrt{|\det(\bar{E})|}}
\end{equation}
with the extrinsic-curvature tensor $\bar{K}_{ab}$. These variables were first
introduced by Schwinger \cite{Schwinger} and are closely related to the
canonical fields commonly used in models of loop quantum gravity, going back
to \cite{SymmRed,IsoCosmo,SphSymmHam}.

In the isotropic slicing, we have
\begin{equation}\label{eq:Homogeneous variables}
    \bar K_a^i = \bar k \delta_a^i, \quad \bar E_i^a = \bar p \delta_i^a
\end{equation}
and
\begin{equation}\label{eq:PB homogeneous}
    \left\{\bar k,\bar p\right\}=\frac{\kappa}{3V_0}
\end{equation}
where $\kappa = 8\pi G$ and $V_0$ is the coordinate volume of a spatial
avaraging region in which the geometry can be considered approximately
homogeneous. In what follows, we will assume $\bar{p}>0$, using right-handed
triads.

The dynamics is determined by the Hamiltonian constraint with density
\begin{equation}
    \label{HamConstH0}
    {\mathcal H}^{(0)} = -6\bar{k}^2 \sqrt{\bar p}
    + 2\Lambda \bar{p}^{3/2}
\end{equation}
equal to $2\kappa/V_0$ times the matter energy. The resulting condition is
equivalent to the Friedmann equation in canonical variables. The Raychaudhuri
equation is equivalent to Hamilton's equation generated by the constraint
expression
\begin{equation}
  \bar{H}[\bar{N}]=\frac{1}{2\kappa} V_0 \bar{N} {\cal
    H}^{(0)}=\frac{1}{2\kappa}\int{\rm d}^3x \bar{N} {\cal H}^{(0)}
\end{equation}
plus matter terms. In the second version, we integrate over the averaging
region with coordinate volume $V_0$.

\subsection{Perturbations}
\label{Pert}

In the presence of perturbative inhomogeneity, we require that corrections to
the Hamiltonian constraint can in principle be obtained from a full
(non-perturbative) theory. Hence, they should depend only on the full triad
$E^a_i\equiv\bar E^a_i+\delta E^a_i$ rather than on the background
$\bar E^a_i$ and inhomogeneous perturbations $\delta E^a_i$ as distinct
arguments.  This perturbative treatment was performed in
\cite{ScalarGaugeInv,ScalarGaugeInvErratum}, which we follow in this entire
section. 

In general, all space-time metric components are perturbed by inhomogeneous
contributions, including the lapse function $N=\bar{N}+\delta N$, the shift
vector $N^a=\delta N^a$, and the canonical pair
$K_a^i=\bar K_a^i + \delta K_a^i$ and $E^a_i=\bar E_i^a + \delta E_i^a$.  For
scalar modes, which will be our main interest here, the canonical variables
are described by a pair of scalar functions each: We have
\begin{equation}\label{ScalarPerturbations}
\delta K_a^i = \delta_a^i \kappa_1 + \partial_a \partial^i
\kappa_2, \quad \delta E_i^a = \delta_i^a \varepsilon_1 +
\partial_i \partial^a \varepsilon_2\,.
\end{equation}
Here, $\partial_a\partial^a \kappa_2=0$ and
$\partial_a\partial^a \varepsilon_2=0$, such that $\kappa_1$ and
$\varepsilon_1$ correspond to (one third of) the trace of the respective canonical variables,
while $\kappa_2$ and $\varepsilon_2$ parameterize their scalar traceless
contributions. The lapse function is already a spatial scalar, while the shift
vector for scalar modes is the gradient of another scalar function.

In the perturbative context, the independent phase-space
variables are $(\bar k,\bar p)$ and $(\delta K_a^i,\delta E_i^a)$. Since
canonical pairs are independent of one another, the perturbations must be free
of background contributions, such that
\begin{equation} \label{KE0}
  \int{\rm d}^3x \delta K_a^i\delta^a_i=0= \int{\rm d}^3x \delta
  E^a_i\delta_a^i\,.
\end{equation}
Similarly,
\begin{equation}\label{N0}
  \int{\rm d}^3x \delta N=0
\end{equation}
in order to maintain the correct number of gauge transformations.
The non-trivial Poisson brackets between the perturbed contributions are given by
\begin{equation}\label{BasicPBGrav}
    \left\{ \delta K_a^i(x) , \delta E_j^b(y) \right\}= \kappa \delta^i_j \delta_a^b\delta^3(x-y)
    \,,
\end{equation}
while the brackets between background and perturbed variables vanish. (On the
solution space of the second-class constraints (\ref{KE0}), the Poisson
bracket (\ref{BasicPBGrav}) is replaced by the Dirac bracket, which subtracts
$1/V_0$ from $\delta^3(x-y)$.)

In general relativity, an expansion of the full Hamiltonian constraint up to
second order in inhomogeneity takes the form
\begin{equation} \label{ClassPertHamConst0}
H[N] =  \frac{1}{2\kappa}\int_{\Sigma}\mathrm{d}^3x N
{\mathcal H} = H[\bar{N}] + H[\delta N] \,,
\end{equation}
where
\begin{equation}
\label{ClassPertHamConst}
    H[\bar{N}] = \frac{1}{2\kappa}\int \mathrm{d}^3x\bar{N}\left[ {\mathcal H}^{(0)} + {\mathcal H}^{(2)}\right]
    \quad,\quad 
    H[\delta N] =  \frac{1}{2\kappa}\int \mathrm{d}^3x \delta N{\mathcal H}^{(1)}
\end{equation}
in terms of ${\cal H}^{(0)}$ from
(\ref{HamConstH0}) as well as 
\begin{eqnarray} \label{HamConstH1}
    {\mathcal H}^{(1)} &=&
    - 4 \bar{k}\sqrt{\bar{p}} \delta^c_j\delta K_c^j
    -\frac{\bar{k}^2-\Lambda \bar{p}}{\sqrt{\bar{p}}} \delta_c^j\delta E^c_j
    +\frac{2}{\sqrt{\bar{p}}} \partial_c\partial^j\delta E^c_j
~,\\
\label{HamConstH2}
    {\mathcal H}^{(2)} &=&
    \bar{p}^{3/2} \left(\frac{\delta K_c^j\delta K_d^k\delta^c_k\delta^d_j}{\bar{p}}
    - \frac{(\delta K_c^j\delta^c_j)^2}{\bar{p}}\right)
    - \frac{2\bar{k}}{\sqrt{\bar{p}}} \delta E^c_j\delta K_c^j
    \\
    && 
    - \frac{(\bar{k}^2 +\Lambda \bar{p})\sqrt{\bar{p}}}{2} \left(\frac{\delta_c^k\delta_d^j\delta E^c_j\delta E^d_k}{\bar{p}^2}
    - \frac{(\delta_c^j \delta E^c_j)^2}{2\bar{p}^2} \right)
    - \frac{\bar{p}^{3/2}}{2} \frac{\delta^{jk} (\partial_c\delta E^c_j)
       (\partial_d\delta E^d_k)}{\bar{p}^3}\,.\nonumber 
\end{eqnarray}
Using (\ref{KE0}) and (\ref{N0}), the equations
$\int {\rm d}^3 x\bar{N} {\cal H}^{(1)}=0=\int {\rm d}^3 x\delta N {\cal
  H}^{(0)}$ explain the splitting of the perturbed constraint into the two
terms in (\ref{ClassPertHamConst}).

Inhomogeneous perturbations are subject to an independent constraint that
vanishes identically for strictly homogeneous variables:
The second-order
diffeomorphism constraint
\begin{equation} \label{PertDiffConst}
D_c[\delta N^c] = \frac{1}{\kappa}\int_{\Sigma}\mathrm{d}^3x\delta N^c
\left[\bar{p}\partial_c(\delta^d_k \delta K^k_d)
-\bar{p}(\partial_k\delta K^k_c)- \bar{k} \partial_d (\delta_c^k\delta E^d_k)\right] 
\end{equation}
must equal suitable matter contributions. For later purposes, 
we define the density ${\cal D}_c$ by $D_c[\delta N^x]= \kappa^{-1} \int {\rm
  d}^3x\ \delta N^c {\cal D}_c$. Triad variables are, in general, also subject
to a Yang--Mills type Gauss constraint that generates gauge transformations
rotating the triad vectors. This constraint vanishes identically for scalar modes.

The space-time geometry can be constructed from solutions for the canonical
fields. 
To linear order in the fields, the inverse spatial metric is given by
\begin{eqnarray}\label{eq:Perturbed structure function - classical}
    q^{a b} &=& \frac{1}{\bar{p}} \Bigg[ \delta^{a b}
    + \frac{\delta^{aj} \delta E^b_j
    + \delta^{bj} \delta E^a_j}{\bar{p}}
    - \delta^{a b} \frac{\delta_c^j\delta E^c_j}{\bar{p}}
    \Bigg]
    \nonumber\\
    &=:&
    \bar{q}^{a b} + \delta q^{a b}
    \,,
\end{eqnarray}
and hence the spatial metric is
\begin{eqnarray}
    q_{ab} &=& \bar{p} \left[ \delta_{a b} - \delta_{a c} \delta_{b d} \frac{\delta^{cj} \delta E^d_j
    + \delta^{dj} \delta E^c_j}{\bar{p}}
    + \delta_{a b} \frac{\delta_c^j\delta E^c_j}{\bar{p}} \right]
    \nonumber\\
    &=&:
    \bar{q}_{a b} + \delta q_{a b}
    \,.
\end{eqnarray}
The spacetime metric, to first order in perturbations, is given by
\begin{eqnarray}\label{eq:Perurbed metric - classical}
    {\rm d} s^2 &=& - \left(\bar{N} + \delta N\right)^2 {\rm d} t^2
    + \left(\bar{q}_{a b} + \delta q_{a b}\right) \left({\rm d} x^a + \delta N^a {\rm d} t\right) \left({\rm d} x^b + \delta N^b {\rm d} t\right)
    \nonumber\\
    &=& - \bar{N}^2 {\rm d} t^2
    + \bar{q}_{a b} {\rm d} x^a {\rm d} x^b
    - 2 \bar{N} \delta N {\rm d} t^2
    + \delta q_{a b} {\rm d} x^a {\rm d} x^b
    + 2 \bar{q}_{a b} {\rm d} x^a \delta N^b {\rm d} t
    \nonumber\\
    &=:& \bar{g}_{\mu\nu} {\rm d} x^\mu {\rm d} x^\nu + \delta g_{\mu\nu} {\rm d} x^\mu {\rm d} x^\nu
    \,.
\end{eqnarray}
The lapse and shift components  depend on gauge choices.

\subsection{Gauge consistency and covariance}

Classical general relativity is covariant, which means that the space-time
metric obtained from solutions can be changed by coordinate transformations or
slicing changes without affecting values of the line element. In their
canonical form, these transformations must be related to gauge transformations
generated by the constraints via Hamilton's equations. The perturbations
(\ref{HamConstH1}), (\ref{HamConstH2}) and (\ref{PertDiffConst}) clearly
fulfill this condition because they have been obtained by perturbing a
covariant full theory: Solutions for perturbations are covariant with respect
to coordinate transformations that preserve isotropy of the background.

Covariance of the canonical theory can also be demonstrated independently,
without recourse to the full covariant theory from which the constraints have
been obtained by perturbation theory. The general condition can be split into
two important properties: (i) Gauge consistency in the sense that the theory
is subject to gauge transformations compatible with the equations of motion,
and (ii) geometrical covariance in the sense that it is possible to extract an
invariant space-time line element for any given solution. In general, the
first condition does not imply the second one. Gauge consistency ensures that
the theory is not overconstrained by incompatible gauge flows, which is a
prerequisite for the second condition to be formulated. The second condition
then requires additional features that make it possible to relate gauge
transformations to coordinate and slicing changes in a suitable space-time geometry.

\subsubsection{Isotropic background}

Gauge consistency is a standard condition in canonical theories, implemented
by the requirement that the constraints be first class: Their mutual Poisson
brackets must all vanish on the solution space of the constraints. The
background constraints are always first class,
\begin{equation} \label{HHiso}
    \{\bar{H}[\bar{N}_1],\bar{H}[\bar{N}_2]\} = 0
\end{equation}
because of antisymmetry of the Poisson bracket combined with the fact that the
background constraints do not contain spatial derivatives. The latter
property implies the absence of non-zero antisymmetric combinations of the
two lapse functions, such as $N_1\partial_aN_2-N_2\partial_aN_1$.

From the
classical theory, we know that there is a meaningful geometrical structure on
solutions with the inverse spatial metric
\begin{equation}\label{eq:Background structure function - classical}
    \bar{q}^{a b} = \frac{\delta^{a b}}{\bar{p}}
  \end{equation}
  extended to the space-time line element (\ref{FRW}) where
  $a^2=\bar{p}$. This line element is invariant under reparameterizations of
  the time coordinate because $\bar{N}(t){\rm d}t$ is invariant: If we impose
  the gauge transformation
\begin{equation} \label{deltaNbar}
    \delta_{\bar{\epsilon}^0} \bar{N} = \dot{\bar{\epsilon}}^0
\end{equation}
where $\bar{\epsilon}^0$ is a time-dependent function, we have
\begin{equation}
  (\bar{N}+\delta_{\bar{\epsilon}^0}\bar{N}){\rm d}t=\bar{N}{\rm d}t+{\rm
    d}\bar{\epsilon}^0\,.
\end{equation}
Defining an infinitesimal time reparameterization by
$t\mapsto t'= t+\bar{\xi}^0(t)$ with
\begin{equation} \label{xieps0}
  \bar{\xi}^0=\bar{\epsilon}^0/\bar{N}\,,
\end{equation}
the gauge transformation of $\bar{N}{\rm d}t$ is equivalent to the coordinate
change from $t$ to $t'$:
\begin{eqnarray} \label{Nprime}
  \bar{N}(t'){\rm d}t'&=&(\bar{N}(t)+\bar{\xi^0}(t)\dot{\bar{N}}(t))
                          \times(1+\dot{\bar{\xi}}^0(t)){\rm d}t\\
  &\sim&
  (\bar{N}(t)+(\bar{N}(t)\bar{\xi}^0(t))^{\bullet}){\rm d}t =
  (\bar{N}+\delta_{\epsilon^0}\bar{N}){\rm d}t \nonumber
\end{eqnarray}
up to first order in $\bar{\xi}^0$. The specific relationship between the gauge
function $\bar{\epsilon}^0$ and the generator $\bar{\xi}^0$ of infinitesimal coordinate
transformation, as well as the specific gauge transformation
(\ref{deltaNbar}), follow from more general considerations, as discussed
below.

The spatial metric $\bar{p}(t)\delta_{ab}$ changes to
$\bar{p}(t')\delta_{ab}=(\bar{p}(t)+\dot{\bar{p}} \xi^0(t))\delta_{ab}$ by the same coordinate
transformation. Also here, we have equivalence with the canonical gauge transformation
\begin{equation} \label{deltapbar}
  \delta_{\bar{\epsilon}^0} \bar{p} = \{\bar{p},\bar{H}[\bar{\epsilon}^0]\}=\bar{\xi}^0
  \{\bar{p},\bar{H}[N]\}= \dot{\bar{p}}\bar{\xi}^0=\bar{p}(t+\xi^0)-\bar{p}(t)\,.
\end{equation}
The isotropic space-time line element is therefore invariant under time
reparameterizations, irrespective of the phase-space expression of
$\bar{H}[\bar{N}]$ as long as the equation of motion
$\dot{\bar{p}}=\{\bar{p},\bar{H}[N]\}$ is imposed. Moreover, any function
$f(\bar{p})$ replacing $\bar{p}$ in the spatial metric would imply an
invariant line element. (By contrast, the dependence of the space-time metric
on $\bar{N}$ cannot be replaced with an arbitrary function because the
transformations of both $\bar{N}$ and ${\rm d}t$ contribute to
(\ref{Nprime}).) While the isotropic reduction is reparameterization
invariant, this symmetry is not restrictive and does not imply specific
classes of covariant dynamics through $\bar{H}$ or of compatible geometries through the
spatial part of ${\rm d}s^2$ and the function $f(\bar{p})$.

For models of quantum gravity, these conclusions mean that the isotropic (or,
more generally, homogeneous) dynamics and spatial geometry can be modified by
proposed quantum contributions without any restrictions. These so-called
mini-superspace models can be defined and analyzed in many ways, with a large
variety of potential physical effects such as singularity resolution as for
instance in loop quantum cosmology \cite{LivRev}. However, such models are not
reliable from a space-time perspective because they are not subject to any
covariance conditions, in contrast to isotropic models derived from a
covariant action principle that includes inhomogeneous fields.

\subsubsection{Perturbed canonical theory}

If we repeat the canonical analysis of gauge consistency and covariance for
the perturbed constraints, we still obtain a vanishing bracket
\begin{equation}\label{HHpert}
    \{H[\delta N_1], H[\delta N_2]\} = 0\,.
\end{equation}
This equation is valid for any first-order perturbation because the only
derivative term in
(\ref{HamConstH1}) is of second order, while a first-order derivative would be
needed to produce a nontrivial antisymmetric term
$\{H[\delta N_1], H[\delta N_2]\} \propto \delta N_1
\partial_a \delta N_2-\delta N_2 \partial_a \delta N_1$. This outcome
therefore does not
depend on the phase-space expression of $H[\delta N]$.

Nevertheless, these perturbative
contributions to the Hamiltonian constraint are subject to gauge and
covariance conditions because they have non-zero Poisson brackets with the
remaining constraints: Since we now have to include the non-zero
diffeomorphism constraint, an additional Poisson bracket is given by
\begin{equation} \label{HDClassical}
\{H[N], D_a[\delta N^a]\} = - H[\delta N^a\partial_a \delta N] \,.
\end{equation}

Moreover, the perturbative contribution to the Hamiltonian constraint has a
non-trivial Poisson bracket with the background constraint, such that
\begin{eqnarray} \label{HHClassical}
\{H[N_1], H[N_2]\} &=&
\{H[\delta N_1],H[\bar{N}_2]\}+
\{H[\bar{N}_1],H[\delta N_2]\}\\
&=& D_a\left[\bar{q}^{a b} (\bar{N}_2 \partial_b\delta N_1-\bar{N}_1 \partial_b\delta N_2)\right] \,.\nonumber
\end{eqnarray}
Imposing this bracket implies that we are treating background and
perturbations as parts of a single perturbative theory, rather than
independent inhomogeneous fields on a given background. The canonical
formalism therefore distinguishes between a single covariant perturbed theory
on one hand, and a covariant field theory on a fixed background on the
other. In the latter case, the background can still be modified at will, as in
the pure background treatment. In the former case, the background for
perturbations is subject to new conditions based on (\ref{HHClassical}). See
also \cite{NonCovDressed} for a detailed discussion.

\subsubsection{Emergent modified gravity}

The perturbative Poisson brackets are special cases of the
hypersurface-deformation brackets \cite{DiracHamGR,Katz,ADM}
\begin{eqnarray}
    \{ D_a[N^a_1] , D_b [N^b_2] \} &=& - D_a [N_2^b \partial_b N_1^a-N_2^b \partial_b N_2^a]
    \ , \label{DD}
    \\
    \{ H [ N ] , D_a [ N^a]\} &=& - H [ N^b \partial_b N ]
    \ ,\label{HD} \\
    \{ H [ N_1 ] , H [ N_2 ] \} &=& - D_a [ q^{a b} ( N_2 \partial_b N_1 - N_1
                                    \partial_b N_2 )]\,. \label{HH} 
\end{eqnarray}
The last equation, just like (\ref{HHClassical}), has the important property
of containing a structure function given by $q^{ab}$, constituting a
phase-space dependent multiplier of lapse and shift in contrast to just
numerical factors as in the other brackets. Structure functions often
complicate quantization procedures because they lead to ordering questions
between quantizations of these functions and the constraint
operators \cite{NonHerm}. However, they play an important role in the construction of general
covariant theories or new modifications of general relativity. The structure
function is an important part of the gauge conditions because it appears in
Poisson brackets of some of the constraints. In addition, it implies new
conditions on the underlying geometry because the spatial metric (or its
inverse) is now unambiguously linked with properties of the constraints. This
is in contrast to the isotropic reduction where the structure function
disappears from (\ref{HHiso}) and any function $f(\bar{p})$ in the spatial
metric would be compatible with the transformation law (\ref{deltapbar}). In
the presence of inhomogeneity, there is additional information that can be
used to construct compatible space-time geometries, or to conclude that no
consistent geometry exists for a given modification that may otherwise be
gauge compatible.

The structure function also plays an important role in analyzing the
relationship between canonical gauge transformations and space-time coordinate
changes.  In general, the relationship between gauge functions $\epsilon^0$
and $\epsilon^a$, which appear in gauge transformations
$\delta_{\epsilon^0,\epsilon^a}f=\{f,H[\epsilon^0]+D_a[\epsilon^a]\}$ for
phase-space functions $f$, and components of a space-time vector field
$(\xi^t,\xi^a)$ interpreted as an infinitesimal coordinate transformation is
given by
\begin{equation} \label {xita}
    \xi^t = \frac{\epsilon^0}{N} \quad,\quad
    \xi^a = \epsilon^a - \frac{\epsilon^0}{N} N^a\,.
\end{equation}
This relationship directly follows from the equation
\begin{equation}
    \xi^\mu = \xi^t t^\mu + \xi^a s^\mu_a
    = \epsilon^0 n^\mu + \epsilon^a s^\mu_a
\end{equation}
that rewrites a given vector field in the coordinate basis and the normal
basis of a given foliation, respectively \cite{BergmannKomarGroup}. The coordinate vector
$t^{\mu}$ is defined by
\begin{equation} \label{t}
  t^{\mu}=N n^{\mu}+N^as_a^{\mu}
\end{equation}
in the normal basis, using the same spatial basis vectors $s_a^{\mu}$ in both
cases. Equation~(\ref{xita}) is reduced to the single condition (\ref{xieps0})
in the isotropic case.

Consistency of gauge transformations and evolution implies that the lapse
function and shift vector, which appear in the time-evolution vector field
(\ref{t}), must be subject to gauge transformations. However, they are not
phase-space functions since there are no momenta associated with them. (The
classical action does not contain derivatives of these metric components.)
Nevertheless, gauge consistency \cite{LapseGauge,CUP,HypDef} can be used to
derive the gauge transformations
\begin{eqnarray}\label{eq:Off-shell gauge transformations for lapse and shift}
    \delta_\epsilon N &=& \dot{\epsilon}^0 + \epsilon^a \partial_a N - N^a \partial_a \epsilon^0
    \,,\\
    \delta_\epsilon N^a &=& \dot{\epsilon}^a + \epsilon^b \partial_b N^a - N^b \partial_b \epsilon^a + q^{a b} \left(\epsilon^0 \partial_b N - N \partial_b \epsilon^0 \right)
    \nonumber
\end{eqnarray}
for lapse and shift, using the same $\epsilon^0$ and $\epsilon^a$ that
transform the metric and its momenta according to
\begin{equation}
  \delta_{\epsilon}f=\{f,H[\epsilon^0]+D_a[\epsilon^a]\}\,.
\end{equation}
The inverse spatial metric appears in the transformation
$\delta_{\epsilon}N^a$ by virtue of its role as the structure function.

There is one final condition that is important when the constraint expressions
are modified. The structure function is identified with the inverse spatial
metric, which is required for gauge transformations to be equivalent to
deformations of spacelike hypersurfaces within a consistent space-time
geometry. Modified constraints are required to maintain gauge consistency,
such that they form closed Poisson brackets with one another. However, gauge
consistency is necessary but not sufficient for general covariance. In
addition to being closed, the brackets must be of the specific form
(\ref{DD})--(\ref{HH}) in order to define hypersurface deformations in a
consistent 4-dimensional geometry \cite{HypDef}. In particular, (\ref{HH}) must be
maintained in the sense that the structure function may be modified to some
$\tilde{q}^{ab}$ but with no other contributions such as an $H[N]$-term added
to the diffeomorphism constraint on the right. The specific form of
hypersurface-deformation brackets must therefore be maintained in order to
ensure that gauge transformations are equivalent to coordinate transformations
when the constraints and equations of motion are used (on-shell).

Finally, if the structure function is modified, the new $\tilde{q}^{ab}$ as a
phase-space function must be subject to gauge transformations that, in
combination with the lapse and shift transformations, are equivalent to the
tensor transformation law on-shell. This condition is not automatically
fulfilled for a gauge-consistent modification of the classical constraints. As
determined by emergent modified gravity \cite{Higher,HigherCov}, implementing
all gauge and covariance conditions can lead to new theories of modified
gravity that rely on the canonical formulation and have no analog as a
modified action principle. (An action principle assumes that one of the basic
fields, such as the metric or a tetrad, defines the space-time volume
element. This assumption is weakened in emergent modified gravity.)

Covariance of the structure function is an essential condition that has not
been analyzed in previous models of perturbative inhomogeneity, especially in
the context of loop quantum cosmology. Here, we fill in this important lacuna.

\subsubsection{Perturbative cosmology from the perspective of emergent
  modified gravity}

In contrast to the case of unrestricted inhomogeneity, the background
structure function appears in the bracket (\ref{HHClassical}).
The perturbed
theory is therefore more restrictive than the purely isotropic reduction
because it eliminates the freedom to have a covariant space-time metric with
any spatial part $f(\bar{p})\delta_{ab}$. However, the constraint brackets do
not determine the full inverse spatial metric which may be perturbed to
$\bar{q}^{ab}+\delta q^{ab}$.

The complete set of brackets in the perturbative setting is given by
\begin{eqnarray}
    \{D_a[\delta N_1^a], D_b[\delta N_2^b]\} &=& 0
    \,,\label{DDpert}\\
    \{H[N], D_a[\delta N^a]\} &=& - H[\delta N^a\partial_a \delta N]
    \,,\label{HDpert}\\
    \{H[N_1], H[N_2]\} &=& - D_a\left[\bar{q}^{a b} (\bar{N}_2 \partial_b\delta N_1-\bar{N}_1 \partial_b\delta N_2)\right]
    \,.\label{HHpert}
\end{eqnarray}
Moreover, gauge consistency implies the gauge transformations
\begin{eqnarray}\label{eq:Gauge transformation lapse - pert}
    \delta_\epsilon \delta N &=& \delta\dot{\epsilon}^0
    \,,\\
    \delta_\epsilon \delta N^a &=& \delta\dot{\epsilon}^a + \bar{q}^{a b}
                                   (\bar{\epsilon}^0 \partial_b\delta
                                   N-\bar{N} \partial_b\delta \epsilon^0) \label{eq:Gauge transformation shift - pert}
\end{eqnarray}
of lapse and shift perturbations.
The background lapse function transforms according to
\begin{eqnarray}\label{eq:Gauge transformation lapse - background}
    \delta_\epsilon \bar{N} = \dot{\bar{\epsilon}}^0
\end{eqnarray}
as in the isotropic reduction, allowing for the rescaling of time while
preserving the underlying symmetry of the background.

The relationship between gauge functions and perturbed space-time vector
fields $\xi^t = \bar{\xi}^t + \delta \xi^t$ and $\xi^a = \delta \xi^a$ for
coordinate changes is now given by
\begin{equation}
    \bar{\xi}^t = \frac{\bar{\epsilon}^0}{\bar{N}}
    \quad,\quad
    \bar{\xi}^a = 0
        \end{equation}
        for the background contributions and
        \begin{equation}
    \delta\xi^t = \frac{\bar{\epsilon}^0}{\bar{N}} \left(\frac{\delta \epsilon^0}{\bar{\epsilon}^0}-\frac{\delta N}{\bar{N}}\right)
    \quad,\quad
    \delta\bar{\xi}^a = \delta\epsilon^a
    - \frac{\bar{\epsilon}^0}{\bar{N}} \delta N^a
  \end{equation}
  for inhomogeneous perturbations.
These relationships appear in the final covariance condition, which states
that canonical gauge transformations of compatible space-time metric
components  must correspond on-shell to infinitesimal coordinate
transformations of the perturbed metric (\ref{eq:Perurbed metric -
  classical}), given by the Lie derivative $\mathcal{L}_\xi g_{\mu\nu}$:
\begin{eqnarray}
    \delta_\epsilon g_{\mu\nu} \big|_{\rm O.S.} = \mathcal{L}_\xi g_{\mu\nu} \big|_{\rm O.S.}
    \,.
\end{eqnarray}
As shown in \cite{HigherCov}, the normal part of this condition can be reduced to
\begin{eqnarray}
    \{q^{ab} , H [N]\} \big|_{\rm O.S.}
    = \{q^{ab} , H [\epsilon^0]\} \big|_{\rm O.S.}\,,
\end{eqnarray}
from which we obtain
\begin{eqnarray}
    \frac{\partial \{q^{ab} , H [\epsilon^0]\}}{\partial (\partial_{c_1}\epsilon^0)} \bigg|_{\rm O.S.} = \frac{\partial \{q^{ab} , H [\epsilon^0]\}}{\partial (\partial_{c_1}\partial_{c_2}\epsilon^0)} \bigg|_{\rm O.S.}
    =\dots = 0\,.
\end{eqnarray}

Using our perturbed variables, this condition, to first order in perturbations, is given by
\begin{eqnarray}\label{eq:Cov cond - 1 - classical}
    \frac{\partial \{\delta q^{ab} , H^{(1)} [\delta \epsilon^0]\}}{\partial (\partial_{c_1}\delta\epsilon^0)} \bigg|_{\rm O.S.} = \frac{\partial \{\delta q^{ab} , H^{(1)} [\delta \epsilon^0]\}}{\partial (\partial_{c_1}\partial_{c_2}\delta\epsilon^0)} \bigg|_{\rm O.S.}
    = \dots = 0\,.
\end{eqnarray}
The inhomogenous dependence of $\delta q^{ab}$ is only through $\delta E^a_i$,
but not its derivatives.  It therefore follows that the first-order conditions
(\ref{eq:Cov cond - 1 - classical}) are satisfied because (\ref{HamConstH1})
does not contain derivatives of $\delta K_a^i$.

The spatial part of the covariance condition takes the form
\begin{eqnarray}
    \delta_{\vec \epsilon}\, q^{a b}
    &=& \mathcal{L}_{\vec\xi}\, g^{a b}
    \nonumber\\
    &=&
    \epsilon^c \partial_c q^{a b}
    - q^{ca} \partial_{c} \epsilon^b
    - q^{c b} \partial_{c} \epsilon^a\label{eq:Spatial cov cond}
\end{eqnarray}
which, again to first order, becomes
\begin{eqnarray}
    \{q^{ab} , D_c [\delta \epsilon^c]\}
    =
    - \bar{q}^{c a} \partial_{c} \delta\epsilon^b
    - \bar{q}^{c b} \partial_{c} \delta\epsilon^a\,.\label{eq:Spatial cov cond
  - red} 
\end{eqnarray}
This condition is indeed satisfied by (\ref{PertDiffConst}).

While the perturbation of the structure function does not appear directly in
the constraint brackets (\ref{HHpert}) to the given order, it plays an
important role in the covariance condition of the space-time metric. This
condition will be analyzed in detail in the remainder of this paper.

\subsection{Background (vacuum) Dirac observable}
\label{sec:Background (vacuum) Dirac observable}

When looking for general classes of modified theories that could be used for
reliable effective descriptions, it is often useful to reduce the freedom by
imposing additional conditions beyond what basic requirements such as general
covariance would imply. We will be making use of a condition that stipulates
the existence of a certain Dirac observable in allowed modifications.

Classically, writing the background Hamiltonian constraint (\ref{HamConstH0}) as
\begin{equation}\label{eq:H0 Dirac obs}
    {\mathcal H}^{(0)}[\bar{N}] = V_0 \bar{N} \left[\frac{1}{2\kappa}\left(-6\bar{k}^2 \sqrt{\bar p}
    + 2\Lambda \bar{p}^{3/2}\right) + \frac{\bar{\pi}^2}{2\bar{p}^{3/2}}
    + \bar{p}^{3/2} V (\bar{\varphi})\right]
    = V_0 {\mathcal H}^{(0)} \bar{N}
  \end{equation}
  assuming scalar matter $\bar{\varphi}$ with a potential $V(\bar{\varphi})$,
  we may consider a potential Dirac observable ${\cal D}(\bar{k},\bar{p},\bar{\varphi},\bar{\pi})$
  such that $\{{\cal D},{\mathcal H}^{(0)}[\bar{N}]\}|_{\rm O.S.}=0$.  Since
  there is only one constraint in the background theory, this condition can be
  written as
\begin{eqnarray}
    \{{\cal D},{\mathcal H}^{(0)}[\bar{N}]\} = \alpha {\mathcal H}^{(0)} \bar{N}
\end{eqnarray}
for some phase-space function $\alpha(\bar{k},\bar{p},\bar{\varphi},\bar{\pi})$.

The Poisson bracket can be evaluated as
\begin{eqnarray}
    \{{\cal D},{\mathcal H}^{(0)}[\bar{N}]\} &=& \left[ \frac{\kappa}{3} \frac{\partial {\cal D}}{\partial \bar{k}} \frac{\partial {\mathcal H}^{(0)}}{\partial \bar{p}} - \frac{\kappa}{3} \frac{\partial {\cal D}}{\partial \bar{p}} \frac{\partial {\mathcal H}^{(0)}}{\partial \bar{k}}
    + \frac{\partial {\cal D}}{\partial \bar{\varphi}} \frac{\partial {\mathcal H}^{(0)}}{\partial \bar{\pi}} - \frac{\partial {\cal D}}{\partial \bar{\pi}} \frac{\partial {\mathcal H}^{(0)}}{\partial \bar{\varphi}} \right] \bar{N}
    \nonumber\\
    &=& \bar{N}\Bigg[ - \frac{1}{2\kappa} \left( \frac{\kappa}{3} \frac{1}{2\bar{p}} \frac{\partial {\cal D}}{\partial \bar{k}}
    - \frac{\kappa}{3} \frac{2}{\bar{k}} \frac{\partial {\cal D}}{\partial \bar{p}} \right) 6 \bar{k}^2 \sqrt{\bar p}
    + \frac{\kappa}{3} \frac{3}{2\bar{p}} \frac{\partial {\cal D}}{\partial \bar{k}} \frac{1}{2\kappa} 2 \Lambda \bar{p}^{3/2}
    \nonumber\\
    &&+ \frac{\kappa}{3} \frac{3}{2} \frac{1}{\bar{p}} \frac{\partial {\cal D}}{\partial \bar{k}} \left( - \frac{\bar{\pi}^2}{2\bar{p}^{3/2}}
    + \bar{p}^{3/2} V \right)
    + \frac{\partial {\cal D}}{\partial \bar{\varphi}} \frac{\bar{\pi}}{\bar{p}^{3/2}}
    - \frac{\partial {\cal D}}{\partial \bar{\pi}} \bar{p}^{3/2} V' \Bigg]
\end{eqnarray}
where $V'=\partial V/\partial \bar{\varphi}$. Therefore, a comparison with (\ref{eq:H0 Dirac obs}) shows that a Dirac observable can exist only in vacuum.
In the presence of a cosmological constant, the condition requires
\begin{eqnarray}
     - \frac{2}{\bar{k}} \frac{\partial {\cal D}}{\partial \bar{p}}
    = \frac{1}{\bar{p}} \frac{\partial {\cal D}}{\partial \bar{k}}
    \,,
\end{eqnarray}
or
\begin{equation}
    {\cal D} = {\cal D} \left(\frac{\bar{k}}{\sqrt{\bar{p}}}\right)\,.
  \end{equation}
We therefore obtain the background Dirac observable
\begin{equation}
    {\cal T} = \frac{3}{\Lambda} \frac{\bar{k}}{\sqrt{\bar{p}}}
\end{equation}
in vacuum, provided there is a non-zero cosmological constant.

If $\Lambda=0$, then we instead require
\begin{equation}
    \frac{1}{2\bar{p}} \frac{\partial {\cal D}}{\partial \bar{k}}
    - \frac{2}{\bar{k}} \frac{\partial {\cal D}}{\partial \bar{p}} = \alpha
\end{equation}
for some arbitrary function $\alpha$. In this case, the Dirac observable takes the
more complicated form
\begin{equation}
    {\cal T} = \tau (\bar{k}^4)\bar{p}+\int {\rm d} z\; \frac{2\bar{k}^4\bar{p} \alpha(z,\bar{k}^4\bar{p}/z^4)}{z}\,.
\end{equation}
However, on-shell we have $\bar{k}=0$, and hence the observable becomes trivial on
dynamical solutions. We will, therefore, only consider the former case when we
impose the existence of a non-trivial background Dirac observable in the modified theory.

\section{Perturbative emergent modified gravity: Vacuum}
\label{s:Vacuum}

Any canonical analysis requires the technical derivation of large numbers of
Poisson brackets. Applied in gauge and covariance conditions, these brackets
then result in coupled partial differential equations for the terms that may
contribute to consistent Hamiltonian constraints. This process leads to new
perturbative theories of emergent modified gravity, including versions that
can be used to clarify the issue of general covariance in cosmological  models of loop
quantum gravity. In the remainder of the main part of this paper, we will
summarize and discuss several resulting models and their dynamical
equations. For readers who are interested in reproducing the detailed
derivations or in finding alternative theories based on varying assumptions,
we collect underlying Poisson brackets as well as relevant partial derivatives
of the constraints in our appendices.

\subsection{Anomaly freedom}

Gauge and covariance conditions imply coupled non-linear partial differential
equations for the Hamiltonian and diffeomorphism constraints as phase-space
functions. In order to make them tractable, it is useful to apply the analysis
not to generic phase-space functions but rather to expressions that generalize the
classical constraints by inserting certain coefficient functions with a
restricted dependence on the phase-space variables. This step is also
necessary from a physical perspective: For reliable predictions from modified
theories, we should make sure that all terms are included which are allowed by the
conditions and are of the same order in derivatives or in perturbative
quantities.

In this way, the result will be a suitable effective theory that models all
competing effects obeying specific symmetries, up to a given order in a
derivative or other perturbative expansion. Standard higher-curvature actions
organize modifications to general relativity in a curvature expansion, which
in most cases introduces higher derivatives in each term. Such contributions
are often problematic for instance because of instabilities
\cite{OstrogradskiProblem}. Emergent modified gravity has shown, so far in several
symmetry-reduced contexts, that canonical gravity allows covariant
modifications even at the classical order of derivatives. In our cosmological
analysis, we will also assume that potential modified constraints to leading
order do not introduce higher derivatives, and thus work with the classical
phase space outlined in the preceding section. In the perturbative treatment,
we will continue to work with constraints of up to second order in inhomogeneity.

As in previous studies of emergent modified gravity, we will maintain the
classical form of the diffeomorphism constraint. Therefore, spatial
transformations will not be modified and we have a well-defined 3-dimensional
integration measure on a spatial manifold $\Sigma$, as well as standard
density weights for the fields. This property allows us to use a smeared
modified Hamiltonian constraint of the form
\begin{equation}\label{eq:Vacuum constraint ansatz-EMG}
   \tilde{H}_{\rm grav}[N] =  \frac{1}{2\kappa}\int_{\Sigma}\mathrm{d}^3x\;
   N \tilde{\cal H}_{\rm grav}\,.
\end{equation}
The modified Hamiltonian density $\tilde{\cal H}_{\rm grav}$ is a function of
the background variables $\bar{p}$ and $\bar{k}$ as well as the perturbations
$\delta E^a_i$ and $\delta K_b^j$. These are phase-space degrees of freedom,
but they no longer retain their classical geometrical meaning as components of
the densitized triad and extrinsic curvature. Rather, the corresponding
geometrical quantities will be determined by the emergent line element that
results from the following gauge and covariance analysis. In general, the
emergent space-time metric may be a function of all the phase-space variables
including $\bar{k}$ and $\delta K_b^j$. We first present the vacuum case and
then discuss how matter terms may be included from the same general perspective.

\subsubsection{Parameterized constraints}

According to our assumptions, we maintain the classical derivative order and
spatial manifold structure. These conditions restrict the number of
perturbation terms that can generically appear in a modified Hamiltonian
density. In particular, we do not include terms that have spatial derivatives
of $\delta K_b^j$ (as in the classical constraint), and spatial indices have
to be fully contracted in each term. These conditions give us the following
general forms of perturbative contributions to the Hamiltonian constraint:
We write the background constraint in the form
\begin{equation}
\label{HamConstH0-EMG}
    \tilde{\mathcal H}^{(0)}_{\rm grav} = - 6 \sqrt{\bar p} {\cal K}^{(0)}
\end{equation}
with a general phase-space dependent background function ${\cal K}^{(0)}$. (We
pull out an explicit factor of $\sqrt{\bar{p}}$ for convenience.) Classically,
${\cal K}^{(0)}=\bar{k}^2 - \Lambda \bar{p}/3$. 

The first-order perturbation takes the form
\begin{equation}
\label{HamConstH1-EMG}
    \tilde{\mathcal H}^{(1)}_{\rm grav} =
    \sqrt{\bar{p}} \left[- 4 {\cal K}^{(1)}_1 \delta^c_j\delta K_c^j
    - {\cal K}^{(1)}_2 \frac{\delta_c^j\delta E^c_j}{\bar{p}}
    + 2 {\cal K}^{(1)}_3 \frac{\partial^j\partial_c\delta E^c_j}{\bar{p}} \right]
\end{equation}
with three phase-space dependent background functions with classical limits
${\cal K}^{(1)}_1=\bar{k}$, ${\cal K}^{(1)}_2=\bar{k}^2-\Lambda \bar{p}$ and
${\cal K}^{(1)}_3=1$.  In the last term,
$\partial^j\partial_c\delta E^c_j=\delta^{bk}\partial_b\partial_c\delta
E^c_k$. We write the second-order perturbation as
\begin{eqnarray}  
\label{HamConstH2-EMG}
    \tilde{\mathcal H}^{(2)}_{\rm grav} &=&
    \sqrt{\bar{p}} \Bigg[ {\cal K}^{(2)}_1 \delta K_c^j\delta K_d^k\delta^c_k\delta^d_j
    - {\cal K}^{(2)}_2 (\delta K_c^j\delta^c_j)^2
    - 2 {\cal K}^{(2)}_3 \delta K_c^j \frac{\delta E^c_j}{\bar{p}}
    \nonumber\\
    && \quad 
    - \frac{1}{2} \left({\cal K}^{(2)}_4 \frac{\delta_c^k\delta_d^j\delta E^c_j\delta E^d_k}{\bar{p}^2}
    - {\cal K}^{(2)}_5 \frac{(\delta_c^j \delta E^c_j)^2}{2\bar{p}^2} \right)
    - \frac{{\cal K}^{(2)}_6}{2} \frac{\delta^{jk} (\partial_c\delta E^c_j) (\partial_d\delta E^d_k)}{\bar{p}^2} \Bigg]
\end{eqnarray}
with six functions of $\bar{k}$ and $\bar{p}$, given in the classical case by
${\cal K}^{(2)}_3 = \bar{k}$,
${\cal K}^{(2)}_4={\cal K}^{(2)}_5=\bar{k}^2+\Lambda \bar{p}$, and
${\cal K}^{(2)}_1={\cal K}^{(2)}_2={\cal K}^{(2)}_6=1$. Since this
perturbation is multiplied by the background lapse function $\bar{N}$, all
second-order derivative terms of $\delta E^a_i$ are (for scalar perturbations)
equivalent to the last term in $\tilde{\mathcal H}^{(2)}_{\rm grav}$ upon
integrating by parts.

We impose that the modified Hamiltonian constraint, together with the
unmodified diffeomorphism constraint, preserves anomaly-free
hypersurface-deformation brackets of the form
\begin{eqnarray}
    \{ D_a [ N^a_1] , D_b [ N^b_2 ] \} &=& - D_a [N_2^b \partial_b N_1^a-N_2^b \partial_b N_2^a]
    \ , \label{DD - mod}
    \\
    \{ \tilde{H}_{\rm grav} [ N ] , D_a [ N^a]\} &=& - \tilde{H}_{\rm grav} [\delta N^b \partial_b \delta N ]
    \ ,\label{HD - mod} \\
    \{ \tilde{H}_{\rm grav} [ N_1 ] , \tilde{H}_{\rm grav} [ N_2 ] \} &=& - D_a [ \bar{\tilde{q}}^{a b} ( \bar{N}_2 \partial_b \delta N_1 - \bar{N}_1 \partial_b \delta N_2 )] \label{HH - mod}
\end{eqnarray}
with possibly modified
$\bar{\tilde{q}}^{ab}\neq \bar{q}^{ab}=\bar{p}^{-1}\delta^{ab}$. This new
function determines the underlying background geometry of space-time. Further
assumptions will be necessary to derive the metric perturbation
$\delta \tilde{q}_{ab}$, subject to covariance conditions, since this quantity
does not appear in the structure functions for constraints up to second
perturbative order.  Our analysis will determine restrictions on the
${\cal K}$-functions implied by these conditions.

\subsubsection{Results}

As discussed in detail in Appendix~\ref{a:Deriv}, the hypersurface-deformation
brackets result in the conditions
\begin{eqnarray}
    \label{eq:K^2_2=K^2_1 - simp}
    {\cal K}^{(2)}_2&=&{\cal K}^{(2)}_1
    \,,\\
    \label{eq:dK^0/dk - simp}
    {\cal K}^{(2)}_3 &=&
    \frac{\partial {\cal K}^{(0)}}{\partial \bar{k}}
    - \bar{k} {\cal K}^{(2)}_1
    \,,\\
    \label{eq:dK^0/dp - simp}
    {\cal K}^{(2)}_4
    &=& 2 \bar{k} {\cal K}^{(2)}_3
    -{\cal K}^{(0)}
    - 2 \bar{p} \frac{\partial {\cal K}^{(0)}}{\partial \bar{p}}
    \,,\\
    \label{eq:K^2_5=K^2_4 - simp}
    {\cal K}^{(2)}_5&=&{\cal K}^{(2)}_4
    \,,\\
    \label{eq:HH cond E - simp}
    0 &=& 2 \bar{p} \frac{\partial {\cal K}^{(1)}_1}{\partial \bar{p}} \frac{\partial {\cal K}^{(0)}}{\partial \bar{k}}
    - \frac{\partial {\cal K}^{(1)}_1}{\partial \bar{k}} \left({\cal K}^{(0)}
    + 2 \bar{p} \frac{\partial {\cal K}^{(0)}}{\partial \bar{p}}\right)
    \nonumber\\
    &&
    - {\cal K}^{(1)}_1 \frac{\partial {\cal K}^{(0)}}{\partial \bar{k}}
    +3 {\cal K}^{(0)} {\cal K}^{(2)}_1
    \,,\\
     \label{eq:HH=D - simp}
    0 &=&
    \frac{\partial {\cal K}^{(0)}}{\partial \bar{k}} \left({\cal K}^{(1)}_3 \frac{\partial {\cal K}^{(0)}}{\partial \bar{k}}
    + 2 \bar{p} \frac{\partial {\cal K}^{(1)}_3}{\partial \bar{p}}\right)
    - \frac{\partial {\cal K}^{(1)}_3}{\partial \bar{k}} \left({\cal K}^{(0)}
    + 2 \bar{p} \frac{\partial {\cal K}^{(0)}}{\partial \bar{p}}\right)
    \nonumber\\
    &&
    - 2 {\cal K}^{(1)}_1 {\cal K}^{(2)}_6
    \,,\\
    \label{eq:K12 - simp}
    {\cal K}^{(1)}_2 &=& 3 {\cal K}^{(0)}
    -2 \bar{k} {\cal K}^{(1)}_1
    \,.
\end{eqnarray}
The first four equations easily eliminate four of the six second-order functions,
while the remaining ones in this subset can be obtained from the next two
equations because ${\cal K}^{(0)}$ and ${\cal K}_1^{(1)}$ cannot vanish
identically owing to the required classical limit. The final equation then
implies one of the three first-order functions. The background function ${\cal
  K}^{(0)}$ and two first-order functions, ${\cal K}_1^{(1)}$ and ${\cal
  K}_3^{(1)}$, are free according to this solution procedure. However, if
additional assumptions are made, it may be more convenient to treat some of
the other functions as free and solve the conditions in a different way.

The background structure function implied by the anomaly-free brackets is given by
\begin{equation}\label{eq:Background structure function - EMG}
    \bar{\tilde{q}}^{ab} = {\cal K}^{(1)}_3 {\cal K}^{(2)}_1 \frac{\delta^{ab}}{\bar p}
    \,.
\end{equation}
When constructing modified vacuum theories, it is therefore possible to choose
the background dynamics ($\tilde{{\cal H}}_{\rm grav}^{(0)}$ through
${\cal K}^{(0)}$) and independently the background geometry
($\bar{\tilde{q}}^{ab}$ through the free ${\cal K}_3^{(1)}$), both with an
arbitrary dependence on $\bar{p}$ and $\bar{k}$.

\subsection{Covariance conditions and the metric perturbation}

Anomaly freedom determines a candidate for the background metric which, like
any background quantity, automatically satisfies the covariance conditions
(\ref{eq:Cov cond - 1 - classical}) and (\ref{eq:Spatial cov cond - red}) to
first order in inhomogeneity. 

Since the perturbation of the structure function cannot be obtained from the
constraint brackets, we formulate the symmetric ansatz
\begin{eqnarray} \label{MetricAnsatz}
    \delta \tilde{q}^{ab} &=& \frac{1}{\bar{p}} \Bigg[ Q_1^{(E)}  \frac{\delta^{aj} \delta E^b_j
    + \delta^{bj} \delta E^a_j}{\bar{p}}
    - Q_2^{(E)} \delta^{a b} \frac{\delta_c^j\delta E^c_j}{\bar{p}}
    - Q_3^{(E)} \delta^{a b} \frac{\partial^k\partial_c\delta E^c_k}{\bar{p}}
    \nonumber\\
    &&\quad
    + Q_1^{(K)} \delta^{a b} \delta^c_j\delta K_c^j
    + Q_2^{(K)} \left(\delta^{ac} \delta^b_j
    + \delta^{bc} \delta^{a}_j\right) \delta K_c^j
    \Bigg]
\end{eqnarray}
following the derivative structure assumed for the parameterized
constraints. The five functions $Q_1^{(E)}$, $Q_2^{(E)}$, $Q_3^{(E)}$,
$Q_1^{(K)}$ and $Q_2^{(K)}$ may depend on the background variables $\bar{p}$
and $\bar{k}$. Classically, $Q_1^{(E)}=Q_2^{(E)}=1$ while the remaining
functions are zero. This expression is subject to covariance conditions.

The relevant contribution to the normal gauge transformation of $\delta\tilde{q}^{ab}$ is given by
\begin{eqnarray}
    \{\delta \tilde{q}^{ab} , \tilde{H}^{(1)}_{\rm grav}[\delta \epsilon^0]\}
    &=& Q^{ab} \delta \epsilon^0
    + \frac{\delta^{a b}}{\bar{p}^{3/2}} 
    \left( Q_1^{(K)} {\cal K}^{(1)}_3-2 Q_3^{(E)} {\cal K}^{(1)}_1
        \right)(\delta^{cd}\partial_c\partial_d\delta \epsilon^0) 
    \nonumber\\
    &&+ Q_2^{(K)} \frac{2{\cal K}^{(1)}_3}{\sqrt{\bar{p}}} (\delta^{ac}\delta^{bd}\partial_c\partial_d\delta \epsilon^0)\,,
\end{eqnarray}
where $Q^{ab}$ is some phase-space function whose explicit expression we will
not need for now. (We  will compute it below when coupling scalar matter.)
The normal covariance condition (\ref{eq:Cov cond - 1 - classical}),
\begin{eqnarray}\label{eq:Cov cond - 1 - EMG}
    \frac{\partial \{\delta \tilde{q}^{ab} , \tilde{\cal H}^{(1)}_{\rm grav} [\delta \epsilon^0]\}}{\partial (\partial_{c_1}\delta\epsilon^0)} \bigg|_{\rm O.S.} = \frac{\partial \{\delta \tilde{q}^{ab} , \tilde{\cal H}^{(1)}_{\rm grav} [\delta \epsilon^0]\}}{\partial (\partial_{c_1}\partial_{c_2}\delta\epsilon^0)} \bigg|_{\rm O.S.}
    = \dots = 0
\end{eqnarray}
applied to the emergent metric, therefore implies
\begin{equation}
    Q_2^{(K)} = 0
\end{equation}
and
\begin{equation}
  Q_3^{(E)}= Q_1^{(K)} \frac{{\cal K}^{(1)}_3}{2{\cal K}^{(1)}_1}\,.
\end{equation}

The spatial covariance condition (\ref{eq:Spatial cov cond - red}) now requires
\begin{equation}\label{eq:Spatial cov cond - red - EMG}
  \{\tilde{q}^{ab} , D_c [\delta \epsilon^c]\} =
  - {\cal K}^{(1)}_3 {\cal K}^{(2)}_1 \frac{\delta^{ca}\partial_{c} \delta\epsilon^b+\delta^{cb}\partial_{c} \delta\epsilon^a}{\bar p}\,,
\end{equation}
using the background spatial metric $\tilde{q}^{ab}$ implied by the
hypersurface-deformation brackets.  A direct computation shows,
\begin{eqnarray}\label{eq:background spatial diffeo}
    &&\{\bar{\tilde{q}}^{ab} , D_c [\delta \epsilon^c]\}\\
    &=&
    \delta^{ab} \left(\frac{\partial}{\partial \bar{p}}\left(\frac{{\cal K}^{(1)}_3 {\cal K}^{(2)}_1}{\bar p}\right) \partial_d (\delta_c^k\delta E^d_k)
    + \frac{\partial}{\partial \bar{k}}\left(\frac{{\cal K}^{(1)}_3 {\cal K}^{(2)}_1}{\bar p}\right) \left(\partial_c(\delta^d_k \delta K^k_d)-\partial_k\delta K^k_c\right)\right) \frac{\delta \epsilon^c}{3}
   \nonumber
\end{eqnarray}
(which is of second order and therefore does not contribute to the leading
term (\ref{eq:Spatial cov cond - red - EMG})),
and
\begin{eqnarray}\label{eq:Pert spatial diffeo}
    \{\delta\tilde{q}^{ab} , D_c [\delta \epsilon^c]\}
    &=&
    2 \left(Q_1^{(E)}-Q_2^{(E)}+\bar{k}Q_1^{(K)}\right) \frac{\delta^{ab}}{\bar{p}} \partial_c \delta \epsilon^c
    - Q_1^{(E)} \frac{\delta^{ca}\partial_{c} \delta\epsilon^b+\delta^{cb}\partial_{c} \delta\epsilon^a}{\bar p}
    \,.
\end{eqnarray}
Hence, to first order, (\ref{eq:Spatial cov cond - red - EMG}) requires
\begin{equation}
    Q_1^{(E)}={\cal K}^{(1)}_3 {\cal K}^{(2)}_1
\end{equation}
and
\begin{equation}
Q_2^{(E)} = Q_1^{(E)}+\bar{k}Q_1^{(K)} \,.
\end{equation}

The perturbation of the structure function therefore becomes
\begin{eqnarray}\label{eq:Metric perturbation}
    \delta \tilde{q}^{ab} &=& \frac{1}{\bar{p}} \left[ {\cal K}^{(1)}_3 {\cal K}^{(2)}_1 \left( \frac{\delta^{aj} \delta E^b_j + \delta^{bj} \delta E^a_j}{\bar{p}}
    - \delta^{a b} \frac{(\delta_d^k\delta E^d_k)}{\bar{p}}\right)\right.
    \nonumber\\
    &&\left.\quad
    + \delta^{a b} Q_1^{(K)} \left(\delta^d_k\delta K_d^k - \frac{{\cal K}^{(1)}_3}{2{\cal K}^{(1)}_1} \frac{\delta^{jk}\partial_j\partial_d\delta E^d_k}{\bar{p}}
    - \bar{k} \frac{(\delta_d^k\delta E^d_k)}{\bar{p}}\right)
    \right]\,.
\end{eqnarray}
There is one remaining function, $Q_1^{(K)}$, that is undetermined by the covariance
conditions, up to compatibility with the classical limit $Q_1^{(K)}\to0$.  The
perturbation of the emergent metric is not fully determined by the covariance
conditions even if we choose a specific modified Hamiltonian
constraint. Unambiguous vacuum models therefore require additional
phenomenological conditions. In addition, we will soon see that coupling
matter in a covariant way in general imposes additional conditions also on the
gravitational modification functions.

For later applications, we note the expression
\begin{equation}\label{eq:Metric perturbation contraction}
    \bar{\tilde{q}}_{ab} \delta \tilde{q}^{ab} =
    - \frac{\delta_d^k\delta E^d_k}{\bar{p}}
    + 3 \frac{Q_1^{(K)}}{{\cal K}^{(1)}_3 {\cal K}^{(2)}_1} \left(\delta^d_k\delta K_d^k - \frac{{\cal K}^{(1)}_3}{2{\cal K}^{(1)}_1} \frac{\partial^k\partial_d\delta E^d_k}{\bar{p}}
    - \bar{k} \frac{\delta_d^k\delta E^d_k}{\bar{p}}\right)
    \,.
\end{equation}

\subsection{Background (vacuum) Dirac observable}

A background function ${\cal D}(\bar{k},\bar{p})$ is a vacuum Dirac observable if
\begin{eqnarray}
    \{{\cal D},{\mathcal H}^{(0)}_{\rm grav}[\bar{N}]\} \propto {\mathcal H}^{(0)}_{\rm grav} \bar{N}
\end{eqnarray}
where
\begin{equation}
    \{{\cal D},\tilde{\cal H}^{(0)}_{\rm grav}[\bar{N}]\}     = - \bar{N} \sqrt{\bar p}\; \Bigg[ \frac{\partial {\cal D}}{\partial \bar{k}} \left( \frac{{\cal K}^{(0)}}{2\bar{p}} + \frac{\partial {\cal K}^{(0)}}{\partial \bar{p}} \right)
    - \frac{\partial {\cal D}}{\partial \bar{p}} \frac{\partial {\cal K}^{(0)}}{\partial \bar{k}}
    \Bigg]\,.
\end{equation}
The existence of such an observable therefore requires
\begin{eqnarray}
    \frac{\partial {\cal D}}{\partial \bar{k}} \left( \frac{{\cal K}^{(0)}}{2\bar{p}} + \frac{\partial {\cal K}^{(0)}}{\partial \bar{p}} \right)
    - \frac{\partial {\cal D}}{\partial \bar{p}} \frac{\partial {\cal K}^{(0)}}{\partial \bar{k}} = \alpha {\cal K}^{(0)}
\end{eqnarray}
for some arbitrary function $\alpha(\bar{k},\bar{p})$.
This equation is satisfied for any background modification function of the form
\begin{equation}
    {\cal K}^{(0)} = S(\bar{k},\bar{p}) - \Pi(\bar{p})
  \end{equation}
  if
\begin{equation}
    {\cal D} = \frac{S(\bar{k},\bar{p})}{\Pi(\bar{p})}\,.
\end{equation}
We recover the classical expressions in the limit $S\to\bar{k}^2$, $\Pi\to \Lambda \bar{p}/3$.

\section{Perfect fluid}
\label{s:Fluid}

As an example of matter couplings, we first consider a perfect fluid in
emergent modified gravity. In this case, a simple option is minimal coupling
which can be achieved by replacing the inverse spatial metric with the
emergent one in relevant canonical matter expressions. Following
\cite{EmergentFluid}, the diffeomorphism constraint has a contribution from a
timelike perfect fluid given by
\begin{equation}
    {\cal H}^{\rm PF}_a =
    P^{(T)} \partial_a T
    + P^{(Z)}_i \partial_a Z^i
    \,,
    \label{eq:Dust contributions to diffeomorphism constraint}
\end{equation}
and the Hamiltonian constraint  by
\begin{equation}
    {\cal H}^{\rm PF} =
    \sqrt{(P^{(T)})^2 + \tilde{q}^{a b} {\cal H}^{\rm PF}_a {\cal H}^{\rm PF}_b}
    - \sqrt{\det \tilde{q}} P
    \,,
    \label{eq:Dust contributions to Hamiltonian constraint}
\end{equation}
where $P$ is the pressure function.

We adapt this formulation to the perturbative treatment by defining
\begin{eqnarray}
    T &=& \bar{T} + \delta T\,,\\
    Z^i &=& \bar{Z}^i + \delta Z^i\,,\\
    P^{(T)} &=& \bar{P}^{(T)} + \delta P^{(T)}\,,\\
    P^{(Z)}_i &=& \bar{P}^{(Z)}_i + \delta P^{(Z)}_i\,,
\end{eqnarray}
where the barred quantities are the background variables and the rest are perturbations.
The assumption of a homogeneous background allows us to fix a
background fluid frame such that $\partial_c \bar{T}=0$, $\partial_c Z^i=\delta^i_c$ and
$\bar{P}^{(Z)}_i = 0$.  To second order, the constraint
contributions from the perfect fluid become
\begin{eqnarray}
    \label{eq:Dust contributions to diffeomorphism constraint - pert}
    D^{\rm PF}_a &=& \int {\rm d}^3 x\ \left(\bar{P}^{(T)} \partial_c \delta T
    + \delta P^{(Z)}_i \delta_c^i\right) \delta \epsilon^c
    \,,\\
    \label{eq:PF Hamiltonian constraint - pert 0 - timelike}
    {\cal H}^{(0){\rm PF}} &=&
    \bar{P}^{(T)}
    - \sqrt{\det \bar{\tilde{q}}} \bar{P}
    \,,\\
    \label{eq:PF Hamiltonian constraint - pert 1 - timelike}
    {\cal H}^{(1){\rm PF}} &=&
    \delta P^{(T)}
    - \sqrt{\det \bar{\tilde{q}}} \left(\delta P - \bar{P} \frac{\bar{\tilde{q}}_{ab} \delta \tilde{q}^{ab}}{2}\right)
    \,,\\
    \label{eq:PF Hamiltonian constraint - pert 2 - timelike}
    {\cal H}^{(2){\rm PF}} &=&
    \frac{\bar{\tilde{q}}^{a b} {\cal D}^{\rm PF}_a {\cal D}^{\rm PF}_b}{2 \bar{P}^{(T)}}
    - \sqrt{\det \bar{\tilde{q}}} \left(\bar{P} \delta_{(2)}- \frac{\bar{\tilde{q}}_{ab} \delta \tilde{q}^{ab}}{2}\delta P \right)
    \,,
\end{eqnarray}
where $\bar{P}$ and $\delta P$ are, respectively, the background value and perturbation of
the pressure function, and an expansion of the square-root determinant to
second order,
\begin{eqnarray}
    \sqrt{\det \tilde{q}} &=& \sqrt{\det \bar{\tilde{q}}} \left(1
    - \frac{\bar{\tilde{q}}_{ab}\delta\tilde{q}^{ab}}{2}
    + \frac{1}{4}\left(\left(\bar{\tilde{q}}_{ab}\delta\tilde{q}^{ab}\right)^2-\bar{\tilde{q}}_{ac}\bar{\tilde{q}}_{bd}\delta \tilde{q}^{ab}\delta \tilde{q}^{cd}\right)
    \right)
    \nonumber\\
    &=:& \sqrt{\det \bar{\tilde{q}}} \left(1
    - \frac{\bar{\tilde{q}}_{ab}\delta\tilde{q}^{ab}}{2}
    + \delta_{(2)}
    \right)
    \,,
\end{eqnarray}
defines the expression $\delta_{(2)}$.

The gauge consistency of the constraints follows directly from the definition
of the emergent metric through the structure function. A perfect fluid
therefore does not lead to additional restrictions of the gravitational
modification functions.

\section{Scalar matter}
\label{s:Scalar}

Coupling scalar matter implies several additional terms in the Hamiltonian constraint
that may be modified by new independent background functions. Gauge and
covariance conditions then provide partial differential equations that couple
these modification functions to the previous ones introduced for vacuum
gravity. A priori, it is not clear whether these new modification functions
lead to more freedom in constructing modified theories, or could, in
combination with gauge and covariance conditions, further restrict consistent
geometries that can be coupled to dynamical matter. Our analysis sheds light
on this question.

\subsection{Classical constraint and conditions}

A canonical scalar field can be split into background variables with Poisson bracket
\begin{equation}
    \{\bar{\varphi},\bar{\pi}\}= \frac{1}{V_0}
\end{equation}
and perturbations with brackets
\begin{equation}
    \{\delta \varphi(x),\delta \pi(y)\}= \delta^3(x-y)\,.
  \end{equation}
As in the case of gravitational variables, we avoid double counting the
background variables by imposing the conditions $\int{\rm
  d}^3x\delta\varphi=0=\int{\rm d}^3x\delta\pi$. On the solution space of
these conditions, the Dirac bracket is obtained by subtracting $1/V_0$ from $\delta^3(x-y)$.
  
The contributions of scalar matter to the classical constraints are given by
\begin{equation}
  \vec{D}_{\rm scalar}[\vec{N}]= \int{\rm d}^3x N^a\pi\partial_a \varphi
  = \int{\rm d}^3x \delta N^a\bar{\pi}\partial_a \delta\varphi\,,
\end{equation}
for the diffeomorphism constraint, and
\begin{equation}
  H_{\varphi}[N] =
  \int{\rm d}^3x \bar{N} \left({\cal H}^{(0)}_\varphi + {\cal H}^{(2)}_\varphi\right)
  + \int{\rm d}^3x \delta N {\cal H}^{(1)}_\varphi
\end{equation}
for the Hamiltonian constraint, with the background contribution
\begin{equation}
  {\cal H}^{(0)}_\varphi =
  \frac{\bar{\pi}^2}{2\bar{p}^{3/2}}
  + \bar{p}^{3/2} V (\bar{\varphi})\,,
\end{equation}
first-order perturbation
\begin{equation}
  {\cal H}^{(1)}_\varphi =
  - \frac{\bar{\pi}^2}{2\bar{p}^{3/2}} \frac{\delta_c^j\delta E^c_j}{2\bar{p}}
  +\frac{\bar{\pi}\delta\pi}{\bar{p}^{3/2}}
  + \bar{p}^{3/2}\left(\frac{\partial V}{\partial \varphi} \delta\varphi + V (\bar{\varphi}) \frac{\delta_c^j\delta E^c_j}{2\bar{p}}\right)
\end{equation}
and second-order perturbation
\begin{eqnarray}
  {\cal H}^{(2)}_\varphi &=&
  \frac{\delta\pi^2}{2 \bar{p}^{3/2}}
  - \frac{\bar{\pi}\delta\pi}{2\bar{p}^{3/2}} \frac{\delta_c^j\delta E^c_j}{\bar{p}}
  + \frac{\bar{p}^{3/2}}{2} \left( \frac{1}{\bar{p}} (\partial^b \delta\varphi) (\partial_b \delta\varphi)
  + \frac{\partial^2 V}{(\partial \varphi)^2} \delta\varphi^2
  + \frac{\delta_c^j\delta E^c_j}{\bar{p}} \frac{\partial V}{\partial \varphi} \delta\varphi\right)
  \nonumber\\
  &&
  + \frac{\bar{\pi}^2}{8\bar{p}^{3/2}} \left( \frac{\delta_c^k\delta_d^j\delta E^c_j\delta E^d_k}{\bar{p}^2}
    + \frac{(\delta_d^k \delta E^d_k)^2}{2\bar{p}^2}\right)
  \nonumber\\
  &&
  + \bar{p}^{3/2} V (\bar{\varphi}) \left( \frac{(\delta^j_c\delta E^c_j)^2}{8\bar{p}^2}-\frac{\delta^k_c\delta^j_d\delta E^c_j\delta E^d_k}{4\bar{p}^2}\right)\,.
\end{eqnarray}

The covariance condition for scalar matter is given by
\begin{equation}
    \delta_\epsilon \varphi |_{\rm O.S.} = \mathcal{L}_\xi \varphi |_{\rm O.S.}
    \,,
\end{equation}
which can be simplified, in the perturbed context, to the series of conditions \cite{EmergentScalar}
\begin{eqnarray}
    \frac{\partial
  \{\delta\varphi,H_\varphi^{(1)}[\delta\epsilon^0]\}}{\partial
  (\partial_{a_1}\delta\epsilon^0)} = \frac{\partial
  \{\delta\varphi,H_\varphi^{(1)}[\delta\epsilon^0]\}}{\partial
  (\partial_{a_1}\partial_{a_2}\delta\epsilon^0)} = \dots = 0\,. 
\end{eqnarray}
Using $\{\delta\varphi,H_\varphi^{(1)}[\delta\epsilon^0]\}=\partial
H_\varphi^{(1)}[\delta\epsilon^0]/\partial \delta \pi$, the condition becomes 
\begin{eqnarray}\label{eq:Covariance cond scalar - pert}
    \frac{\partial {\cal H}_\varphi^{(1)}}{\partial (\partial_{a_1} \delta\pi)}=\frac{\partial {\cal H}_\varphi^{(1)}}{\partial (\partial_{a_1} \partial_{a_2} \delta\pi)}=\dots=0\,.
\end{eqnarray}
Therefore, scalar matter is covariant if and only if ${\cal H}_\varphi^{(1)}$
does not contain  derivatives of the scalar momentum perturbation $\delta \pi$.

When the potential vanishes, this system has a symmetry generator for the free field given by
\begin{eqnarray}\label{eq:Scalar symm gen}
    G[\alpha]=\int{\rm d}^3x \alpha \pi
    = \int{\rm d}^3x \alpha \bar{\pi}\,,
\end{eqnarray}
where $\alpha$ is a constant.  This $G[\alpha]$ Poisson commutes with the
diffeomorphism constraint because the latter is independent of
$\bar{\varphi}$.  On the other hand, commutation with the Hamiltonian
constraint requires it to be independent of $\bar{\varphi}$, hence it commutes
only for $V=0$. The existence of such a quantity may be used as a restriction
on possible modified theories.

\subsection{Modified scalar matter: Free field}
\label{sec:Modified scalar matter: Free field}

Scalar matter introduces several new terms to generic Hamiltonian constraints,
even within an effective theory that maintains the classical derivative order
and second-order perturbative inhomogeneity in the constraints. We therefore
first consider the case of a free field and introduce terms implied by a
scalar potential in a second step.

We formulate the ansatz
\begin{equation}\label{eq:Free scalar constraint ansatz}
  \tilde{H}_{\varphi}[N] =
  \int{\rm d}^3x \bar{N} \left(\tilde{\cal H}^{(0)}_\varphi + \tilde{\cal H}^{(2)}_\varphi\right)
  + \int{\rm d}^3x \delta N \tilde{\cal H}^{(1)}_\varphi
\end{equation}
with the background contribution
\begin{equation}
  \tilde{\cal H}^{(0)}_\varphi =
  {\cal P}_1\frac{\bar{\pi}^2}{2\bar{p}^{3/2}}
  + \Phi_0 \bar{\pi}
  \,,
\end{equation}
the first-order perturbation
\begin{eqnarray}
  \tilde{\cal H}^{(1)}_\varphi &=&
  - {\cal P}_2 \frac{\bar{\pi}^2}{2\bar{p}^{3/2}} 
  \frac{\delta^j_c \delta E^c_j}{2\bar{p}}
  + {\cal P}_3 \frac{\bar{\pi}\delta\pi}{\bar{p}^{3/2}}
  + \Phi_1 \delta \pi
  + \Phi_2 \bar{\pi} \delta^j_c \delta E^c_j
  + \Phi_3 \bar{\pi} \delta_j^c \delta K_c^j
  \nonumber\\
  &&
  + \Theta_1 \bar{\pi} \delta^{ab} \partial_a \partial_b (\delta^j_c \delta E^c_j)
  + \Theta_{2} \bar{\pi} \frac{\partial_c \partial^j\delta E^c_j}{2 \bar{p}}
  \,,
\end{eqnarray}
and the second-order perturbation
\begin{eqnarray}
  \tilde{\cal H}^{(2)}_\varphi &=&
  {\cal P}_4 \frac{\delta\pi^2}{2 \bar{p}^{3/2}}
  - {\cal P}_5 \frac{\bar{\pi}\delta\pi}{\bar{p}^{3/2}} \frac{\delta^j_c\delta E^c_j}{2\bar{p}}
  + {\cal P}_6 \frac{\bar{\pi}^2}{2\bar{p}^{3/2}} \frac{\delta^k_c\delta^j_d\delta E^c_j\delta E^d_k}{4\bar{p}^2}
  + {\cal P}_7 \frac{\bar{\pi}^2}{2\bar{p}^{3/2}} \frac{(\delta^j_c\delta E^c_j)^2}{8\bar{p}^2}
  \nonumber\\
  &&
  + {\cal P}_8 \frac{\sqrt{\bar{p}}}{2} \delta^{a b} (\partial_a \delta\varphi) (\partial_b \delta\varphi)
  + \Phi \frac{\bar{\pi}^2}{2\bar{p}^{3/2}} \delta K_c^j \frac{\delta E^c_j}{2 \bar{p}}
  \nonumber\\
  &&
  + \Phi_4 \delta\pi \frac{\delta^j_c \delta E^c_j}{2\bar{p}}
  + \Phi_5 \delta\pi\delta_j^c\delta K_c^j
  + \Phi_6 \bar{\pi} \frac{\delta^k_c\delta^j_d\delta E^c_j\delta E^d_k}{4\bar{p}^2}
  + \Phi_7 \bar{\pi} \frac{(\delta^j_c\delta E^c_j)^2}{8\bar{p}^2}
  \nonumber\\
  &&
  + \Phi_8 \bar{\pi} \delta K_c^j \frac{\delta E^c_j}{2 \bar{p}}
  + \Phi_9 \bar{\pi} \delta^d_k \delta K_d^k \frac{\delta_c^j \delta E^c_j}{2 \bar{p}}
  + \Phi_{10} \bar{\pi} (\delta^c_j K_c^j)^2
  \nonumber\\
  &&
  + \left(\Theta_3 \bar{\pi} + \Theta_6 \frac{\bar{\pi}^2}{2\bar{p}^{3/2}} \right) \frac{\delta^{jk} (\partial_c \delta E^c_j) (\partial_d \delta E^d_k)}{2 \bar{p}}\nonumber\\
&&  + \left(\Theta_4 \bar{\pi} + \Theta_7 \frac{\bar{\pi}^2}{2\bar{p}^{3/2}} \right)
  \frac{\delta^{ab} \delta_{cd} \delta^{jk} (\partial_a \delta E^c_j) (\partial_b \delta E^d_k)}{2 \bar{p}}
  \nonumber\\
  &&
  + \left(\Theta_5 \bar{\pi} + \Theta_8 \frac{\bar{\pi}^2}{2\bar{p}^{3/2}} \right) \frac{\delta^{ab}\partial_a  (\delta_c^j \delta E^c_j) \partial_b (\delta_d^k \delta E^d_k)}{2 \bar{p}}
  \nonumber\\
  &&
  + \left(\Phi_{11} \frac{\sqrt{\bar{p}}}{2} + \Theta_{9} \bar{\pi} \right) (\partial^j \delta\varphi) (\partial_c \delta E^c_j)
  \,.
\end{eqnarray}
The modification functions $\Phi$ and $\Phi_0$ as well as $\Phi_I$, $\Theta_I$ and ${\cal P}_I$ with
$I=1,2,...$ depend on the background variables $\bar{k}$ and $\bar{p}$ and
have classical limits $\Phi,\Phi_0,\Phi_I,\Theta_I\to0$ while ${\cal P}_I\to1$.

We have not included derivatives of the scalar momentum $\delta \pi$ because
the covariance condition (\ref{eq:Covariance cond scalar - pert}) requires
their absence.  We will also require the existence of the symmetry generator
for the free field (when $V\to0$) given by (\ref{eq:Scalar symm gen}).
Commutation of this symmetry generator with the Hamiltonian constraint
requires that the undetermined functions do not depend on $\bar{\varphi}$,
hence the condition that they depend on $\bar{k}$ and $\bar{p}$ only.  The
addition of a potential, which turns out to be highly nontrivial, will be
performed in a second step.

Appendix~\ref{a:scalar} collects the relevant Poisson brackets and an analysis
of the resulting conditions on modification functions. As a result, the case
of free scalar matter is characterized as follows: All functions except for ${\cal P}_1$,
${\cal P}_2$, $\Phi_0$ and $\Theta_9$ can be obtained in terms of
these functions by simple manipulations: The
remaining ${\cal P}$-functions are determined by
\begin{eqnarray}\label{eq:P3 - summary}
    {\cal P}_3 &=& \frac{1}{2}({\cal P}_1 + {\cal P}_2)
    \,,\\
    \label{eq:P5=P4}
    {\cal P}_5 &=& {\cal P}_4
    \,,\\
    {\cal P}_7 &=& 2{\cal P}_4-{\cal P}_6
    \,,\\
    {\cal P}_6 &=& {\cal P}_1 + \frac{2}{3}\left(\bar{k} \frac{\partial{\cal P}_1}{\partial \bar{k}}-\bar{p}\frac{\partial{\cal P}_1}{\partial \bar{p}}\right)
    \,,\label{eq:P6 - summary}
\end{eqnarray}
as well as
\begin{eqnarray}
    \frac{6}{\bar{p}} {\cal K}^{(1)}_1 {\cal P}_4 &=& 
    \frac{\partial {\cal K}^{(0)}}{\partial \bar{k}} \left(
    \frac{3{\cal P}_1}{2\bar{p}}-\frac{\partial{\cal P}_1}{\partial \bar{p}}
    + \frac{3{\cal P}_2}{2\bar{p}} - \frac{\partial{\cal P}_2}{\partial \bar{p}} \right)\nonumber\\
&&    +
    \left( \frac{{\cal K}^{(0)}}{2\bar{p}} 
    +
    \frac{\partial {\cal K}^{(0)}}{\partial \bar{p}}\right)
    \left(
    \frac{\partial{\cal P}_1}{\partial \bar{k}}
    +
    \frac{\partial{\cal P}_2}{\partial \bar{k}}
    \right)
    \,,\label{eq:P4 - summary}
    \\
    \frac{({\cal P}_1 + {\cal P}_2)}{2} {\cal P}_8 &=&  {\cal K}^{(1)}_3 {\cal K}^{(2)}_1- 2 \bar{p}^{3/2} {\cal K}^{(1)}_1 {\Theta}_9
    \,.\label{eq:P8 - summary}
\end{eqnarray}

The remaining $\Phi$-functions are determined by
\begin{equation}
  \Phi_1=\Phi_3=\Phi_4=\Phi_5=\Phi_9=\Phi_{10}=\Phi_{11}=0
\end{equation}
as well as
\begin{eqnarray}
    \label{eq:Phi - summary}
    \Phi
    &=& 
    \frac{2}{3} \frac{\partial{\cal P}_1}{\partial \bar{k}}
    \,,\\
    {\Phi}_2 &=& \frac{1}{2 \bar{p}} {\Phi}_0
    \,,\label{eq:Phi2 - summary}
    \\
    \Phi_6
    &=& 
    \frac{2 \bar{p}}{3} \frac{\partial{\Phi}_0}{\partial \bar{p}}- \frac{2 \bar{k}}{3} \frac{\partial{\Phi}_0}{\partial \bar{k}}
    \,,\label{eq:Phi6 - summary}
    \\
    \Phi_7 &=&  - \Phi_6
    \,,\label{eq:Phi7 - summary}
    \\
    \Phi_8 &=& \frac{2}{3} \frac{\partial{\Phi}_0}{\partial \bar{k}}
    \,.\label{eq:Phi8 - summary}
\end{eqnarray}
The remaining $\Theta$-functions are determined by
\begin{equation}
  \Theta_1=\Theta_2=\Theta_4=\Theta_5=\Theta_7=\Theta_8=0
\end{equation}
as well as 
\begin{eqnarray}
    {\Theta_3} &=&
    \frac{1}{6{\cal K}^{(1)}_1} \left[
    \left(\frac{\partial {\cal K}^{(1)}_3}{\partial \bar{p}}
    +
    \frac{{\cal K}^{(1)}_3}{2\bar{p}}
    \right) \frac{\partial {\Phi}_0}{\partial \bar{k}} -
    \frac{\partial {\cal K}^{(1)}_3}{\partial \bar{k}} \frac{\partial{\Phi}_0}{\partial \bar{p}}
    \right]
    \,,\label{eq:Theta3 - summary}
\end{eqnarray}
and
\begin{eqnarray}
     6{\cal K}^{(1)}_1 {\Theta_6}&=&
     \frac{\partial{\cal P}_1}{\partial \bar{k}} \left(\frac{\partial {\cal K}^{(1)}_3}{\partial \bar{p}} + \frac{{\cal K}^{(1)}_3}{2\bar{p}}\right)
     -
     \
    \left( \frac{\partial{\cal P}_1}{\partial \bar{p}}
    -\frac{3 {\cal P}_1}{2\bar{p}}\right) \frac{\partial {\cal K}^{(1)}_3}{\partial \bar{k}} 
    \nonumber \\
    &&\qquad
    -3
    {\kappa}
    \sqrt{\bar{p}}
    ({\cal P}_1 + {\cal P}_2){\Theta}_9
    \,,\label{eq:Theta6 - summary}
\end{eqnarray}

The function $\Theta_9$ remains completely free, while the three functions
${\cal P}_1$, ${\cal P}_2$ and $\Phi_0$ are related to one another and some of
the ${\cal K}$-functions by the six differential equations
\begin{eqnarray}%
    0&=& 
    \frac{\partial{\cal P}_1}{\partial \bar{k}}
    \left(
    \frac{\partial{\cal P}_2}{\partial \bar{p}}
    -
    \frac{3{\cal P}_2}{2\bar{p}}
    \right)
    -
    \frac{\partial{\cal P}_2}{\partial \bar{k}}
    \left(
    \frac{\partial{\cal P}_1}{\partial \bar{p}}
    -
    \frac{3{\cal P}_1}{2\bar{p}}
    \right)
    \,,\label{eq:Subset1 - summary}
    \\
    0 &=& 
    \left( \frac{\partial{\cal P}_1}{\partial \bar{p}} - \frac{3{\cal P}_1}{2\bar{p}} + \frac{\partial{\cal P}_2}{\partial \bar{p}}-\frac{3{\cal P}_2}{2\bar{p}}\right) \frac{\partial{\Phi}_0}{\partial \bar{k}}
    - \left( \frac{\partial{\cal P}_1}{\partial \bar{k}} + \frac{\partial{\cal P}_2}{\partial \bar{k}} \right) \frac{\partial{\Phi}_0}{\partial \bar{p}}
    \,,\label{eq:Subset2 - summary}
    \\
    0&=&
    \frac{1}{3} \left( {\cal K}^{(1)}_1 - 2\bar{p}
    \frac{\partial {\cal K}^{(1)}_1}{\partial \bar{p}}\right) \frac{\partial {\Phi}_0}{\partial \bar{k}}
    + \frac{2\bar{p}}{3} \frac{\partial {\cal K}^{(1)}_1}{\partial \bar{k}} \frac{\partial {\Phi}_0}{\partial \bar{p}}
    - {\cal K}^{(2)}_1 {\Phi}_0
    \,,\label{eq:Subset3 - summary}
    \\
    0&=&
    \left(\frac{1}{6} \frac{\partial {\cal K}^{(1)}_2}{\partial \bar{k}} - \frac{1}{2} \frac{\partial {\cal K}^{(0)}}{\partial \bar{k}} \right) \frac{\partial {\Phi}_0}{\partial \bar{p}}
    + \left[\frac{1}{2} \left(
    \frac{{\cal K}^{(0)}}{2\bar{p}} + \frac{\partial {\cal K}^{(0)}}{\partial \bar{p}}
    \right)
    - \frac{{\cal K}^{(1)}_2}{6\bar{p}}
    + \frac{1}{6} \left(\frac{{\cal K}^{(1)}_2}{2\bar{p}}-\frac{\partial {\cal K}^{(1)}_2}{\partial \bar{p}} \right)\right] \frac{\partial {\Phi}_0}{\partial \bar{k}}
    \nonumber\\
    && + \frac{1}{2\bar{p}} \left(\frac{\partial {\cal K}^{(0)}}{\partial \bar{k}} - {\cal K}^{(2)}_3\right) \Phi_0
    -
    \frac{{\cal K}^{(1)}_1}{3\bar{p}}
    \left(
    \bar{k}\frac{\partial {\cal P}_1}{\partial \bar{k}}
    -
    \bar{p}\frac{\partial {\cal P}_1}{\partial \bar{p}}
    \right)
    \,,\label{eq:Subset4 - summary}
    \\
    0 &=&
    \frac{3 {\cal P}_2}{2\bar{p}} {\cal K}^{(2)}_1
    - \left(\frac{3 {\cal P}_1}{2\bar{p}}-\frac{\partial {\cal P}_1}{\partial \bar{p}}\right) \frac{\partial {\cal K}^{(1)}_1}{\partial \bar{k}}
    - \frac{\partial {\cal P}_1}{\partial \bar{k}} \left(\frac{\partial {\cal K}^{(1)}_1}{\partial \bar{p}} - \frac{{\cal K}^{(1)}_1}{2 \bar{p}} \right)
    \,,\label{eq:Subset5 - summary}
    \\
    0&=&
    \frac{\partial {\cal K}^{(0)}}{\partial \bar{k}} \left( \bar{p}\frac{\partial{\cal P}_2}{\partial \bar{p}}
    - \frac{5 {\cal P}_2}{2} \right)
    - \left( \frac{{\cal K}^{(0)}}{2} + \bar{p}\frac{\partial {\cal K}^{(0)}}{\partial \bar{p}}\right) \frac{\partial{\cal P}_2}{\partial \bar{k}}
    + {\cal P}_2 {\cal K}^{(2)}_3
    \nonumber\\
    &&
    - \frac{1}{3} \left(
    \frac{\partial{\cal P}_1}{\partial \bar{k}} 
    \left( \bar{p}\frac{\partial {\cal K}^{(1)}_2}{\partial \bar{p}} +\frac{{\cal K}^{(1)}_2}{2}\right)
    -
     \left( 
     \bar{p}\frac{\partial {\cal P}_1}{\partial \bar{p}} - \frac{3{\cal P}_1}{2}
     \right)
    \frac{\partial {\cal K}^{(1)}_2}{\partial \bar{k}} \right)
    \nonumber\\
    &&-
    \frac{2}{3} {\cal K}^{(1)}_1
    \left(
    \bar{k}\frac{\partial{\cal P}_1}{\partial \bar{k}}
    -
    \bar{p}\frac{\partial{\cal P}_1}{\partial \bar{p}}
    \right)
    - {\cal K}^{(1)}_1{\cal P}_1
      +
    \frac{{2\kappa}}{3}\bar{p}
    \frac{\partial {\Phi}_0}{\partial \bar{k}}
    {\Phi}_0 
    \nonumber\\
    && 
    - \bar{p} 
    \frac{\partial {\cal K}^{(0)}}{\partial \bar{k}} \left(
    \frac{\partial{\cal P}_1}{\partial \bar{p}}
    - \frac{3{\cal P}_1}{2\bar{p}}
    + \frac{\partial{\cal P}_2}{\partial \bar{p}}
    - \frac{3{\cal P}_2}{2\bar{p}} \right)
    +\bar{p}
    \left( \frac{{\cal K}^{(0)}}{2\bar{p}} 
    +
    \frac{\partial {\cal K}^{(0)}}{\partial \bar{p}}\right)
    \left(
    \frac{\partial{\cal P}_1}{\partial \bar{k}}
    +
    \frac{\partial{\cal P}_2}{\partial \bar{k}}
    \right)
    \,,\label{eq:Subset6 - summary}
\end{eqnarray}

The constraint ansatz for free scalar matter, (\ref{eq:Free scalar constraint
  ansatz}), introduced 30 undetermined functions, 26 of which can be solved
for in a straightforward manner by using the independent anomaly freedom
equations (\ref{eq:P3 - summary})--(\ref{eq:Theta6 - summary}). The remaining
four functions are subject to six differential equations (\ref{eq:Subset1 -
  summary})--(\ref{eq:Subset6 - summary}). However, the scalar contribution is
not uniquely determined since one function, $\Theta_9$, is free. It is not
straightforward to evaluate the remaining freedom because the conditions
(\ref{eq:Subset1 - summary})--(\ref{eq:Subset6 - summary}) are partial
differential equations whose solutions have integration functions.

For given gravitational modifications with fixed ${\cal K}$-functions,
equations (\ref{eq:Subset3 - summary})--(\ref{eq:Subset6 - summary}) are
linear partial differential equations (if $\Phi_0^2$ is used instead of
$\Phi_0$). If a solution for $\Phi_0$ can be found, equation (\ref{eq:Subset2
  - summary}) is then a linear partial differential equation relating ${\cal
  P}_1$ and ${\cal P}_2$. The first equation, (\ref{eq:Subset1 - summary}), is
non-linear. Since there are six partial differential equations for three
functions of two variables, one may expect that the dependence of these
functions on $\bar{p}$ and $\bar{k}$ is completely determined since
integration constants are fixed by the classical limit. However, the partial
non-linear nature of the equations could complicate this behavior. 

Alternatively, one may combine the gravitational and spatial parts and
simultaneously solve for the corresponding modification functions. In this
case, all equations (\ref{eq:Subset1 - summary})--(\ref{eq:Subset6 - summary})
are non-linear.  In this interpretation, the final set of equations also
demonstrates that the gravitational part is more restricted in the presence of
scalar matter because some of the ${\cal K}$-functions appear in the new
conditions.

In all cases, the free scalar system has the symmetry generator
\begin{equation}
    G[\alpha]=\int{\rm d}^3x \alpha \pi
    = \int{\rm d}^3x \alpha \bar{\pi}\,,
\end{equation}
where $\alpha$ is a constant.  This $G[\alpha]$ commutes automatically with
the constraints because they are independent of $\bar{\varphi}$ in the free
case.

\subsection{Scalar matter with potential}

In order to include a potential we formulate the ansatz
\begin{equation}\label{eq:Scalar potential constraint ansatz}
    \tilde{H}_{V}[N] =
  \int{\rm d}^3x \bar{N} \left(\tilde{\cal H}^{(0)}_V + \tilde{\cal H}^{(2)}_V\right)
  + \int{\rm d}^3x \delta N \tilde{\cal H}^{(1)}_V
\end{equation}
 for its constraint contribution, with the background contribution
\begin{equation}
  \label{HamConstH0-EMG-Potential}
  \tilde{\cal H}^{(0)}_V =
  \left({\cal P}_{V1}+ \Theta_{V1} \bar{\pi}\right) \bar{p}^{3/2}\,,
\end{equation}
the first-order perturbation
\begin{equation}
  \label{HamConstH1-EMG-Potential}
  \tilde{\cal H}^{(1)}_V =
  \left({\cal P}_{dV1}+ \Theta_{dV1} \bar{\pi}\right) \bar{p}^{3/2} \delta\varphi
  + \left({\cal P}_{V2}+ \Theta_{V2} \bar{\pi}\right) \bar{p}^{3/2} \frac{\delta^j_c \delta E^c_j}{2 \bar{p}}
  + \Phi_{V1} \bar{p}^{3/2} \delta^d_k \delta K_d^k
`\end{equation}
and the second-order perturbation
\begin{eqnarray}
  \label{HamConstH2-EMG-Potential}
  \tilde{\cal H}^{(2)}_V &=&
  \bar{p}^{3/2} \left( \left({\cal P}_{V3}+ \Theta_{V3} \bar{\pi}\right) 
  \frac{(\delta^j_c\delta E^c_j)^2}{8\bar{p}^2}
  - \left({\cal P}_{V4}+ \Theta_{V4} \bar{\pi}\right) \frac{\delta^k_c\delta^j_d\delta E^c_j\delta E^d_k}{4\bar{p}^2}\right)
  \nonumber\\
  &&
  + \left({\cal P}_{ddV}+ \Theta_{ddV} \bar{\pi}\right) \frac{\bar{p}^{3/2}}{2} \delta\varphi^2
  \nonumber\\
  &&
  + \left({\cal P}_{dV2}+ \Theta_{dV2} \bar{\pi}\right) \bar{p}^{3/2} \delta\varphi\frac{\delta^j_c\delta E^c_j}{2\bar{p}}
  + \Phi_{V2} \bar{p}^{3/2} \frac{\delta^j_c\delta E^c_j}{2\bar{p}} \delta^d_k \delta K^k_d
  \nonumber\\
  &&
  + \Phi_{V3} \bar{p}^{3/2} \frac{\delta E^d_k \delta K^k_d}{2\bar{p}}
  + \Phi_{V4} \bar{p}^{3/2} \delta\varphi \delta^d_k \delta K^k_d
  \nonumber\\
  && + \Phi_{V5} (\delta_j^c\delta K_c^j)^2 + \Phi_{V6} \delta_k^c\delta_j^d\delta K_c^j\delta K_d^k + \Phi_{V7} \delta^{cd}\delta_{jk}\delta K_c^j\delta K_d^k
  \,.
\end{eqnarray}
Here, ${\cal P}_{VI}$, ${\cal P}_{dVI}$, ${\cal P}_{ddV}$, $\Theta_{VI}$,
$\Theta_{dVI}$, $\Theta_{ddV}$ and
$\Phi_{VI}$  for $I=1,2,\dots$ are functions of $\bar{p}$, $\bar{k}$, and $\bar{\varphi}$ with
the classical limits ${\cal P}_{VI}\to V(\bar{\varphi})$,
${\cal P}_{VI}\to \partial V/ \partial \bar{\varphi}$,
${\cal P}_{ddV}\to \partial^2 V/ \partial \bar{\varphi}^2$, and
$\Phi_{VI},\Theta_{VI},\Theta_{dVI},\Theta_{ddV}\to0$ for a given classical scalar
potential $V(\bar{\varphi})$.

Imposing
anomaly freedom for this potential contribution, added to
the full constraints, we obtain a large set of independent equations. Some of
them can be simplified using the conditions $\Phi_1=\Phi_3=\Phi_4=\Phi_5=0$
observed for free scalar matter. The resulting equations for $\Phi_{VI}$ take
the compact form
\begin{eqnarray}
    \Phi_{V1}&=&0
    \,,\\
    \label{eq:HV,D - 4 - summary}
    \Phi_{V2} &=&0
    \,,\\
    \label{eq:HV,D - 5 - summary}
    \Phi_{V3} &=& \frac{2}{3} 
    \frac{\partial {\cal P}_{V1}}{\partial \bar{k}}
    \,,\\
    \Phi_{V4} &=& \Phi_{V5} = \Phi_{V6} = \Phi_{V7} = 0
    \,.
\end{eqnarray}
The equations for $\Phi_{V1}$, $\Phi_{V2}$, and $\Phi_{V3}$ are initially more complicated but can be simplified using the last equation.
This step also simplifies the remaining equations,
\begin{eqnarray}
    \label{eq:HV,D - 10 - summary}
    {\cal P}_{V2} &=&
    {\cal P}_{V1}
    \,,\\
    \label{eq:HV,D - 2 - summary}
    {\cal P}_{V4} = {\cal P}_{V3} &=& {\cal P}_{V1}
    + \frac{2}{3} \left(\bar{p} \frac{\partial {\cal P}_{V1}}{\partial \bar{p}} - \bar{k} \frac{\partial {\cal P}_{V1}}{\partial \bar{k}}\right)
\end{eqnarray}
for ${\cal P}_{VI}$ in terms of ${\cal P}_{V1}$,
\begin{eqnarray}
    \label{eq:HV,D - 8 - summary}
    \frac{\partial {\Theta}_{V1}}{\partial \bar{k}}
    &=&
    0
    \,,\\
    \label{eq:HV,D - 11 - summary}
    {\Theta}_{V1} &=& {\Theta}_{V2}
    \,,\\
    \label{eq:HV,D - 6 - summary}
     {\Theta}_{V4} = {\Theta}_{V3} &=& {\Theta}_{V1}
    + \frac{2\bar{p}}{3} \frac{\partial {\Theta}_{V1}}{\partial \bar{p}}
\end{eqnarray}
for $\Theta_{VI}$ in terms of $\Theta_{V1}(\bar{p},\bar{\varphi})$,
\begin{eqnarray}
{\cal P}_{dV1} &=&
\frac{\kappa \bar{p}^{3/2}}{3} \frac{1}{{\cal P}_4} \left( 
    \frac{\partial {\cal P}_{V1}}{\partial \bar{k}}
    \frac{\partial{\Phi}_1}{\partial \bar{p}}
    - \frac{\partial{\Phi}_1}{\partial \bar{k}}
    \left(\frac{3 {\cal P}_{V1}}{2\bar{p}} + \frac{\partial {\cal P}_{V1}}{\partial \bar{p}} \right)
    \right)
\nonumber \\
&&
+ \frac{1}{2} \frac{\mathcal{P}_1 + \mathcal{P}_2}{{\cal P}_4}
   \frac{\partial \mathcal{P}_{V1}}{\partial \bar{\varphi}}
       \,,\\
    \label{eq:HV,D - 3 - summary}
    {\cal P}_{dV2} &=& \frac{\partial {\cal P}_{V1}}{\partial \bar{\varphi}}
\end{eqnarray}
for ${\cal P}_{dVI}$,
\begin{eqnarray}
-{\cal P}_{ddV} {\cal P}_{3} &=&
% Bracket 1: 1 contribution
\bar{p}^2 \biggl( \left( \frac{{\cal K}^{(0)}}{2\bar{p}} + \frac{\partial {\cal K}^{(0)}}{\partial \bar{p}} \right) \frac{\partial{\Theta}_{dV1}}{\partial\bar{k}} 
- \frac{\partial {\cal K}^{(0)}}{\partial \bar{k}} \left( \frac{3{\Theta}_{dV1}}{2\bar{p}} + \frac{\partial{\Theta}_{dV1}}{\partial\bar{p}} \right) \biggr) \nonumber\\
% Bracket 3: 3 contributions
&&+ \frac{\kappa \bar{p}^{3/2}}{3} \biggl( \frac{\partial {\Phi}_{0}}{\partial \bar{k}} \left( \frac{3 {\cal P}_{dV1}}{2\bar{p}} + \frac{\partial{\cal P}_{dV1}}{\partial\bar{p}} \right) 
+ \left( \frac{3 {\Phi}_{0}}{2\bar{p}} - \frac{\partial {\Phi}_{0}}{\partial \bar{p}} \right) \frac{\partial{\cal P}_{dV1}}{\partial\bar{k}} \biggr) \nonumber\\
&&- {\cal P}_{1} \frac{\partial {\cal P}_{dV1}}{\partial \bar{\varphi}}
- \bar{p}^{3/2} {\Phi}_{0} \frac{\partial {\Theta}_{dV1}}{\partial \bar{\varphi}}
+
\bar{p}^3
\frac{\partial {\Theta}_{V1}}{\partial \bar{\varphi}}
{\Theta}_{dV1}
\delta \varphi
\nonumber\\
% Bracket 5: 4 contributions
&&+ \frac{\kappa \bar{p}^3}{3} \Biggl( \frac{\partial {\Theta}_{V1}}{\partial \bar{k}} \left( \frac{3 {\cal P}_{dV1}}{2\bar{p}} + \frac{\partial{\cal P}_{dV1}}{\partial\bar{p}} \right) 
- \left( \frac{3 {\Theta}_{V1}}{2\bar{p}} + \frac{\partial {\Theta}_{V1}}{\partial \bar{p}} \right) \frac{\partial{\cal P}_{dV1}}{\partial\bar{k}} \Biggr)  \nonumber \\
&&+ \frac{\kappa \bar{p}^3}{3} \Biggl( \frac{\partial {\cal P}_{V1}}{\partial \bar{k}} \left( \frac{3 {\Theta}_{dV1}}{2\bar{p}} + \frac{\partial{\Theta}_{dV1}}{\partial\bar{p}} \right) 
- \left( \frac{3 {\cal P}_{V1}}{2\bar{p}} + \frac{\partial {\cal P}_{V1}}{\partial \bar{p}} \right) \frac{\partial{\Theta}_{dV1}}{\partial\bar{k}} \Biggr)
\nonumber \\
&&
% Bracket 7: 1 contribution
+ 3\bar{p} {\cal K}^{(1)}_1 {\Theta}_{dV2}
% Bracket 9: 2 contribution
- 3 \kappa \bar{p}^{3/2} {\Phi}_{2} \Phi_{V4}
- \bar{p}^3 \frac{\partial \Theta_{dV1}}{\partial \bar{\varphi}} {\Theta}_{V1}
+ \bar{p}^3 \frac{\partial \Theta_{V1}}{\partial \bar{\varphi}} {\Theta}_{dV1}
\nonumber\\
&&
% Bracket 10: 1 contribution
+ \kappa \bar{p}^2 \frac{3{\Theta}_{V2}}{2} \Phi_{V4}
\,.
\end{eqnarray}
for ${\cal P}_{ddV}$,
\begin{eqnarray}
- {\Theta}_{ddV} {\cal P}_3  &=&
% Bracket 3: 3 contributions
\frac{\kappa}{6} \biggl[\frac{\partial {\cal P}_{1}}{\partial \bar{k}} \left( \frac{3 {\cal P}_{dV1}}{2\bar{p}} + \frac{\partial{\cal P}_{dV1}}{\partial\bar{p}} \right) 
+ \left( \frac{3 {\cal P}_{1}}{2\bar{p}} - \frac{\partial {\cal P}_{1}}{\partial \bar{p}} \right) \frac{\partial{\cal P}_{dV1}}{\partial\bar{k}} \biggr] \nonumber\\'
&&+ \frac{\kappa \bar{p}^{3/2}}{3} \biggl[ \frac{\partial {\Phi}_{0}}{\partial \bar{k}} \left( \frac{3 {\Theta}_{dV1}}{2\bar{p}} + \frac{\partial{\Theta}_{dV1}}{\partial\bar{p}} \right) 
+ \left( \frac{3 {\Phi}_{0}}{2\bar{p}} - \frac{\partial {\Phi}_{0}}{\partial \bar{p}} \right) \frac{\partial{\Theta}_{dV1}}{\partial\bar{k}} \biggr] \nonumber\\
&&- {\cal P}_{1} \frac{\partial {\Theta}_{dV1}}{\partial \bar{\varphi}}
% Bracket 9: 2 contribution
- \frac{3 {\kappa}}{4 \bar{p}} {\cal P}_2 \Phi_{V4}
+ \bar{p}^3
\frac{\partial {\Phi}_{V0}}{\partial \bar{\varphi}}
{\Theta}_{dV1}
\nonumber \\
% Bracket 5: 1 contribution
&&+\frac{\kappa}{3}\bar{p}^3
\biggl[
\frac{\partial {\Theta}_{V1}}{\partial \bar{k}} \left( \frac{3 {\Theta}_{dV1}}{2\bar{p}} + \frac{\partial{\Theta}_{dV1}}{\partial\bar{p}}\right)
    - \left(\frac{3 {\Theta}_{V1}}{2\bar{p}} + \frac{\partial {\Theta}_{V1}}{\partial \bar{p}} \right) \frac{\partial{\Theta}_{dV1}}{\partial\bar{k}} 
    \biggr]
% new contribution from Hv0 , Hv1 bracket
\nonumber
\,.
\end{eqnarray}
for $\Theta_{ddV}$, and
\begin{eqnarray}
 {\cal P}_{4}{\Theta}_{dV1}
 &=&
 \frac{\kappa}{3} \Biggl[
    \frac{\partial {\cal P}_{V1}}{\partial \bar{k}}
    \Biggl(
    \frac{\partial {\cal P}_3}{\partial \bar{p}}
    -
    \frac{3{\cal P}_{3}}{2\bar{p}}
\Biggr)
-\frac{\partial{\cal P}_3}{\partial \bar{k}}
\Biggl(
       \frac{3{\cal P}_{V1}}{2\bar{p}}
       +\frac{\partial {\cal P}_{V1}}{\partial \bar{p}}
     \Biggr)
\Biggr]
\nonumber\\
&&+\frac{\kappa \bar{p}^{3/2}}{3} \left[ \frac{\partial {\Theta}_{V1}}{\partial \bar{k}}
    \frac{\partial{\Phi}_1}{\partial \bar{p}}
    - \frac{\partial{\Phi}_1}{\partial \bar{k}}
    \left(\frac{3{\Theta}_{V1}}{2\bar{p}} + \frac{\partial {\Theta}_{V1}}{\partial \bar{p}} \right)
    \right]
\nonumber\\
&&+ \frac{\mathcal{P}_1 + \mathcal{P}_2}{2}
   \frac{\partial \Theta_{V1}}{\partial \bar{\varphi}}
% Bracket 8: One contribution
+ \frac{3{\kappa}}{2\bar{p}}{\cal P}_{4}{\Phi}_{V1}
\,,
\end{eqnarray}
\begin{eqnarray}
-{\Theta}_{dV2} \frac{{\cal P}_3}{2\,\bar{p}} &=&
\frac{\kappa}{3} \left( 
\frac{\partial {\Theta}_{V1}}{\partial \bar{k}}
\frac{\partial{\Phi}_2}{\partial \bar{p}}
- \frac{\partial{\Phi}_2}{\partial \bar{k}}
\left(\frac{3 {\Theta}_{V1}}{2\bar{p}} 
+ \frac{\partial {\Theta}_{V1}}{\partial \bar{p}} \right)\right)
\nonumber\\
%Bracket 3 contribution
&&+ \frac{\kappa}{12\bar{p}}\,\biggl[
\frac{\partial{\cal P}_1}{\partial \bar{k}}
\left(\frac{{\cal P}_{V2}}{2\bar{p}}
+ \frac{\partial {\cal P}_{V2}}{\partial \bar{p}}\right)
+\left(\frac{\partial{\cal P}_1}{\partial \bar{p}}
 - \frac{3\,{\cal P}_1}{2\bar{p}}\right)\,
\frac{\partial {\cal P}_{V2}}{\partial\bar{k}}
\biggr]
\nonumber\\
%Bracket 4 contribution
&&- \frac{\kappa}{12\bar{p}}\biggl[
\frac{\partial{\cal P}_2}{\partial \bar{k}}
\left(\frac{3{\cal P}_{V1}}{2\bar{p}}
+ \frac{\partial {\cal P}_{V1}}{\partial \bar{p}}\right)
+ \frac{\partial {\cal P}_{V1}}{\partial \bar{k}}
\left(\frac{\partial {\cal P}_2}{\partial \bar{p}}
- \frac{5\,{\cal P}_2}{2\bar{p}}\right)
\biggr]
\nonumber\\
&&- \frac{{\cal P}_2}{2\bar{p}} \frac{\partial {\Theta}_{V1}}{\partial \bar{\varphi}}
%Bracket 8 and 9 contributions
+ \frac{{\cal P}_4}{2\bar{p}} {\Theta}_{dV1}
- \kappa \Phi_{V3} \frac{{\cal P}_2}{8\bar{p}^2}
\nonumber \\
&&
+\frac{3\kappa}{8\bar{p}^2}\left(\frac{{\cal P}_{6}}{3} -2{\cal P}_4\right) {\Phi}_{V1}
-\frac{{\cal P}_1}{2\bar{p}} \frac{\partial {\Theta}_{V2}}{\partial \bar{\varphi}}
\nonumber\\
&&
+ \frac{\kappa\sqrt{\bar{p}}}{6}\left( \frac{\partial{\Phi}_0}{\partial \bar{k}} \left( \frac{{\Theta}_{V2}}{2\bar{p}} + \frac{\partial{\Theta}_{V2}}{\partial\bar{p}} \right)
+ \left( \frac{\partial{\Phi}_0}{\partial \bar{p}}
- \frac{3 {\Phi}_0}{2\bar{p}}\right) \frac{\partial{\Theta}_{V2}}{\partial\bar{k}} \right)
\nonumber \\
&&+
\frac{\kappa\bar{p}^2}{6}
\biggl[
\frac{\partial {\Theta}_{V1}}{\partial \bar{k}} \left( \frac{{\Theta}_{V2}}{2\bar{p}} + \frac{\partial{\Theta}_{V2}}{\partial\bar{p}} \right)
- \left(\frac{3 {\Theta}_{V1}}{2\bar{p}} + \frac{\partial {\Theta}_{V1}}{\partial \bar{p}} \right) \frac{\partial{\Theta}_{V2}}{\partial\bar{k}} 
\biggr]
\nonumber \\
&&
+
\frac{\kappa}{8\bar{p}^2} {\cal P}_{V2}{\Phi}
+
\frac{\kappa}{6\sqrt{\bar{p}}}\frac{\partial \Phi_0}{\partial \bar{k}}
{\Theta}_{V2}
\,.
\end{eqnarray}
for $\Theta_{dVI}$. If these  equations have been solved,
\begin{equation}
    \label{eq:HV,D - 7 - summary}
    \frac{\partial {\Theta}_{V1}}{\partial \bar{\varphi}}=   {\Theta}_{dV2}
\end{equation}
restricts the $\bar{\varphi}$-dependence of $\Theta_{V1}$, while 
\begin{eqnarray}
% (Bracket 2) from \{\tilde{H}^{(0)}_{V}, \tilde{H}^{(1)}_{\rm grav}\}
0 &=&
 \,\frac{2\bar{p}^2}{3}\;\Biggl[
    \left(\frac{3\Theta_{V1}}{2\bar{p}}
          +
          \frac{\partial \Theta_{V1}}{\partial \bar{p}}\right)
    \frac{\partial {\cal K}^{(1)}_1}{\partial \bar{k}}
    -
    \frac{\partial \Theta_{V1}}{\partial \bar{k}}\,
    \left(\frac{\partial {\cal K}^{(1)}_1}{\partial \bar{p}}
          +
          \frac{{\cal K}^{(1)}_1}{2\,\bar{p}}\right)
\Biggr]
% (Bracket 4) from \{\tilde{H}^{(0)}_{V}, \tilde{H}^{(1)}_{\varphi}\}
+\Phi_{V4}{\cal P}_3
\,.\nonumber
\end{eqnarray}
can be used to restrict the $\bar{p}$-dependence (since ${\cal K}_1^{(1)}$ has
to depend on $\bar{k}$ for the correct classical limit).
The equations
\begin{eqnarray}
0 &=&
% 2) from \{\tilde{H}^{(0)}_{V},\tilde{H}^{(1)}_{\rm grav}\}, no pi
\frac{2\bar{p}^2}{3}\biggl[
  \left(\frac{3{\cal P}_{V1}}{2\bar{p}}
        + \frac{\partial {\cal P}_{V1}}{\partial \bar{p}}\right)
  \,\frac{\partial {\cal K}^{(1)}_1}{\partial \bar{k}}
  -
  \frac{\partial {\cal P}_{V1}}{\partial \bar{k}}
  \left(\frac{\partial {\cal K}^{(1)}_1}{\partial \bar{p}}
        + \frac{{\cal K}^{(1)}_1}{2\,\bar{p}}\right)
\biggr]
\nonumber\\
&&\quad
% 4) from \{\tilde{H}^{(0)}_V,\tilde{H}^{(1)}_\varphi\}, no pi
+ \bar{p} {\cal K}^{(1)}_1 \Phi_{V3}
% 6) from \{\tilde{H}^{(2)}_{\rm grav},\tilde{H}^{(1)}_V\}_\delta
- \bar{p} {\cal P}_{V2} {\cal K}^{(2)}_1
\,.
\end{eqnarray}
\begin{eqnarray}
0 &=&
\frac{\bar{p}}{2} \left[ \left( \frac{{\cal K}^{(0)}}{2\bar{p}} + \frac{\partial {\cal K}^{(0)}}{\partial \bar{p}}\right) \frac{\partial{\cal P}_{V2}}{\partial\bar{k}}
- \frac{\partial {\cal K}^{(0)}}{\partial \bar{k}} \left( \frac{3{\cal P}_{V2}}{2\bar{p}} + \frac{\partial{\cal P}_{V2}}{\partial\bar{p}} \right) \right]
\nonumber \\
&&+ \frac{\bar{p}}{6} \left( \left(\frac{3 {\cal P}_{V1}}{2\bar{p}} + \frac{\partial {\cal P}_{V1}}{\partial \bar{p}} \right) \frac{\partial {\cal K}^{(1)}_2}{\partial \bar{k}}
- \frac{\partial {\cal P}_{V1}}{\partial \bar{k}} \left(\frac{\partial {\cal K}^{(1)}_2}{\partial \bar{p}} - \frac{{\cal K}^{(1)}_2}{2 \bar{p}}\right) \right)
\nonumber \\
&&+ \frac{\kappa\bar{p}}{6}
\left( \frac{\partial {\cal P}_{V1}}{\partial \bar{k}} \left( \frac{{\cal P}_{V2}}{2\bar{p}} + \frac{\partial{\cal P}_{V2}}{\partial\bar{p}} \right)
- \left(\frac{3 {\cal P}_{V1}}{2\bar{p}} + \frac{\partial {\cal P}_{V1}}{\partial \bar{p}} \right) \frac{\partial{\cal P}_{V2}}{\partial\bar{k}} \right)
\nonumber \\
&&-\frac{\sqrt{\bar{p}}}{2}
{\Phi}_0 \frac{\partial {\cal P}_{V2}}{\partial \bar{\varphi}}
+ \frac{1}{2}
\left({\cal K}^{(1)}_1 {\cal P}_{V3}
- {\cal K}^{(2)}_3 {\cal P}_{V2}
-\frac{{\cal K}^{(1)}_2}{2} \Phi_{V3} \right)
\nonumber \\
&&
+ \frac{{\kappa}\bar{p}}{4} {\cal P}_{V2} \Phi_{V3}
+ \bar{p}^{3/2} \frac{\partial {\cal P}_{V1}}{\partial \bar{\varphi}} {\Phi}_2
+
\bar{p}^2 \frac{\partial {\cal P}_{V1}}{\partial \bar{\varphi}} {\Theta}_{V2}
-\frac{\bar{p}^2}{2}
\frac{\partial {\cal P}_{V2}}{\partial \bar{\varphi}}
{\Theta}_{V1}
\end{eqnarray}
and
\begin{eqnarray}
0 &=&
\frac{\bar{p}}{3} \frac{\partial {\cal P}_{V1}}{\partial \bar{k}} \left(\frac{\partial {\cal K}^{(1)}_3}{\partial \bar{p}} - \frac{{\cal K}^{(1)}_3}{2 \bar{p}}\right)
- \frac{\bar{p}}{3} \left(\frac{3 {\cal P}_{V1}}{2\bar{p}} + \frac{\partial {\cal P}_{V1}}{\partial \bar{p}} \right) \frac{\partial {\cal K}^{(1)}_3}{\partial \bar{k}}
+\frac{1}{2} {\cal K}^{(1)}_3 \Phi_{V3}
\,.
\end{eqnarray}
restrict the dependence on $\bar{p}$, $\bar{k}$ and $\bar{\varphi}$ of the
remaining free function related to potential terms ${\cal P}_{VI}$.

There are six final equations required for anomaly freedom,
\begin{eqnarray}
0
&=&
-\frac{\partial{\cal P}_2}{\partial \bar{k}} 
    \left(\frac{3 {\Theta}_{V1}}{2\bar{p}}
    + \frac{\partial {\Theta}_{V1}}{\partial \bar{p}}\right)
    - \frac{\partial {\Theta}_{V1}}{\partial \bar{k}}
    \left(\frac{\partial{\cal P}_2}{\partial \bar{p}}
    - \frac{5{\cal P}_2}{2\bar{p}} \right)
\nonumber \\
&&
+\frac{\partial{\cal P}_1}{\partial \bar{k}}
\left(\frac{{\Theta}_{V2}}{2\bar{p}}
+ \frac{\partial {\Theta}_{V2}}{\partial \bar{p}}\right)
+\left(\frac{\partial{\cal P}_1}{\partial \bar{p}}
 - \frac{3{\cal P}_1}{2\bar{p}}\right)
\frac{\partial {\Theta}_{V2}}{\partial\bar{k}}
+
\frac{3 {\Theta}_{V2}}{8\bar{p}} {\Phi}
\end{eqnarray}
\begin{eqnarray}
0 = \frac{\partial{\cal P}_3}{\partial \bar{k}}
    \left(\frac{3 {\Theta}_{V1}}{2\bar{p}} + \frac{\partial {\Theta}_{V1}}{\partial \bar{p}} \right)
\end{eqnarray}
\begin{eqnarray}
0 =
- \frac{\bar{p}}{3} \left(\frac{3\Theta_{V1}}{2\,\bar{p}}
+ \frac{\partial \Theta_{V1}}{\partial \bar{p}}\right)
\frac{\partial {\cal K}^{(1)}_3}{\partial \bar{k}}
+\frac{\sqrt{\bar{p}}}{2}{\Theta}_{2}
\frac{\partial {\Theta}_{V1}}{\partial \bar{\varphi}}
\end{eqnarray}
\begin{eqnarray}
0 &=&
\frac{\bar{p}}{3}\,\biggl[
\frac{\partial \Theta_{V1}}{\partial \bar{k}}
\Bigl(\frac{\partial {\cal K}^{(1)}_3}{\partial \bar{p}}
- \frac{{\cal K}^{(1)}_3}{2\bar{p}}\Bigr)
-
\Bigl(\frac{3\,\Theta_{V1}}{2\,\bar{p}}
+ \frac{\partial \Theta_{V1}}{\partial \bar{p}}\Bigr)
\frac{\partial {\cal K}^{(1)}_3}{\partial \bar{k}}
\biggr]
\,.
\end{eqnarray}
\begin{eqnarray}
0 &=&\bar{p}^2 \left( \left( \frac{{\cal K}^{(0)}}{2\bar{p}} + \frac{\partial {\cal K}^{(0)}}{\partial \bar{p}}\right) \frac{\partial{\cal P}_{dV1}}{\partial\bar{k}}
    - \frac{\partial {\cal K}^{(0)}}{\partial \bar{k}} \left( \frac{3 {\cal P}_{dV1}}{2\bar{p}} + \frac{\partial{\cal P}_{dV1}}{\partial\bar{p}}\right) \right) \nonumber\\
    &&+ \frac{\kappa \bar{p}^3}{3}
 \left(\frac{\partial {\cal P}_{V1}}{\partial \bar{k}} \left( \frac{3 {\cal P}_{dV1}}{2\bar{p}} + \frac{\partial{\cal P}_{dV1}}{\partial\bar{p}}\right)
    - \left(\frac{3 {\cal P}_{V1}}{2\bar{p}} + \frac{\partial {\cal P}_{V1}}{\partial \bar{p}} \right) \frac{\partial{\cal P}_{dV1}}{\partial\bar{k}} \right) \nonumber \\
    &&-
    \bar{p}^{3/2}
    {\Phi}_0 \frac{\partial{\cal P}_{dV1}}{\partial \bar{\varphi}}
    +\bar{p}^3\biggl(
    \frac{\partial {\cal P}_{V1}}{\partial \bar{\varphi}}
    {\Theta}_{dV1}
    -
    \frac{\partial {\cal P}_{dV1}}{\partial \bar{\varphi}}  {\Theta}_{V1}\biggl)
    \nonumber \\
    &&+3 \bar{p} {\cal P}_{dV2} {\cal K}^{(1)}_1
    +
    \bar{p}^3 
    \left(
    \frac{\partial {\cal P}_{V1}}{\partial \bar{\varphi}}
    {\Theta}_{dV1}
    -
    \frac{\partial {\cal P}_{dV1}}{\partial \bar{\varphi}}
    {\Theta}_{V1}
    \right)
\end{eqnarray}
and
\begin{eqnarray}
0 &=&
\frac{\bar{p}}{2} \left[ \left( \frac{{\cal K}^{(0)}}{2\bar{p}} + \frac{\partial {\cal K}^{(0)}}{\partial \bar{p}}\right) \frac{\partial{\Theta}_{V2}}{\partial\bar{k}}
- \frac{\partial {\cal K}^{(0)}}{\partial \bar{k}} \left( \frac{3{\Theta}_{V2}}{2\bar{p}} + \frac{\partial{\Theta}_{V2}}{\partial\bar{p}} \right) \right] 
\nonumber \\
&&
+\frac{{\bar{p}}}{6} \left(\frac{3}{2\bar{p}} {\Theta}_{V1} + \frac{\partial {\Theta}_{V1}}{\partial \bar{p}} \right) \frac{\partial {\cal K}^{(1)}_2}{\partial \bar{k}}
- \frac{\bar{p}}{6} \frac{\partial {\Theta}_{V1}}{\partial \bar{k}} \left(\frac{\partial {\cal K}^{(1)}_2}{\partial \bar{p}} - \frac{{\cal K}^{(1)}_2}{2 \bar{p}}\right)
\nonumber \\
&&
+ \frac{\kappa\sqrt{\bar{p}}}{6}\left( \frac{\partial{\Phi}_0}{\partial \bar{k}} \left( \frac{{\cal P}_{V2}}{2\bar{p}} + \frac{\partial{\cal P}_{V2}}{\partial\bar{p}} \right)
+ \left( \frac{\partial{\Phi}_0}{\partial \bar{p}}
- \frac{3 {\Phi}_0}{2\bar{p}}\right) \frac{\partial{\cal P}_{V2}}{\partial\bar{k}} \right)
\nonumber \\
&&
+ \frac{\kappa}{3} \left(
\frac{\partial {\cal P}_{V1}}{\partial \bar{k}} \frac{\partial{\Phi}_2}{\partial \bar{p}}
-\frac{\partial{\Phi}_2}{\partial \bar{k}}
\left(\frac{3}{2\bar{p}} {\cal P}_{V1} 
+ \frac{\partial {\cal P}_{V1}}{\partial \bar{p}} \right)
\right)
\nonumber \\
&&-\frac{{\cal P}_1}{2\bar{p}} \frac{\partial {\cal P}_{V2}}{\partial \bar{\varphi}}
+ \bar{p}^{3/2} 
{\Phi}_2 \frac{\partial {\Theta}_{V1}}{\partial \bar{\varphi}}
- \frac{{\cal P}_2}{2\bar{p}} \frac{\partial {\cal P}_{V1}}{\partial \bar{\varphi}}
+
\bar{p}^2 \frac{\partial {\Theta}_{V1}}{\partial \bar{\varphi}} {\Theta}_{V2}
\nonumber \\
&&+ \frac{\kappa}{2\bar{p}}
\left( \frac{\partial {\Theta}_{V1}}{\partial \bar{k}} \left( \frac{{\cal P}_{V2}}{2\bar{p}} + \frac{\partial{\cal P}_{V2}}{\partial\bar{p}} \right)
- \left(\frac{3 {\Theta}_{V1}}{2\bar{p}} + \frac{\partial {\Theta}_{V1}}{\partial \bar{p}} \right) \frac{\partial{\cal P}_{V2}}{\partial\bar{k}} \right)
\nonumber \\
&&+ \frac{\kappa}{2\sqrt{\bar{p}}} \left[\left({\cal P}_{6} + 2 {\cal P}_1 - 3 {\cal P}_{3}
- 6 \bar{k}^2 \Phi_{10} \right) \frac{{\Phi}_{V1}}{2}
+
\frac{1}{6}\frac{\partial \Phi_0}{\partial \bar{k}}
{\cal P}_{V2}\right]
\nonumber \\
&&- \frac{\kappa}{2} \sqrt{\bar{p}} \Phi_{2} \Phi_{V3}
+ \frac{{\cal P}_3}{2\bar{p}} {\cal P}_{dV2}
+ \frac{{\cal P}_4}{2\bar{p}} {\cal P}_{dV1}
+ \frac{1}{2} {\cal K}^{(1)}_1{\Theta}_{V3}
-\frac{1}{4}{\cal K}^{(1)}_2{\Theta}_{V5}
\nonumber \\
&&
+ \bar{p}^{3/2}
{\Phi}_2 \frac{\partial {\Theta}_{V1}}{\partial \bar{\varphi}}
-\frac{\bar{p}^2}{2}
\frac{\partial {\Theta}_{V2}}{\partial \bar{\varphi}}
{\Theta}_{V1}
\end{eqnarray}
in ascending order of complexity. These equations can be used to impose
further conditions on the potential-independent terms of the scalar
contribution. The first four of them hold automatically if $\Theta_{V1}$ and
$\Theta_{dV1}$ vanish, which may be used as an assumption that ensures an even
dependence of the Hamiltonian constraint on momentum components. (Some of the
previous equations suggested for solutions of these coefficients then impose
consistency conditions on the other functions.)

\section{Discussion}
\label{s:Discussion}

Our effective Hamiltonian constraints, in full generality up to second order
in perturbations and with the classical derivative structure, include a large
number of free functions.  The gravitational part introduced ten
${\cal K}$-functions depending on $\bar{k}$ and $\bar{p}$. The
potential-independent terms for scalar matter introduced thirty functions
$\Phi$, $\Phi_0$, $\Phi_I$, $\Theta_I$, and ${\cal P}_I$ depending on
$\bar{k}$ and $\bar{p}$, and the potential terms introduced another eighteen
functions ${\cal P}_{VI}$, $\Theta_{VI}$, $\Phi_{VI}$ $\Phi_{dVI}$,
$\Theta_{dVI}$, $\Phi_{ddV}$ and $\Theta_{ddV}$ depending on $\bar{p}$,
$\bar{k}$, and $\bar{\varphi}$.

The dependence of these functions on the background phase-space variables is
restricted by the requirement of anomaly freedom as well as compatibility with
the classical limit given by ${\cal K}^{(0)}\to\bar{k}^2 - \Lambda \bar{p}/3$,
${\cal K}^{(1)}_1, {\cal K}^{(2)}_3 \to \bar{k}$,
${\cal K}^{(1)}_2\to\bar{k}^2-\Lambda \bar{p}$,
${\cal K}^{(2)}_4,{\cal K}^{(2)}_5\to\bar{k}^2+\Lambda \bar{p}$, as well as
${\cal K}^{(1)}_3,{\cal K}^{(2)}_1,{\cal K}^{(2)}_2,{\cal K}^{(2)}_6\to1$ for
the gravitational coefficients, $\Phi,\Phi_0,\Phi_I,\Theta_I\to0$ as well as
${\cal P}_I\to1$  for
coefficients in the potential-independent matter terms, and
${\cal P}_{VI}\to V(\bar{\varphi})$,
${\cal P}_{dVI}\to \partial V/ \partial \bar{\varphi}$,
${\cal P}_{ddV}\to \partial^2 V/ \partial \bar{\varphi}^2$, as well as
$\Phi_{VI},\Phi_{dVI},\Theta_{VI}\to0$, for $I=1,2,\dots$.

The vacuum case (\ref{eq:Vacuum constraint ansatz-EMG}) implies only seven
independent equations (\ref{eq:K^2_2=K^2_1 - simp})--(\ref{eq:K12 - simp}) for
ten free ${\cal K}$-functions.  However, the equations for scalar modification
functions are much more restrictive and also include some of the
${\cal K}$-functions which may impose additional restrictions on the
latter. Given the nature of these conditions, some of which are non-linear
partial differential equations, it is not easy to count the number of
independent conditions in order to see whether the complete system is
overdetermined or underdetermined for given background modifications, or
whether the classical dynamics may be unique. Below, we present a non-trivial
example in order to show that non-classical models are possible.

We recall that the background structure function is given by
\begin{equation}\label{eq:Background structure function - summary}
    \bar{\tilde{q}}^{ab} = {\cal K}^{(1)}_3 {\cal K}^{(2)}_1 \frac{\delta^{ab}}{\bar p}
    \,.
\end{equation}
After imposing vacuum conditions, the perturbation to the structure function is of the form
\begin{eqnarray}\label{eq:Metric perturbation - summary}
    \delta \tilde{q}^{ab} &=& \frac{1}{\bar{p}} \left[ {\cal K}^{(1)}_3 {\cal K}^{(2)}_1 \left( \frac{\delta^{aj} \delta E^b_j + \delta^{bj} \delta E^a_j}{\bar{p}}
    - \delta^{a b} \frac{(\delta_d^k\delta E^d_k)}{\bar{p}}\right)\right.
    \nonumber\\
    &&\left.\quad
    + \delta^{a b} Q_1^{(K)} \left(\delta^d_k\delta K_d^k - \frac{{\cal K}^{(1)}_3}{2{\cal K}^{(1)}_1} \frac{\delta^{jk}\partial_j\partial_d\delta E^d_k}{\bar{p}}
    - \bar{k} \frac{(\delta_d^k\delta E^d_k)}{\bar{p}}\right)
    \right]\,,
\end{eqnarray}
with a function $Q_1^{(K)}(\bar{k},\bar{p})$ whose classical limit is
$Q_1^{(K)}\to0$. This function is unrestricted in vacuum, but it turns out to  vanish for
consistent couplings of scalar matter with a potential which requires
\begin{eqnarray}
    Q_1^{(K)} = 0
\end{eqnarray}
(unless $\Theta_6\not=0$; see App.~\ref{eq:Covariance cond - scalar potential}).
The emergent line
element is given by
\begin{eqnarray}\label{eq:Line-element - summary}
    {\rm d} s^2 = - \left(\bar{N}^2+2\bar{N}\delta N\right) {\rm d}t^2
    + 2\bar{\tilde{q}}_{ab} \delta N^b {\rm d}x^a{\rm d}t
    + \left(\bar{\tilde{q}}_{ab}+\delta\tilde{q}_{ab}\right) {\rm d}x^a{\rm d}x^b\,.
\end{eqnarray}

As a non-trivial example, it is possible to reproduce (and generalize) the
ansatz considered in \cite{ScalarHol} by restricting some of our modification
functions by
\begin{eqnarray}
    {\cal K}^{(0)} \!\!&=&\!\! \frac{\sin^2(\lambda \bar{k})}{\lambda^2}\,,\\
    {\cal P}_1 \!\!&=&\!\! {\cal P}_2 ={\cal P}_3 = {\cal P}_4 = {\cal P}_5 = {\cal P}_6 = {\cal P}_7 = 1
    \,,\\
    {\cal P}_{VI}\!\!&=&\!\! V(\bar{\varphi})
    \quad,\quad (I=1,2,3,4)
    \\
    {\cal P}_{VI} \!\!&=&\!\! \partial V/ \partial \bar{\varphi}
    \,,
    \quad,\quad (I=1,2)\\
    {\cal P}_{ddV} \!\!&=&\!\! \partial^2 V / \partial \bar{\varphi}^2\,,
\end{eqnarray}
with a function $\lambda(\bar{p})=\sqrt{\Delta} \bar{p}^\beta$ defined by two
constant parameters $\Delta$ and $\beta$. We set
$\Theta_I=\Phi=\Phi_I=\Theta_{VI}=\Theta_{dVI}=\Theta_{ddV}=\Phi_{VI}=0$ for
all $I$.  The remaining modification functions are written in the form
\begin{eqnarray}
    {\cal K}^{(1)}_1 &=& \frac{\sin(s_1\lambda \bar{k})}{s_1\lambda} + \alpha_1
    \\
    {\cal K}^{(1)}_2 &=& \frac{\sin^2(\lambda \bar{k})}{\lambda^2} + \alpha_2
    \\
    {\cal K}^{(1)}_3 &=& 1 + \alpha_3
    \\
    {\cal K}^{(2)}_1 &=& 1+\alpha_4
    \\
    {\cal K}^{(2)}_2 &=& 1+\alpha_5
    \\
    {\cal K}^{(2)}_3 &=& \frac{\sin(s_2\lambda \bar{k})}{s_2\lambda}+\alpha_6
    \\
    {\cal K}^{(2)}_4 &=& \frac{\sin^2(\lambda \bar{k})}{\lambda^2} + \alpha_7
    \\
    {\cal K}^{(2)}_5 &=& \frac{\sin^2(\lambda \bar{k})}{\lambda^2} + \alpha_8
    \\
    {\cal K}^{(2)}_6 &=& 1 + \alpha_9\,,\\
    {\cal P}_8 &=& 1 + \alpha_{10}\,,
\end{eqnarray}
where $s_1$ and $s_2$ are nonzero integers to be determined, and the
$\alpha_I$ are functions of $\bar{k}$.

  In terms of the $\alpha$-functions, the conditions of anomaly-freedom reduce
  to the fourteen equations,
\begin{eqnarray}
    0 &=& - 2 \frac{\sin(2\lambda \bar{k})}{2\lambda}
    + \bar{k} (1+\alpha_4) + \frac{\sin(s_2\lambda \bar{k})}{s_2\lambda}+\alpha_6
    \,,\\
    0 &=&
    2 \bar{k} \left(\frac{\sin(s_1\lambda \bar{k})}{s_1\lambda} + \alpha_1\right)
    + \alpha_2
    - 2 \frac{\sin^2(\lambda \bar{k})}{\lambda^2}
    \,,\\
    0 &=&\alpha_5-\alpha_4
    \,,\\
    0 &=& 
    - \frac{\sin^2(\lambda \bar{k})}{\lambda^2}
    - \bar{p} \frac{\partial}{\partial \bar{p}} \left(\frac{\sin^2(\lambda \bar{k})}{\lambda^2}\right)
    - \frac{1}{2} \alpha_7
    + \bar{k} \left(\frac{\sin(s_2\lambda \bar{k})}{s_2\lambda}
    + \alpha_6\right)\
    \,,\\
    0&=&\alpha_8-\alpha_7
    \,,\\
    0 &=& \alpha_5-\alpha_4
    \,,\\
    \label{eq:A6}
    0 &=& 
    (1+\alpha_9) \left(\frac{\sin(s_1\lambda \bar{k})}{s_1\lambda} + \alpha_1\right)
    - (1 + \alpha_3) \left(\frac{\sin(s_2\lambda \bar{k})}{s_2\lambda}+\alpha_6\right)
    + \frac{\sin(2\lambda \bar{k})}{2\lambda} \left(1 + \alpha_3\right)
    \\
    &&
    - 2 \frac{\sin(2\lambda \bar{k})}{2\lambda} \bar{p} \frac{\partial \alpha_3}{\partial \bar{p}}
    + \frac{1}{2}\left( \frac{\sin^2(\lambda \bar{k})}{\lambda^2} + 2\bar{p} \frac{\partial}{\partial \bar{p}}\left(\frac{\sin^2(\lambda \bar{k})}{\lambda^2}\right)\right) \frac{\partial \alpha_3}{\partial \bar{k}}
    - \bar{k} (1 + \alpha_3) (1+\alpha_5)
    \,,\nonumber\\
    0 &=&
    4 \frac{\sin(2\lambda \bar{k})}{2\lambda}\bar{p} \frac{\partial}{\partial \bar{p}}\left(\frac{\sin(s_1\lambda \bar{k})}{s_1\lambda} + \alpha_1\right)
    - \left( \frac{\sin^2(\lambda \bar{k})}{\lambda^2} + 2\bar{p} \frac{\partial}{\partial \bar{p}}\left(\frac{\sin^2(\lambda \bar{k})}{\lambda^2}\right) \right) \frac{\partial}{\partial \bar{k}}\left(\frac{\sin(s_1\lambda \bar{k})}{s_1\lambda} + \alpha_1\right)
    \nonumber\\
    &&
    + \left( 1 + \frac{3}{2} \alpha_5 - \frac{1}{2} \alpha_4\right) \left(\frac{\sin^2(\lambda \bar{k})}{\lambda^2} + \alpha_2\right)
    - 2 \left(\frac{\sin(s_2\lambda \bar{k})}{s_2\lambda}+\alpha_6\right) \left(\frac{\sin(s_1\lambda \bar{k})}{s_1\lambda} + \alpha_1\right)
    \nonumber\\
    &&
    + 2 \frac{\sin(2\lambda \bar{k})}{2\lambda}\left(\frac{\sin(s_1\lambda \bar{k})}{s_1\lambda} + \alpha_1\right)
    \,,\\
    0 &=&
    \frac{1}{2} \left(\frac{\sin(s_2\lambda \bar{k})}{s_2\lambda}+\alpha_6\right) \left(\frac{\sin^2(\lambda \bar{k})}{\lambda^2} + \alpha_2\right)
    - \left(\frac{\sin(s_1\lambda \bar{k})}{s_1\lambda} + \alpha_1\right) \left(\frac{\sin^2(\lambda \bar{k})}{\lambda^2} + \alpha_7\right)
    \nonumber\\
    &&
    + \frac{3}{2} \left(\frac{\sin(s_1\lambda \bar{k})}{s_1\lambda} + \alpha_1\right) \left(\frac{\sin^2(\lambda \bar{k})}{\lambda^2} + \alpha_8\right)
    - \frac{1}{2} \frac{\sin(2\lambda \bar{k})}{2\lambda}\left(\frac{\sin^2(\lambda \bar{k})}{\lambda^2} + \alpha_2\right)
    \nonumber\\
    &&
    + \frac{\sin(2\lambda \bar{k})}{2\lambda} \bar{p} \frac{\partial}{\partial \bar{p}}\left(\frac{\sin^2(\lambda \bar{k})}{\lambda^2} + \alpha_2\right)
    \nonumber\\
    &&
    - \frac{1}{4}\left( \frac{\sin^2(\lambda \bar{k})}{\lambda^2} + 2\bar{p} \frac{\partial}{\partial \bar{p}}\left(\frac{\sin^2(\lambda \bar{k})}{\lambda^2}\right)\right) \frac{\partial}{\partial \bar{k}}\left(\frac{\sin^2(\lambda \bar{k})}{\lambda^2} + \alpha_2\right)
    \,,\\
    0 &=& \frac{\partial \alpha_3}{\partial \bar{k}}
    \,,\\
    0 &=& \frac{\sin(2\lambda \bar{k})}{2\lambda} - \frac{\sin(s_1\lambda \bar{k})}{s_1\lambda} - \alpha_1
    \,,\\
    0 &=& - \frac{\partial}{\partial\bar{k}} \left(\frac{\sin(s_1\lambda \bar{k})}{s_1\lambda}+\alpha_1\right)
    + \frac{3}{2} \left(1+\alpha_5\right)
    - \frac{1}{2} \left(1+\alpha_4\right)
    \,,\\
    0 &=& - \frac{1}{2} \frac{\partial}{\partial\bar{k}} \left(\frac{\sin(\lambda \bar{k})}{\lambda}+\alpha_2\right)
    + 5 \left(\frac{\sin(s_1\lambda \bar{k})}{s_1\lambda}+\alpha_1\right)
    - 5 \frac{\sin(2\lambda \bar{k})}{2\lambda}
    + \frac{\sin(s_2\lambda \bar{k})}{s_2\lambda}
    + \alpha_6
    \,,\\
  0 &=& \frac{1}{2} \frac{\partial}{\partial\bar{k}} \left(\frac{\sin(\lambda \bar{k})}{\lambda}+\alpha_2\right)
    + \frac{\sin(s_1\lambda \bar{k})}{s_1\lambda}+\alpha_1
    - \frac{\sin(2\lambda \bar{k})}{2\lambda}
    - \frac{\sin(s_2\lambda \bar{k})}{s_2\lambda}
    - \alpha_6\,.
\end{eqnarray}
The same set had been obtained in \cite{ScalarHol}, with the solutions
\begin{eqnarray}
    \alpha_1 &=& \frac{\sin(2\lambda \bar{k})}{2\lambda}-\frac{\sin(s_1\lambda \bar{k})}{s_1\lambda}
    \\
    \alpha_2 &=& 2 \frac{\sin^2(\lambda \bar{k})}{\lambda^2} - 2 \bar{k} \frac{\sin(2\lambda \bar{k})}{2\lambda}
    \\
    \alpha_3 &=& 0
    \\
    \alpha_4 &=& \cos(2\lambda \bar{k}) - 1
    \\
    \alpha_5 &=& \cos(2\lambda \bar{k}) - 1
    \\
    \alpha_6 &=& 2 \frac{\sin(2\lambda \bar{k})}{2\lambda} - \frac{\sin(s_2\lambda \bar{k})}{s_2\lambda} - \bar{k} \cos(2\lambda\bar{k})
    \\
    \alpha_7 &=& -4\frac{\sin^2(\lambda \bar{k})}{\lambda^2} + 6 \bar{k} \frac{\sin(2\lambda \bar{k})}{2\lambda}-2\bar{k}^2\cos(2\lambda \bar{k})
    \\
    \alpha_8 &=& -4\frac{\sin^2(\lambda \bar{k})}{\lambda^2} + 6 \bar{k} \frac{\sin(2\lambda \bar{k})}{2\lambda}-2\bar{k}^2\cos(2\lambda \bar{k})
    \\
    \alpha_9 &=& 0
    \\
    \alpha_{10} &=& \cos(2\lambda \bar{k})-1
\end{eqnarray}
for arbitrary integers $s_1$ and $s_2$. (Corresponding terms cancel in the
definition of ${\cal K}^{(1)}_1$ and ${\cal K}^{(2)}_3$, resulting in a
constraint that is independent of $s_1$ and $s_2$.) Moreover, the exponent in
$\lambda(\bar{p})=\sqrt{\Delta}\; \bar{p}^\beta$ is restricted by the
anomaly-freedom equations for this choice of modification functions to equal
$\beta=-1/2$, such that
\begin{equation}\label{eq:mubar scheme - LQC}
    \lambda = \sqrt{\frac{\Delta}{\bar{p}}}\,.
  \end{equation}

  Our further analysis differs crucially from \cite{ScalarHol}, which did not
  consider an emergent metric and instead worked with the classical
  expression, $q^{ab}=|\bar{p}|^{-1}\delta^{ab}$. This omission means that the
  underlying theory was not covariant, and it has important cosmological
  implications.  Given the solutions for anomaly-free modifications, the
  background structure function (\ref{eq:Background structure function - EMG})
  can be written as
\begin{equation}\label{eq:Background structure function - LQC - 1}
    \bar{\tilde{q}}^{ab} = (1+\alpha_3)(1+\alpha_4) \frac{\delta^{ab}}{\bar p}
 = \frac{\cos(2\lambda \bar{k})}{\bar p} \delta^{ab}
\end{equation}
and turns out to be different from the classical inverse metric.

These results demonstrate several important properties: First, the large set
of equations found in the main part of this paper has non-classical
solutions. Secondly, we confirm that the modifications constructed in
\cite{ScalarHol} do correspond to a covariant system, even though only
anomaly-freedom but not covariance had been checked in that paper. However,
covariance requires using the emergent metric, which had not been done in the
cosmological analysis of \cite{ScalarHol}. This paper used the structure
function to conclude, following \cite{Action}, that there is signature change
at large curvature where the structure function changes sign. But by working
with the classical rather than emergent metric for solutions, the
corresponding cosmological analysis was not based on a covariant geometry.

Going beyond \cite{ScalarHol}, our effective Hamiltonians are more general in
that they allow an arbitrary function ${\cal K}^{(0)}$ and several other new
terms in the constraint, which are important for a reliable treatment in the
spirit of effective field theory. In particular, singularity resolution in
loop quantum cosmology should be re-analyzed from our new perspective.
Another implication of relevance for loop quantum cosmology is the fact that
the fixed value of the holonomy parameter (\ref{eq:mubar scheme - LQC}) can be
relaxed to arbitrary functions $\lambda(\bar{p})$ in our system due to the
additional terms. The lattice refinement scheme \cite{InhomLattice} is therefore not determined by
anomaly-freedom and covariance.

An analysis of geometrical properties such as singularity resolution must be
based on the emergent line element
\begin{equation} \label{dsiso}
    {\rm d}s^2=-\bar{N}{\rm d} t^2 + \frac{\bar{p}}{\cos(2\lambda\bar{k})}
    \delta_{ab} {\rm d}x^a {\rm d}x^b \,.
\end{equation} 
The cosine introduces new divergences of metric components at
$\lambda\bar{k}=\pi/4$, which defines the moment of signature change where the
structure function changes sign. For generic solutions, $\bar{p}$ is still
finite at this time, implying a significant difference between the
classical-type scale factor $\sqrt{\bar{p}}$ (which never reaches zero in the
modified background dynamics of loop quantum cosmology) and the actual
geometrical scale factor
\begin{equation}
  \tilde{a}=\sqrt{\frac{\bar{p}}{\cos(2\lambda\bar{k})}}
\end{equation}
determined by the compatible line element (\ref{dsiso}). This expression
diverges when $\lambda\bar{k}$ approaches $\pi/4$. The physical nature of this
singularity and implications for loop quantum cosmology, as well as a new
nonsingular alternative, are discussed in \cite{EMGCosmo}.

\section*{Acknowledgements}

This work was supported in part by NSF grant PHY-2206591.

\begin{appendix}

\section{Vacuum}
\label{a:Deriv}

We collect detailed derivations of Poisson brackets of the constraint and
implications for consistent modifications in this and the following appendices.

In the vacuum case, we use the general constraint expressions
(\ref{HamConstH0-EMG})--(\ref{HamConstH2-EMG}) and impose
hypersurface-deformation brackets as well as the covariance condition on the parametrized
perturbed spatial metric (\ref{MetricAnsatz}).

\subsection{$\{\tilde{H},D\}$ bracket}

Since the bracket (\ref{HD - mod}) contains only
$\delta N^b \partial_b \delta N$ terms, the contribution
$\{\tilde{H}_{\rm grav}[\bar{N}] ,\vec{D}[\vec{N}]\} = \{\tilde{H}^{(0)}_{\rm
  grav}[\bar{N}], \vec{D}[\vec{N}] \}+\{\tilde{H}^{(2)}_{\rm grav}[\bar{N}],
\vec{D}[\vec{N}] \}$ must vanish.  The latter two give, using the brackets
computed in Appendix~\ref{a:Brackets-modified},
\begin{eqnarray}
    \{\tilde{H}^{(0)}_{\rm grav}[\bar{N}], \vec{D}[\vec{N}] \}
    &=& - \frac{\sqrt{\bar p} \bar{N}}{16 \pi G} \int {\rm d}^3x\ \delta N^c \Bigg[
    \left( \frac{{\cal K}^{(0)}}{\bar{p}}
    + 2 \frac{\partial {\cal K}^{(0)}}{\partial \bar{p}} \right) \delta_c^k\partial_d \delta E^d_k
    \\
    &&\qquad
    + 2 \frac{\partial {\cal K}^{(0)}}{\partial \bar{k}} \left(\partial_c(\delta^d_k \delta K^k_d)-\partial_k\delta K^k_c\right) \Bigg]\nonumber
    \,,\\
    \{\tilde{H}^{(2)}_{\rm grav}[\bar{N}], \vec{D}[\vec{N}] \}
    &=& - \frac{\sqrt{\bar p} \bar{N}}{16\pi G} \int {\rm d}^3x\ \delta N^c \Bigg[ 2 \bar{k} {\cal K}^{(2)}_1 \partial_k\delta K_c^k
    - 2 \bar{k} {\cal K}^{(2)}_2 \partial_c (\delta^d_k \delta K_d^k)
    \nonumber\\
    && \quad 
    + 2 {\cal K}^{(2)}_3 \left( \partial_k \delta K_c^k
    - \partial_c (\delta_k^d \delta K_d^k)
    - \frac{\bar{k}}{\bar{p}} \delta_c^k \partial_d \delta E^d_k\right)
    \nonumber\\
    && \quad 
    + \frac{{\cal K}^{(2)}_4}{\bar{p}} \left(- \partial_c (\delta_d^k \delta E^d_k) + \delta_c^k \partial_d \delta E^d_k\right)
    + \frac{{\cal K}^{(2)}_5}{\bar{p}} \partial_c (\delta_d^k \delta E^d_k) \Bigg]
    \nonumber
\end{eqnarray}

Since this must hold for arbitrary $\delta N^c$, the vanishing of their sum implies
\begin{eqnarray}
    0 &=&
    \frac{1}{\bar{p}} \left[ {\cal K}^{(2)}_4
    + {\cal K}^{(0)}
    - 2 \bar{k} {\cal K}^{(2)}_3
    + 2 \bar{p} \frac{\partial {\cal K}^{(0)}}{\partial \bar{p}} 
    \right] \delta_c^k \partial_d \delta E^d_k
    + \frac{1}{\bar{p}} \left[ {\cal K}^{(2)}_5 - {\cal K}^{(2)}_4 \right] \partial_c (\delta_d^k \delta E^d_k)
    \nonumber\\
    &&
    - 2 \left[\frac{\partial {\cal K}^{(0)}}{\partial \bar{k}} - \bar{k} {\cal K}^{(2)}_1 - {\cal K}^{(2)}_3 \right] \partial_k\delta K_c^k
    + 2 \left[ \frac{\partial {\cal K}^{(0)}}{\partial \bar{k}} - \bar{k} {\cal K}^{(2)}_2 - {\cal K}^{(2)}_3 \right] \partial_c(\delta^d_k \delta K^k_d)
    \nonumber
    \,.
\end{eqnarray}
Because the ${\cal K}$ functions are independent of the perturbations, the terms in square brackets must vanish independently, yielding the restrictions
\begin{eqnarray}
    \bar{p} \frac{\partial {\cal K}^{(0)}}{\partial \bar{p}} &=& \bar{k} {\cal K}^{(2)}_3 - \frac{{\cal K}^{(0)}+{\cal K}^{(2)}_4}{2}
    \,,\label{eq:dK^0/dp}\\
    \frac{\partial {\cal K}^{(0)}}{\partial \bar{k}} &=& \bar{k} {\cal K}^{(2)}_1 + {\cal K}^{(2)}_3
    \,,\label{eq:dK^0/dk}\\
    {\cal K}^{(2)}_5&=&{\cal K}^{(2)}_4
    \,,\label{eq:K^2_5=K^2_4}\\
    {\cal K}^{(2)}_2&=&{\cal K}^{(2)}_1
    \,.\label{eq:K^2_2=K^2_1}
\end{eqnarray}

On the other hand,
\begin{eqnarray}
    \{\tilde{H}^{(1)}_{\rm grav}[\delta N], D_c[\delta N^c]\} &=&
    \frac{\sqrt{\bar{p}}}{16 \pi G} \left[4 {\cal K}^{(1)}_1 \bar{k}
    + 2 {\cal K}^{(1)}_2 \right] \delta N^b \partial_b \delta N
    \,.
\end{eqnarray}
Using the above, the bracket (\ref{HD - mod}) becomes simply $\{\tilde{H}^{(1)}_{\rm grav}[\delta N], D_c[\delta N^c] \}=- \tilde{H}^{(0)}_{\rm grav} [\delta N^b \partial_b \delta N ]$.
For this to hold for arbitrary $\delta N$ and $\delta N^a$ we obtain the equation
\begin{eqnarray}\label{eq:HD bracket cond}
    {\cal K}^{(0)} = \frac{2 \bar{k} {\cal K}^{(1)}_1+{\cal K}^{(1)}_2}{3}
    \,.
\end{eqnarray}

\subsection{$\{\tilde{H},\tilde{H}\}$ bracket}

We will now divide (\ref{HamConstH2-EMG}) into parts.
For what follows we use the subindices $\bar{A}$ and $\delta$ in the brackets to denote that only the background and perturbed variables, respectively, are used in the bracket's functional derivatives.

First, because of antisymmetry we have $\{\tilde{H}^{(0)}[\bar{N}_1],\tilde{H}^{(0)}[\bar{N}_2]\}=0$, $\{\tilde{H}^{(2)}[\bar{N}_1],\tilde{H}^{(2)}[\bar{N}_2]\}=0$, and $\{\tilde{H}^{(0)}[\bar{N}_1],\tilde{H}^{(2)}[\bar{N}_2]\}+\{\tilde{H}^{(0)}[\bar{N}_2],\tilde{H}^{(2)}[\bar{N}_1]\}=0$.
Also notice that $\{\tilde{H}^{(1)}[\delta N_1],\tilde{H}^{(1)}[\delta N_2]\}_{\bar{A}}$ and $\{\tilde{H}^{(2)}[\bar{N}_1],\tilde{H}^{(1)}[\delta N_2]\}_{\bar{A}}$ are of fourth order in perturbations and hence may be neglected.
Furthermore, $\{\tilde{H}^{(0)}[\bar{N}_1],\tilde{H}^{(1)}[\delta N_2]\}_{\delta}=0$ because $\tilde{\cal H}^{(0)}$ does not depend on perturbations by definition.
On the other hand, a direct computation shows
\begin{eqnarray}
    \{\tilde{H}^{(1)}_{\rm grav}[\delta N_1],\tilde{H}^{(1)}_{\rm grav}[\delta N_2]\}_{\delta} &=& \kappa \int {\rm d}^3x \left[\frac{\delta \tilde{H}^{(1)}_{\rm grav}[\delta N_1]}{\delta (\delta K_a^i)}\frac{\delta \tilde{H}^{(1)}_{\rm grav}[\delta N_2]}{\delta (\delta E^a_i)}
    -\frac{\delta \tilde{H}^{(1)}_{\rm grav}[\delta N_1]}{\delta (\delta E^a_i)}\frac{\delta \tilde{H}^{(1)}_{\rm grav}[\delta N_2]}{\delta (\delta K_a^i)}\right]
    \nonumber\\
    &=&\frac{2 {\cal K}^{(1)}_1 {\cal K}^{(1)}_3}{\kappa} \int {\rm d}^3x \left[
    (\partial_a\delta N_2) (\partial^a \delta N_1) - (\partial_a\delta N_1) (\partial^a \delta N_2)\right]
    \nonumber\\
    &=&0
    \,.
\end{eqnarray}

Using all this, the bracket (\ref{HamConstH2-EMG}) is reduced to
\begin{eqnarray}
    \{\tilde{H}^{(0)}_{\rm grav}[\bar{N}_1],\tilde{H}^{(1)}_{\rm grav}[\delta N_2]\}_{\bar{A}}
    + \{\tilde{H}^{(2)}_{\rm grav}[\bar{N}_1],\tilde{H}^{(1)}_{\rm grav}[\delta N_2]\}_{\delta}
    \quad&&
    \nonumber\\
    + \{\tilde{H}^{(0)}_{\rm grav}[\bar{N}_2],\tilde{H}^{(1)}_{\rm grav}[\delta N_1]\}_{\bar{A}}
    + \{\tilde{H}^{(2)}_{\rm grav}[\bar{N}_2],\tilde{H}^{(1)}_{\rm grav}[\delta N_1]\}_{\delta}
    &&
    \nonumber\\
    = - D_a^{\rm grav} \left[ \tilde{\bar q}^{a b} ( \bar{N}_2 \partial_b \delta N_1 - \bar{N}_1 \partial_b \delta N_2 )]\right]\,,&&
\end{eqnarray}
or, simply,
\begin{eqnarray}\label{eq:Reduced HH bracket}
    \{\tilde{H}^{(0)}_{\rm grav}[\bar{N}_1],\tilde{H}^{(1)}_{\rm grav}[\delta N_2]\}_{\bar{A}}
    + \{\tilde{H}^{(2)}_{\rm grav}[\bar{N}_1],\tilde{H}^{(1)}_{\rm grav}[\delta N_2]\}_{\delta} = D_a^{\rm grav} \left[ \tilde{\bar q}^{a b} \bar{N}_1 \partial_b \delta N_2\right]
    \,.
\end{eqnarray}

Therefore, the relevant brackets are
\begin{eqnarray}
    &&\{\tilde{H}^{(0)}_{\rm grav}[\bar{N}_1],\tilde{H}^{(1)}_{\rm grav}[\delta N_2]\}_{\bar{A}}
    = \frac{\kappa}{3 V_0} \left[\frac{\partial \tilde{H}^{(0)}_{\rm grav}[\bar{N}_1]}{\partial \bar{k}} \frac{\partial \tilde{H}^{(1)}_{\rm grav}[\delta N_2]}{\partial \bar{p}}
    - \frac{\partial \tilde{H}^{(0)}_{\rm grav}[\bar{N}_1]}{\partial \bar{p}} \frac{\partial \tilde{H}^{(1)}_{\rm grav}[\delta N_2]}{\partial \bar{k}} \right]
    \nonumber\\
    &&= \int \frac{{\rm d}^3 x}{2\kappa} \Bigg[ 4 \left( \frac{\partial {\cal K}^{(0)}}{\partial \bar{k}} \left(\frac{{\cal K}^{(1)}_1}{2} + \bar{p} \frac{\partial {\cal K}^{(1)}_1}{\partial \bar{p}}\right)
    - \frac{\partial {\cal K}^{(1)}_1}{\partial \bar{k}} \left( \frac{{\cal K}^{(0)}}{2} + \bar{p} \frac{\partial {\cal K}^{(0)}}{\partial \bar{p}}\right) \right) \delta^d_k\delta K_d^k
    \nonumber\\
    &&\qquad
    - \left( \frac{\partial {\cal K}^{(0)}}{\partial \bar{k}} \left(\frac{{\cal K}^{(1)}_2}{2} - \bar{p} \frac{\partial {\cal K}^{(1)}_2}{\partial \bar{p}}\right)
    + \left( \frac{{\cal K}^{(0)}}{2} + \bar{p} \frac{\partial {\cal K}^{(0)}}{\partial \bar{p}}\right) \frac{\partial {\cal K}^{(1)}_2}{\partial \bar{k}} \right) \frac{\delta_d^k\delta E^d_k}{\bar{p}} \Bigg] \bar{N}_1
    \delta N_2
    \nonumber\\
    &&
    - \int \frac{{\rm d}^3 x}{\kappa} \left(\frac{\partial {\cal K}^{(0)}}{\partial \bar{k}} \left(\frac{{\cal K}^{(1)}_3}{2}-\bar{p} \frac{\partial {\cal K}^{(1)}_3}{\partial \bar{p}}\right)
    + \left( \frac{{\cal K}^{(0)}}{2} + \bar{p} \frac{\partial {\cal K}^{(0)}}{\partial \bar{p}}\right) \frac{\partial {\cal K}^{(1)}_3}{\partial \bar{k}} \right) \delta^k_c(\partial_d\delta E^d_k) \frac{\delta^{ce}}{\bar{p}} \bar{N}_1 \partial_e \delta N_2
    \nonumber
    \nonumber\\
    &&\to \int \frac{{\rm d}^3 x}{2\kappa} \Bigg[ 2\bar{k}^2\delta^d_k\delta K_d^k
    - 2 \bar{k}^3\frac{\delta_d^k\delta E^d_k}{\bar{p}} \Bigg] \bar{N}_1
    \delta N_2
    - \int \frac{{\rm d}^3 x}{\kappa} \bar{k} \delta^k_c(\partial_d\delta E^d_k) \frac{\delta^{ce}}{\bar{p}} \bar{N}_1 \partial_e \delta N_2
\end{eqnarray}
and
\begin{eqnarray}
    &&\!\!\!\!
    \{\tilde{H}^{(2)}_{\rm grav}[\bar{N}_1],\tilde{H}^{(1)}_{\rm grav}[\delta N_2]\}_{\delta} 
    = \kappa \int {\rm d}^3x \left[\frac{\delta \tilde{H}^{(2)}_{\rm grav}[\bar{N}_1]}{\delta (\delta K_a^i)}\frac{\delta \tilde{H}^{(1)}_{\rm grav}[\delta N_2]}{\delta (\delta E^a_i)}
    -\frac{\delta \tilde{H}^{(2)}_{\rm grav}[\bar{N}_1]}{\delta (\delta E^a_i)}\frac{\delta \tilde{H}^{(1)}_{\rm grav}[\delta N_2]}{\delta (\delta K_a^i)}\right]
    \nonumber\\
    &&= \int \frac{{\rm d}^3x}{2\kappa} \Bigg[
    \left({\cal K}^{(1)}_2 \left( 3 {\cal K}^{(2)}_2 - {\cal K}^{(2)}_1\right)
    - 4 {\cal K}^{(1)}_1 {\cal K}^{(2)}_3
    \right) \delta^d_k \delta K_d^k
    \nonumber\\
    &&\qquad
    + \left({\cal K}^{(1)}_2 {\cal K}^{(2)}_3
    + {\cal K}^{(1)}_1 \left(3 {\cal K}^{(2)}_5-2{\cal K}^{(2)}_4\right)\right) \frac{\delta_d^k\delta E^d_k}{\bar{p}}
    \Bigg] \bar{N}_1 \delta N_2
    \nonumber\\
    &&\qquad
    + \int \frac{{\rm d}^3x}{\kappa} \Bigg[
    {\cal K}^{(1)}_3 \left( {\cal K}^{(2)}_2 \bar{p} \partial_c(\delta^d_k\delta K_d^k)
    - {\cal K}^{(2)}_1 \bar{p} (\partial_k \delta K_c^k) \right)
    \nonumber\\
    &&\qquad
    + \left({\cal K}^{(1)}_3 {\cal K}^{(2)}_3
    - {\cal K}^{(1)}_1 {\cal K}^{(2)}_6\right) \delta^{k}_c (\partial_d\delta E^d_k) \Bigg] \frac{\delta^{ce}}{\bar{p}} \bar{N}_1 (\partial_e\delta N_2)
    \,.
\end{eqnarray}

The condition (\ref{eq:Reduced HH bracket}) then implies the following:
The terms multiplying $(\delta^d_k \delta K_d^k)\bar{N}_1 \delta N_2$ and $(\delta_d^k \delta E^d_k)\bar{N}_1 \delta N_2$ must vanish independently because they cannot contribute to the diffeomorphism constraint, respectively implying
\begin{eqnarray}\label{eq:HH cond K}
    0 &=&
    \left(\frac{{\cal K}^{(1)}_1}{2} + \bar{p} \frac{\partial {\cal K}^{(1)}_1}{\partial \bar{p}}\right) \frac{\partial {\cal K}^{(0)}}{\partial \bar{k}}
    - \left( \frac{{\cal K}^{(0)}}{2} + \bar{p} \frac{\partial {\cal K}^{(0)}}{\partial \bar{p}}\right) \frac{\partial {\cal K}^{(1)}_1}{\partial \bar{k}}
    \nonumber\\
    &&
    + \frac{{\cal K}^{(1)}_2 \left( 3 {\cal K}^{(2)}_2 - {\cal K}^{(2)}_1\right) - 4 {\cal K}^{(1)}_1 {\cal K}^{(2)}_3}{4}
    \,,
\end{eqnarray}
and
\begin{eqnarray}\label{eq:HH cond E}
    0 &=& \left(\frac{{\cal K}^{(1)}_2}{2} - \bar{p} \frac{\partial {\cal K}^{(1)}_2}{\partial \bar{p}}\right) \frac{\partial {\cal K}^{(0)}}{\partial \bar{k}}
    + \left( \frac{{\cal K}^{(0)}}{2} + \bar{p} \frac{\partial {\cal K}^{(0)}}{\partial \bar{p}}\right) \frac{\partial {\cal K}^{(1)}_2}{\partial \bar{k}}
    \nonumber\\
    &&
    - {\cal K}^{(1)}_2 {\cal K}^{(2)}_3
    - {\cal K}^{(1)}_1 \left(3 {\cal K}^{(2)}_5-2{\cal K}^{(2)}_4\right)
    \,.
\end{eqnarray}

The rest of the terms must combine into reproducing the diffeomorphsim
constraint, hence the conditions
\begin{eqnarray}
    {\cal K}^{(2)}_2 &=& {\cal K}^{(2)}_1
\end{eqnarray}
(which is equivalent to (\ref{eq:K^2_2=K^2_1})) and
\begin{eqnarray}\label{eq:HH=D}
    0&=&-\left(\frac{{\cal K}^{(1)}_3}{2}-\bar{p} \frac{\partial {\cal K}^{(1)}_3}{\partial \bar{p}}\right) \frac{\partial {\cal K}^{(0)}}{\partial \bar{k}}
    - \left( \frac{{\cal K}^{(0)}}{2} + \bar{p} \frac{\partial {\cal K}^{(0)}}{\partial \bar{p}}\right) \frac{\partial {\cal K}^{(1)}_3}{\partial \bar{k}}
    \nonumber\\
    &&
    + {\cal K}^{(1)}_3 {\cal K}^{(2)}_3
    - {\cal K}^{(1)}_1 {\cal K}^{(2)}_6
    + \bar{k} {\cal K}^{(1)}_3 {\cal K}^{(2)}_1
\end{eqnarray}
while the background structure function is determined by
\begin{equation}\label{eq:Background structure function - EMGApp}
    \bar{\tilde{q}}^{ab} = {\cal K}^{(1)}_3 {\cal K}^{(2)}_1 \frac{\delta^{ab}}{\bar p}
    \,.
\end{equation}

\subsection{Summary}

Imposing the $\{\tilde{H},D\}$ bracket determines the conditions
(\ref{eq:dK^0/dp})-(\ref{eq:K^2_2=K^2_1}) and (\ref{eq:HD bracket cond}).
Imposing the $\{\tilde{H},\tilde{H}\}$ bracket determines the conditions
(\ref{eq:HH cond K})-(\ref{eq:HH=D}) which, using the $\{\tilde{H},D\}$
bracket conditions, further:
Using (\ref{eq:K^2_5=K^2_4}) and (\ref{eq:K^2_2=K^2_1}) to eliminate
${\cal K}^{(2)}_2$ and ${\cal K}^{(2)}_5$, and solving (\ref{eq:HD bracket
  cond}) to substitute ${\cal K}^{(1)}_2$, it can be shown that equations
(\ref{eq:HH cond K}) and (\ref{eq:HH cond E}) are equivalent to each
other. The result is given by equations (\ref{eq:K^2_2=K^2_1 - simp})--(\ref{eq:K12
  - simp}) written in the main text.

\section{Scalar matter}

\subsection{Vanishing potential}
\label{a:scalar}

\subsubsection{$\{\tilde{H}_\varphi,D\}$ bracket}

This bracket requires
\begin{eqnarray}
    \label{eq:H1H2 - HD - Scalar}
    \{\tilde{H}^{(0)}_\varphi[\bar{N}], \vec{D}[\vec{N}] \}+\{\tilde{H}^{(2)}_\varphi[\bar{N}], \vec{D}[\vec{N}] \} &=& 0\,,\\
    \label{eq:H1D - HD - Scalar}
    \{\tilde{H}^{(1)}_\varphi[\delta N], \vec{D}[\vec{N}] \} &=& - \tilde{H}^{(0)}_\varphi \left[\delta N^b \partial_b \delta N\right]
    \,.
\end{eqnarray}

Using the brackets computed in Appendix~\ref{a:Brackets-modified} we obtain
\begin{eqnarray}
    \{\tilde{H}^{(0)}_\varphi[\bar{N}], \vec{D}[\vec{N}] \} &=&
    \int {\rm d}^3 x \bar{N} \delta N^c \left[
    \left(\frac{2\bar{p}}{3}\frac{\partial{\cal P}_1}{\partial \bar{p}}-{\cal P}_1 \right)\frac{\bar{\pi}^2}{2\bar{p}^{3/2}} \frac{\delta_c^k\partial_d \delta E^d_k}{2\bar{p}}
    \right.
    \nonumber\\
    &&\left.
    \qquad
    + \frac{1}{3} \frac{\partial{\cal P}_1}{\partial \bar{k}} \frac{\bar{\pi}^2}{2\bar{p}^{3/2}} \left(\partial_c(\delta^d_k \delta K^k_d)-\partial_k\delta K^k_c\right)
    + \frac{2\bar{p}}{3}\frac{\partial{\Phi}_0}{\partial \bar{p}}\bar{\pi} \frac{\delta_c^k\partial_d \delta E^d_k}{2\bar{p}}
    \right.
    \nonumber\\
    &&\left.
    \qquad
    + \frac{1}{3} \frac{\partial{\Phi}_0}{\partial \bar{k}} \bar{\pi} \left(\partial_c(\delta^d_k \delta K^k_d)-\partial_k\delta K^k_c\right)
    \right]
\end{eqnarray}
and
\begin{eqnarray}
    \{\tilde{H}^{(2)}_\varphi[\bar{N}], \vec{D}[\vec{N}] \} &=&
    \int{\rm d}^3x \bar{N} \delta N^c \left[
    \left({\cal P}_5-{\cal P}_4\right) \frac{\bar{\pi}}{\bar{p}^{3/2}} \partial_c \delta\pi
    \right.
    \\
    &&\left.
    + \left(2{\cal P}_5-{\cal P}_6-{\cal P}_7\right) \frac{\bar{\pi}^2}{2\bar{p}^{3/2}} \frac{\partial_c (\delta^k_d\delta E^d_k)}{2\bar{p}}
    + \left({\cal P}_6-\bar{k} \Phi\right)\frac{\bar{\pi}^2}{2\bar{p}^{3/2}} \frac{\delta_c^k \partial_d \delta E^d_k}{2\bar{p}}
    \right.
    \nonumber\\
;    &&\left.
    -\frac{\Phi}{2} \frac{\bar{\pi}^2}{2\bar{p}^{3/2}} \left(\partial_c(\delta^d_k\delta K_d^k)
    - \partial_k \delta K_c^k\right)
    - ({\Phi}_4 + \bar{k}{\Phi}_5)\partial_c\delta{\pi}
    \right.
    \nonumber\\
    &&\left.
    - ({\Phi}_4 +{\Phi}_6 +{\Phi}_7 + \bar{k}{\Phi}_9)\bar{\pi}\frac{\partial_c (\delta^k_d\delta E^d_k)}{2\bar{p}}
    \right.
    \nonumber\\
    &&\left.
    - ({\Phi}_5 +{\Phi}_9+2\bar{k}{\Phi}_{10})\bar{\pi}\partial_c \left(\delta^d_k\delta K^k_d\right)
    - ({\Phi}_6 + \bar{k}{\Phi}_8)\bar{\pi}\frac{\delta^k_c \partial_d \delta E^d_k}{2\bar{p}}
    \right.
    \nonumber\\
     &&\left.
    -\frac{{\Phi}_8}{2}\bar{\pi}\left(\partial_c (\delta^d_k\delta K^k_d) - \partial_k \delta K^k_c\right)
    +
    \left(\Theta_4 \bar{\pi} + \Theta_7 \frac{\bar{\pi}^2}{2\bar{p}^{3/2}} \right)
    \delta^{ab} \delta_{dc} \delta^{kj} (\partial_j \partial_b\partial_a \delta E^d_k)
    \right.
    \nonumber\\
    &&\left.
    +
    \left(\frac{2}{3}{\Theta}_4 \bar{\pi}
    + \Theta_7 
    \frac{\bar{\pi}^2}{3\bar{p}^{3/2}}
    +
    2{\Theta}_5 \bar{\pi}
    + \Theta_8 
    \frac{\bar{\pi}^2}{\bar{p}^{3/2}}
    \right)
    \delta^{ab}
    \partial_c 
    \partial_a
    \partial_b
    (\delta^k_d\delta E^d_k)
    \right]
    \,.
\end{eqnarray}
Therefore, the condition (\ref{eq:H1H2 - HD - Scalar}) requires, upon some simplification,
\begin{eqnarray} \label{eq:HDconditions}
    {\cal P}_5 &=& {\cal P}_4
    \,,\\
    {\cal P}_7 &=& -{\cal P}_6 + 2{\cal P}_5
    \,,\\
    \label{eq:Phi_3-P1}
    {\cal P}_6 &=& {\cal P}_1 + \frac{2}{3}\left(\bar{k} \frac{\partial{\cal P}_1}{\partial \bar{k}}-\bar{p}\frac{\partial{\cal P}_1}{\partial \bar{p}}\right)
    \,,\\
    \label{eq:Phi-P1}
    \Phi
    &=& 
    \frac{2}{3} \frac{\partial{\cal P}_1}{\partial \bar{k}}
    \,,\\
     \Phi_4
    &=& 
   -{ \bar{k}}\Phi_5
   \,,\\
     \Phi_7
    &=& 
 -\Phi_4 - \Phi_6  - { \bar{k}}\Phi_9
    \,,\\
    \Phi_9
    &=& 
 -\Phi_5  - 2{ \bar{k}}\Phi_{10}
    \,,\\
     \Phi_6
    &=& 
  \frac{2 \bar{p}}{3} \frac{\partial{\Phi}_0}{\partial \bar{p}}- { \bar{k}}\Phi_8
    \,,\\
    \Phi_8
    &=& 
  \frac{2}{3} \frac{\partial{\Phi}_0}{\partial \bar{k}}\,,
\end{eqnarray}
as well as
\begin{equation}
    {\Theta}_4
    = 
    {\Theta}_7
    = 
    {\Theta}_5
    = 
    {\Theta}_8
    =0
    \,.\label{eq:HDconditions-Theta8}
\end{equation}

On the other hand,
\begin{eqnarray}
    \{\tilde{H}^{(1)}_\varphi[\delta N], D_c[\delta N^c]\} &=&
    - \int{\rm d}^3x \delta N^c \partial_c \delta N \Bigg[ \left(2 {\cal P}_3 - {\cal P}_2\right) \frac{\bar{\pi}^2}{2\bar{p}^{3/2}}
     + \left(\Phi_1+2\bar{p}\Phi_2+\bar{k}\Phi_3\right) \bar{\pi}\Bigg]
    \nonumber\\
    && -
    \int{\rm d}^3x 
    (\partial^b \partial_b \delta N^c) (\partial_c \delta N) 2\bar{p}\Theta_1 \bar{\pi}
    \,.
\end{eqnarray}
Inserting this into the condition (\ref{eq:H1D - HD - Scalar}), we obtain
\begin{eqnarray}
    \label{eq:Phi_1-P1}
    {\cal P}_3 &=& \frac{1}{2}({\cal P}_1 + {\cal P}_2) \,,\\
    \Phi_1&=&\Phi_0 -2\bar{p}\Phi_2-\bar{k}\Phi_3 \,,\\
    \Theta_1&=&0
    \,.\label{eq:Theta1=0}
\end{eqnarray}

The constraint contributions therefore simplify into
\begin{eqnarray}
  \tilde{H}_{\varphi}[N] &=&
  \int{\rm d}^3x \bar{N} \left(\tilde{\cal H}^{(0)}_\varphi + \tilde{\cal H}^{(2)}_\varphi\right)
  + \int{\rm d}^3x \delta N \tilde{\cal H}^{(1)}_\varphi
  \\
  \label{HamConstH0-EMG-scalar}
  \tilde{\cal H}^{(0)}_\varphi &=&
  {\cal P}_1 \frac{\bar{\pi}^2}{2\bar{p}^{3/2}}
  + \Phi_0 \bar{\pi}
  \,,
  \\
  \label{HamConstH1-EMG-scalar}
  \tilde{\cal H}^{(1)}_\varphi &=&
  - {\cal P}_2 \frac{\bar{\pi}^2}{2\bar{p}^{3/2}} 
  \frac{\delta^j_c \delta E^c_j}{2\bar{p}}
  + ({\cal P}_1 + {\cal P}_2) \frac{\bar{\pi}\delta\pi}{2\bar{p}^{3/2}}
  + (\Phi_0 -2\bar{p}\Phi_2-\bar{k}\Phi_3) \delta \pi
  \,\nonumber\\ 
  &\,& +\Phi_2 \bar{\pi} \delta^j_c \delta E^c_j
  + \Phi_3 \bar{\pi} \delta_j^c \delta K_c^j
  + \Theta_{2} \bar{\pi} \frac{\partial_c \partial^j\delta E^c_j}{2 \bar{p}}
  \,,
  \\
  \label{HamConstH2-EMG-scalar}
  \tilde{\cal H}^{(2)}_\varphi &=&
  {\cal P}_4 \frac{\delta\pi^2}{2 \bar{p}^{3/2}}
  - {\cal P}_4 \frac{\bar{\pi}\delta\pi}{\bar{p}^{3/2}} \frac{\delta^j_c\delta E^c_j}{2\bar{p}}
  + \left(\frac{2}{3}\left[\bar{k}\frac{\partial  {\cal P}_1}{\partial \bar{k}} - \bar{p}\frac{\partial  {\cal P}_1}{\partial \bar{p}}\right]+{\cal P}_1 \right) \frac{\bar{\pi}^2}{2\bar{p}^{3/2}} \frac{\delta^k_c\delta^j_d\delta E^c_j\delta E^d_k}{4\bar{p}^2}
   \nonumber\\
  &&
  -\left(\frac{2}{3}\left(\bar{k}\frac{\partial  {\cal P}_1}{\partial \bar{k}} - \bar{p}\frac{\partial  {\cal P}_1}{\partial \bar{p}}\right)+{\cal P}_1 -2{\cal P}_4\right)  \frac{\bar{\pi}^2}{2\bar{p}^{3/2}} \frac{(\delta^j_c\delta E^c_j)^2}{8\bar{p}^2} 
  + {\cal P}_8 \frac{\sqrt{\bar{p}}}{2} \delta^{a b} (\partial_a \delta\varphi) (\partial_b \delta\varphi)
  \nonumber\\
  &&
  + \frac{2}{3}\frac{\partial  {\cal P}_1}{\partial \bar{k}} \frac{\bar{\pi}^2}{2\bar{p}^{3/2}} \delta K_c^j \frac{\delta E^c_j}{2 \bar{p}}
  - \bar{k}\Phi_5 \delta\pi \frac{\delta^j_c \delta E^c_j}{2\bar{p}}
  + \Phi_5 \delta\pi\delta_j^c\delta K_c^j
  \nonumber\\
  &&
  - \frac{2}{3}\left(\bar{k}\frac{\partial  {\cal P}_1}{\partial \bar{k}} - \bar{p}\frac{\partial  {\cal P}_1}{\partial \bar{p}}\right) \bar{\pi} \frac{\delta^k_c\delta^j_d\delta E^c_j\delta E^d_k}{4\bar{p}^2}
  +\left(\frac{2}{3}\left(\bar{k}\frac{\partial  {\cal P}_1}{\partial \bar{k}} - \bar{p}\frac{\partial  {\cal P}_1}{\partial \bar{p}}\right)+ 2\bar{k}^2 \Phi_{10}\right) \bar{\pi} \frac{(\delta^j_c\delta E^c_j)^2}{8\bar{p}^2}
  \nonumber\\
  &&
  +\frac{2}{3}\frac{\partial \Phi_0}{\partial \bar{k}} \bar{\pi} \delta K_c^j \frac{\delta E^c_j}{2 \bar{p}}
  - (\Phi_5 + 2 \bar{k}\Phi_{10})\bar{\pi} \delta^d_k \delta K_d^k \frac{\delta_c^j \delta E^c_j}{2 \bar{p}}
  + \Phi_{10} \bar{\pi} (\delta^c_j K_c^j)^2
  \nonumber\\
  &&
  + \left(\Theta_3 \bar{\pi} + \Theta_6 \frac{\bar{\pi}^2}{2\bar{p}^{3/2}} \right) \frac{\delta^{jk} (\partial_c \delta E^c_j) (\partial_d \delta E^d_k)}{2 \bar{p}}
  \nonumber\\
  &&
  + \left(\Phi_{11} \frac{\sqrt{\bar{p}}}{2} + \Theta_9 \bar{\pi} \right) (\partial^j \delta\varphi) (\partial_c \delta E^c_j)
  \,,
\end{eqnarray}
where ${\cal P}_4$ and $\Phi$ are given by (\ref{eq:Phi_3-P1}) and (\ref{eq:Phi-P1}), respectively.

\subsubsection{$\{\tilde{H},\tilde{H}\}$ bracket}
Here we assume that $\tilde{\cal H}_{\rm grav}$ is independent of the scalar matter variables $\varphi$ and $\pi$.
We leave the full treatment for future works.

The full bracket reads
\begin{eqnarray}
    \{\tilde{H}[N_1],\tilde{H}[N_2]\} &=&
    \{\tilde{H}^{(0)}[\bar{N}_1],\tilde{H}^{(1)}[\delta N_2]\}_{\bar{A}}
    + \{\tilde{H}^{(1)}[\delta N_1],\tilde{H}^{(0)}[\bar{N}_2]\}_{\delta}
    \nonumber\\
    &&
    + \{\tilde{H}^{(2)}[\bar{N}_1],\tilde{H}^{(1)}[\delta N_2]\}_{\delta}
    + \{\tilde{H}^{(1)}[\delta N_1],\tilde{H}^{(2)}[\bar{N}_2]\}_{\delta}
    \nonumber\\
    &&
    + \{\tilde{H}^{(1)}[\delta N_1],\tilde{H}^{(1)}[\delta N_2]\}_{\delta}
    \,.
\end{eqnarray}
Using $\{ \tilde{H}_{\rm grav} [ N_1 ] , \tilde{H}_{\rm grav} [ N_2 ] \}=-D^{\rm grav}_a[ \tilde{\bar q}^{a b} ( \bar{N}_2 \partial_b \delta N_1 - \bar{N}_1 \partial_b \delta N_2 )]$, the right-hand-side of (\ref{HH - mod}) implies the conditions
\begin{eqnarray}\label{eq:HH bracket cond - 1}
    &&\{\tilde{H}^{(1)}_{\rm grav}[\delta N_1],\tilde{H}^{(1)}_\varphi[\delta N_2]\}_{\delta}
    + \{\tilde{H}^{(1)}_\varphi[\delta N_1],\tilde{H}^{(1)}_{\rm grav}[\delta N_2]\}_{\delta}
    \nonumber\\
    &&
    + \{\tilde{H}^{(1)}_\varphi[\delta N_1],\tilde{H}^{(1)}_\varphi[\delta N_2]\}_{\delta}
    = 0\,,
\end{eqnarray}
and 
\begin{equation}
    B[\bar{N_1},\delta N_2] - B[\bar{N_2},\delta N_1]
    =- \int {\rm d}^x \bar{\pi} (\partial_a \delta\varphi) \tilde{\bar q}^{a b} ( \bar{N}_2 \partial_b \delta N_1 - \bar{N}_1 \partial_b \delta N_2 )
\end{equation}
where
\begin{eqnarray}
    B[\bar{N_1},\delta N_2] &=& \{\tilde{H}^{(0)}_{\rm grav}[\bar{N}_1],\tilde{H}^{(1)}_\varphi[\delta N_2]\}_{\bar{A}}
    + \{\tilde{H}^{(0)}_\varphi[\bar{N}_1],\tilde{H}^{(1)}_{\rm grav}[\delta N_2]\}_{\bar{A}}
    \\
    &&
    + \{\tilde{H}^{(0)}_\varphi[\bar{N}_1],\tilde{H}^{(1)}_\varphi[\delta N_2]\}_{\bar{A}}
    \nonumber\\
    &&
    + \{\tilde{H}_{\rm grav}^{(2)}[\bar{N}_1],\tilde{H}_\varphi^{(1)}[\delta N_2]\}_{\delta}
    + \{\tilde{H}_\varphi^{(2)}[\bar{N}_1] ,\tilde{H}_{\rm grav}^{(1)}[\delta N_2]\}_{\delta}
    \nonumber\\
    &&
    + \{\tilde{H}_\varphi^{(2)}[\bar{N}_1] ,\tilde{H}_\varphi^{(1)}[\delta N_2]\}_{\delta}
    \,,\nonumber
\end{eqnarray}
or, since the antisymmetrization $(\bar{N}_1, \delta N_2) \leftrightarrow (\bar{N}_2, \delta N_1)$ cannot cancel any term, we simply get the condition
\begin{equation}\label{eq:HH bracket cond - 2}
    B[\bar{N_1},\delta N_2]
    = \int {\rm d}x\; \bar{\pi} (\partial_a \delta\varphi) \tilde{\bar q}^{a b}\bar{N}_1 \partial_b \delta N_2\,.
\end{equation}

A direct computation shows
\begin{eqnarray}
    \{\tilde{H}^{(1)}_{\rm grav}[\delta N_1],\tilde{H}^{(1)}_\varphi[\delta N_2]\}_{\delta} &=&
    \int{\rm d}^3x \kappa \left(\frac{\delta \tilde{H}^{(1)}_{\rm grav}[\delta N_1]}{\delta K_a^i} \frac{\delta \tilde{H}^{(1)}_\varphi[\delta N_2]}{\delta E^a_i} - \frac{\delta \tilde{H}^{(1)}_{\rm grav}[\delta N_1]}{\delta E^a_i} \frac{\delta \tilde{H}^{(1)}_\varphi[\delta N_2]}{\delta K_a^i}\right)
    \nonumber\\
    &=&
    \int{\rm d}^3x
    \delta N_1 \delta N_2\left[
    \frac{3}{2} {\cal P}_2  \frac{\bar{\pi}^2}{\bar{p}^2}
    {\cal K}^{(1)}_1
    + \left(\frac{3{\cal K}^{(1)}_2}{2\sqrt{\bar{p}}}{\Phi}_3
    - 6 {\Phi}_2 \sqrt{\bar{p}}\right)
    \bar{\pi}
    \right]
    \\
    &&+
    \int{\rm d}^3x
    (\partial_a\delta N_1) (\partial^a\delta N_2)\left[
    -\frac{\bar{\pi}}{\sqrt{\bar{p}}}
   \left(
    {\cal K}^{(1)}_3{\Phi}_3
    +{\cal K}^{(1)}_1{\Theta}_3
    \right)
    \right] \nonumber
    \,,
\end{eqnarray}
\begin{eqnarray}
    \{\tilde{H}^{(1)}_\varphi[\delta N_1],\tilde{H}^{(1)}_\varphi[\delta N_2]\}_{\delta} &=& 
    0
    \,,
\end{eqnarray}
and hence (\ref{eq:HH bracket cond - 1}) is automatically satisfied upon antisymmetrization.

To evaluate (\ref{eq:HH bracket cond - 2}) we use the computations from App.~\ref{sec:Scalar matter contribution - app}.
The $\bar{N}_1\delta N_2$ terms cannot contribute to the right-hand-side of (\ref{eq:HH bracket cond - 2}) and must therefore vanish.
Such terms can be separated into powers of $\bar{\pi}$ and into perturbation terms that must vanish independently since the undetermined functions can only depend on $\bar{k}$ and $\bar{p}$.
The term $(\bar{N}_1\delta N_2)\bar{\pi}^0 {\delta \pi}$ implies the equation
\begin{eqnarray}\label{eq:pi0deltapi}
    0&=&
    - \sqrt{\bar{p}} \frac{\partial {\cal K}^{(0)}}{\partial \bar{k}}
    \left(
    \frac{\partial {\Phi}_0}{\partial \bar{p}}
    - 2 {\Phi}_2
    - 2 \bar{p}
    \frac{{\partial \Phi}_2}{\partial \bar{p}}
    -\bar{k}
    \frac{\partial {\Phi}_3}{\partial \bar{p}}
    \right)
    \nonumber \\
    &&+ \sqrt{\bar{p}} \left(
    \frac{{\cal K}^{(0)}}{2\bar{p}}
    + \frac{\partial {\cal K}^{(0)}}{\partial \bar{p}}
    \right)
    \left(\frac{\partial {\Phi}_0}{\partial \bar{k}}
    - 2 \bar{p}
    \frac{{\partial \Phi}_2}{\partial \bar{k}}
    - {\Phi}_3
    -\bar{k}
    \frac{\partial {\Phi}_3}{\partial \bar{k}}
    \right)
    \nonumber\\
    &&- \frac{3}{\sqrt{\bar{p}}} \left( \frac{{\cal K}^{(1)}_2}{2}
    + \bar{k} {\cal K}^{(1)}_1 \right) {\Phi}_5\,.
\end{eqnarray}
The term $(\bar{N}_1\delta N_2)\bar{\pi}^1 {\delta \pi}$ implies the equation
\begin{eqnarray}\label{eq:pi1deltapi}
    0&=&
    -\frac{1}{2\bar{p}} \frac{\partial {\cal K}^{(0)}}{\partial \bar{k}} \left(
    \frac{\partial{\cal P}_1}{\partial \bar{p}}
    +
    \frac{\partial{\cal P}_2}{\partial \bar{p}}
    - \frac{3{\cal P}_1}{2\bar{p}}
    - \frac{3{\cal P}_2}{2\bar{p}} \right)
    + 
    \frac{1}{2\bar{p}}
    \left( \frac{{\cal K}^{(0)}}{2\bar{p}} 
    + 
    \frac{\partial {\cal K}^{(0)}}{\partial \bar{p}}\right)
    \left(
    \frac{\partial{\cal P}_1}{\partial \bar{k}}
    +
    \frac{\partial{\cal P}_2}{\partial \bar{k}}
    \right)    \nonumber\\
    &&
    - \frac{3 {\cal P}_4}{\bar{p}^2} {\cal K}^{(1)}_1
+
    \frac{\kappa}{3}
    \frac{{\partial \Phi}_0}{\partial \bar{k}}
    \left(
    \frac{{\partial \Phi}_0}{\partial \bar{p}}
    -2 {\Phi}_2
    - 2\bar{p}\frac{{\partial \Phi}_2}{\partial \bar{p}}
    - \bar{k}\frac{{\partial \Phi}_3}{\partial \bar{p}}
    \right)
    \nonumber\\
    &&-
    \frac{\kappa}{3}
    \frac{{\partial \Phi}_0}{\partial\bar{p}}
    \left(
    \frac{{\partial \Phi}_0}{\partial \bar{k}}
    - 2\bar{p}\frac{{\partial \Phi}_2}{\partial \bar{k}}
    - {\Phi}_3
    - \bar{k}\frac{{\partial \Phi}_3}{\partial \bar{k}}
    \right)
    \nonumber\\
    &&+ 3 {\kappa} \left({\Phi}_2 {\Phi}_5
    + \frac{\bar{k}}{2\bar{p}} {\Phi}_3^2\right)\,.
\end{eqnarray}
The term $(\bar{N}_1\delta N_2)\bar{\pi}^2 {\delta \pi}$ implies the equation
\begin{eqnarray}\label{eq:pi2deltapi}
    0&=&
    \frac{1}{6} \left[
    \frac{\partial{\Phi}_0}{\partial \bar{k}}
    \left(
    \frac{\partial{\cal P}_1}{\partial \bar{p}}
    +
    \frac{\partial{\cal P}_2}{\partial \bar{p}}
    -
    \frac{3{\cal P}_1}{2\bar{p}}
    -
    \frac{3{\cal P}_2}{2\bar{p}}
    \right)
    -
    \frac{\partial{\Phi}_0}{\partial \bar{p}}
    \left(
    \frac{\partial{\cal P}_1}{\partial \bar{k}}
    +
    \frac{\partial{\cal P}_2}{\partial \bar{k}}
    \right)
    \right]
    \nonumber\\
    &&-
    \frac{3 {\cal P}_2}{4 \bar{p}} {\Phi}_5
    +
    \frac{3{\cal P}_4}{2 \bar{p}} {\Phi}_3\,.
\end{eqnarray}
The term $(\bar{N}_1\delta N_2)\bar{\pi}^3 {\delta \pi}$ implies the equation
\begin{eqnarray}\label{eq:pi3deltapi}
    0&=&
    \frac{1}{2}
    \frac{\partial{\cal P}_1}{\partial \bar{k}}
    \left(
    \frac{\partial{\cal P}_1}{\partial \bar{p}}
    +
    \frac{\partial{\cal P}_2}{\partial \bar{p}}
    -
    \frac{3{\cal P}_1}{2\bar{p}}
    -
    \frac{3{\cal P}_2}{2\bar{p}}
    \right)
    -
    \frac{1}{2}
    \left(
    \frac{\partial{\cal P}_1}{\partial \bar{p}}
    -
    \frac{3{\cal P}_1}{2 \bar{p}}
    \right)
    \left(
    \frac{\partial{\cal P}_1}{\partial \bar{k}}
    +
    \frac{\partial{\cal P}_2}{\partial \bar{k}}
    \right)
    \nonumber\\
    &&+
    \frac{\partial{\cal P}_1}{\partial \bar{k}}
    \left(
    \frac{{\partial \Phi}_0}{\partial \bar{p}}
    -2 {\Phi}_2
    - 2\bar{p}\frac{{\partial \Phi}_2}{\partial \bar{p}}
    - \bar{k}\frac{{\partial \Phi}_3}{\partial \bar{p}}
    \right)
    \nonumber\\
    &&-
    \left(
    \frac{\partial{\cal P}_1}{\partial \bar{p}}
    -
    \frac{3{\cal P}_1}{2 \bar{p}}
    \right)
    \left(
    \frac{{\partial \Phi}_0}{\partial \bar{k}}
    - 2\bar{p}\frac{{\partial \Phi}_2}{\partial \bar{k}}
    -{\Phi}_3
    - \bar{k}\frac{{\partial \Phi}_3}{\partial \bar{k}}
    \right)\,.
\end{eqnarray}.
The term $(\bar{N}_1\delta N_2)\bar{\pi}^1 
{\delta}^k_d
{\delta E^d_k}$ implies the equation
\begin{eqnarray}\label{eq:pi1deltaE}
    0&=&
    - \bar{p} \frac{\partial {\cal K}^{(0)}}{\partial \bar{k}}
    \frac{{\partial \Phi}_2}{\partial \bar{p}}
    + \bar{p} \left(
    \frac{{\cal K}^{(0)}}{2\bar{p}}
    + \frac{\partial {\cal K}^{(0)}}{\partial \bar{p}}\right)
    \frac{{\partial \Phi}_2}{\partial \bar{k}}
    \nonumber \\
    &&
    -\frac{1}{6}
    \frac{\partial {\Phi}_0}{\partial \bar{k}}
    \left(
    \frac{\partial {\cal K}^{(1)}_2}{\partial \bar{p}}
    -
    \frac{{\cal K}^{(1)}_2}{2\bar{p}}
    \right)
    +
    \frac{1}{6}
    \frac{\partial {\Phi}_0}{\partial \bar{p}}
    \frac{\partial {\cal K}^{(1)}_2}{\partial \bar{k}}
    \nonumber\\
    &&
    - {\cal K}^{(2)}_3 {\Phi}_2
    - \frac{{\cal K}^{(2)}_4}{4 \bar{p}} {\Phi}_3
    - \frac{{\cal K}^{(1)}_2}{6\bar{p}}
    \frac{\partial {\Phi}_0}{\partial \bar{k}}
    + \frac{3{\cal K}^{(1)}_2}{4\bar{p}}\left(
    {\Phi}_5
    + 2\bar{k} {\Phi}_{10}\right)
    \nonumber\\
    &&-
    \frac{{\cal K}^{(1)}_1}{3\bar{p}}
    \left(
    \bar{k}\frac{\partial {\cal P}_1}{\partial \bar{k}}
    -
    \bar{p}\frac{\partial {\cal P}_1}{\partial \bar{p}}
    \right)
    - \frac{3}{2\bar{p}} \bar{k}^2 {\cal K}^{(1)}_1 {\Phi}_{10}\,.
\end{eqnarray}

The $(\bar{N}_1\delta N_2)\bar{\pi}^2 \delta_d^k\delta E^d_k$ term implies the equation
\begin{eqnarray}\label{eq:pi2deltaE}
    0&=&
    \frac{\partial {\cal K}^{(0)}}{\partial \bar{k}} \left( \bar{p}\frac{\partial{\cal P}_2}{\partial \bar{p}}
    - \frac{5 {\cal P}_2}{2} \right)
    - \left( \frac{{\cal K}^{(0)}}{2} + \bar{p}\frac{\partial {\cal K}^{(0)}}{\partial \bar{p}}\right) \frac{\partial{\cal P}_2}{\partial \bar{k}}
    + {\cal P}_2 {\cal K}^{(2)}_3
    \nonumber\\
    &&
    - \frac{1}{3} \left(
    \frac{\partial{\cal P}_2}{\partial \bar{k}} 
    \left( \bar{p}\frac{\partial {\cal K}^{(1)}_2}{\partial \bar{p}} - \frac{{\cal K}^{(1)}_2}{2}\right)
    -
     \left( 
     \bar{p}\frac{\partial {\cal P}_1}{\partial \bar{p}} - \frac{3{\cal P}_1}{2}
     \right)
    \frac{\partial {\cal K}^{(1)}_2}{\partial \bar{k}} \right)
    \nonumber\\
    &&
    \frac{4\kappa}{3}\bar{p}^3
    \left(
    \frac{\partial {\Phi}_0}{\partial \bar{k}}
    \frac{\partial {\Phi}_2}{\partial \bar{p}}
    -
    \frac{\partial {\Phi}_0}{\partial \bar{p}}
    \frac{\partial {\Phi}_2}{\partial \bar{k}}
    \right)
    -
    \frac{{\cal K}^{(1)}_2}{3}\frac{\partial {\cal P}_1}{\partial \bar{k}}
    \nonumber\\
    &&
    - \frac{2}{3} {\cal K}^{(1)}_1
    \left(
    \bar{k}\frac{\partial{\cal P}_1}{\partial \bar{k}}
    -
    \bar{p}\frac{\partial{\cal P}_1}{\partial \bar{p}}
    \right)
    -{\cal K}^{(1)}_1
    \left(
    {\cal P}_1
    -
    6{\cal P}_4
    \right)
    \nonumber\\
    &&
    \frac{{4\kappa}}{3}\bar{p}^2
    \frac{\partial {\Phi}_0}{\partial \bar{k}}
    {\Phi}_2 
    +
    6{\kappa} \bar{p}
    \left(
    {\Phi}_5 + 2\bar{k}{\Phi}_{10}
    \right)
    {\Phi}_2
    -
    \frac{2{\kappa}}{3} \bar{p}
    \left(
    \bar{k}\frac{\partial{\cal P}_1}{\partial \bar{k}}
    -
    \bar{p}\frac{\partial{\cal P}_1}{\partial \bar{p}}
    \right) {\Phi}_3
    \nonumber\\
    &&
    -6{\kappa} \bar{p} \bar{k}^2  {\Phi}_{3} {\Phi}_{10}\,.
\end{eqnarray}
The $(\bar{N}_1\delta N_2)\bar{\pi}^3 \delta_d^k\delta E^d_k$ term implies the equation
\begin{eqnarray}\label{eq:pi3deltaE}
    0&=&
    - \frac{1}{12}
    \frac{\partial {\Phi}_0}{\partial \bar{k}} 
    \left(\frac{\partial{\cal P}_2}{\partial \bar{p}}
    - \frac{5{\cal P}_2}{2\bar{p}} \right)
    +\frac{1}{12}
    \frac{\partial {\Phi}_0}{\partial \bar{p}}
    \frac{\partial{\cal P}_2}{\partial \bar{k}}
    +\bar{p}\frac{\partial {\cal P}_1}{\partial \bar{k}}
    \frac{\partial {\Phi}_2}{\partial \bar{p}}
    +
    \frac{1}{6}
    \left(
    \bar{p}\frac{\partial {\cal P}_1}{\partial \bar{p}}
    -
    \frac{3{\cal P}_1}{2}
    \right)
    \frac{\partial {\Phi}_2}{\partial \bar{p}}
    \nonumber\\
    &&+
    \frac{1}{6}\frac{\partial {\cal P}_1}{\partial \bar{p}} {\Phi}_2
    -\frac{{\cal P}_2}{12\bar{p}}
    \frac{\partial {\Phi}_0}{\partial \bar{k}}
    +
    \frac{3{\cal P}_2}{8\bar{p}}
    \left(
    {\Phi}_5 + 2\bar{k}{\Phi}_{10}
    \right)
     \nonumber\\
     &&
     + \frac{1}{\bar{p}} \left[\frac{1}{12}
     \left(
     \bar{k}\frac{\partial {\cal P}_1}{\partial \bar{k}}
     -
     \bar{p}\frac{\partial {\cal P}_1}{\partial \bar{p}}
     \right)
     + \frac{{\cal P}_1}{8}
     - \frac{3 {\cal P}_4}{4}\right] {\Phi}_3\,.
\end{eqnarray}
The $(\bar{N}_1\delta N_2)\bar{\pi}^4 \delta_d^k\delta E^d_k$ term implies the equation
\begin{eqnarray}\label{eq:pi4deltaE}
    0 &=&
    \frac{\partial {\cal P}_1}{\partial \bar{k}} \left( \bar{p}\frac{\partial{\cal P}_2}{\partial \bar{p}}
    - \frac{3{\cal P}_2}{2} \right)
    + \left( \bar{p}\frac{\partial {\cal P}_1}{\partial \bar{p}} - \frac{3{\cal P}_1}{2}\right) \frac{\partial{\cal P}_2}{\partial \bar{k}}\,.
\end{eqnarray}
The $(\bar{N}_1\delta N_2)\bar{\pi}^1 \delta_k^d\delta K^k_d$ term implies the equation
\begin{eqnarray}\label{eq:pi1deltaK}
    0&=&
    -
    \bar{p}
    \frac{\partial {\cal K}^{0}}{\partial \bar{k}}
    \frac{{\partial \Phi}_3}{\partial \bar{p}}
    +
    \bar{p}
    \bigg(
    \frac{{\cal K}^{0}}{2 \bar{p}}
    +
    \frac{\partial {\cal K}^{0}}{\partial \bar{p}}
    \bigg)
    \frac{{\partial \Phi}_3}{\partial \bar{k}}
    -
    \frac{2\bar{p}}{3}
    \frac{\partial {\Phi}_0}{\partial \bar{k}}
    \left(
    \frac{\partial {\cal K}^{(1)}_1}{\partial \bar{p}}
    +
    \frac{{\cal K}^{(1)}_1}{2 \bar{p}}
    \right)
    +
    \frac{2\bar{p}}{3}
    \frac{\partial {\Phi}_0}{\partial \bar{p}}
    \frac{\partial {\cal K}^{(1)}_1}{\partial \bar{k}}
    \nonumber\\
    &&-
    2 \bar{p} {\cal K}^{(2)}_1 {\Phi}_2
    + {\cal K}^{(2)}_3 {\Phi}_3
    + \frac{2}{3} {\cal K}^{(1)}_1 \frac{\partial {\Phi}_0}{\partial \bar{k}}
    - 3 {\cal K}^{(1)}_1 \left(
    {\Phi}_5 + 2\bar{k} {\Phi}_{10}
    \right)\,.
\end{eqnarray}
The $(\bar{N}_1\delta N_2)\bar{\pi}^2 \delta_k^d\delta K^k_d$ term implies the equation
\begin{eqnarray}\label{eq:pi2deltaK}
    0&=&
    -\frac{1}{3 \bar{p}}
    \frac{\partial {\cal P}_1}{\partial \bar{k}}
    \left(
    \frac{\partial {\cal K}^{(1)}_1}{\partial \bar{p}}
    +
    \frac{{\cal K}^{(1)}_1}{2 \bar{p}}
    \right)
    +
    \frac{1}{3 \bar{p}}
    \frac{\partial {\cal K}^{(1)}_1 }{\partial \bar{k}}
    \left(
    \frac{\partial {\cal P}_1}{\partial \bar{p}}
    -
    \frac{3 {\cal P}_1}{2\bar{p}}
    \right)
    +
    \frac{\partial {\cal P}_1}{\partial \bar{k}}
    \frac{{\cal K}^{(1)}_1}{3\bar{p}^2}
    \nonumber\\
    &&
    + \frac{{\cal P}_2}{2\bar{p}^2} {\cal K}^{(2)}_1
    + \frac{\kappa}{3}
    \left( \frac{\partial {\Phi}_0}{\partial \bar{k}} \frac{\partial {\Phi}_3}{\partial \bar{p}}
    - \frac{\partial {\Phi}_0}{\partial \bar{p}}\frac{\partial {\Phi}_3}{\partial \bar{k}}\right)
    + 6 {\kappa} {\Phi}_2 {\Phi}_{10}
    \nonumber
    \\
    &&
    -\frac{{\kappa}}{3\bar{p}}
    {\Phi}_3 \frac{\partial {\Phi}_0}{\partial \bar{k}}
    + \frac{3{\kappa}}{2\bar{p}}
    {\Phi}_3 \left(
    {\Phi}_5 + 2 \bar{k}{\Phi}_{10}
    \right)\,.
\end{eqnarray}
The $(\bar{N}_1\delta N_2)\bar{\pi}^3 \delta_k^d\delta K^k_d$ term implies the equation
\begin{eqnarray}\label{eq:pi3deltaK}
    0&=&
    \bar{p}\frac{\partial {\cal P}_1}{\partial \bar{k}}
    \frac{\partial {\Phi}_3}{\partial \bar{p}}
    - \bar{p}\left(\frac{\partial {\cal P}_1}{\partial \bar{p}}
    - \frac{3 {\cal P}_1}{2 \bar{p}}\right) \frac{\partial {\Phi}_3}{\partial \bar{k}}
    - 9 {\cal P}_2 {\Phi}_{10}
    - {\Phi}_3 \frac{\partial {\cal P}_1}{\partial \bar{k}}\,.
\end{eqnarray}
The $(\bar{N}_1\partial_a\delta N_2)\bar{\pi}^1 {\partial}^a {\delta}{\pi}$ term implies the equation
\begin{equation}\label{eq:pi1Partialdeltapi}
\left(
{\Theta}_1
+
\frac{{\Theta}_2}{2\bar{p}}
\right)
{\Phi}_5
=0\,.
\end{equation}
The $(\bar{N}_1\partial_a\delta N_2)\bar{\pi}^0 {\partial}^a {\delta}{\varphi}$ term implies the equation
\begin{eqnarray}\label{eq:pi0Partialdeltaphi}
\sqrt{p}
{\cal K}^{(1)}_1
\Phi_{11} - {\cal P}_8 \left(
{\Phi}_0 - 2 \bar{p}{\Phi}_2 -\bar{k}{\Phi}_3
\right) = 0\,.
\end{eqnarray}
The $(\bar{N}_1\partial_a\delta N_2)\bar{\pi}^1 {\partial}^a {\delta}{\varphi}$ term implies the equation
\begin{eqnarray}\label{eq:pi1Partialdeltaphi}
\frac{{\cal P}_1 + {\cal P}_2}{2\bar{p}}{\cal P}_8
-
\frac{\sqrt{\bar{p}}}{2}
{\Phi}_3
\Phi_{11}
-
2\sqrt{\bar{p}} {\cal K}^{(1)}_1 {\Theta}_9
= \frac{{\cal K}^{(1)}_3 {\cal K}^{(2)}_1}{\bar{p}}\,.
\end{eqnarray}
The $(\bar{N}_1\partial_a\delta N_2)\bar{\pi}^2 {\partial}^a {\delta}{\varphi}$ term implies the equation
\begin{equation} \label{eq:pi2partialdeltaphi}
    {\Theta}_9 \Phi_3 = 0\,.
\end{equation}
The $(\bar{N}_1\partial_c\delta N_2) \bar{\pi} \delta^{cd} \partial_k\delta K^k_d$ term implies
\begin{eqnarray}\label{eq:pi1PartialdeltaK}
2\sqrt{\bar{p}}{\cal K}^{(2)}_1
\left({\Theta}_1
+ \frac{{\Theta}_2}{2\bar{p}}
\right)
+ \frac{{\cal K}^{(1)}_3}{\sqrt{\bar{p}}} {\Phi}_{10} = 0\,.
\end{eqnarray}
The $(\bar{N}_1\partial_c\delta N_2) \bar{\pi}^0\delta^{ck} \partial_d\delta E^d_k$ term implies
\begin{eqnarray}\label{eq:pi0PartialdeltaE}
\Phi_{11}
\left(
{\Phi}_0 - 2 \bar{p}{\Phi}_2 -\bar{k}{\Phi}_3
\right)=0\,.
\end{eqnarray}
The $(\bar{N}_1\partial_c\delta N_2) \bar{\pi}^1\delta^{ck} \partial_d\delta E^d_k$ term implies
\begin{eqnarray}\label{eq:pi1PartialdeltaE}
    0&=&
    \frac{{\cal K}^{(2)}_6}{2\bar{p}^{3/2}} {\Phi}_3
    -
    \frac{{\cal K}^{(1)}_3}{3 \bar{p}^{3/2}}
    \frac{\partial {\Phi}_0}{\partial \bar{k}}
    + \frac{{\cal K}^{(1)}_3}{2\bar{p}^{3/2}} \left(
    {\Phi}_5 + 2\bar{k} {\Phi}_{10}
    \right)
    \\
    &&
    -
    \frac{1}{3 \sqrt{\bar{p}}}
    \left[
    \frac{\partial{\Phi}_0}{\partial \bar{k}} \left(\frac{\partial {\cal K}^{(1)}_3}{\partial \bar{p}} - \frac{{\cal K}^{(1)}_3}{2\bar{p}}\right) - \frac{\partial{\Phi}_0}{\partial \bar{p}} \frac{\partial {\cal K}^{(1)}_3}{\partial \bar{k}}
    \right]
    \nonumber\\
    \nonumber \\
    &&
    - \frac{2{\cal K}^{(1)}_1}{\sqrt{\bar{p}}}
    {\Theta_3} 
    - \frac{{\cal K}^{(2)}_3}{\sqrt{\bar{p}}}
    \bigg(
    {\Theta}_1
    +
    \frac{{\Theta}_2}{2\bar{p}}
    \bigg)
    \nonumber \\
    &&
    + \frac{\partial {\cal K}^{(0)}}{\partial \bar{k}} \frac{\partial \Theta_1}{\partial \bar{p}}
    - \left( \frac{{\cal K}^{(0)}}{2\bar{p}} + \frac{\partial {\cal K}^{(0)}}{\partial \bar{p}}\right) \frac{\partial \Theta_1}{\partial \bar{k}}
    \nonumber \\
    &&
    +
    \frac{1}{2\bar{p}}
    \Bigg[\frac{\partial {\cal K}^{(0)}}{\partial \bar{k}} 
    \left(\frac{\partial \Theta_2}{\partial \bar{p}}
    - \frac{\Theta_2}{\bar{p}}
    \right)
    - \left( \frac{{\cal K}^{(0)}}{2\bar{p}} + \frac{\partial {\cal K}^{(0)}}{\partial \bar{p}}\right)
    \frac{\partial \Theta_2}{\partial \bar{k}} \Bigg]
    \nonumber \\
    &&
    -\kappa
    \left[
    \frac{{\cal P}_1 + {\cal P}_2}{4 \bar{p}} {\cal P}_9
    +
    {\Theta}_9
    \left(
    {\Phi}_0 
    - 2 \bar{p}{\Phi}_2 
    -\bar{k}{\Phi}_3
    \right)
    \right]\,.
    \nonumber
\end{eqnarray}
The $(\bar{N}_1\partial_c\delta N_2) \bar{\pi}^2 \delta^{ck}\partial_d\delta E^d_k$ term implies
\begin{eqnarray}\label{eq:pi2PartialdeltaE}
    0&=&
    -
    \frac{1}{6\bar{p}^2}
    \left[
    \frac{\partial{\cal P}_1}{\partial \bar{k}} \left(\frac{\partial {\cal K}^{(1)}_3}{\partial \bar{p}} - \frac{{\cal K}^{(1)}_3}{2\bar{p}}\right) - \left( \frac{\partial{\cal P}_1}{\partial \bar{p}}
    - \frac{3 {\cal P}_1}{2\bar{p}}\right) \frac{\partial {\cal K}^{(1)}_3}{\partial \bar{k}}
    \right]
    - \frac{{\cal K}^{(1)}_3}{4\bar{p}^3} \Phi
    \nonumber\\
    &&
    + \left(\frac{1}{3\bar{p}}\frac{\partial {\Phi}_0}{\partial \bar{k}}- \frac{{\Phi}_5}{2 \bar{p}}
    +
    2 \left(1-\frac{\bar{k}}{2\bar{p}}\right){\Phi}_{10} \right)
    \bigg(
    {\Theta}_1
    +
    \frac{{\Theta}_2}{2\bar{p}}
    \bigg)
    \nonumber\\\
    &&-
    \frac{{\cal K}^{(1)}_1}
    {\bar{p}^2}
    {\Theta_6}
    +
    \frac{\Phi_3}
    {\bar{p}}
    {\Theta_3}
    -\frac{{\cal P}_1 + {\cal P}_2}{2\bar{p}^{3/2}}{\Theta}_9\,.
\end{eqnarray}
The $(\bar{N}_1\partial_c\delta N_2) \bar{\pi}^3 \delta^{ck} \partial_d\delta E^d_k$ term implies
\begin{eqnarray}\label{eq:pi3PartialdeltaE}
0 =\frac{{\Phi}}{2} \left( {\Theta}_1
    + \frac{{\Theta}_2}{2\bar{p}} \right)
    - {\Phi}_3 \Theta_6\,.
\end{eqnarray}

\subsubsection{Spacetime covariance}

Consider the modified spacetime metric with background structure function (\ref{eq:Background structure function - EMG}) and its perturbation (\ref{eq:Metric perturbation}).
In the presence of a matter contribution of the constraint given by (\ref{HamConstH0-EMG-scalar})-(\ref{HamConstH2-EMG-scalar}), the covariance condition further requires
\begin{eqnarray}\label{eq:Cov cond - 1 - EMG - scalar matter}
    \frac{\partial \{\delta \tilde{q}^{ab}_{(1)} , \tilde{\cal H}_\varphi^{(1)} [\delta \epsilon^0]\}}{\partial (\partial_{c_1}\delta\epsilon^0)} \bigg|_{\rm O.S.} = \frac{\partial \{\delta \tilde{q}^{ab}_{(1)} , \tilde{\cal H}_\varphi^{(1)} [\delta \epsilon^0]\}}{\partial (\partial_{c_1}\partial_{c_2}\delta\epsilon^0)} \bigg|_{\rm O.S.}
    = \dots = 0\,.
\end{eqnarray}
Using our ansatz we can directly compute
\begin{eqnarray}
    \{\delta \tilde{q}^{ab}_{(1)} , \tilde{\cal H}_\varphi^{(1)} [\delta \epsilon^0]\} &=&
    \kappa \frac{\delta^{ab}}{\bar{p}} \left[ {\cal K}^{(1)}_3 {\cal K}^{(2)}_1 \frac{\bar{\pi}}{\bar{p}}
    + 3 Q_1^{(K)} \left(
    \left( \bar{p}\Phi_2 + \bar{k}\right) \frac{\bar{\pi}}{\bar{p}}
    - \frac{1}{2 \bar{p}}{\cal P}_2 \frac{\bar{\pi}^2}{2\bar{p}^{3/2}}
    \right) \right] \delta \epsilon^0
    \nonumber\\
    &&
    + \kappa \frac{\delta^{a b}}{\bar{p}} Q_1^{(K)} \left(\frac{{\cal K}^{(1)}_3}{2{\cal K}^{(1)}_1} \Phi_3
    + 3 \bar{p} \Theta_1
    + \frac{\Theta_2}{2}
    \right) \frac{\bar{\pi}}{\bar{p}} \partial^d \partial_d \delta \epsilon^0
    \,.
\end{eqnarray}
The spacetime covariance condition, therefore, requires that either
\begin{equation} \label{QK1}
    Q_1^{(K)}=0
\end{equation}
or
\begin{equation}
    \frac{{\cal K}^{(1)}_3}{2{\cal K}^{(1)}_1} \Phi_3
    + 3 \bar{p} \Theta_1
    + \frac{\Theta_2}{2} = 0\,.
  \end{equation}
  Here, $\Theta_1=0$ by virtue of (\ref{eq:Theta1=0}), and $\Phi_3=0$ follows
  from (\ref{eq:pi2partialdeltaphi}) as shown below in Section~\ref{a:Simp2},
  which then implies $\Theta_2=0$ as shown in Section~\ref{a:Simp1}. 
Therefore, $Q_1^{(K)}$ is not restricted at this point.

\subsubsection{Simplification: ${\Phi}_3 = 0$ case}
\label{a:Simp1}

We now show that some of the conditions are redundant, simplifying several of them in the process.

From (\ref{eq:pi0PartialdeltaE}) we have that 
\begin{equation} \label{eq:Phi0}
{\Phi}_0 = 2 \bar{p}{\Phi}_2 +\bar{k}{\Phi}_3\,.
\end{equation}
Using this result to simplify (\ref{eq:pi0Partialdeltaphi}) then gives us that
\begin{equation} \label{eq:P9}
\Phi_{11} = 0\,.
\end{equation}
Using (\ref{eq:Phi0}) to simplify (\ref{eq:pi0deltapi}), then implies that
\begin{equation} \label{eq:Phi5}
{\Phi}_5 = 0\,.
\end{equation}
Continuing to simplify our equations in this way we have from (\ref{eq:pi2partialdeltaphi}) that either $\Phi_3$ or $\Theta_9$ must vanish.
This freedom of choice results in two inequivalent anomaly-free systems.
We choose
\begin{equation} \label{eq:Phi3}
{\Phi}_3 = 0\,,
\end{equation}
while the case $\Theta_9=0$ is discussed in the next Subsection.

Using (\ref{eq:Phi3}) to simplify (\ref{eq:pi3deltaK}) results in
\begin{equation} \label{eq:Phi10}
{\Phi}_{10} = 0\,.
\end{equation}
Using (\ref{eq:Phi10}) together with (\ref{eq:Theta1=0}) to simplify (\ref{eq:pi1PartialdeltaK}) then gives us
\begin{equation} \label{eq:Theta2}
{\Theta}_2 = 0\,.
\end{equation}
Using (\ref{eq:Phi3}) and (\ref{eq:Phi5}) simplifies (\ref{eq:pi2deltapi}) to
\begin{equation} \label{eq:pi2deltapi2}
\frac{\partial{\Phi}_0}{\partial \bar{p}}
    \left(
    \frac{\partial{\cal P}_1}{\partial \bar{k}}
    +
    \frac{\partial{\cal P}_2}{\partial \bar{k}}
    \right) = 
    \frac{\partial{\Phi}_0}{\partial \bar{k}}
    \left(\frac{\partial{\cal P}_1}{\partial \bar{p}}
    - \frac{3{\cal P}_1}{2\bar{p}}
    \frac{\partial{\cal P}_2}{\partial \bar{p}}
    - \frac{3{\cal P}_2}{2\bar{p}}\right)
\end{equation}
Simplifying (\ref{eq:pi4deltaE}) gives us the equation
\begin{equation}\label{eq:pi4deltaE2}
    0 = 
    \frac{\partial{\cal P}_1}{\partial \bar{k}}
    \left(
    \frac{\partial{\cal P}_2}{\partial \bar{p}}
    -
    \frac{3{\cal P}_2}{2\bar{p}}
    \right)
    -
    \frac{\partial{\cal P}_2}{\partial \bar{k}}
    \left(
    \frac{\partial{\cal P}_1}{\partial \bar{p}}
    -
    \frac{3{\cal P}_1}{2\bar{p}}
    \right)
\end{equation}
Using (\ref{eq:Phi0}), (\ref{eq:Phi5}), (\ref{eq:Phi3}) and (\ref{eq:Phi10}) to simplify (\ref{eq:pi3deltaE}) we get the same expression given by (\ref{eq:pi2deltapi2}), so (\ref{eq:pi3deltaE}) is a redundant equation.
Using (\ref{eq:Phi0}) to simplify (\ref{eq:pi3deltapi}) we get the same expression given by (\ref{eq:pi4deltaE2}), so (\ref{eq:pi3deltapi}) is a redundant equation.
Using (\ref{eq:Phi0}), (\ref{eq:Phi5}), (\ref{eq:Phi3}) and (\ref{eq:pi2deltapi2}) we get that (\ref{eq:pi1deltaK}) is a redundant equation. Using (\ref{eq:Phi_1-P1}) we have that ${\Theta}_1 = 0$, which together with (\ref{eq:Theta2}) and (\ref{eq:Phi10}) we get that (\ref{eq:pi1PartialdeltaK}) is a redundant equation.
Using (\ref{eq:Phi_1-P1}), ${\Theta}_1 = 0$, together with (\ref{eq:Theta2}) and (\ref{eq:Phi3}) we get that (\ref{eq:pi3PartialdeltaE}) is a redundant equation.
Using (\ref{eq:Phi5}) we get that (\ref{eq:pi1Partialdeltapi}) is a redundant equation.

Finally, using all the above, we can simplify the rest of the anomaly freedom equations from the $\{\tilde{H},\tilde{H}\}$ bracket, which are reduced to
\begin{eqnarray}\label{eq:pi1deltapi2}
    0&=&
    -\bar{p} 
    \frac{\partial {\cal K}^{(0)}}{\partial \bar{k}} \left(
    \frac{\partial{\cal P}_1}{\partial \bar{p}}
    +
    \frac{\partial{\cal P}_2}{\partial \bar{p}}
    - \frac{3{\cal P}_1}{2\bar{p}}
    - \frac{3{\cal P}_2}{2\bar{p}} \right)
    +\bar{p}
    \left( \frac{{\cal K}^{(0)}}{2\bar{p}} 
    +
    \frac{\partial {\cal K}^{(0)}}{\partial \bar{p}}\right)
    \left(
    \frac{\partial{\cal P}_1}{\partial \bar{k}}
    +
    \frac{\partial{\cal P}_2}{\partial \bar{k}}
    \right)
    \nonumber\\
    &&
    - 6 {\cal P}_4 {\cal K}^{(1)}_1
\end{eqnarray}
\begin{eqnarray}\label{eq:pi1deltaE2}
    0&=&
    -
    \frac{\partial {\cal K}^{(0)}}{\partial \bar{k}}
    \frac{{\partial \Phi}_2}{\partial \bar{p}}\bar{p}
    +
    \left(
    \frac{{\cal K}^{(0)}}{2\bar{p}}
    +
    \frac{\partial {\cal K}^{(0)}}{\partial \bar{p}}
    \right)
    \frac{{\partial \Phi}_2}{\partial \bar{k}}
    \bar{p}
    \nonumber \\
    &&
    -\frac{1}{6}
    \frac{\partial {\Phi}_0}{\partial \bar{k}}
    \left(
    \frac{\partial {\cal K}^{(1)}_2}{\partial \bar{p}}
    -
    \frac{{\cal K}^{(1)}_2}{2\bar{p}}
    \right)
    +
    \frac{1}{6}
    \frac{\partial {\Phi}_0}{\partial \bar{p}}
    \frac{\partial {\cal K}^{(1)}_2}{\partial \bar{k}}
    \nonumber\\
    &&- {\Phi}_2 {\cal K}^{(2)}_3
    -
    \frac{{\cal K}^{(1)}_2}{6\bar{p}}
    \frac{\partial {\Phi}_0}{\partial \bar{k}}
    -
    \frac{{\cal K}^{(1)}_1}{3\bar{p}}
    \left(
    \bar{k}\frac{\partial {\cal P}_1}{\partial \bar{k}}
    -
    \bar{p}\frac{\partial {\cal P}_1}{\partial \bar{p}}
    \right)
\end{eqnarray}

\begin{eqnarray}\label{eq:pi2deltaE2}
    0&=&
    \frac{\partial {\cal K}^{(0)}}{\partial \bar{k}} \left( \bar{p}\frac{\partial{\cal P}_2}{\partial \bar{p}}
    - \frac{5 {\cal P}_2}{2} \right)
    - \left( \frac{{\cal K}^{(0)}}{2} + \bar{p}\frac{\partial {\cal K}^{(0)}}{\partial \bar{p}}\right) \frac{\partial{\cal P}_2}{\partial \bar{k}}
    + {\cal P}_2 {\cal K}^{(2)}_3
    \nonumber\\
    &&
    - \frac{1}{3} \left(
    \frac{\partial{\cal P}_1}{\partial \bar{k}} 
    \left( \bar{p}\frac{\partial {\cal K}^{(1)}_2}{\partial \bar{p}} +\frac{{\cal K}^{(1)}_2}{2}\right)
    -
     \left( 
     \bar{p}\frac{\partial {\cal P}_1}{\partial \bar{p}} - \frac{3{\cal P}_1}{2}
     \right)
    \frac{\partial {\cal K}^{(1)}_2}{\partial \bar{k}} \right)
    \nonumber\\
    &&
    - \frac{2}{3} {\cal K}^{(1)}_1
    \left(
    \bar{k}\frac{\partial{\cal P}_1}{\partial \bar{k}}
    -
    \bar{p}\frac{\partial{\cal P}_1}{\partial \bar{p}}
    \right)
    -{\cal K}^{(1)}_1
    \left(
    {\cal P}_1
    -
    6{\cal P}_4
    \right)
      +
    \frac{{2\kappa}}{3}\bar{p}
    \frac{\partial {\Phi}_0}{\partial \bar{k}}
    {\Phi}_0
\end{eqnarray}

\begin{eqnarray}\label{eq:pi1deltaK2}
    0&=&
    -
    \frac{2\bar{p}}{3}
    \frac{\partial {\Phi}_0}{\partial \bar{k}}
    \frac{\partial {\cal K}^{(1)}_1}{\partial \bar{p}}
    +
    \frac{1}{3}
    \frac{\partial {\Phi}_0}{\partial \bar{k}}
    {\cal K}^{(1)}_1
    +
    \frac{2\bar{p}}{3}
    \frac{\partial {\Phi}_0}{\partial \bar{p}}
    \frac{\partial {\cal K}^{(1)}_1}{\partial \bar{k}}
    -
    {\Phi}_0 {\cal K}^{(2)}_1
\end{eqnarray}

\begin{eqnarray}\label{eq:pi2deltaK2}
    0&=&
    -\frac{1}{3 \bar{p}}
    \frac{\partial {\cal P}_1}{\partial \bar{k}}
    \left(
    \frac{\partial {\cal K}^{(1)}_1}{\partial \bar{p}}
    +
    \frac{{\cal K}^{(1)}_1}{2 \bar{p}}
    \right)
    +
    \frac{1}{3 \bar{p}}
    \frac{\partial {\cal K}^{(1)}_1 }{\partial \bar{k}}
    \left(
    \frac{\partial {\cal P}_1}{\partial \bar{p}}
    -
    \frac{3 {\cal P}_1}{2\bar{p}}
    \right)
    +
    \frac{\partial {\cal P}_1}{\partial \bar{k}}
    \frac{{\cal K}^{(1)}_1}{3\bar{p}^2}
    +
    \frac{{\cal P}_2{\cal K}^{(2)}_1}{2\bar{p}^2}
    \nonumber\\
\end{eqnarray}

\begin{eqnarray}\label{eq:pi1PartialdeltaE2}
    0&=&
    \frac{{\cal K}^{(1)}_3}{\bar{p}}
    \frac{\partial {\Phi}_0}{\partial \bar{k}}
    -
    \frac{\partial{\Phi}_0}{\partial \bar{k}}
    \frac{\partial {\cal K}^{(1)}_3}{\partial \bar{p}}
    +
    \frac{\partial{\Phi}_0}{\partial \bar{p}} \frac{\partial {\cal K}^{(1)}_3}{\partial \bar{k}}
    - 
    6{\cal K}^{(1)}_1{(\Theta_3 + \Theta_4)}
\end{eqnarray}

\begin{eqnarray}\label{eq:pi2PartialdeltaE2}
    0&=&
    \frac{\partial{\cal P}_1}{\partial \bar{k}} \left(\frac{\partial {\cal K}^{(1)}_3}{\partial \bar{p}} - \frac{3{\cal K}^{(1)}_3}{2\bar{p}}\right) - \left( \frac{\partial{\cal P}_1}{\partial \bar{p}}
    - \frac{3 {\cal P}_1}{2\bar{p}}\right) \frac{\partial {\cal K}^{(1)}_3}{\partial \bar{k}}
    \nonumber\\\
    &&
    - 3\sqrt{\bar{p}}
    \left[
    2{\cal K}^{(1)}_1 {\Theta_6 }
    +({\cal P}_1 + {\cal P}_2){\Theta}_9
    \right]
\end{eqnarray}

\begin{eqnarray}\label{eq:pi1Partialdeltaphi2}
\frac{({\cal P}_1 + {\cal P}_2)}{2\bar{p}}{\cal P}_8
-
2\sqrt{\bar{p}}{\Theta}_9 {\cal K}^{(1)}_1
&=& \frac{{\cal K}^{(1)}_3 {\cal K}^{(2)}_1}{\bar{p}}\,.
\end{eqnarray}

\subsubsection{Simplification: $\Theta_9 = 0$ case}
\label{a:Simp2}

We now analyze here the consequences of resolving (\ref{eq:pi2partialdeltaphi}), ${\Theta}_9 {\Phi}_3 = 0$ by choosing instead ${\Theta}_9=0$ while keeping ${\Phi}_3 \neq 0$.

The conditions ${\Phi}_{11} = 0$ (\ref{eq:P9}) and ${\Phi}_5 = 0$ (\ref{eq:Phi5}) still hold exactly as before. Consequently, Eq.~(\ref{eq:pi0PartialdeltaE}), ${\Phi}_0= 2 \bar{p}{\Phi}_2 +\bar{k}{\Phi}_3$, remains unchanged.
Also (\ref{eq:pi3deltaK}) remains unchanged,
\begin{equation} \label{eq:Alt_pi3deltaK_constraint}
 \bar{p}\frac{\partial{\cal P}_{1}}{\partial\bar{k}}\frac{\partial{\Phi}_{3}}{\partial\bar{p}}-\bar{p}\left(\frac{\partial{\cal P}_{1}}{\partial\bar{p}}-\frac{3{\cal P}_{1}}{2\bar{p}}\right)\frac{\partial{\Phi}_{3}}{\partial\bar{k}}-9{\Phi}_{10}{\cal P}_{2}-{\Phi}_{3}\frac{\partial{\cal P}_{1}}{\partial\bar{k}} = 0
 \,.
\end{equation}
The equations (\ref{eq:pi2deltapi2}) and (\ref{eq:pi4deltaE2}) remain unchanged:
\begin{equation} \label{eq:pi2deltapi2-alt}
\frac{\partial{\Phi}_0}{\partial \bar{p}}
    \left(
    \frac{\partial{\cal P}_1}{\partial \bar{k}}
    +
    \frac{\partial{\cal P}_2}{\partial \bar{k}}
    \right) = 
    \frac{\partial{\Phi}_0}{\partial \bar{k}}
    \left(\frac{\partial{\cal P}_1}{\partial \bar{p}}
    - \frac{3{\cal P}_1}{2\bar{p}}
    + \frac{\partial{\cal P}_2}{\partial \bar{p}}
    - \frac{3{\cal P}_2}{2\bar{p}}\right)\,,
\end{equation}
\begin{equation}\label{eq:pi4deltaE2-alt}
    0 = 
    -\frac{\partial{\cal P}_1}{\partial \bar{k}}
    \left(
    \frac{\partial{\cal P}_2}{\partial \bar{p}}
    -
    \frac{3{\cal P}_2}{2\bar{p}}
    \right)
    +
    \frac{\partial{\cal P}_2}{\partial \bar{k}}
    \left(
    \frac{\partial{\cal P}_1}{\partial \bar{p}}
    -
    \frac{3{\cal P}_1}{2\bar{p}}
    \right)
\end{equation}

We note that while Eq.~(\ref{eq:pi4deltaE2-alt}) has a simple solution
${\cal P}_2={\cal P}_1$, the rest of the equations are very complicated to
solve generally if ${\cal P}_1$ depends on $\bar{k}$. Consequently, using
compatibility with the classical limit ${\cal P}_1,{\cal P}_2\to 1$, the cases
of interest will be given by the special solution
\begin{eqnarray}
\label{eq:dPdk_zero}
    \frac{\partial{\cal P}_1}{\partial \bar{k}} = \frac{\partial{\cal P}_2}{\partial \bar{k}} = 0\,.
\end{eqnarray}
Using this, Eq.~(\ref{eq:pi2deltapi2-alt}), and compatibility with the classical limit of ${\cal P}_1$ and ${\cal P}_2$, implies that
\begin{equation}
    \frac{\partial \Phi_0}{\partial \bar{k}} = 0\,.
\end{equation}
It follows that (\ref{eq:pi2deltapi}) implies
\begin{eqnarray}\label{eq:pi2deltapi3}
    {\Phi}_3 = 0
    \,,
\end{eqnarray}
a direct contradiction with the central assumption $\Phi_3 \neq 0$.  If
${\cal P}_1$ and ${\cal P}_2$ are allowed to depend on $\bar{k}$, this
contradiction would potentially be avoided, but the equations remain
challenging to solve.

\subsection{Scalar matter potential}
\label{sec:Scalar matter potential}

\subsubsection{$\{\tilde{H}_V,D\}$ bracket}

This bracket requires
\begin{eqnarray}
    \label{eq:H0H2 - HD - Potential}
    \{\tilde{H}^{(0)}_V[\bar{N}], \vec{D}[\vec{N}] \}+\{\tilde{H}^{(2)}_V[\bar{N}], \vec{D}[\vec{N}] \} &=& 0\,,\\
    \label{eq:H1D - HD - Potential}
    \{\tilde{H}^{(1)}_V[\delta N], \vec{D}[\vec{N}] \} &=& - \tilde{H}^{(0)}_V \left[\delta N^b \partial_b \delta N\right]
    \,.
\end{eqnarray}

Using the brackets computed in the App.~\ref{sec:Scalar potential contributions} we get
\begin{eqnarray}
    \{\tilde{H}^{(0)}_V[\bar{N}], \vec{D}[\vec{N}] \} &=&
    \int {\rm d}^3 x \bar{N} \delta N^c 
    \left[
    \left(
    {\cal P}_{V1} + \frac{2 \bar{p}}{3} \frac{\partial {\cal P}_{V1}}{\partial \bar{p}}\right) 
    \bar{p}^{3/2} \frac{\delta_c^k\partial_d \delta E^d_k}{2\bar{p}}
    \right.
    \\
    &&\left.
    \qquad
    + \bar{p}^{3/2}
    \frac{\partial {\cal P}_{V1}}{\partial \bar{\varphi}} 
    \partial_c \delta \varphi
    + \frac{1}{3} \frac{\partial {\cal P}_{V1}}{\partial \bar{k}} \bar{p}^{3/2}\left(\partial_c(\delta^d_k \delta K^k_d)-\partial_k\delta K^k_c\right)
    \right.
    \nonumber\\
    &&
    \left.\qquad
    + \left(
    {\Theta}_{V1} + \frac{2 \bar{p}}{3} \frac{\partial {\Theta}_{V1}}{\partial \bar{p}}\right) 
    \bar{p}^{3/2}
    \bar{\pi}
    \frac{\delta_c^k\partial_d \delta E^d_k}{2\bar{p}}
    \right.
    \nonumber\\
    &&\left.
    \qquad
    + \bar{p}^{3/2}
    \frac{\partial {\Theta}_{V1}}{\partial \bar{\varphi}} 
    \bar{\pi}
    \partial_c \delta \varphi
    + \frac{1}{3} \frac{\partial {\Theta}_{V1}}{\partial \bar{k}} \bar{p}^{3/2}
    \bar{\pi}
    \left(\partial_c(\delta^d_k \delta K^k_d)-\partial_k\delta K^k_c\right)
    \right]
    \,,\nonumber
    \\
    \{\tilde{H}^{(2)}_V[\bar{N}], \vec{D}[\vec{N}] \} &=&
    \int{\rm d}^3x \bar{N} \delta N^c \Bigg[
    - \left( {\cal P}_{V4} + \bar{k} \Phi_{V3} \right) \bar{p}^{3/2} \frac{\delta_c^k \partial_d \delta E^d_k}{2\bar{p}}
    - {\Theta}_{V4} \bar{\pi}
    \bar{p}^{3/2} \frac{\delta_c^k \partial_d \delta E^d_k}{2\bar{p}}
    \nonumber\\
    &&\qquad
    - \left({\cal P}_{dV2} + \bar{k} \Phi_{V4}\right) \bar{p}^{3/2}
    \partial_c \delta\varphi
    - {\Theta}_{dV2} \bar{\pi} \bar{p}^{3/2}
    \partial_c \delta\varphi
    \nonumber\\
    &&\qquad
    - \left( 
    \Phi_{V2}
    + \frac{1}{2} \Phi_{V3}
    +\frac{2\bar{k}}{\bar{p}^{3/2}} {\Phi}_{V5}
    \right) \bar{p}^{3/2}
    \partial_c (\delta_k^d \delta K_d^k)
    + 
    \left(
    \frac{1}{2} \Phi_{V3}
    -\frac{2\bar{k}}{\bar{p}^{3/2}}{\Phi}_{V6}
    \right)
    \bar{p}^{3/2} \partial_k \delta K_c^k
    \nonumber\\
    &&\qquad
    + \left( {\cal P}_{V4}-{\cal P}_{V3}
    - \bar{k} \Phi_{V2} \right) \bar{p}^{3/2} \frac{\partial_c (\delta^k_d\delta E^d_k)}{2\bar{p}}
    -2\bar{k} {\Phi}_{V7} \delta_{kc}{\partial}^d \delta K_d^k
    \nonumber\\
    &&\qquad
    + \left( {\Theta}_{V4}-{\Theta}_{V3} \right)
    \bar{\pi}
    \bar{p}^{3/2} \frac{\partial_c (\delta^k_d\delta E^d_k)}{2\bar{p}}
    + \partial_d T^d
    \Bigg]
    \,.
\end{eqnarray}
The condition (\ref{eq:H0H2 - HD - Potential}) therefore requires
\begin{eqnarray}
    {\cal P}_{V1} + \frac{2 \bar{p}}{3} \frac{\partial {\cal P}_{V1}}{\partial \bar{p}} - {\cal P}_{V4} - \bar{k} \Phi_{V3} &=& 0
    \,,\\
    {\Theta}_{V1} + \frac{2 \bar{p}}{3} \frac{\partial {\Theta}_{V1}}{\partial \bar{p}} - {\Theta}_{V4} &=& 0
    \,,\\
    \frac{\partial {\cal P}_{V1}}{\partial \bar{\varphi}} 
    - {\cal P}_{dV2} 
    - \bar{k} \Phi_{V4} &=&0
    \,,\\
    \frac{\partial {\Theta}_{V1}}{\partial \bar{\varphi}} 
    - {\Theta}_{dV2} &=&0
    \,,\\
    \frac{1}{3} \frac{\partial {\cal P}_{V1}}{\partial \bar{k}}
    - \Phi_{V2}
    - \frac{1}{2} \Phi_{V3} 
    -\frac{2\bar{k}}{\bar{p}^{3/2}} {\Phi}_{V5}
    &=&0
    \,,\\
    \frac{1}{3} \frac{\partial {\Theta}_{V1}}{\partial \bar{k}}
     &=&0
    \,,\\
    \frac{1}{2} \Phi_{V3}
    -\frac{2\bar{k}}{\bar{p}^{3/2}} {\Phi}_{V6}
    - \frac{1}{3} \frac{\partial {\cal P}_{V1}}{\partial \bar{k}} 
    &=& 0
    \,,\\
    {\cal P}_{V4}-{\cal P}_{V3}
    - \bar{k} \Phi_{V2} &=& 0
    \,,\\
    {\Phi}_{V7} &=& 0
    \,,\\
    {\Theta}_{V4}-{\Theta}_{V3} &=& 0
    \,.
\end{eqnarray}
Combining them, they simplify to
\begin{eqnarray}
    \label{eq:HV,D - 1}
    {\cal P}_{V4} &=& {\cal P}_{V3}
    \,,\\
    \label{eq:HV,D - 2}
    {\cal P}_{V4} &=& {\cal P}_{V1}
    + \frac{2}{3} \left(\bar{p} \frac{\partial {\cal P}_{V1}}{\partial \bar{p}} - \bar{k} \frac{\partial {\cal P}_{V1}}{\partial \bar{k}}\right)
    \,,\\
    \label{eq:HV,D - 3}
    \frac{\partial {\cal P}_{V1}}{\partial \bar{\varphi}} 
    &=& {\cal P}_{dV2} - \bar{k} \Phi_{V4}
    \,,\\
    \label{eq:HV,D - 4}
    {\Phi}_{V2}
     &=& \frac{2\bar{k}}{\bar{p}^{3/2}}
     \left(
     {\Phi}_{V5} - {\Phi}_{V6}
     \right)
    \,,\\
    \label{eq:HV,D - 5}
    \Phi_{V3}
    &=&
    \frac{2}{3} 
    \frac{\partial {\cal P}_{V1}}{\partial \bar{k}}
    +
    \frac{4\bar{k}}{\bar{p}^{3/2}} {\Phi}_{V6}
    \,,\\
    \label{eq:HV,D - 6}
    \frac{\partial {\Theta}_{V1}}{\partial \bar{p}}
    &=&
    \frac{3}{2\bar{p}} 
    \left(
    {\Theta}_{V4}
    -
    {\Theta}_{V1}
    \right)
    \,,\\
    \label{eq:HV,D - 7}
    \frac{\partial {\Theta}_{V1}}{\partial \bar{\varphi}}
    &=&
    {\Theta}_{dV2}
    \,,\\
    \label{eq:HV,D - 8}
    \frac{\partial {\Theta}_{V1}}{\partial \bar{k}}
    &=&
    0
    \,,\\
    \label{eq:HV,D - 9}
    {\Theta}_{V4} &=& {\Theta}_{V3}
    \,.
    \,,\\
    \label{eq:HV,D - 10}
    {\Phi}_{V7} &=& 0
    \,.
\end{eqnarray}

On the other hand,
\begin{eqnarray}
    \{\tilde{H}^{(1)}_V[\delta N], D_c[\delta N^c]\} =
    - \int{\rm d}^3x\; (\delta N^b \partial_b \delta N) \left(
    {\cal P}_{V2} + \bar{k} \Phi_{V1} 
    + {\Theta}_{V2} \bar{\pi}
    \right) \bar{p}^{3/2}
    \,.
\end{eqnarray}
Plugging this into the condition (\ref{eq:H1D - HD - Potential}) we obtain
\begin{eqnarray}
    \label{eq:HV,D - 10}
    {\cal P}_{V2} &=&
    {\cal P}_{V1}
    - \bar{k} \Phi_{V1}
    \,
    \\
    \label{eq:HV,D - 11}
    {\Theta}_{V1} &=& {\Theta}_{V2}
\end{eqnarray}

\subsubsection{$\{\tilde{H},\tilde{H}\}$ bracket}
Here we assume that $\tilde{\cal H}_{\rm grav}$ is independent of the scalar matter variables $\varphi$ and $\pi$.
We leave the full treatment for future works.

The full bracket reads
\begin{eqnarray}
    \{\tilde{H}[N_1],\tilde{H}[N_2]\} &=&
    \{\tilde{H}^{(0)}[\bar{N}_1],\tilde{H}^{(1)}[\delta N_2]\}_{\bar{A}}
    + \{\tilde{H}^{(1)}[\delta N_1],\tilde{H}^{(0)}[\bar{N}_2]\}_{\delta}
    \nonumber\\
    &&
    + \{\tilde{H}^{(2)}[\bar{N}_1],\tilde{H}^{(1)}[\delta N_2]\}_{\delta}
    + \{\tilde{H}^{(1)}[\delta N_1],\tilde{H}^{(2)}[\bar{N}_2]\}_{\delta}
    \nonumber\\
    &&
    + \{\tilde{H}^{(1)}[\delta N_1],\tilde{H}^{(1)}[\delta N_2]\}_{\delta}
    \,.
\end{eqnarray}
Using $\{ \tilde{H}_{\rm grav} [ N_1 ]+\tilde{H}_\varphi [ N_1 ] , \tilde{H}_{\rm grav} [ N_2 ] + \tilde{H}_\varphi [ N_2 ] \}=-D_a[ \tilde{\bar q}^{a b} ( \bar{N}_2 \partial_b \delta N_1 - \bar{N}_1 \partial_b \delta N_2 )]$, the right-hand-side of (\ref{HH - mod}) implies the conditions
\begin{eqnarray}\label{eq:HH bracket cond - 1 - V}
    \{\tilde{H}^{(1)}_{\rm grav}[\delta N_1],\tilde{H}^{(1)}_V[\delta N_2]\}_{\delta}
    + \{\tilde{H}^{(1)}_V[\delta N_1],\tilde{H}^{(1)}_{\rm grav}[\delta N_2]\}_{\delta}&&
    \nonumber\\
    + \{\tilde{H}^{(1)}_\varphi[\delta N_1],\tilde{H}^{(1)}_V[\delta N_2]\}_{\delta}
    + \{\tilde{H}^{(1)}_V[\delta N_1],\tilde{H}^{(1)}_\varphi[\delta N_2]\}_{\delta}&&
    \nonumber\\
    + \{\tilde{H}^{(1)}_V[\delta N_1],\tilde{H}^{(1)}_V[\delta N_2]\}_{\delta}
    &=& 0\,,
\end{eqnarray}
and 
\begin{equation}
    B_V[\bar{N_1},\delta N_2] - B_V[\bar{N_2},\delta N_1]
    = 0\,,
\end{equation}
where
\begin{eqnarray}
    B_V[\bar{N_1},\delta N_2] &=& \{\tilde{H}^{(0)}_{\rm grav}[\bar{N}_1],\tilde{H}^{(1)}_V[\delta N_2]\}_{\bar{A}}
    + \{\tilde{H}^{(0)}_V[\bar{N}_1],\tilde{H}^{(1)}_{\rm grav}[\delta N_2]\}_{\bar{A}}
    \nonumber\\
    &&
    + \{\tilde{H}^{(0)}_\varphi[\bar{N}_1],\tilde{H}^{(1)}_V[\delta N_2]\}_{\bar{A}}
    + \{\tilde{H}^{(0)}_V[\bar{N}_1],\tilde{H}^{(1)}_\varphi[\delta N_2]\}_{\bar{A}}
    \nonumber\\
    &&
    + \{\tilde{H}^{(0)}_V[\bar{N}_1],\tilde{H}^{(1)}_V[\delta N_2]\}_{\bar{A}}
    \nonumber\\
    &&
    + \{\tilde{H}^{(2)}_{\rm grav}[\bar{N}_1],\tilde{H}^{(1)}_V[\delta N_2]\}_\delta
    + \{\tilde{H}^{(2)}_V[\bar{N}_1],\tilde{H}^{(1)}_{\rm grav}[\delta N_2]\}_\delta
    \nonumber\\
    &&
    + \{\tilde{H}^{(2)}_\varphi[\bar{N}_1],\tilde{H}^{(1)}_V[\delta N_2]\}_\delta
    + \{\tilde{H}^{(2)}_V[\bar{N}_1],\tilde{H}^{(1)}_\varphi[\delta N_2]\}_\delta
    \nonumber\\
    &&
    + \{\tilde{H}^{(2)}_V[\bar{N}_1],\tilde{H}^{(1)}_V[\delta N_2]\}_\delta
    \,,
\end{eqnarray}
or, since the antisymmetry $\bar{N}_1, \delta N_2 \leftrightarrow \bar{N}_2, \delta N_1$ cannot cancel any term, we simply get the condition
\begin{equation}\label{eq:HH bracket cond - 2 - V}
    B_V[\bar{N_1},\delta N_2]
    = 0\,.
\end{equation}

From the derivative terms of each constraint contribution it is easy to see that we get
\begin{eqnarray}
    \{\tilde{H}^{(1)}_{\rm grav}[\delta N_1],\tilde{H}^{(1)}_V[\delta N_2]\}_{\delta} &=&
    \int{\rm d}^3x
    \delta N_1 \delta N_2 G_{1}
    \nonumber\\
    &&
    + \int{\rm d}^3x \delta^{ab}(\partial_a\delta N_1)(\partial_b\delta N_2) G_2
    \,,\\
    \{\tilde{H}^{(1)}_\varphi[\delta N_1],\tilde{H}^{(1)}_V[\delta N_2]\}_\delta &=& \delta N_1 \delta N_2 G_3 
    \,,\\
    \{\tilde{H}^{(1)}_V[\delta N_1],\tilde{H}^{(1)}_V[\delta N_2]\}_\delta &=& 0
    \,,
\end{eqnarray}
where $G_1$, $G_2$, and $G_3$ are some functions of the background phase space variables, and hence (\ref{eq:HH bracket cond - 1 - V}) is automatically satisfied.

The equation $B_V[\bar{N}_1,\delta N_2]=0$ can be obtained by summing the brackets computed in the appendix~\ref{sec:Scalar potential contributions} and setting it to zero.
As before, since all the undetermined functions depend only on $\bar{k}$ and $\bar{p}$, $B_V[\bar{N}_1,\delta N_2]$ can be separated in terms multiplying different phase space variables that must then vanish independently.
The vanishing of the $\bar{\pi}^{0}\partial_j\partial^j\delta\varphi$ term implies
\begin{equation}
    {\kappa}\frac{\bar{p}^2}{2}\Phi_{V1}{\Phi}_{11}
    + \bar{p}\Phi_{V4}{\cal K}^{(1)}_3=0\,.
\end{equation}
The vanishing of the $\bar{\pi}\delta_j^c\partial_c\partial^j\delta\varphi$ term implies
\begin{equation}
    \Phi_{V1}{\Theta}_{9} =0\,.
  \end{equation}
  However, $\Phi_{11}$ is already zero because of (\ref{eq:P9}) and ${\Theta}_{9}$ is nonzero.  Therefore, the two terms reduce to the implications that
\begin{eqnarray}\label{eq:HV,HV - 1}
    \Phi_{V4} = 0\,.
\end{eqnarray}
\begin{eqnarray}\label{eq:HV,HV - 2}
    \Phi_{V1} = 0\,.
\end{eqnarray}
The $(\bar{N}_1\delta N_2)\bar{\pi}^0 {\delta \varphi}$ term implies the equation
\begin{eqnarray}
0 &=&\bar{p}^2\left( \left( \frac{{\cal K}^{(0)}}{2\bar{p}} + \frac{\partial {\cal K}^{(0)}}{\partial \bar{p}}\right) \frac{\partial{\cal P}_{dV1}}{\partial\bar{k}}
    - \frac{\partial {\cal K}^{(0)}}{\partial \bar{k}} \left( \frac{3 {\cal P}_{dV1}}{2\bar{p}} + \frac{\partial{\cal P}_{dV1}}{\partial\bar{p}}\right) \right)  \nonumber\\
    &&+ \frac{\kappa \bar{p}^3}{3}
 \left(\frac{\partial {\cal P}_{V1}}{\partial \bar{k}} \left( \frac{3 {\cal P}_{dV1}}{2\bar{p}} + \frac{\partial{\cal P}_{dV1}}{\partial\bar{p}}\right)
    - \left(\frac{3 {\cal P}_{V1}}{2\bar{p}} + \frac{\partial {\cal P}_{V1}}{\partial \bar{p}} \right) \frac{\partial{\cal P}_{dV1}}{\partial\bar{k}} \right) \nonumber \\
    &&-
    \bar{p}^{3/2}
    {\Phi}_0 \frac{\partial{\cal P}_{dV1}}{\partial \bar{\varphi}}
    +\bar{p}^3\biggl(
    \frac{\partial {\cal P}_{V1}}{\partial \bar{\varphi}}
    {\Theta}_{dV1}
    -
    \frac{\partial {\cal P}_{dV1}}{\partial \bar{\varphi}}  {\Theta}_{V1}\biggl)
    \nonumber \\
    &&+\bar{p}^2 
    \left(
    \frac{3}{\bar{p}} {\cal K}^{(1)}_1 {\cal P}_{dV2}
    - 
    \frac{3}{2 \bar{p}} {\cal K}^{(1)}_2  \Phi_{V4}
    \right)
    -
    \frac{3\kappa\bar{p}^2}{2}
    {\cal P}_{dV2} \Phi_{V1}
    \nonumber \\
    &&+\kappa \bar{p}^2 
    \left( 
    \frac{3{\cal P}_{V2}}{2} \Phi_{V4}
    - \frac{3 {\cal P}_{dV2}}{2} \Phi_{V1} \right)
    \nonumber \\
    &&
    +
    \bar{p}^3 
    \left(
    \frac{\partial {\cal P}_{V1}}{\partial \bar{\varphi}}
    {\Theta}_{dV1}
    -
    \frac{\partial {\cal P}_{dV1}}{\partial \bar{\varphi}}
    {\Theta}_{V1}
    \right)
\end{eqnarray}
The term $(\bar{N}_1\delta N_2)\bar{\pi}^1 {\delta \varphi}$ implies the equation
\begin{eqnarray}
0 &=&
% Bracket 1: 1 contribution
\bar{p}^2 \biggl( \left( \frac{{\cal K}^{(0)}}{2\bar{p}} + \frac{\partial {\cal K}^{(0)}}{\partial \bar{p}} \right) \frac{\partial{\Theta}_{dV1}}{\partial\bar{k}} 
- \frac{\partial {\cal K}^{(0)}}{\partial \bar{k}} \left( \frac{3{\Theta}_{dV1}}{2\bar{p}} + \frac{\partial{\Theta}_{dV1}}{\partial\bar{p}} \right) \biggr) \nonumber\\
% Bracket 3: 3 contributions
&&+ \frac{\kappa \bar{p}^{3/2}}{3} \biggl( \frac{\partial {\Phi}_{0}}{\partial \bar{k}} \left( \frac{3 {\cal P}_{dV1}}{2\bar{p}} + \frac{\partial{\cal P}_{dV1}}{\partial\bar{p}} \right) 
+ \left( \frac{3 {\Phi}_{0}}{2\bar{p}} - \frac{\partial {\Phi}_{0}}{\partial \bar{p}} \right) \frac{\partial{\cal P}_{dV1}}{\partial\bar{k}} \biggr) \nonumber\\
&&- {\cal P}_{1} \frac{\partial {\cal P}_{dV1}}{\partial \bar{\varphi}}
- \bar{p}^{3/2} {\Phi}_{0} \frac{\partial {\Theta}_{dV1}}{\partial \bar{\varphi}}
+ \bar{p}^3 \frac{\partial {\Theta}_{V1}}{\partial \bar{\varphi}}
{\Theta}_{dV1}
\delta \varphi
\nonumber\\
% Bracket 5: 4 contributions
&&+ \frac{\kappa \bar{p}^3}{3} \Biggl( \frac{\partial {\Theta}_{V1}}{\partial \bar{k}} \left( \frac{3 {\cal P}_{dV1}}{2\bar{p}} + \frac{\partial{\cal P}_{dV1}}{\partial\bar{p}} \right) 
- \left( \frac{3 {\Theta}_{V1}}{2\bar{p}} + \frac{\partial {\Theta}_{V1}}{\partial \bar{p}} \right) \frac{\partial{\cal P}_{dV1}}{\partial\bar{k}} \Biggr)  \nonumber \\
&&+ \frac{\kappa \bar{p}^3}{3} \Biggl( \frac{\partial {\cal P}_{V1}}{\partial \bar{k}} \left( \frac{3 {\Theta}_{dV1}}{2\bar{p}} + \frac{\partial{\Theta}_{dV1}}{\partial\bar{p}} \right) 
- \left( \frac{3 {\cal P}_{V1}}{2\bar{p}} + \frac{\partial {\cal P}_{V1}}{\partial \bar{p}} \right) \frac{\partial{\Theta}_{dV1}}{\partial\bar{k}} \Biggr)
\nonumber \\
&&
% Bracket 7: 1 contribution
+ 3 \bar{p} {\cal K}^{(1)}_1 {\Theta}_{dV2}
% Bracket 9: 2 contribution
- 3 \kappa \bar{p}^{3/2} {\Phi}_{2} \Phi_{V4}
- \bar{p}^3 \frac{\partial \Theta_{dV1}}{\partial \bar{\varphi}} {\Theta}_{V1}
+ \bar{p}^3 \frac{\partial \Theta_{V1}}{\partial \bar{\varphi}} {\Theta}_{dV1}
\nonumber\\
&&+ 
{\cal P}_{ddV} {\cal P}_{3}
% Bracket 10: 1 contribution
+ \frac{3\kappa \bar{p}^2}{2} \left( 
{\Theta}_{V2} \Phi_{V4} 
- {\Theta}_{dV2} \Phi_{V1} 
\right)
\nonumber \\
&&-
\frac{3\kappa\bar{p}^2}{2}
{\Theta}_{dV2} \Phi_{V1}
\,.
\end{eqnarray}
The term $(\bar{N}_1\delta N_2)\bar{\pi}^2 {\delta \varphi}$ implies the equation
\begin{eqnarray}
0 &=&
% Bracket 3: 3 contributions
\frac{\kappa}{6} \biggl( \frac{\partial {\cal P}_{1}}{\partial \bar{k}} \left( \frac{3 {\cal P}_{dV1}}{2\bar{p}} + \frac{\partial{\cal P}_{dV1}}{\partial\bar{p}} \right) 
+ \left( \frac{3 {\cal P}_{1}}{2\bar{p}} - \frac{\partial {\cal P}_{1}}{\partial \bar{p}} \right) \frac{\partial{\cal P}_{dV1}}{\partial\bar{k}} \biggr) \nonumber\\'
&&+ \frac{\kappa \bar{p}^{3/2}}{3} \biggl( \frac{\partial {\Phi}_{0}}{\partial \bar{k}} \left( \frac{3 {\Theta}_{dV1}}{2\bar{p}} + \frac{\partial{\Theta}_{dV1}}{\partial\bar{p}} \right) 
+ \left( \frac{3 {\Phi}_{0}}{2\bar{p}} - \frac{\partial {\Phi}_{0}}{\partial \bar{p}} \right) \frac{\partial{\Theta}_{dV1}}{\partial\bar{k}} \biggr) \nonumber\\
&& 
- {\cal P}_{1} \frac{\partial {\Theta}_{dV1}}{\partial \bar{\varphi}}
% Bracket 9: 2 contribution
-\frac{3 {\kappa}}{4 \bar{p}} {\cal P}_2 \Phi_{V4}
+ {\Theta}_{ddV} {\cal P}_3 \nonumber \\
% Bracket 5: 1 contribution
&&+\frac{\kappa\bar{p}^3}{3}
\left(\frac{\partial {\Theta}_{V1}}{\partial \bar{k}} \left( \frac{3 {\Theta}_{dV1}}{2\bar{p}} + \frac{\partial{\Theta}_{dV1}}{\partial\bar{p}}\right)
    - \left(\frac{3 {\Theta}_{V1}}{2\bar{p}} + \frac{\partial {\Theta}_{V1}}{\partial \bar{p}} \right) \frac{\partial{\Theta}_{dV1}}{\partial\bar{k}} \right)\,. \nonumber
\end{eqnarray}
The term $(\bar{N}_1\delta N_2)\bar{\pi}^3 {\delta \varphi}$ implies the equation
\begin{eqnarray}
0 =
\frac{\partial{\cal P}_1}{\partial \bar{k}} \left( \frac{3 {\Theta}_{dV1}}{2\bar{p}} + \frac{\partial{\Theta}_{dV1}}{\partial\bar{p}}\right)
    + \left(\frac{3 {\cal P}_1}{2\bar{p}}-\frac{\partial{\cal P}_1}{\partial \bar{p}}\right) \frac{\partial{\Theta}_{dV1}}{\partial\bar{k}} \,.
\end{eqnarray}
The term $(\bar{N}_1\delta N_2)\bar{\pi}^0 {\delta \pi}$ implies the equation
\begin{eqnarray}
0 &=&
\frac{\kappa \bar{p}^{3/2}}{3} \left( 
    \frac{\partial {\cal P}_{V1}}{\partial \bar{k}}
    \frac{\partial{\Phi}_1}{\partial \bar{p}}
    -\frac{\partial{\Phi}_1}{\partial \bar{k}}
    \left(\frac{3 {\cal P}_{V1}}{2\bar{p}} + \frac{\partial {\cal P}_{V1}}{\partial \bar{p}} \right)
    \right)
\nonumber \\
&&+\frac{\mathcal{P}_1 + \mathcal{P}_2}{2}
\frac{\partial \mathcal{P}_{V1}}{\partial \bar{\varphi}}
+\frac{3 \kappa \sqrt{\bar{p}}}{2} \bar{k} \Phi_{V1}\Phi_5
-{\cal P}_4\,{\cal P}_{dV1}.
\end{eqnarray}
The term $(\bar{N}_1\delta N_2)\bar{\pi}^1 {\delta \pi}$ implies the equation
\begin{eqnarray}
0 &=&
% Bracket 4: Three contributions
\frac{\kappa}{3} \Biggl(
    \frac{\partial {\cal P}_{V1}}{\partial \bar{k}}
    \Biggl(
    \frac{\partial {\cal P}_3}{\partial \bar{p}}
    -
    \frac{3{\cal P}_{3}}{2\bar{p}}
\Biggr)
-\frac{\partial{\cal P}_3}{\partial \bar{k}}
\Biggl(
       \frac{3{\cal P}_{V1}}{2\bar{p}}
       +\frac{\partial {\cal P}_{V1}}{\partial \bar{p}}
     \Biggr)
\Biggr)
\nonumber\\
&&+\frac{\kappa \bar{p}^{3/2}}{3} \left( 
    -\frac{\partial{\Phi}_1}{\partial \bar{k}}
    \left(\frac{3}{2\bar{p}} {\Theta}_{V1} + \frac{\partial {\Theta}_{V1}}{\partial \bar{p}} \right)
    + \frac{\partial {\Theta}_{V1}}{\partial \bar{k}}
    \frac{\partial{\Phi}_1}{\partial \bar{p}}
    \right)
\nonumber\\
&&+ \frac{\mathcal{P}_1 + \mathcal{P}_2}{2}
   \frac{\partial \Theta_{V1}}{\partial \bar{\varphi}}
% Bracket 8: One contribution
-
{\cal P}_{4}{\Theta}_{dV1}
+\frac{3{\kappa}}{2\bar{p}} {\cal P}_{4}{\Phi}_{V1}
\end{eqnarray}

The term $(\bar{N}_1\delta N_2)\bar{\pi}^2 {\delta \pi}$ implies the equation
\begin{eqnarray}
0 =
% Bracket 3: 3 contributions
    -\frac{\partial{\cal P}_3}{\partial \bar{k}}
    \left(\frac{3 {\Theta}_{V1}}{2\bar{p}} + \frac{\partial {\Theta}_{V1}}{\partial \bar{p}} \right)
    + \frac{\partial {\Theta}_{V1}}{\partial \bar{k}} \left(\frac{\partial{\cal P}_3}{\partial \bar{p}}
    - \frac{3{\cal P}_3}{2\bar{p}} \right)
\end{eqnarray}

The term $(\bar{N}_1\delta N_2) \bar{\pi}^0 \delta^k_d\delta E^d_k$ implies the equation
\begin{eqnarray}
0 &=&
 \frac{\bar{p}}{2}  \left[ \left( \frac{{\cal K}^{(0)}}{2\bar{p}} + \frac{\partial {\cal K}^{(0)}}{\partial \bar{p}}\right) \frac{\partial{\cal P}_{V2}}{\partial\bar{k}}
- \frac{\partial {\cal K}^{(0)}}{\partial \bar{k}} \left( \frac{3{\cal P}_{V2}}{2\bar{p}} + \frac{\partial{\cal P}_{V2}}{\partial\bar{p}} \right) \right]
\nonumber \\
&&+\frac{\bar{p}}{2} \left( \frac{1}{3} \left(\frac{3}{2\bar{p}} {\cal P}_{V1} + \frac{\partial {\cal P}_{V1}}{\partial \bar{p}} \right) \frac{\partial {\cal K}^{(1)}_2}{\partial \bar{k}}
- \frac{1}{3} \frac{\partial {\cal P}_{V1}}{\partial \bar{k}} \left(\frac{\partial {\cal K}^{(1)}_2}{\partial \bar{p}} - \frac{{\cal K}^{(1)}_2}{2 \bar{p}}\right) \right)
\nonumber \\
&&-\frac{\sqrt{\bar{p}}}{2} {\Phi}_0 \frac{\partial {\cal P}_{V2}}{\partial \bar{\varphi}}
-
\frac{\kappa \bar{p}}{4}{\cal P}_{V3} \Phi_{V1}
\nonumber \\
&&+ \frac{\kappa\bar{p}}{6}
\left( \frac{\partial {\cal P}_{V1}}{\partial \bar{k}} \left( \frac{{\cal P}_{V2}}{2\bar{p}} + \frac{\partial{\cal P}_{V2}}{\partial\bar{p}} \right)
- \left(\frac{3 {\cal P}_{V1}}{2\bar{p}} + \frac{\partial {\cal P}_{V1}}{\partial \bar{p}} \right) \frac{\partial{\cal P}_{V2}}{\partial\bar{k}} \right)
\nonumber \\
&&+ \frac{1}{2}
\left(
- \frac{\Phi_{V1}}{2} {\cal K}^{(2)}_4
- {\cal P}_{V2} {\cal K}^{(2)}_3
+{\cal P}_{V3} {\cal K}^{(1)}_1
-\frac{\Phi_{V3}}{2} {\cal K}^{(1)}_2 \right)
\nonumber \\
&&
+ \frac{\sqrt{\bar{p}}}{2}\bar{k} {\Phi}_5 {\cal P}_{dV1}
+
\frac{{\kappa}\bar{p}}{2}\left(\Phi_{V3} \frac{{\cal P}_{V2}}{2}
- \Phi_{V1} \frac{{\cal P}_{V3}}{2}
\right)
\nonumber \\
&&+ \bar{p}^{3/2}
\left({\Phi}_2 \frac{\partial {\cal P}_{V1}}{\partial \bar{\varphi}} \right)
+
\bar{p}^2 \frac{\partial {\cal P}_{V1}}{\partial \bar{\varphi}} {\Theta}_{V2}
-\frac{\bar{p}^2}{2}
\frac{\partial {\cal P}_{V2}}{\partial \bar{\varphi}}
{\Theta}_{V1}
\end{eqnarray}
The term $(\bar{N}_1\delta N_2)\bar{\pi}^1 \delta^k_d\delta E^d_k$ implies the equation
\begin{eqnarray}
0 &=&
\left[ \left( \frac{{\cal K}^{(0)}}{2\bar{p}} + \frac{\partial {\cal K}^{(0)}}{\partial \bar{p}}\right) \frac{\partial{\Theta}_{V2}}{\partial\bar{k}}
- \frac{\partial {\cal K}^{(0)}}{\partial \bar{k}} \left( \frac{3{\Theta}_{V2}}{2\bar{p}} + \frac{\partial{\Theta}_{V2}}{\partial\bar{p}} \right) \right] \frac{\bar{p}}{2} 
\nonumber \\
&&
+\frac{{\bar{p}}}{6} \left(\frac{3 {\Theta}_{V1}}{2\bar{p}} + \frac{\partial {\Theta}_{V1}}{\partial \bar{p}} \right) \frac{\partial {\cal K}^{(1)}_2}{\partial \bar{k}}
- \frac{\bar{p}^2}{6\bar{p}} \frac{\partial {\Theta}_{V1}}{\partial \bar{k}} \left(\frac{\partial {\cal K}^{(1)}_2}{\partial \bar{p}} - \frac{{\cal K}^{(1)}_2}{2 \bar{p}}\right)
\nonumber \\
&&
+ \frac{\kappa \sqrt{\bar{p}}}{6}\left( \frac{\partial{\Phi}_0}{\partial \bar{k}} \left( \frac{{\cal P}_{V2}}{2\bar{p}} + \frac{\partial{\cal P}_{V2}}{\partial\bar{p}} \right)
+ \left( \frac{\partial{\Phi}_0}{\partial \bar{p}}
- \frac{3 {\Phi}_0}{2\bar{p}}\right) \frac{\partial{\cal P}_{V2}}{\partial\bar{k}} \right) 
\nonumber \\
&&-\frac{{\cal P}_1}{2\bar{p}} \frac{\partial {\cal P}_{V2}}{\partial \bar{\varphi}}
+ \frac{\kappa}{3} \left( 
-\frac{\partial{\Phi}_2}{\partial \bar{k}}
\left(\frac{3 {\cal P}_{V1}}{2\bar{p}} 
+ \frac{\partial {\cal P}_{V1}}{\partial \bar{p}} \right)
+ \frac{\partial {\cal P}_{V1}}{\partial \bar{k}} \frac{\partial{\Phi}_2}{\partial \bar{p}}
\right)
\nonumber \\
&&+ \bar{p}^{3/2} 
{\Phi}_2 \frac{\partial {\Theta}_{V1}}{\partial \bar{\varphi}}
- \frac{{\cal P}_2}{2\bar{p}} \frac{\partial {\cal P}_{V1}}{\partial \bar{\varphi}}
+
\bar{p}^2 \frac{\partial {\Theta}_{V1}}{\partial \bar{\varphi}} {\Theta}_{V2}
\nonumber \\
&&+ \frac{\kappa}{2\bar{p}}
\left( \frac{\partial {\Theta}_{V1}}{\partial \bar{k}} \left( \frac{{\cal P}_{V2}}{2\bar{p}} + \frac{\partial{\cal P}_{V2}}{\partial\bar{p}} \right)
- \left(\frac{3 {\Theta}_{V1}}{2\bar{p}} + \frac{\partial {\Theta}_{V1}}{\partial \bar{p}} \right) \frac{\partial{\cal P}_{V2}}{\partial\bar{k}} \right)
\nonumber \\
&&+\frac{\kappa}{4 \sqrt{\bar{p}}} \left({\cal P}_{6} - {\cal P}_1\right) \Phi_{V1}
- \frac{3 \kappa}{4\bar{p}^{1/2}}
\left(
\left({\cal P}_{6} - {\cal P}_1\right)
+ 2 \bar{k}^2 \Phi_{10}
\right) {\Phi}_{V1}
+
\frac{\kappa}{6\sqrt{\bar{p}}}\frac{\partial \Phi_0}{\partial \bar{k}}
{\cal P}_{V2}
\nonumber \\
&&-\kappa \bar{p}^{3/2}\Phi_{2} \frac{\Phi_{V3}}{2\bar{p}}
+ \frac{{\cal P}_3}{2\bar{p}} {\cal P}_{dV2}
+\frac{\sqrt{\bar{p}}}{2}
\bar{k}{\Phi}_{5}{\Theta}_{dV1}
-\frac{1}{4}
\Phi_{V1} {\cal K}^{(2)}_4
\nonumber \\
&&- \frac{\kappa \bar{p}}{2} 
\Phi_{V1} \frac{{\Theta}_{V3}}{2}
+ \frac{{\cal P}_4}{2\bar{p}} {\cal P}_{dV1}
+
\frac{1}{2} {\cal K}^{(1)}_1 {\Theta}_{V3}
-
\frac{\kappa \bar{p}}{4}{\Theta}_{V3} \Phi_{V1}
\nonumber \\
&&
+ \bar{p}^{3/2}
\left({\Phi}_2 \frac{\partial {\Theta}_{V1}}{\partial \bar{\varphi}} \right)
-\bar{p}^2
\frac{\partial {\cal P}_{V2}}{\partial \bar{\varphi}}
{\Phi}_{V0}
-\frac{\bar{p}^2}{2}
\frac{\partial {\Theta}_{V2}}{\partial \bar{\varphi}}
{\Theta}_{V1}
\end{eqnarray}
The term $(\bar{N}_1\delta N_2)\bar{\pi}^2 \delta^k_d\delta E^d_k$ implies the equation
\begin{eqnarray}
0 &=&
\frac{\kappa}{3} \left( 
-\frac{\partial{\Phi}_2}{\partial \bar{k}}
\left(\frac{3 {\Theta}_{V1}}{2\bar{p}} 
+ \frac{\partial {\Theta}_{V1}}{\partial \bar{p}} \right)
+ \frac{\partial {\Theta}_{V1}}{\partial \bar{k}}
\frac{\partial{\Phi}_2}{\partial \bar{p}}\right)
\nonumber\\
%Bracket 3 contribution
&&+ \frac{\kappa}{12\bar{p}}\,\biggl[
\frac{\partial{\cal P}_1}{\partial \bar{k}}
\Bigl(\frac{{\cal P}_{V2}}{2\bar{p}}
+ \frac{\partial {\cal P}_{V2}}{\partial \bar{p}}\Bigr)
+\Bigl(\frac{\partial{\cal P}_1}{\partial \bar{p}}
 - \frac{3\,{\cal P}_1}{2\bar{p}}\Bigr)\,
\frac{\partial {\cal P}_{V2}}{\partial\bar{k}}
\biggr]
\nonumber\\
%Bracket 4 contribution
&&+ \frac{\kappa}{12\bar{p}}\,\biggl[
-\frac{\partial{\cal P}_2}{\partial \bar{k}}
\Bigl(\frac{3{\cal P}_{V1}}{2\bar{p}}
+ \frac{\partial {\cal P}_{V1}}{\partial \bar{p}}\Bigr)
-\frac{\partial {\cal P}_{V1}}{\partial \bar{k}}
\Bigl(\frac{\partial {\cal P}_2}{\partial \bar{p}}
- \frac{5\,{\cal P}_2}{2\bar{p}}\Bigr)
\biggr]
\nonumber\\
&&- \frac{{\cal P}_2}{2\bar{p}} \frac{\partial {\Theta}_{V1}}{\partial \bar{\varphi}}
%Bracket 8 and 9 contributions
- \kappa \frac{{\cal P}_2}{8\bar{p}^2} \Phi_{V3}
+ \frac{{\cal P}_3}{2\bar{p}}{\Theta}_{dV2}
+ \frac{{\cal P}_4}{2\bar{p}} {\Theta}_{dV1}
\nonumber \\
&&
+{3\kappa}\left({\cal P}_{6} -2{\cal P}_4\right) \frac{{\Phi}_{V1}}{8\bar{p}^2}
-\frac{\kappa {\cal P}_6}{4\bar{p}^2}\Phi_{V1}
-\frac{{\cal P}_1}{2\bar{p}} \frac{\partial {\Theta}_{V2}}{\partial \bar{\varphi}}
\nonumber\\
&&
+ \frac{\kappa\sqrt{\bar{p}}}{6}\left( \frac{\partial{\Phi}_0}{\partial \bar{k}} \left( \frac{{\Theta}_{V2}}{2\bar{p}} + \frac{\partial{\Theta}_{V2}}{\partial\bar{p}} \right)
+ \left( \frac{\partial{\Phi}_0}{\partial \bar{p}}
- \frac{3 {\Phi}_0}{2\bar{p}}\right) \frac{\partial{\Theta}_{V2}}{\partial\bar{k}} \right)
\nonumber \\
&&+
\frac{\kappa}{6}\bar{p}^2
\biggl[
\frac{\partial {\Theta}_{V1}}{\partial \bar{k}} \left( \frac{{\Theta}_{V2}}{2\bar{p}} + \frac{\partial{\Theta}_{V2}}{\partial\bar{p}} \right)
- \left(\frac{3 {\Theta}_{V1}}{2\bar{p}} + \frac{\partial {\Theta}_{V1}}{\partial \bar{p}} \right) \frac{\partial{\Theta}_{V2}}{\partial\bar{k}} 
\biggr]
\nonumber \\
&&
+
\frac{\kappa{\Phi}}{8\bar{p}^2} {\cal P}_{V2}
+
\frac{\kappa}{6\bar{p}^{1/2}}\frac{\partial \Phi_0}{\partial \bar{k}}
{\Theta}_{V2}
\,.
\end{eqnarray}
The term $(\bar{N}_1\delta N_2)\bar{\pi}^3 \delta^k_d\delta E^d_k$ implies the equation
\begin{eqnarray}
0
&=&
-\frac{\partial{\cal P}_2}{\partial \bar{k}} 
    \left(\frac{3 {\Theta}_{V1}}{2\bar{p}}
    + \frac{\partial {\Theta}_{V1}}{\partial \bar{p}}\right)
    - \frac{\partial {\Theta}_{V1}}{\partial \bar{k}}
    \left(\frac{\partial{\cal P}_2}{\partial \bar{p}}
    - \frac{5{\cal P}_2}{2\bar{p}} \right)
\nonumber \\
&&
+\frac{\partial{\cal P}_1}{\partial \bar{k}}
\Bigl(\frac{{\Theta}_{V2}}{2\bar{p}}
+ \frac{\partial {\Theta}_{V2}}{\partial \bar{p}}\Bigr)
+\Bigl(\frac{\partial{\cal P}_1}{\partial \bar{p}}
 - \frac{3\,{\cal P}_1}{2\bar{p}}\Bigr)\,
\frac{\partial {\Theta}_{V2}}{\partial\bar{k}}
+
\frac{3 {\Theta}_{V2}}{8\bar{p}} {\Phi}
\end{eqnarray}
The term $(\bar{N}_1\delta N_2)\bar{\pi}^0 \delta^d_k \delta K^k_d$ implies the equation
\begin{eqnarray}
0 &=&
% 1) from \{\tilde{H}^{(0)}_{\rm grav},\tilde{H}^{(1)}_V\}
\bar{p}^2\biggl[
  \biggl(\frac{{\cal K}^{(0)}}{2\bar{p}}
        + \frac{\partial {\cal K}^{(0)}}{\partial \bar{p}}\biggr)\,
  \frac{\partial\Phi_{V1}}{\partial\bar{k}}
  -
  \frac{\partial {\cal K}^{(0)}}{\partial \bar{k}}\,
  \biggl(\frac{3}{2\bar{p}}\,\Phi_{V1}
        + \frac{\partial\Phi_{V1}}{\partial\bar{p}}\biggr)
\biggr]
\nonumber\\
% 2) from \{\tilde{H}^{(0)}_{V},\tilde{H}^{(1)}_{\rm grav}\}, no pi
&&\quad
+ \frac{2\bar{p}^2}{3}\biggl[
  \biggl(\frac{3\,{\cal P}_{V1}}{2\bar{p}}
        + \frac{\partial {\cal P}_{V1}}{\partial \bar{p}}\biggr)
  \,\frac{\partial {\cal K}^{(1)}_1}{\partial \bar{k}}
  -
  \frac{\partial {\cal P}_{V1}}{\partial \bar{k}}
  \biggl(\frac{\partial {\cal K}^{(1)}_1}{\partial \bar{p}}
        + \frac{{\cal K}^{(1)}_1}{2\,\bar{p}}\biggr)
\biggr]
\nonumber\\
% 3) from \{\tilde{H}^{(0)}_\varphi,\tilde{H}^{(1)}_V\}, no pi
&&\quad
- \bar{p}^{3/2} \Phi_{0}\,\frac{\partial \Phi_{V1}}{\partial \bar{\varphi}}
% 4) from \{\tilde{H}^{(0)}_V,\tilde{H}^{(1)}_\varphi\}, no pi
+ \bar{p}
  \Phi_{V3}\,{\cal K}^{(1)}_1
% 9) from \{\tilde{H}^{(2)}_{V},\tilde{H}^{(1)}_{V}\}_\delta
- \bar{p}^2 \kappa\,\frac{\Phi_{V1}\,\Phi_{V3}}{2}
\nonumber\\
% 5) from \{\tilde{H}^{(0)}_V,\tilde{H}^{(1)}_V\}, line w/out pi
&&\quad
+ \bar{p}^3\,\frac{\kappa}{3}\,\biggl[
  \frac{\partial {\cal P}_{V1}}{\partial \bar{k}}
  \Bigl(\frac{3\,\Phi_{V1}}{2\bar{p}}
        + \frac{\partial\Phi_{V1}}{\partial\bar{p}}\Bigr)
  - \Bigl(\frac{3\,{\cal P}_{V1}}{2\bar{p}}
          + \frac{\partial {\cal P}_{V1}}{\partial \bar{p}}\Bigr)
    \frac{\partial\Phi_{V1}}{\partial\bar{k}}
\biggr]
\nonumber\\
&&\quad
% 6) from \{\tilde{H}^{(2)}_{\rm grav},\tilde{H}^{(1)}_V\}_\delta
+ \bar{p}^2\,\bigl[
  \bigl(\frac{\Phi_{V1}}{\bar{p}}\,{\cal K}^{(2)}_3
        - \frac{{\cal P}_{V2}}{\bar{p}}\,{\cal K}^{(2)}_1\bigr)
\bigr]
- \bar{p}^3
  \frac{\partial \Phi_{V1}}{\partial \bar{\varphi}}\,\Theta_{V1}
-
\bar{p}^{3/2} \Phi_0 \Phi_{V1}
\nonumber\\
% 7) from \{\tilde{H}^{(2)}_{V},\tilde{H}^{(1)}_{\rm grav}\}_\delta
&&\quad
-{\Phi}_5 {\cal P}_{dV1} \bar{p}^{3/2}\,
+ \bar{p}^{3/2}
  \Phi_{3}\,\frac{\partial {\cal P}_{V1}}{\partial \bar{\varphi}}
\nonumber\\
&&\quad
+ \bar{p}^2 {\kappa} {\cal P}_{V2} (3\Phi_{V5} + \Phi_{V6})
- \bar{p} {\cal K}^{(1)}_2 (3\Phi_{V5} + \Phi_{V6})\,.
\end{eqnarray}

The term $(\bar{N}_1\delta N_2)\bar{\pi}^1 \delta^d_k \delta K^k_d$ implies the equation
\begin{eqnarray}
% (Bracket 2) from \{\tilde{H}^{(0)}_{V}, \tilde{H}^{(1)}_{\rm grav}\}
0 &=&
\bar{p}^2 \,\frac{2}{3}\;\Biggl[
    \Bigl(\frac{3\,\Theta_{V1}}{2\bar{p}}
          +
          \frac{\partial \Theta_{V1}}{\partial \bar{p}}\Bigr)\,
    \frac{\partial {\cal K}^{(1)}_1}{\partial \bar{k}}
    -
    \frac{\partial \Theta_{V1}}{\partial \bar{k}}\,
    \Bigl(\frac{\partial {\cal K}^{(1)}_1}{\partial \bar{p}}
          +
          \frac{{\cal K}^{(1)}_1}{2\,\bar{p}}\Bigr)
\Biggr]
% (Bracket 3) from \{\tilde{H}^{(0)}_{\varphi}, \tilde{H}^{(1)}_{V}\}
\nonumber\\
&&\quad
+\frac{\kappa\bar{p}^{3/2}}{3}\,\Biggl[
  \frac{\partial \Phi_{0}}{\partial \bar{k}}
  \Bigl(\frac{3\,\Phi_{V1}}{2\bar{p}}
        + \frac{\partial \Phi_{V1}}{\partial \bar{p}}\Bigr)
  +
  \Bigl(\frac{\partial \Phi_{0}}{\partial \bar{p}}
        - \frac{3\,\Phi_{0}}{2\bar{p}}\Bigr)\,
  \frac{\partial \Phi_{V1}}{\partial \bar{k}}
\Biggr]
% (Bracket 4) from \{\tilde{H}^{(0)}_{V}, \tilde{H}^{(1)}_{\varphi}\}
\nonumber\\
&&\quad
+\,\frac{\kappa}{3}\,\Biggl[
  -\,\frac{\partial \Phi_{3}}{\partial \bar{k}}
   \Bigl(\frac{3}{2\bar{p}}\, {\cal P}_{V1}
         + \frac{\partial {\cal P}_{V1}}{\partial \bar{p}}\Bigr)
  +
  \frac{\partial {\cal P}_{V1}}{\partial \bar{k}}
  \,\frac{\partial \Phi_{1}}{\partial \bar{p}}
\Biggr]
% (Bracket 5) from \{\tilde{H}^{(0)}_{V}, \tilde{H}^{(1)}_{V}\}, the bar{\pi} piece
-\frac{\kappa\bar{\bar{p}}}{3}\frac{\partial \Phi_0}{\partial \bar{k}}\Phi_{V1}
\nonumber\\
&&\quad
+\frac{\kappa\bar{p}^3}{3}\,\Biggl[
  \frac{\partial \Theta_{V1}}{\partial \bar{k}}
  \left(\frac{3\,\Phi_{V1}}{2\bar{p}}
        + \frac{\partial \Phi_{V1}}{\partial \bar{p}}\right)
  -
  \left(\frac{3\,\Theta_{V1}}{2\bar{p}}
        + \frac{\partial \Theta_{V1}}{\partial \bar{p}}\right)\,
  \frac{\partial \Phi_{V1}}{\partial \bar{k}}
\Biggr]
% (Bracket 10) from \{\tilde{H}^{(2)}_{\varphi}, \tilde{H}^{(1)}_{V}\}_\delta
\nonumber\\
&&\quad
+ \bar{p}^{3/2} {\Phi}_{3}
\frac{\partial {\Theta}_{V1}}{\partial \bar{\varphi}}
+ \frac{\kappa\sqrt{\bar{p}}}{2} \left[ 3 \bigl(\Phi_{5} + 2\bar{k}\Phi_{10}\bigr)
%
% (Bracket 11) from \{\tilde{H}^{(2)}_{V}, \tilde{H}^{(1)}_{\varphi}\}_\delta
- \bar{p}^{3/2}{\Phi}_{V3} \right] \Phi_{V1}
+\bar{p} {\cal K}^{(2)}_3 \Phi_{V1}
\nonumber\\
&&\quad
+{\cal P}_3\Phi_{V4}
- \bar{p}^{3/2} {\Phi}_5 {\Theta}_{dV1}
- {\cal P}_1 \frac{\partial \Phi_{V1}}{\partial \bar{\varphi}}
+
\frac{\kappa\bar{p}^{3/2}}{3}
\biggl[
\left(
\frac{3 \Phi_{V1}}{2\bar{p}}
+
\frac{\partial \Phi_{V1}}{\partial \bar{p}}
\right)
\frac{\partial {\Phi}_{0}}{\partial \bar{k}}
-
\frac{\partial {\Phi}_{0}}{\partial \bar{p}}
\frac{\partial {\Phi}_{V1}}{\partial \bar{k}}
\biggr]
\nonumber\\
&&\quad
+ 2{\kappa}\bar{p}^{3/2}{\Phi}_{2} (3\Phi_{V5} + \Phi_{V6})
- \bar{p}{\cal K}^{(2)}_1{\Theta}_{V2}
+ {\kappa} \bar{p}^2 (3\Phi_{V5} + \Phi_{V6}) {\Theta}_{V2}
- {\cal P}_1 \Phi_{V1}
\,.
\end{eqnarray}
The term $(\bar{N}_1\delta N_2)\bar{\pi}^2 \delta^d_k \delta K^k_d$ implies the equation
\begin{eqnarray}
0 &=&
%(1) from \{\tilde{H}^{(0)}_\varphi,\tilde{H}^{(1)}_V\}
\frac{\kappa}{6}\;\Biggl[
   \frac{\partial{\cal P}_1}{\partial \bar{k}}
   \left(\frac{3\,\Phi_{V1}}{2\bar{p}}
         + \frac{\partial\Phi_{V1}}{\partial \bar{p}}\right)
+
   \left(\frac{\partial{\cal P}_1}{\partial \bar{p}}
         - \frac{3\,{\cal P}_1}{2\bar{p}}\right)\,
   \frac{\partial\Phi_{V1}}{\partial \bar{k}}
\Biggr]\;
\nonumber\\
%(2) from \{\tilde{H}^{(0)}_V,\tilde{H}^{(1)}_V\}, the bar{\pi}^2 piece
&&\quad
+\bar{p}^3\,\frac{\kappa}{3}\,\Biggl[
  \frac{\partial {\Theta}_{V1}}{\partial \bar{k}}
  \frac{\partial \Phi_{3}}{\partial \bar{p}}
  -
  \left(\frac{3\,{\Theta}_{V1}}{2\bar{p}}
        + \frac{\partial {\Theta}_{V1}}{\partial \bar{p}}\right)\,
  \frac{\partial\Phi_{3}}{\partial \bar{k}}
\Biggr]
\nonumber\\
%(3) from \{\tilde{H}^{(2)}_\varphi,\tilde{H}^{(1)}_V\}_\delta
&&\quad
-\kappa\frac{\,\Phi_{V1}}{6\,\bar{p}}\,
   \frac{\partial {\cal P}_1}{\partial \bar{k}}
-{\kappa}\frac{{\cal P}_2}{2\bar{p}}
(
3\Phi_{V5} + \Phi_{V6}
)
\,.
\end{eqnarray}
The term 
$\delta^{jk}\partial_j\partial_c\delta E^c_k$
implies the equation
\begin{eqnarray}
0 &=&
\frac{\bar{p}}{3} \frac{\partial {\cal P}_{V1}}{\partial \bar{k}} \left(\frac{\partial {\cal K}^{(1)}_3}{\partial \bar{p}} - \frac{{\cal K}^{(1)}_3}{2 \bar{p}}\right)
- \frac{\bar{p}}{3} \left(\frac{3 {\cal P}_{V1}}{2\bar{p}} + \frac{\partial {\cal P}_{V1}}{\partial \bar{p}} \right) \frac{\partial {\cal K}^{(1)}_3}{\partial \bar{k}}
\nonumber\\
&&+\frac{1}{2}\Phi_{V3} {\cal K}^{(1)}_3
-\frac{1}{2}
\Phi_{V1} {\cal K}^{(2)}_6
+
\frac{\sqrt{\bar{p}}}{2}{\Theta}_{2}
\frac{\partial {\cal P}_{V1}}{\partial \bar{\varphi}}\,.
\end{eqnarray}
The term 
$\bar{\pi} \delta^{jk}\partial_j\partial_c\delta E^c_k$
implies the equation
\begin{eqnarray}
0 &=&
\frac{\bar{p}}{3}\,\biggl[
\frac{\partial \Theta_{V1}}{\partial \bar{k}}
\left(\frac{\partial {\cal K}^{(1)}_3}{\partial \bar{p}}
- \frac{{\cal K}^{(1)}_3}{2\bar{p}}\right)
-
\left(\frac{3\,\Theta_{V1}}{2\,\bar{p}}
+ \frac{\partial \Theta_{V1}}{\partial \bar{p}}\right)
\frac{\partial {\cal K}^{(1)}_3}{\partial \bar{k}}
\biggr]
\nonumber \\
&&+ \frac{\sqrt{\bar{p}}}{2}{\Theta}_{2}
\frac{\partial {\Theta}_{V1}}{\partial \bar{\varphi}}
+ {\kappa} \sqrt{\bar{p}}
{\Phi}_{V1}
{\Theta}_{3}\,.
\end{eqnarray}

The term 
$\bar{\pi}^2 \delta^{jk}\partial_j\partial_c\delta E^c_k$
implies the equation
\begin{equation}
\label{eq:PhiV1}
0 = {\Phi}_{V1} {\Theta}_{6}\,.
\end{equation}

The term 
$\delta_{jk}\partial^j \partial^c\delta K_c^k$
implies the equation
\begin{equation}
0 = 2\bar{p}
{\Phi}_{V6}
{\cal K}^{(1)}_3
\end{equation}

The term 
$\delta^c_k \partial^a \partial_a \delta K_c^k$
implies the equation
\begin{equation}
0 = 2\bar{p} {\Phi}_{V5} {\cal K}^{(1)}_3
\end{equation}

\subsubsection{Spacetime covariance}
\label{eq:Covariance cond - scalar potential}

Consider the modified spacetime metric with background structure function (\ref{eq:Background structure function - EMG}) and its perturbation (\ref{eq:Metric perturbation}).
In the presence of a scalar matter potential contribution of the constraint, the covariance condition further requires
\begin{eqnarray}\label{eq:Cov cond - 1 - EMG - V}
    \frac{\partial \{\delta \tilde{q}^{ab}_{(1)} , \tilde{\cal H}_V^{(1)} [\delta \epsilon^0]\}}{\partial (\partial_{c_1}\delta\epsilon^0)} \bigg|_{\rm O.S.} = \frac{\partial \{\delta \tilde{q}^{ab}_{(1)} , \tilde{\cal H}_V^{(1)} [\delta \epsilon^0]\}}{\partial (\partial_{c_1}\partial_{c_2}\delta\epsilon^0)} \bigg|_{\rm O.S.}
    = \dots = 0\,.
\end{eqnarray}
Using our ansatz this becomes
\begin{eqnarray}
    \{\delta \tilde{q}^{ab}_{(1)} , \tilde{\cal H}_V^{(1)} [\delta \epsilon^0]\}
    &=&
    \kappa \frac{\delta^{a b}}{\bar{p}} \left[ Q_1^{(K)} \frac{3{\cal P}_{V2}}{2}
    + \left( {\cal K}^{(1)}_3 {\cal K}^{(2)}_1
    + 3 \bar{k} Q_1^{(K)}
    \right) \Phi_{V1}\right] \sqrt{\bar{p}}\;\delta \epsilon^0
    \nonumber\\
    &&\qquad
    + \kappa \frac{\delta^{a b}}{\bar{p}} Q_1^{(K)} \Phi_{V1} \frac{{\cal K}^{(1)}_3}{2{\cal K}^{(1)}_1} \sqrt{\bar{p}}\; \partial^d\partial_d \delta \epsilon^0
    \,.
\end{eqnarray}
The covariance condition, therefore, requires either
\begin{equation} \label{Q1K}
    Q_1^{(K)} = 0
\end{equation}
or
\begin{equation}
    \Phi_{V1} = 0\,.
\end{equation}
The latter is already satisfied by the anomaly-freedom equation
(\ref{eq:PhiV1}) unless $\Theta_6=0$. If $\Theta_6=0$, $\Phi_{V1}$ may be
non-zero, in which case (\ref{Q1K}) is required.
 
\section{Classical system}
\label{a:Brackets-classical}

\subsection{Diffeomorphism constraint}

The brackets between the diffeomorphism constraint and the basic variables are
\begin{eqnarray}\label{eq:Basic D brackets}
    \{\bar{p} , D_c[\delta N^c]\} &=&
    \int \frac{{\rm d}^3 x}{V_0}\frac{\delta N^c}{3} \partial_d (\delta_c^k\delta E^d_k)
    \\
    \{\bar{k} , D_c[\delta N^c]\} &=&
    \int \frac{{\rm d}^3 x}{V_0} \frac{\delta N^c}{3} \left(\partial_c(\delta^d_k \delta K^k_d)-\partial_k\delta K^k_c\right)
    \\
    \{\delta K^i_a,D_c[\delta N^c]\}
    &=& \bar{k} \delta_c^i \partial_a \delta N^c
    \,,\\
    \{\delta E^a_i,D_c[\delta N^c]\}
    &=& \bar{p} \left(\delta_i^a \partial_c \delta N^c - \partial_i \delta N^a\right)
    \,,\\
    \{\bar{\varphi},D_c[\delta N^c]\}
    &=& \int \frac{{\rm d}^3 x}{V_0} \delta N^c \partial_c \delta \varphi\,,\\
    \{\bar{\pi},D_c[\delta N^c]\}
    &=& 0
    \,,\\
    \{\delta \varphi,D_c[\delta N^c]\}
    &=& 0
    \,,\\
    \{\delta\pi,D_c[\delta N^c]\}
    &=& \bar{\pi} \partial_c \delta N^c
    \,,
\end{eqnarray}
It follows that
\begin{eqnarray}
    \{\delta^a_i\delta K^i_a,D_c[\delta N^c]\}
    &=& \bar{k} \partial_c \delta N^c
    \,,\\
    \{\delta_a^i\delta E^a_i,D_c[\delta N^c]\}
    &=& 2 \bar{p} \partial_c \delta N^c
    \,,\\
    \{\partial_a\delta E^a_i,D_c[\delta N^c]\}
    &=&
    0
\end{eqnarray}
and
\begin{eqnarray}
    \{\delta K_c^j\delta K_d^k\delta^c_k\delta^d_j,D_c[\delta N^c]\}
    &=& 2 \bar{k} \delta K_c^j \partial_j \delta N^c
    \\
    &=& - 2 \bar{k} \delta N^c \partial_j \delta K_c^j + \partial_d R^d
    \,,\nonumber\\
    \{\delta E^c_j\delta E^d_k\delta_c^k\delta_d^j,D_c[\delta N^c]\}
    &=& 2 \delta E^c_j \bar{p} \left(\delta_c^j \partial_e \delta N^e - \delta_d^j \partial_c \delta N^d\right)
    \\
    &=& - 2 \bar{p} \delta N^e \partial_e (\delta_c^j \delta E^c_j) + 2 \bar{p} \delta N^d \delta_d^j \partial_c \delta E^c_j + \partial_d T^d
    \,,\nonumber\\
    \{\delta K_a^i\delta E^a_i,D_c[\delta N^c]\}
    &=& \delta K_a^i \bar{p} \left(\delta_i^a \partial_c \delta N^c - \partial_i \delta N^a\right)
    + \delta E^a_i\bar{k} \delta_c^i \partial_a \delta N^c
    \\
    &=&
    \bar{p} \delta N^c \partial_k \delta K_c^k
    - \bar{p} \delta N^c \partial_c (\delta_k^d \delta K_d^k)
    - \bar{k} \delta N^c \delta_c^k \partial_d \delta E^d_k
    + \partial_d Y^d
    \,.\nonumber
\end{eqnarray}

Furthermore, for any function $F(\bar{k},\bar{p})$ we get
\begin{eqnarray}\label{eq:F(k,p) D bracket}
    \{F , D_c[\delta N^c]\} &=&
    \frac{\delta N^c}{3} \left[\frac{\partial F}{\partial \bar{k}} \left(\partial_c(\delta^d_k \delta K^k_d)-\partial_k\delta K^k_c\right) 
    + \frac{\partial F}{\partial \bar{p}} \delta_c^k(\partial_d \delta E^d_k)\right]
    =: F_c [\delta N^c]
    \,,
\end{eqnarray}
and hence
\begin{eqnarray}
    \{F , D_c[\delta N^c]\} &=&
    \frac{\delta N^c}{3} \left[\frac{\partial F}{\partial \bar{k}} \left(\partial_c(\delta^d_k \delta K^k_d)-\partial_k\delta K^k_c\right) 
    + \frac{\partial F}{\partial \bar{p}} \delta_c^k(\partial_d \delta E^d_k)\right]
    \\
    &=&
    \frac{1}{3} \left[\frac{\partial F}{\partial \bar{k}} \left(-\delta^d_k \delta K^k_d\partial_c\delta N^c+\delta K^k_c\partial_k\delta N^c\right) 
    - \frac{\partial F}{\partial \bar{p}} \delta_c^k\delta E^d_k(\partial_d \delta N^c)\right]
    + \partial_c Z^c
    \,,\nonumber
\end{eqnarray}

\subsection{Hamiltonian constraint}
\subsubsection{Derivatives}

\begin{eqnarray}
    \frac{\partial H_{\rm grav}^{(0)} [\bar{N}]}{\partial \bar{k}} &=&
    - \frac{V_0}{2 \kappa} \bar{N} 12 \sqrt{\bar p} \bar{k}
    \,,\\
    \frac{\partial H_{\rm grav}^{(0)} [\bar{N}]}{\partial \bar{p}} &=&
    - \frac{V_0}{2 \kappa} \bar{N} 6 \sqrt{\bar p} \frac{\bar{k}^2-\Lambda\bar{p}}{2\bar{p}}
    \,,
\end{eqnarray}
\begin{eqnarray}
    \frac{\partial H_{\rm grav}^{(2)} [\bar{N}]}{\partial \bar{k}} &=&
    \int \frac{{\rm d}^3x}{16 \pi G} 
    \bar{N} \sqrt{\bar{p}} \Bigg[
    - 2 \delta K_c^j \frac{\delta E^c_j}{\bar{p}}
    - \bar{k} \left( \frac{\delta_c^k\delta_d^j\delta E^c_j\delta E^d_k}{\bar{p}^2}
    - \frac{(\delta_c^j \delta E^c_j)^2}{2\bar{p}^2} \right)
    \Bigg]
    \,,\\
    \frac{\partial H_{\rm grav}^{(2)} [\bar{N}]}{\partial \bar{p}} &=&
    \int \frac{{\rm d}^3x}{16 \pi G} 
    \bar{N} \sqrt{\bar{p}} \Bigg[ \frac{1}{2\bar{p}} \delta K_c^j\delta K_d^k\delta^c_k\delta^d_j
    - \frac{1}{2\bar{p}} (\delta K_c^j\delta^c_j)^2
    + \frac{\bar{k}}{\bar{p}} \delta K_c^j \frac{\delta E^c_j}{\bar{p}}
    \nonumber\\
    && \quad 
    + \frac{\bar{k}^2- \Lambda \bar{p}}{4\bar{p}} \left( \frac{\delta_c^k\delta_d^j\delta E^c_j\delta E^d_k}{\bar{p}^2}
    - \frac{(\delta_c^j \delta E^c_j)^2}{2\bar{p}^2}\right)
    + \frac{1}{4\bar{p}} \frac{\delta^{jk} (\partial_c\delta E^c_j) (\partial_d\delta E^d_k)}{\bar{p}^2} \Bigg]
    \,,\\
    \frac{\delta H_{\rm grav}^{(2)} [\bar{N}]}{\delta (\delta K_a^i)} &=& \frac{\bar{N} \sqrt{\bar{p}}}{2\kappa} \left( 2 \delta_k^a \delta^d_i \delta K_d^k
    - 2 \delta^d_k\delta K_d^k \delta_i^a
    - 2 \bar{k} \frac{\delta E^a_i}{\bar{p}} \right)
    \,\\
    \frac{\delta H_{\rm grav}^{(2)} [\bar{N}]}{\delta (\delta E^a_i)} &=& \frac{\bar{N} \sqrt{\bar{p}}}{2\kappa} \Bigg( 
    - 2 \bar{k} \frac{\delta K_a^i}{\bar{p}} 
    - \left(\bar{k}^2+\Lambda \bar{p}\right)\left( \frac{\delta^i_c\delta_a^j\delta E^c_j}{\bar{p}^2}
    - \frac{\delta_c^j \delta E^c_j}{2\bar{p}^2} \delta^i_a\right)
    + \frac{\delta^{ij} (\partial_a \partial_c\delta E^c_j)}{\bar{p}^2}\Bigg)
\end{eqnarray}
and
\begin{eqnarray}
    \frac{\partial H_{\rm grav}^{(1)} [\delta N]}{\partial \bar{p}} &=& \frac{1}{2\kappa} \int{\rm d}^3 x \delta N \sqrt{\bar{p}} \left(- 2 \frac{\bar{k}}{\bar{p}} \delta^d_k\delta K_d^k
    + \frac{\bar{k}^2+\Lambda \bar{p}}{2 \bar{p}} \frac{\delta_d^k\delta E^d_k}{\bar{p}}
    - \frac{1}{\bar{p}} \frac{\delta^{jk}\partial_j\partial_c\delta E^c_k}{\bar{p}} \right)
    \,\\
    \frac{\partial H_{\rm grav}^{(1)} [\delta N]}{\partial \bar{k}} &=& \frac{1}{2\kappa} \int{\rm d}^3 x \delta N \sqrt{\bar{p}} \left(- 4 \delta^c_j\delta K_c^j
    - 2 \bar{k} \frac{\delta_c^j\delta E^c_j}{\bar{p}}
    \right)
    \,,\\
    \frac{\delta H_{\rm grav}^{(1)} [\delta N]}{\delta (\delta K_a^i)} &=& 
    - \frac{\delta N}{2 \kappa} 4 \sqrt{\bar{p}} \bar{k} \delta^a_i
    \,\\
    \frac{\delta H_{\rm grav}^{(1)} [\delta N]}{\delta(\delta E^a_i)} &=& 
    - \frac{\delta N}{16\pi G}
    \frac{\bar{k}^2-\Lambda \bar{p}}{\sqrt{\bar{p}}} \delta_a^i
    + \frac{(\partial^i\partial_a\delta N)}{16\pi G} \frac{2}{\sqrt{\bar{p}}}
\end{eqnarray}

\textbf{Classical scalar matter:}

\begin{eqnarray}
    \frac{\partial H_\varphi^{(0)} [\bar{N}]}{\partial \bar{k}} &=&
    0
    \,,\\
    \frac{\partial H_\varphi^{(0)} [\bar{N}]}{\partial \bar{p}} &=& \frac{3V_0}{2}\bar{N} \left(- \frac{\bar{\pi}^2}{2\bar{p}^{5/2}}
    + \sqrt{\bar{p}}\; V\right)
    \,,\\
    \frac{\partial H_\varphi^{(0)} [\bar{N}]}{\partial \bar{\varphi}} &=&
    V_0\bar{N}\bar{p}^{3/2} \frac{\partial V}{\partial \bar{\varphi}}
    \,,\\
    \frac{\partial H_\varphi^{(0)} [\bar{N}]}{\partial \bar{\pi}} &=& V_0\bar{N}\frac{\bar{\pi}}{\bar{p}^{3/2}}\,.
\end{eqnarray}

\begin{eqnarray}
    \frac{\delta H_\varphi^{(2)} [\bar{N}]}{\delta (\delta K_a^i)} &=& 0
    \,,\\
    \frac{\delta H_\varphi^{(2)} [\bar{N}]}{\delta (\delta E^a_i)} &=& \bar{N} \Bigg[
    - \frac{\bar{\pi}\delta\pi}{\bar{p}^{3/2}} \frac{\delta^i_a}{2\bar{p}}
    + \frac{\bar{\pi}^2}{2\bar{p}^{3/2}} \left(\frac{(\delta^k_d\delta E^d_k) \delta^i_a}{4\bar{p}^2}+\frac{\delta^k_a\delta^i_d\delta E^d_k}{2\bar{p}^2}\right)
    \nonumber\\
    &&
    + \bar{p}^{3/2} V \left(\frac{(\delta^k_d\delta E^d_k) \delta^i_a}{4\bar{p}^2}-\frac{\delta^k_a\delta^i_d\delta E^d_k}{2\bar{p}^2}\right)
    + \bar{p}^{3/2}\frac{\partial V}{\partial \varphi} \delta\varphi\frac{\delta^i_a}{2\bar{p}}
    \Bigg]
    \,,\\
    \frac{\delta H_\varphi^{(2)} [\bar{N}]}{\delta (\delta \varphi)} &=& \bar{N} \Bigg[ 
    \bar{p}^{3/2} \frac{\partial^2 V}{(\partial \varphi)^2} \delta\varphi
    + \bar{p}^{3/2}\frac{\partial V}{\partial\varphi} \frac{\delta^j_c\delta E^c_j}{2\bar{p}}
    - \sqrt{\bar{p}} \delta^{a b} (\partial_a \partial_b \delta\varphi)
    \Bigg]
    \,,\\
    \frac{\delta H_\varphi^{(2)} [\bar{N}]}{\delta (\delta \pi)} &=& \bar{N} \Bigg[
    \frac{\delta\pi}{\bar{p}^{3/2}}
    - \frac{\bar{\pi}}{\bar{p}^{3/2}} \frac{\delta^j_c\delta E^c_j}{2\bar{p}}
    \Bigg]
    \,.
\end{eqnarray}

\begin{eqnarray}
    \frac{\partial H_\varphi^{(1)} [\delta N]}{\partial \bar{k}} &=&
    0
    \,,\\
    \frac{\partial H_\varphi^{(1)} [\delta N]}{\partial \bar{p}} &=&
    \int {\rm d}^3x 
    \delta N \frac{1}{\bar{p}} \Bigg[ 
    \frac{5}{2} \frac{\bar{\pi}^2}{2\bar{p}^{3/2}} \frac{\delta^j_c \delta E^c_j}{2\bar{p}}
    - \frac{3}{2} \frac{\bar{\pi}\delta\pi}{\bar{p}^{3/2}}
    + \bar{p}^{3/2} \left( \frac{3}{2} \frac{\partial V}{\partial \varphi} \delta\varphi
    + \frac{1}{2} V (\bar{\varphi}) \frac{\delta^j_c \delta E^c_j}{2 \bar{p}}\right)  \Bigg]
    \,,\\
    \frac{\partial H_\varphi^{(1)} [\delta N]}{\partial \bar{\varphi}} &=&
    \int {\rm d}^3 x \delta N \left( \bar{p}^{3/2} \frac{\partial^2 V}{(\partial \varphi)^2} \delta\varphi
    + \sqrt{\bar{p}} \frac{\partial V}{\partial \varphi} \frac{\delta^k_d \delta E^d_k}{2}\right)
    \,,\\
    \frac{\partial H_\varphi^{(1)} [\delta N]}{\partial \bar{\pi}} &=& 
    \int {\rm d}^3 x \delta N\left( \frac{\delta\pi}{\bar{p}^{3/2}}
    - \frac{\delta^k_d \delta E^d_k}{2\bar{p}^{5/2}} \bar{\pi} \right)
    \,,\\
    \frac{\delta H_\varphi^{(1)} [\delta N]}{\delta (\delta K_a^i)} &=& 0
    \,,\\
    \frac{\delta H_\varphi^{(1)} [\delta N]}{\delta (\delta E^a_i)} &=& \delta N \delta^i_a \Bigg[ 
    - \frac{\bar{\pi}^2}{2\bar{p}^{3/2}} \frac{1}{2\bar{p}}
    + \bar{p}^{3/2} \frac{1}{2 \bar{p}} V
    \Bigg]
    \,,\\
    \frac{\delta H_\varphi^{(1)} [\delta N]}{\delta (\delta \varphi)} &=& \delta N \bar{p}^{3/2} \frac{\partial V}{\partial \varphi}
    \,,\\
    \frac{\delta H_\varphi^{(1)} [\delta N]}{\delta (\delta \pi)} &=& \delta N \frac{\bar{\pi}}{\bar{p}^{3/2}}
    \,.
\end{eqnarray}

\subsubsection{Scalar matter brackets}

\begin{eqnarray}
    \{H^{(1)}_{\rm grav}[\delta N_1],H^{(1)}_\varphi[\delta N_2]\}_{\delta} &=&
    \int{\rm d}^3x
    \delta N_1 \delta N_2 \left( - \bar{p} \bar{k} V
    + \frac{\bar{k}}{2\bar{p}^2} \bar{\pi}^2 \right)
    \,,
\end{eqnarray}
\begin{eqnarray}
    \{H^{(1)}_\varphi[\delta N_1],H^{(1)}_\varphi[\delta N_2]\}_{\delta} &=& 0
    \,.
\end{eqnarray}
Thus,
\begin{eqnarray}
    \{H^{(1)}_{\rm grav}[\delta N_1],H^{(1)}_\varphi[\delta N_2]\}_{\delta} + \{H^{(1)}_{\rm grav}[\delta N_2],H^{(1)}_\varphi[\delta N_1]\}_{\delta}&&
    \nonumber\\
    + \{H^{(1)}_\varphi[\delta N_1],H^{(1)}_\varphi[\delta N_2]\}_{\delta} = 0&&
\end{eqnarray}

Also,
\begin{eqnarray}
    \{H^{(0)}_{\rm grav}[\bar{N}_1],H^{(1)}_\varphi[\delta N_2]\}_{\bar{A}} &=& \frac{\kappa}{3V_0} \frac{\partial H^{(0)}_{\rm grav}[\bar{N}_1]}{\partial\bar{k}} \frac{\partial H^{(1)}_\varphi[\delta N_2]}{\partial\bar{p}}
    \\
    &=& \int {\rm d}^3x \bar{N}_1 \delta N_2 \Bigg[ 
    - 3 \bar{p} \bar{k} \frac{\partial V}{\partial \varphi} \delta\varphi
    - \frac{\bar{k} V}{2} \delta^k_d \delta E^d_k
    + \frac{3 \bar{k}}{\bar{p}^2} \delta\pi \bar{\pi}
    - \frac{5 \bar{k}}{4\bar{p}^3} \delta^j_c \delta E^c_j \bar{\pi}^2
    \Bigg]\nonumber
\end{eqnarray}
\begin{eqnarray}
    \{H^{(0)}_\varphi[\bar{N}_1],H^{(1)}_{\rm grav}[\delta N_2]\}_{\bar{A}} &=& - \frac{\kappa}{3V_0} \frac{\partial H^{(0)}_\varphi[\bar{N}_1]}{\partial\bar{p}} \frac{\partial H^{(1)}_{\rm grav}[\delta N_2]}{\partial\bar{k}}
    \\
    &=& \int{\rm d}^3 x \bar{N}_1 \delta N_2 \left[\bar{p} V\left( \delta^d_k\delta K_d^k
    + \bar{k} \frac{\delta_d^k\delta E^d_k}{2\bar{p}}\right)
    - \left( \frac{\delta^d_k\delta K_d^k}{2\bar{p}^2}
    + \bar{k} \frac{\delta_d^k\delta E^d_k}{4\bar{p}^3}\right) \bar{\pi}^2 \right]
    \nonumber
\end{eqnarray}
\begin{eqnarray}
    \{H^{(0)}_\varphi[\bar{N}_1],H^{(1)}_\varphi[\delta N_2]\}_{\bar{A}} &=& \frac{1}{V_0} \left(\frac{\partial H^{(0)}_\varphi[\bar{N}_1]}{\partial\bar{\varphi}} \frac{\partial H^{(1)}_\varphi[\delta N_2]}{\partial\bar{\pi}}
    - \frac{\partial H^{(0)}_\varphi[\bar{N}_1]}{\partial\bar{\pi}} \frac{\partial H^{(1)}_\varphi[\delta N_2]}{\partial\bar{\varphi}}\right)
    \\
    &=& \int {\rm d}^3 x \bar{N}_1 \delta N_2 \left[ \frac{\partial V}{\partial \bar{\varphi}} \delta\pi
    - \left( \frac{\partial^2 V}{(\partial \varphi)^2} \delta\varphi
    + \frac{\partial V}{\partial \varphi} \frac{\delta^k_d \delta E^d_k}{\bar{p}}\right) \bar{\pi} \right]
    \nonumber
\end{eqnarray}

\begin{eqnarray}
    \{H^{(2)}_{\rm grav}[\bar{N}_1],H^{(1)}_\varphi[\delta N_2]\}_\delta &=& \int{\rm d}^3x \left[\kappa\frac{\delta H^{(2)}_{\rm grav}[\bar{N}_1]}{\delta \delta K_a^i} \frac{\delta H^{(1)}_\varphi[\delta N_2]}{\delta \delta E^a_i}\right]
    \\
    &=& \int{\rm d}^3x \bar{N}_1 \delta N_2 \left( 2 \delta^d_k\delta K_d^k
    + \bar{k} \frac{\delta^k_d \delta E^d_k}{\bar{p}} \right) \left( 
    - \frac{\bar{p}}{2} V
    + \frac{\bar{\pi}^2}{4\bar{p}^2}
    \right)
    \,,
    \nonumber
\end{eqnarray}
\begin{eqnarray}
    \{H^{(2)}_\varphi[\bar{N}_1],H^{(1)}_{\rm grav}[\delta N_2]\}_\delta &=&
    - \kappa \int{\rm d}^3x \frac{\delta H^{(2)}_\varphi[\bar{N}_1]}{\delta \delta E^a_i} \frac{\delta H^{(1)}_{\rm grav}[\delta N_2]}{\delta \delta K_a^i}
    \\
    &=&
    \int{\rm d}^3x \bar{N}_1 \delta N_2 \Bigg[
    \frac{\bar{k} V}{2} \delta^k_d\delta E^d_k
    + 3 \bar{p} \bar{k} \frac{\partial V}{\partial \varphi} \delta\varphi
    - \frac{3 \bar{k}}{\bar{p}^2} \delta\pi \bar{\pi}
    + \frac{5 \bar{k}}{4\bar{p}^3} \delta^k_d\delta E^d_k \bar{\pi}^2
    \Bigg]
    \,,\nonumber
\end{eqnarray}

\begin{eqnarray}
    \{H^{(2)}_\varphi[\bar{N}_1],H^{(1)}_\varphi[\delta N_2]\}_\delta &=&
    \int{\rm d}^3x \left(\frac{\delta H^{(2)}_\varphi[\bar{N}_1]}{\delta \delta \varphi} \frac{\delta H^{(1)}_\varphi[\delta N_2]}{\delta \delta \pi}
    - \frac{\delta H^{(2)}_\varphi[\bar{N}_1]}{\delta \delta \pi} \frac{\delta H^{(1)}_\varphi[\delta N_2]}{\delta \delta \varphi}\right)
    \nonumber\\
    &=&
    \int{\rm d}^3x \bar{N}_1 \delta N_2 \Bigg[
    - \frac{\partial V}{\partial \varphi} \delta\pi
    + \left( 
    \frac{\partial^2 V}{(\partial \varphi)^2} \delta\varphi
    + \frac{\partial V}{\partial\varphi} \frac{\delta^j_c\delta E^c_j}{\bar{p}}
    \right) \bar{\pi} \Bigg]
    \nonumber\\
    &&
    + \int{\rm d}^3x (\bar{N}_1 \partial_c \delta N_2) \frac{\delta^{ce}}{\bar{p}} \bar{\pi}\partial_e \delta\varphi
    \,.
\end{eqnarray}

We therefore get
\begin{eqnarray}
    B[\bar{N_1},\delta N_2] &=& 
    \int{\rm d}^3x (\bar{N}_1 \partial_c \delta N_2) \frac{\delta^{ce}}{\bar{p}} \bar{\pi}\partial_e \delta\varphi
    \,,
\end{eqnarray}
and hence the constraint algebra is anomaly free.

\section{Modified Hamiltonian constraint}
\label{a:Brackets-modified}

\subsection{Gravitational contribution}

Here we will use the gravitational contribution to the modified Hamiltonian constraint ansatz (\ref{HamConstH0-EMG})-(\ref{HamConstH2-EMG}).

The relevant (functional) derivatives of the constraints are
\begin{eqnarray}
    \frac{\partial\tilde{H}_{\rm grav}^{(0)} [\bar{N}]}{\partial \bar{k}} &=&
    - \frac{V_0}{2\kappa} \bar{N} 6 \sqrt{\bar p} \frac{\partial {\cal K}^{(0)}}{\partial \bar{k}}
    \,,\\
    \frac{\partial\tilde{H}_{\rm grav}^{(0)} [\bar{N}]}{\partial \bar{p}} &=&
    - \frac{V_0}{2\kappa} \bar{N} 6 \sqrt{\bar p} \left( \frac{{\cal K}^{(0)}}{2\bar{p}} + \frac{\partial {\cal K}^{(0)}}{\partial \bar{p}}\right)
    \,,\\
    \frac{\delta \tilde{H}_{\rm grav}^{(0)} [\bar{N}]}{\delta (\delta K_a^i)} &=& 0
    \,,\\
    \frac{\delta \tilde{H}_{\rm grav}^{(0)} [\bar{N}]}{\delta (\delta E^a_i)} &=& 0\,.
\end{eqnarray}
\begin{eqnarray}
    \frac{\partial\tilde{H}_{\rm grav}^{(2)} [\bar{N}]}{\partial \bar{k}} &=&
    \int \frac{{\rm d}^3x}{2\kappa} 
    \bar{N} \sqrt{\bar{p}} \Bigg[ \frac{\partial{\cal K}^{(2)}_1}{\partial \bar{k}} \delta K_c^j\delta K_d^k\delta^c_k\delta^d_j
    - \frac{\partial{\cal K}^{(2)}_2}{\partial \bar{k}} (\delta K_c^j\delta^c_j)^2
    - 2 \frac{\partial{\cal K}^{(2)}_3}{\partial \bar{k}} \delta K_c^j \frac{\delta E^c_j}{\bar{p}}
    \nonumber\\
    && \quad 
    - \frac{1}{2} \left(\frac{\partial{\cal K}^{(2)}_4}{\partial \bar{k}} \frac{\delta_c^k\delta_d^j\delta E^c_j\delta E^d_k}{\bar{p}^2}
    - \frac{\partial{\cal K}^{(2)}_5}{\partial \bar{k}} \frac{(\delta_c^j \delta E^c_j)^2}{2\bar{p}^2} \right)
    - \frac{1}{2} \frac{\partial {\cal K}^{(2)}_6}{\partial \bar{k}} \frac{\delta^{jk} (\partial_c\delta E^c_j) (\partial_d\delta E^d_k)}{\bar{p}^2} \Bigg]
    \nonumber\\
    \,,
    \\
    \frac{\partial\tilde{H}_{\rm grav}^{(2)} [\bar{N}]}{\partial \bar{p}} &=&
    \int \frac{{\rm d}^3x}{2\kappa} 
    \bar{N} \sqrt{\bar{p}} \Bigg[ \left(\frac{{\cal K}^{(2)}_1}{2\bar{p}} + \frac{\partial{\cal K}^{(2)}_1}{\partial \bar{p}}\right) \delta K_c^j\delta K_d^k\delta^c_k\delta^d_j
    - \left(\frac{{\cal K}^{(2)}_2}{2\bar{p}} +\frac{\partial{\cal K}^{(2)}_2}{\partial \bar{p}} \right) (\delta K_c^j\delta^c_j)^2
    \nonumber\\
    && \quad 
    - 2 \left( \frac{\partial{\cal K}^{(2)}_3}{\partial \bar{p}} - \frac{{\cal K}^{(2)}_3}{2\bar{p}}\right) \delta K_c^j \frac{\delta E^c_j}{\bar{p}}
    \nonumber\\
    && \quad 
    - \frac{1}{2} \left( \frac{\partial{\cal K}^{(2)}_4}{\partial \bar{p}} - \frac{{\cal K}^{(2)}_4}{2\bar{p}} \right) \frac{\delta_c^k\delta_d^j\delta E^c_j\delta E^d_k}{\bar{p}^2}
    + \frac{1}{2} \left(\frac{\partial{\cal K}^{(2)}_5}{\partial \bar{p}} - \frac{{\cal K}^{(2)}_5}{2\bar{p}} \right) \frac{(\delta_c^j \delta E^c_j)^2}{2\bar{p}^2}
    \nonumber\\
    && \quad 
    - \frac{1}{2} \left(\frac{\partial {\cal K}^{(2)}_6}{\partial \bar{p}} - \frac{{\cal K}^{(2)}_6}{2\bar{p}} \right)\frac{\delta^{jk} (\partial_c\delta E^c_j) (\partial_d\delta E^d_k)}{\bar{p}^2} \Bigg]
    \,,\\
    \frac{\delta \tilde{H}_{\rm grav}^{(2)} [\bar{N}]}{\delta (\delta K_a^i)} &=& \frac{\bar{N} \sqrt{\bar{p}}}{\kappa} \left( {\cal K}^{(2)}_1 \delta_k^a \delta^d_i \delta K_d^k
    - {\cal K}^{(2)}_2 \delta^d_k\delta K_d^k \delta_i^a
    - {\cal K}^{(2)}_3 \frac{\delta E^a_i}{\bar{p}} \right)
    \,\\
    \frac{\delta \tilde{H}_{\rm grav}^{(2)} [\bar{N}]}{\delta (\delta E^a_i)} &=& \frac{\bar{N} \sqrt{\bar{p}}}{2\kappa} \Bigg( 
    - 2 {\cal K}^{(2)}_3 \frac{\delta K_a^i}{\bar{p}} 
    - {\cal K}^{(2)}_4 \frac{\delta^i_c\delta_a^j\delta E^c_j}{\bar{p}^2}
    + {\cal K}^{(2)}_5 \frac{\delta_c^j \delta E^c_j}{2\bar{p}^2} \delta^i_a
    + {\cal K}^{(2)}_6 \frac{\delta^{ij} (\partial_a \partial_c\delta E^c_j)}{\bar{p}^2}\Bigg)
    \nonumber\\
    \
\end{eqnarray}
and
\\
\begin{eqnarray}
    \frac{\partial\tilde{H}_{\rm grav}^{(1)} [\delta N]}{\partial \bar{p}} &=& \int\frac{{\rm d}^3 x}{2\kappa} \delta N \sqrt{\bar{p}} \Bigg[- 4 \left(\frac{\partial {\cal K}^{(1)}_1}{\partial \bar{p}} + \frac{{\cal K}^{(1)}_1}{2 \bar{p}}\right) \delta^c_j\delta K_c^j
    - \left(\frac{\partial {\cal K}^{(1)}_2}{\partial \bar{p}} - \frac{{\cal K}^{(1)}_2}{2 \bar{p}}\right) \frac{\delta_c^j\delta E^c_j}{\bar{p}}
    \nonumber \\
    &&\qquad
    + 2 \left(\frac{\partial {\cal K}^{(1)}_3}{\partial \bar{p}} - \frac{{\cal K}^{(1)}_3}{2 \bar{p}}\right) \frac{\delta^{jk}\partial_j\partial_c\delta E^c_k}{\bar{p}} \Bigg]
\end{eqnarray}
\begin{eqnarray}
    \frac{\partial\tilde{H}_{\rm grav}^{(1)} [\delta N]}{\partial \bar{k}} &=& \int\frac{{\rm d}^3 x}{2\kappa} \delta N \sqrt{\bar{p}} \left[- 4 \frac{\partial {\cal K}^{(1)}_1}{\partial \bar{k}} \delta^c_j\delta K_c^j
    - \frac{\partial {\cal K}^{(1)}_2}{\partial \bar{k}} \frac{\delta_c^j\delta E^c_j}{\bar{p}}
    + 2 \frac{\partial {\cal K}^{(1)}_3}{\partial \bar{k}} \frac{\delta^{jk}\partial_j\partial_c\delta E^c_k}{\bar{p}} \right]
    \,,\\
    \frac{\delta\tilde{H}_{\rm grav}^{(1)} [\delta N]}{\delta (\delta K_a^i)} &=& 
    - \frac{\delta N}{2\kappa} 4 \sqrt{\bar{p}} {\cal K}^{(1)}_1 \delta^a_i
    \,
\end{eqnarray}
\begin{eqnarray}
    \frac{\delta\tilde{H}_{\rm grav}^{(1)} [\delta N]}{\delta(\delta E^a_i)} &=& 
    - \frac{\delta N}{2\kappa}
    \frac{{\cal K}^{(1)}_2}{\sqrt{\bar{p}}} \delta_a^i
    + \frac{(\partial^i\partial_a\delta N)}{2\kappa} \frac{2 {\cal K}^{(1)}_3}{\sqrt{\bar{p}}}
\end{eqnarray}

\subsection{Free scalar matter contributions}
Here we assume that 
${\cal P}_3 = \frac{1}{2}({\cal P}_1+{\cal P}_2)$\,,
$\Phi_1=\Phi_0 -2\bar{p}\Phi_2-\bar{k}\Phi_3$\,
and
$\Theta_{4}=\Theta_{5} = \Theta_{7}=\Theta_{8} = 0$.
\label{sec:Scalar matter contribution - app}
\subsubsection{Derivatives}

The relevant (functional) derivatives of the constraint contributions (\ref{HamConstH0-EMG-scalar})-(\ref{HamConstH2-EMG-scalar}) for scalar matter are
\begin{eqnarray}
    \frac{\partial\tilde{H}^{(0)}_\varphi[\bar{N}]}{\partial \bar{k}} &=&
    V_0 \bar{N} \frac{\partial{\cal P}_1}{\partial \bar{k}}\frac{\bar{\pi}^2}{2\bar{p}^{3/2}}
    + V_0 \bar{N} \frac{\partial \Phi_0}{\partial \bar{k}}\bar{\pi}
    \,,
    \nonumber\\
    \frac{\partial\tilde{H}^{(0)}_\varphi[\bar{N}]}{\partial \bar{p}} &=&
    V_0 \bar{N} \left(\frac{\partial{\cal P}_1}{\partial \bar{p}}
    - \frac{3 {\cal P}_1}{2\bar{p}}\right)\frac{\bar{\pi}^2}{2\bar{p}^{3/2}} + V_0 \bar{N} \frac{\partial \Phi_0}{\partial \bar{p}}\bar{\pi}
    \,,
    \nonumber\\
    \frac{\partial\tilde{H}^{(0)}_\varphi[\bar{N}]}{\partial \bar{\varphi}} &=& 0
    \,,
    \nonumber\\
    \frac{\partial\tilde{H}^{(0)}_\varphi[\bar{N}]}{\partial \bar{\pi}} &=&
    V_0 \bar{N} \left({\cal P}_1\frac{\bar{\pi}}{\bar{p}^{3/2}}
    + \Phi_0
    \right)
    \,,
\end{eqnarray}
\begin{eqnarray}
    \frac{\partial\tilde{H}^{(1)}_\varphi[\delta N]}{\partial \bar{k}} &=&
    \int{\rm d}^3 x \delta N \left[
    - \frac{\partial{\cal P}_2}{\partial \bar{k}} \frac{\bar{\pi}^2}{2\bar{p}^{3/2}} 
    \frac{\delta^j_c \delta E^c_j}{2\bar{p}}
    + \left(\frac{\partial{\cal P}_1}{\partial \bar{k}} + \frac{\partial{\cal P}_2}{\partial \bar{k}}\right)\frac{\bar{\pi}\delta\pi}{2\bar{p}^{3/2}}  
    \right. \nonumber\\
    && \qquad + \left(\frac{\partial{\Phi}_0}{\partial \bar{k}} - 2\bar{p}\frac{\partial{\Phi}_2}{\partial \bar{k}} - {\Phi}_3 - \bar{k}\frac{\partial{\Phi}_3}{\partial \bar{k}}\right)\delta\pi
    \nonumber\\
    && \qquad + \frac{\partial{\Phi}_2}{\partial \bar{k}} \bar{\pi} \delta^j_c \delta E^c_j + \frac{\partial{\Phi}_3}{\partial \bar{k}} \bar{\pi} \delta^c_j \delta K^j_c \, \nonumber\\
    && \left.\qquad + \frac{\partial \Theta_1}{\partial \bar{k}} \bar{\pi} \delta^{ab} 
    \partial_a \partial_b (\delta^j_c \delta E^c_j) 
    + 
    \frac{\partial \Theta_2}{\partial \bar{k}} \bar{\pi} 
    \frac{\partial_c \partial^j \delta E^c_j}{2\bar{p}}
    \right] \,, \nonumber\\
    \frac{\partial\tilde{H}^{(1)}_\varphi[\delta N]}{\partial \bar{p}} &=&
    \int{\rm d}^3 x \delta N \left[
    - \left(\frac{\partial{\cal P}_2}{\partial \bar{p}}
    - \frac{5{\cal P}_2}{2\bar{p}} \right) \frac{\bar{\pi}^2}{2\bar{p}^{3/2}} 
    \frac{\delta^j_c \delta E^c_j}{2\bar{p}}
    + \left(\frac{\partial{\cal P}_1}{\partial \bar{p}}+\frac{\partial{\cal P}_2}{\partial \bar{p}}
    - \frac{3 {\cal P}_1}{2\bar{p}}
    - \frac{3 {\cal P}_2}{2\bar{p}} \right) \frac{\bar{\pi}\delta\pi}{2\bar{p}^{3/2}}\right.
    \nonumber\\
    && \qquad +\left(\frac{\partial{\Phi}_0}{\partial \bar{p}} - 2\Phi_2 - 2\bar{p}\frac{\partial{\Phi}_2}{\partial \bar{p}} - \bar{k}\frac{\partial{\Phi}_3}{\partial \bar{p}}\right)\delta\pi
    \nonumber\\
    && \qquad +\frac{\partial{\Phi}_2}{\partial \bar{p}} \bar{\pi} \delta^j_c \delta E^c_j
    +  \frac{\partial{\Phi}_3}{\partial \bar{p}} \bar{\pi} \delta^c_j \delta K^j_c
    \nonumber\\
    && \qquad\left.
    + \frac{\partial \Theta_1}{\partial \bar{p}} \bar{\pi} \delta^{ab} \partial_a \partial_b \delta^j_c \delta E^c_j + \frac{\partial \Theta_2}{\partial \bar{p}} \bar{\pi} \frac{\partial_c \partial^j \delta E^c_j}{2\bar{p}} - \Theta_2 \bar{\pi} \frac{\partial_c \partial^j \delta E^c_j}{2\bar{p}^2}
    \right]
    \,,
    \nonumber\\
    \frac{\partial\tilde{H}^{(1)}_\varphi[\delta N]}{\partial \bar{\varphi}} &=& 0\,,\nonumber\\
    \frac{\partial\tilde{H}^{(1)}_\varphi[\delta N]}{\partial \bar{\pi}} &=&
    \int{\rm d}^3 x \delta N 
    \left[ 
    - {\cal P}_2 \frac{\bar{\pi}}{\bar{p}^{3/2}} \frac{\delta^j_c \delta E^c_j}{2\bar{p}}
    + ({\cal P}_1+{\cal P}_2) \frac{\delta\pi}{2\bar{p}^{3/2}}
    + \Phi_2 \delta^j_c \delta E^c_j + \Phi_3 \delta^c_j \delta K^j_c
    \right.\nonumber\\
    && \qquad\qquad\qquad\left.
    +\Theta_1 \delta^{ab} \partial_a \partial_b \delta^j_c \delta E^c_j
    +\Theta_2 \frac{\partial_c \partial^j \delta E^c_j}{2\bar{p}}
    \right] \,,
\end{eqnarray}

\begin{eqnarray}
    \frac{\delta\tilde{H}^{(1)}_\varphi[\delta N]}{\delta \delta E_j^c} &=&
    \delta N \delta^j_c\left[
    - \frac{{\cal P}_2}{2 \bar{p}} \frac{\bar{\pi}^2}{2\bar{p}^{3/2}}
    + \Phi_2 \bar{\pi}\right]
    + \delta^j_c \Theta_1 \bar{\pi}\delta^{ab} \partial_a \partial_b \delta N
    +\Theta_2 \bar{\pi}\frac{\partial^j \partial_c \delta N}{2\bar{p}}
    \,,
    \nonumber\\
    \frac{\delta\tilde{H}^{(1)}_\varphi[\delta N]}{\delta \delta K^i_a} &=& 
    \delta N \delta^a_i \Phi_3 \bar{\pi}
    \,,
    \nonumber\\
    \frac{\delta\tilde{H}^{(1)}_\varphi[\delta N]}{\delta \delta \varphi} &=& 0
    \,,
    \nonumber\\
    \frac{\delta\tilde{H}^{(1)}_\varphi[\delta N]}{\delta \delta \pi} &=&
    \delta N 
    \left(
    ({\cal P}_1 + {\cal P}_2 )\frac{\bar{\pi}}{2\bar{p}^{3/2}} 
    + \Phi_0 - 2\bar{p}\Phi_2 - \bar{k}\Phi_3
    \right)
    \,,
\end{eqnarray}
\begin{eqnarray}
    \frac{\delta\tilde{H}^{(2)}_\varphi[\bar{N}]}{\delta \delta E_i^a} &=&
    \bar{N} 
    \left[ 
    - \frac{{\cal P}_4}{2\bar{p}} \frac{\bar{\pi}\delta\pi}{\bar{p}^{3/2}} \delta^i_a
    + \left(\frac{2}{3}\left(\bar{k}\frac{\partial  {\cal P}_1}{\partial \bar{k}} - \bar{p}\frac{\partial  {\cal P}_1}{\partial \bar{p}}\right)+{\cal P}_1 \right)\frac{\bar{\pi}^2}{2\bar{p}^{3/2}} \frac{\delta^i_c\delta^j_a\delta E^c_j}{2\bar{p}^2}
    \right. 
    \nonumber \\
    &&\quad -\left(\frac{2}{3}\left(\bar{k}\frac{\partial  {\cal P}_1}{\partial \bar{k}} - \bar{p}\frac{\partial  {\cal P}_1}{\partial \bar{p}}\right)+{\cal P}_1 -2{\cal P}_4\right)
    \frac{\bar{\pi}^2}{2\bar{p}^{3/2}} \frac{\delta^i_a \delta^j_c\delta E^c_j}{4\bar{p}^2}
    + \frac{1}{3\bar{p}}\frac{\partial  {\cal P}_1}{\partial \bar{k}} \frac{\bar{\pi}^2}{2\bar{p}^{3/2}} \delta K_a^i
    \nonumber \\
    &&\quad -\bar{k} \Phi_5 \frac{\delta\pi}{2\bar{p}} \delta^i_a
    -\frac{2}{3}\left(\bar{k}\frac{\partial  {\cal P}_1}{\partial \bar{k}} - \bar{p}\frac{\partial  {\cal P}_1}{\partial \bar{p}}\right) \bar{\pi}\frac{\delta^i_c\delta^j_a\delta E^c_j}{4\bar{p}^2}
    \nonumber \\
    &&\quad +
    \left(
    \frac{2}{3}\left(\bar{k}\frac{\partial  {\cal P}_1}{\partial \bar{k}} - \bar{p}\frac{\partial  {\cal P}_1}{\partial \bar{p}}\right)
    + 2 \bar{k}^2 \Phi_{10}
    \right)
    \bar{\pi}\frac{\delta^i_a\delta^j_c\delta E^c_j}{4\bar{p}^2}
    + \frac{2}{3}\frac{\partial \Phi_0}{\partial \bar{k}} \frac{\bar{\pi}}{2 \bar{p}}
    \delta K_a^i
    \nonumber \\
    &&\quad \left.
    - (\Phi_5 + 2\bar{k}\Phi_{10}) \delta^c_j \delta K_c^j \frac{\bar{\pi}}{2 \bar{p}} \delta^i_a
    - \left(\Theta_3 \bar{\pi} + \Theta_6 \frac{\bar{\pi}^2}{2\bar{p}^{3/2}}\right)
    \frac{{\delta}^{ik}}{\bar{p}} \partial_a \partial_d
    \delta E^d_k
    \right.
    \nonumber \\
    &&\quad \left.
     -\bigg(
    \Theta_4 \bar{\pi} + \Theta_7 \frac{\bar{\pi}^2}{2\bar{p}^{3/2}}
    \bigg)
    \frac{{\delta}^{cb}}{\bar{p}} 
    {\delta}_{ad}{\delta}^{ik} 
    \partial_c \partial_b
    \delta E^d_k
     -\bigg(
    \Theta_5 \bar{\pi} + \Theta_8
    \frac{\bar{\pi}^2}{2\bar{p}^{3/2}}
    \bigg)
    \frac{\partial^b \partial_b \left(\delta^k_d\delta E^d_k\right)}{\bar{p}} 
    {\delta}^i_a
    \right.
    \nonumber \\
    &&\quad \left.
    -\bigg(
    \Phi_{11} \frac{\sqrt{\bar{p}}}{2}
    +
    \Theta_9 \bar{\pi}
    \bigg)
    \delta^{ac} \delta_{ij} 
    \partial_c \partial^j
    \delta \varphi
    \right] \,,\nonumber
\end{eqnarray}
\begin{eqnarray}
    \frac{\delta\tilde{H}^{(2)}_\varphi[\bar{N}]}{\delta \delta K_a^i} &=&
    \bar{N} \left[
    \frac{\bar{\pi}^2}{2\bar{p}^{3/2}} \Phi \frac{\delta E^a_i}{2 \bar{p}}
    + \Phi_5 \delta\pi \delta^a_i 
    + \frac{2}{3}\frac{\partial \Phi_0}{\partial \bar{k}}\bar{\pi}\frac{\delta E^a_i}{2\bar{p}}
    -(\Phi_5 + 2\bar{k}\Phi_{10})
    \bar{\pi}\frac{\delta^a_i \delta^j_c \delta E^c_j}{2\bar{p}}
    + 2\Phi_{10}\bar{\pi}\delta^a_i\delta^c_j\delta K_c^j
    \right]\,,
    \nonumber\\
    \frac{\delta\tilde{H}^{(2)}_\varphi[\bar{N}]}{\delta \delta \varphi} &=&
    \bar{N} \left[ - {\cal P}_8 \sqrt{\bar{p}} \delta^{a b} (\partial_a \partial_b \delta\varphi)
    -
    \left(\frac{\sqrt{\bar{p}}}{2}\Phi_{11}+\Theta_9 \bar{\pi}\right) \partial^j\partial_c \delta E^c_j \right]\,,
    \nonumber\\
    \frac{\delta\tilde{H}^{(2)}_\varphi[\bar{N}]}{\delta \delta \pi} &=&
    \bar{N}  \left[ {\cal P}_4 \frac{\delta \pi}{\bar{p}^{3/2}} - {\cal P}_4 \frac{\bar{\pi}}{\bar{p}^{3/2}} \frac{\delta^j_c\delta E^c_j}{2\bar{p}}
    -\bar{k}{\Phi}_5 \frac{\delta^j_c \delta E^c_j}{2\bar{p}} + {\Phi}_5 \delta^c_j \delta K^j_c \right] \,.
\end{eqnarray}

\subsubsection{Brackets}
It will be useful to compute
\begin{eqnarray}
    \frac{\kappa}{3 V_0} \frac{\partial\tilde{H}^{(0)}_{\rm grav}[\bar{N}_1]}{\partial\bar{k}} \frac{\partial\tilde{H}^{(1)}_\varphi[\delta N_2]}{\partial \bar{p}} 
    &=& -\int {\rm d}^3 x \, \bar{N}_1 \delta N_2 \frac{\partial {\cal K}^{(0)}}{\partial \bar{k}} 
    \Bigg[ - \frac{1}{4\bar{p}^2} \left(\frac{\partial{\cal P}_2}{\partial \bar{p}} - \frac{5{\cal P}_2}{2\bar{p}} \right) \bar{\pi}^2
    \delta^j_c \delta E^c_j
    \nonumber \\
    &&\quad + \left(\frac{\partial{\cal P}_1}{\partial \bar{p}}+\frac{\partial{\cal P}_2}{\partial \bar{p}} - \frac{3 {\cal P}_1}{2\bar{p}} - \frac{3 {\cal P}_2}{2\bar{p}} \right) \frac{\bar{\pi}\delta\pi}{2 \bar{p}} \nonumber \\
    &&\quad + \sqrt{\bar p} \left(\frac{\partial{\Phi}_0}{\partial \bar{p}} -2\Phi_2 - 2\bar{p}\frac{\partial{\Phi}_2}{\partial \bar{p}} - \bar{k}\frac{\partial{\Phi}_3}{\partial \bar{p}}\right) \delta\pi \nonumber \\
    &&\quad + \sqrt{\bar p}\frac{\partial{\Phi}_2}{\partial \bar{p}} \bar{\pi} \delta^j_c \delta E^c_j +  \sqrt{\bar p} \frac{\partial{\Phi}_3}{\partial \bar{p}} \bar{\pi} \delta^c_j \delta K^j_c  \\
    &&\quad + \frac{\partial \Theta_1}{\partial \bar{p}} \bar{\pi} \delta^{ab} \partial_a \partial_b \delta^j_c \delta E^c_j 
    + \frac{\partial \Theta_2}{\partial \bar{p}} \bar{\pi} \frac{\partial_c \partial^j \delta E^c_j}{2\bar{p}} 
    - \Theta_2 \bar{\pi} \frac{\partial_c \partial^j \delta E^c_j}{2\bar{p}^2}
    \Bigg] \,,\nonumber
\end{eqnarray}
\begin{eqnarray}
    \frac{\kappa}{3 V_0} \frac{\partial\tilde{H}^{(0)}_{\rm grav}[\bar{N}_1]}{\partial\bar{p}} \frac{\partial\tilde{H}^{(1)}_\varphi[\delta N_2]}{\partial \bar{k}} &=&
    -\int {\rm d}^3 x \bar{N}_1 \delta N_2 \left( \frac{{\cal K}^{(0)}}{2\bar{p}} + \frac{\partial {\cal K}^{(0)}}{\partial \bar{p}} \right)
    \nonumber \\
    &&\quad \times
    \Bigg[
    - \frac{\partial{\cal P}_2}{\partial \bar{k}} \frac{\bar{\pi}^2}{2\bar p} \frac{\delta^j_c \delta E^c_j}{2\bar{p}}
    + \left(\frac{\partial{\cal P}_1}{\partial \bar{k}} + \frac{\partial{\cal P}_2}{\partial \bar{k}}\right)
    \frac{\bar{\pi}\delta\pi}{2\bar{p}}
    \nonumber\\
    &&\quad
    + \sqrt{\bar p} \left(
    \frac{\partial{\Phi}_0}{\partial \bar{k}} - 2\bar{p}\frac{\partial{\Phi}_2}{\partial \bar{k}} - {\Phi}_3 - \bar{k}\frac{\partial{\Phi}_3}{\partial \bar{k}}
    \right) \delta\pi
    \nonumber \\
    &&\quad + \sqrt{\bar p} \frac{\partial{\Phi}_2}{\partial \bar{k}} \bar{\pi} \delta^j_c E^c_j
    + \sqrt{\bar p} \frac{\partial{\Phi}_3}{\partial \bar{k}} \bar{\pi} \delta^c_j K^j_c 
    \nonumber \\
    &&\quad + \frac{\partial \Theta_1}{\partial \bar{k}} \bar{\pi}\delta^{ab} \partial_a \partial_b \delta^j_c \delta E^c_j 
    + \frac{\partial \Theta_2}{\partial \bar{k}} \bar{\pi} \frac{\partial_c \partial^j \delta E^c_j}{2\bar{p}} 
    \Bigg]\,,
\end{eqnarray}

\begin{eqnarray}
    \frac{\kappa}{3 V_0}\frac{\partial \tilde{H}^{(0)}_\varphi[\bar{N}_1]}{\partial \bar{k}} \frac{\partial \tilde{H}^{(1)}_{\rm grav}[\delta N_2]}{\partial \bar{p}} &=&
    \int {\rm d}^3 x \bar{N}_1 \delta N_2 \frac{1}{3} \frac{\partial{\cal P}_1}{\partial \bar{k}} \Bigg[- 2 \left(\frac{\partial {\cal K}^{(1)}_1}{\partial \bar{p}} + \frac{{\cal K}^{(1)}_1}{2 \bar{p}}\right) \delta^c_j\delta K_c^j 
    \\
    &&\qquad\qquad\qquad
    - \left(\frac{\partial {\cal K}^{(1)}_2}{\partial \bar{p}} - \frac{{\cal K}^{(1)}_2}{2 \bar{p}}\right) \frac{\delta_c^j\delta E^c_j}{2 \bar{p}}
    \Bigg] \frac{\bar{\pi}^2}{2\bar{p}}
    \nonumber\\
    &&-
    \int {\rm d}^3 x (\bar{N}_1 \partial_j \delta N_2) \frac{1}{3} \frac{\partial{\cal P}_1}{\partial \bar{k}} \left[ \left(\frac{\partial {\cal K}^{(1)}_3}{\partial \bar{p}} - \frac{{\cal K}^{(1)}_3}{2 \bar{p}}\right) \frac{\delta^{jk}\partial_c\delta E^c_k}{2\bar{p}^2} \right] \bar{\pi}^2
    \nonumber\\
    &&-
    \int {\rm d}^3 x (\bar{N}_1 \partial_j \delta N_2) \frac{1}{3} \frac{\partial{\Phi}_0}{\partial \bar{k}} \left[ \left(\frac{\partial {\cal K}^{(1)}_3}{\partial \bar{p}} - \frac{{\cal K}^{(1)}_3}{2 \bar{p}}\right) \frac{\delta^{jk}\partial_c\delta E^c_k}{\sqrt{\bar{p}}} \right] \bar{\pi}
    \,, \nonumber
\end{eqnarray}
\begin{eqnarray}
    \frac{\kappa}{3 V_0}\frac{\partial \tilde{H}^{(0)}_\varphi[\bar{N}_1]}{\partial \bar{p}} \frac{\partial \tilde{H}^{(1)}_{\rm grav}[\delta N_2]}{\partial \bar{k}} &=&
    \int{\rm d}^3 x \bar{N}_1 \delta N_2 \frac{1}{3} \left( \frac{\partial{\cal P}_1}{\partial \bar{p}}
    - \frac{3 {\cal P}_1}{2\bar{p}}\right)
    \left[- \frac{1}{\bar{p}} \frac{\partial {\cal K}^{(1)}_1}{\partial \bar{k}} \delta^c_j\delta K_c^j
    - \frac{\partial {\cal K}^{(1)}_2}{\partial \bar{k}} \frac{\delta_c^j\delta E^c_j}{4 \bar{p}^2} \right] \bar{\pi}^2
    \nonumber\\
    &&
    - \int{\rm d}^3 x (\bar{N}_1 \partial_c \delta N_2) \delta^{ce} \frac{1}{3} \left( \frac{\partial{\cal P}_1}{\partial \bar{p}}
    - \frac{3 {\cal P}_1}{2\bar{p}}\right) \frac{\partial {\cal K}^{(1)}_3}{\partial \bar{k}} \frac{\bar{\pi}^2}{2\bar{p}} \frac{\delta^k_e\partial_d\delta E^d_k}{\bar{p}} \nonumber\\
    && + \int {\rm d}^3 x (\bar{N}_1 \delta N_2) \frac{\sqrt{\bar{p}}}{3} 
    \frac{\partial \Phi_0}{\partial \bar{p}} \left[- 2 \frac{\partial {\cal K}^{(1)}_1}{\partial \bar{k}} \delta^c_j\delta K_c^j
    - \frac{\partial {\cal K}^{(1)}_2}{\partial \bar{k}} \frac{\delta_c^j\delta E^c_j}{2\bar{p}}
    \right]\bar{\pi} \nonumber \\
     &&  - \int{\rm d}^3 x (\bar{N}_1 \partial_c \delta N_2) \frac{\sqrt{\bar{p}}}{3} 
    \frac{\partial \Phi_0}{\partial \bar{p}} \frac{\partial {\cal K}^{(1)}_3}{\partial \bar{k}} \bar{\pi} \frac{\delta^{jk}\partial_j\delta E^c_k}{\bar{p}} \,,
\end{eqnarray}
\begin{eqnarray}
    \frac{\kappa}{3 V_0} \frac{\partial \tilde{H}^{(0)}_\varphi[\bar{N}_1]}{\partial \bar{k}}
    \frac{\partial\tilde{H}^{(1)}_\varphi[\delta N_2]}{\partial \bar{p}} 
    &=&
    \int{\rm d}^3 x  \bar{N}_1\delta N_2 
    \frac{\kappa}{3}
    \left[
    \frac{\partial \Phi_0}{\partial \bar{k}} \bar{\pi}
    + \frac{\partial{\cal P}_1}{\partial \bar{k}}\frac{\bar{\pi}^2}{2\bar{p}^{3/2}}
    \right]
    \nonumber \\
    &&\qquad
    \Bigg[
    - \left(\frac{\partial{\cal P}_2}{\partial \bar{p}}
    - \frac{5{\cal P}_2}{2\bar{p}} \right) \frac{\bar{\pi}^2}{2\bar{p}^{3/2}} 
    \frac{\delta^j_c \delta E^c_j}{2\bar{p}}
    \nonumber \\
    &&\qquad
    + \left(\frac{\partial{\cal P}_1}{\partial \bar{p}}+\frac{\partial{\cal P}_2}{\partial \bar{p}}
    - \frac{3 {\cal P}_1}{2\bar{p}}
    - \frac{3 {\cal P}_2}{2\bar{p}} \right) \frac{\bar{\pi}\delta\pi}{2 \bar{p}^{3/2}}
    \nonumber\\
    &&\qquad +\left(\frac{\partial{\Phi}_0}{\partial \bar{p}} - 2\Phi_2 - 2\bar{p}\frac{\partial{\Phi}_2}{\partial \bar{p}}
    - \bar{k}\frac{\partial{\Phi}_3}{\partial \bar{p}}\right)\delta\pi \\
    &&\qquad +\frac{\partial{\Phi}_2}{\partial \bar{p}} \bar{\pi} \delta^j_c \delta E^c_j
    +  \frac{\partial{\Phi}_3}{\partial \bar{p}} \bar{\pi} \delta^c_j \delta K^j_c 
    \nonumber\\
    &&\qquad + \frac{\partial \Theta_1}{\partial \bar{p}} \pi \delta^{ab} \partial_a \partial_b \delta^j_c \delta E^c_j 
    + \frac{\partial \Theta_2}{\partial \bar{p}} \pi \frac{\partial_c \partial^j \delta E^c_j}{2\bar{p}} 
    - \Theta_2 \pi \frac{\partial_c \partial^j \delta E^c_j}{2\bar{p}^2}
    \Bigg]
    \,, \nonumber
\end{eqnarray}

\begin{eqnarray}
    \frac{\kappa}{3 V_0} \frac{\partial \tilde{H}^{(0)}_\varphi[\bar{N}_1]}{\partial \bar{p}}
    \frac{\partial\tilde{H}^{(1)}_\varphi[\delta N_2]}{\partial \bar{k}} &=&
    \int{\rm d}^3 x \bar{N}_1 \delta N_2 
    \frac{\kappa}{3}
    \left[\frac{\partial \Phi_0}{\partial \bar{p}}\bar{\pi}
    + \left(\frac{\partial{\cal P}_1}{\partial \bar{p}}
    - \frac{3 {\cal P}_1}{2\bar{p}}\right)\frac{\bar{\pi}^2}{2\bar{p}^{3/2}} \right]
    \nonumber \\
    &&\quad 
    \left[
    - \frac{\partial{\cal P}_2}{\partial \bar{k}} \frac{\bar{\pi}^2}{2\bar{p}^{3/2}} 
    \frac{\delta^j_c \delta E^c_j}{2\bar{p}}
    + \left(\frac{\partial{\cal P}_1}{\partial \bar{k}} + \frac{\partial{\cal P}_2}{\partial \bar{k}}\right)\frac{\bar{\pi}\delta\pi}{2\bar{p}^{3/2}}  
    \right. \nonumber\\
    && \qquad + \left(\frac{\partial{\Phi}_0}{\partial \bar{k}} - 2\bar{p}\frac{\partial{\Phi}_2}{\partial \bar{k}} - {\Phi}_3 - \bar{k}\frac{\partial{\Phi}_3}{\partial \bar{k}}\right)\delta\pi 
    \nonumber \\
    && \left.\qquad + \frac{\partial{\Phi}_2}{\partial \bar{k}} \bar{\pi} \delta^j_c \delta E^c_j + \frac{\partial{\Phi}_3}{\partial \bar{k}} \bar{\pi} \delta^c_j \delta K^j_c \right. \nonumber \\
    && \left.\qquad + \frac{\partial \Theta_1}{\partial \bar{k}} \pi \delta^{ab} \partial_a \partial_b \delta^j_c \delta E^c_j 
    + \frac{\partial \Theta_2}{\partial \bar{k}} \pi \frac{\partial_c \partial^j \delta E^c_j}{2\bar{p}} 
    \right].
\end{eqnarray}

Here we use the fact that $\mathcal{K}^{(2)}_1 = \mathcal{K}^{(2)}_2$,
\begin{eqnarray}
    \int{\rm d}^3x \kappa \frac{\delta\tilde{H}^{(2)}_{\rm grav}[\bar{N}_1]}{\delta \delta K_a^i}
    \frac{\delta\tilde{H}^{(1)}_\varphi[\delta N_2]}{\delta \delta E^a_i} &=&
     \int{\rm d}^3x \bar{N}_1 \delta N_2 {\cal P}_2 \left( \frac{{\cal K}^{(2)}_1}{2\bar{p}^2} \delta^d_k \delta K_d^k
    + \frac{{\cal K}^{(2)}_3}{4\bar{p}^3} \delta^k_d \delta E^d_k \right) \bar{\pi}^2
    \\
    &&\quad +
     \int{\rm d}^3x \bar{N}_1 \delta N_2 \sqrt{\bar{p}} {\Phi}_2 \left( {-2\cal K}^{(2)}_1 \delta^d_k \delta K_d^k
    -\frac{{\cal K}^{(2)}_3}{\bar{p}} \delta^k_d \delta E^d_k \right) \bar{\pi}
    \nonumber
    \\
    &&\quad +
     \int{\rm d}^3x \bar{N}_1 
     {\partial}_c
     \delta N_2
     \left( {-2\sqrt{\bar{p}}\cal K}^{(2)}_1 \Theta_1
     -{\cal K}^{(2)}_1\frac{\Theta_2}{\sqrt{\bar{p}}}
     \right) \bar{\pi} \delta^{cd} \partial_k \delta K_d^k
    \nonumber
    \\
    &&\quad +
     \int{\rm d}^3x \bar{N}_1 
     {\partial}_c
     \delta N_2  
     \left( 
     -\frac{{\cal K}^{(2)}_3}{\sqrt{\bar{p}}} \Theta_1
     -\frac{{\cal K}^{(2)}_3}{2\bar{p}^{3/2}} \Theta_2
     \right) \bar{\pi} \delta^{ck} \partial_d \delta E_k^d
    \,,
    \nonumber
\end{eqnarray}

\begin{eqnarray}
    \int{\rm d}^3x \kappa
    \frac{\delta\tilde{H}^{(2)}_{\rm grav}[\bar{N}_1]}{\delta \delta E^a_i}
    \frac{\delta\tilde{H}^{(1)}_\varphi[\delta N_2]}{\delta \delta K_a^i} &=&
    \int{\rm d}^3x \bar{N}_1 \delta N_2   \Bigg( 
    - {\cal K}^{(2)}_3 \frac{\delta^c_j\delta K_c^j}{\sqrt{\bar{p}}} \\
    &&\qquad\qquad\qquad
    + \left(\frac{3{\cal K}^{(2)}_5}{2}-{\cal K}^{(2)}_4\right) \frac{\delta_c^j\delta E^c_j}{2\bar{p}^{3/2}}
    \Bigg)\Phi_3 \bar{\pi}
    \nonumber\\
    &&
    -\int {\rm d}^3x (\bar{N}_1 \partial_c\delta N_2) {\cal K}^{(2)}_6 {\Phi}_3 \frac{\bar{\pi}}{2\bar{p}^{3/2}} {\delta}^{ck}\partial_d \delta E^d_k
    \,,\nonumber
\end{eqnarray}
\begin{eqnarray}
    \int {\rm d}^3x \kappa \frac{\delta \tilde{H}^{(2)}_\varphi[\bar{N}_1]}{\delta K_a^i}\frac{\delta \tilde{H}^{(1)}_{\rm grav}[\delta N_2]}{\delta E^a_i} 
    &=&
    \int \bar{N}_1 \delta N_2 \Bigg[
    - \frac{3}{2 \sqrt{\bar{p}}} {\cal K}^{(1)}_2 {\Phi}_5 \delta \pi
    - 3 {\cal K}^{(1)}_2 \Phi_{10} \frac{\bar{\pi}}{\sqrt{\bar{p}}}\delta_k^d \delta K^k_d
    \nonumber\\
    &&+
    \Bigg(- {\cal K}^{(1)}_2 \Phi  \frac{\bar{\pi}^2}{8 \bar{p}^3} 
    -\frac{2}{3} {\cal K}^{(1)}_2 \frac{\partial \Phi_0}{\partial \bar{k}} \frac{\bar{\pi}}{4\bar{p}^{3/2}}
    + 3 {\cal K}^{(1)}_2 \left(
    \Phi_5 + 2\bar{k}\Phi_{10}
    \right)
    \frac{\bar{\pi}}{4 \bar{p}^{3/2}}\Bigg)\delta_d^k \delta E^d_k
    \Bigg]
    \nonumber\\
    &&
    + \int {\rm d}^3x (\bar{N}_1 \partial_c\delta N_2)\Bigg[
    \Phi  \frac{\bar{\pi}^2}{4 \bar{p}^3}
    +\frac{1}{3}\frac{\partial \Phi_0}{\partial \bar{k}}\frac{\bar{\pi}}{\bar{p}^{3/2}}
    \nonumber\\
    &&\qquad\qquad\qquad\qquad\quad
    -(\Phi_{5} + 2\bar{k}\Phi_{10})\frac{\bar{\pi}}{2\bar{p}^{3/2}}
    \Bigg]
    {\cal K}^{(1)}_3\delta^{ce} \delta^k_e \partial_d\delta E^d_k
    \nonumber\\
    &&
    -\int {\rm d}^3x (\bar{N}_1 \partial_c\delta N_2)
    2\Phi_{10}
    {\cal K}^{(1)}_3 \frac{\bar{\pi}}{\sqrt{\bar{p}}} \delta^{ce} \delta^d_e \partial_k\delta K^k_d
    \,,
\end{eqnarray}
\begin{eqnarray}
    \int {\rm d}^3x \kappa \frac{\delta \tilde{H}^{(2)}_\varphi[\bar{N}_1]}{\delta E^a_i}\frac{\delta \tilde{H}^{(1)}_{\rm grav}[\delta N_2]}{\delta K_a^i} &=&
    \int {\rm d}^3x \bar{N}_1 \delta N_2 
    \Bigg[ \frac{3 {\cal P}_4 {\cal K}^{(1)}_1}{\bar{p}^2} \bar{\pi} \delta\pi
    - {\cal P}_6 \frac{{\cal K}^{(1)}_1}{2\bar{p}^3} \bar{\pi}^2 \delta^k_d\delta E^d_k
    \\
    &&\qquad
    + \frac{3}{4\bar{p}^3} \left({\cal P}_6 - 2 {\cal P}_4\right) {\cal K}^{(1)}_1
    \bar{\pi}^2 \delta^k_d\delta E^d_k
    \nonumber
    \\
    &&\qquad
    -\frac{\partial {\cal P}_1}{\partial \bar{k}}
    \frac{{\cal K}^{(1)}_1}{3 \bar{p}^2}
    \bar{\pi}^2 \delta^d_k\delta K_d^k
    + \frac{3\bar{k}}{\sqrt{\bar{p}}} {\cal K}^{(1)}_1 {\Phi}_5 {\delta \pi}
    \nonumber\\
    &&\qquad
    +\frac{2}{3}\left(\bar{k}\frac{\partial {\cal P}_1}{\partial \bar{k}}
    -\bar{p}\frac{\partial {\cal P}_1}{\partial \bar{p}}\right) 
    {\cal K}^{(1)}_1 {\bar{\pi}} \frac{\delta^k_d E^d_k}{2 \bar{p}^{3/2}} 
    \nonumber
     \\
     &&\qquad
     -3\left(
    \frac{2}{3}\left(\bar{k}\frac{\partial {\cal P}_1}{\partial \bar{k}}
    -\bar{p}\frac{\partial {\cal P}_1}{\partial \bar{p}}\right)
    + 2\bar{k}^2 {\Phi}_{10}\right) 
    {\cal K}^{(1)}_1 {\bar{\pi}} \frac{\delta^k_d E^d_k}{2\bar{p}^{3/2}} 
     \nonumber
     \\
     &&\qquad
     + \left(3 {\cal K}^{(1)}_1 ({\Phi}_5 + 2\bar{k}{\Phi}_{10})
    -\frac{2}{3} {\cal K}^{(1)}_1 \frac{\partial \Phi_0}{\partial \bar{k}} \right) \frac{\bar{\pi}}{\sqrt{\bar{p}}} \delta^d_k\delta K_d^k
     \nonumber
    \Bigg]
    \nonumber\\
    &&
    + \int {\rm d}^3x (\bar{N}_1 \partial_c\delta N_2)
    {\cal K}^{(1)}_1 \Bigg[
    \Theta_3
    \frac{2\bar{\pi} }{\sqrt{\bar{p}}}
    +
    \Theta_6
    \frac{\bar{\pi}^2}{\bar{p}^{5/2}}
    \Bigg]
    \delta^{ck}\partial_d\delta E^d_k
    \nonumber\\
    &&+ \int{\rm d}^3x \bar{N}_1 \delta N_2 
    \bigg(
    \bar{p}\Phi_{11}
    +
    2\sqrt{\bar{p}} \Theta_9 \bar{\pi}
    \bigg)
    {\cal K}^{(1)}_1
    \delta^{ck} \partial_c\partial_d \delta E^d_k
    \,,
    \nonumber
\end{eqnarray}
The relevant brackets are therefore
\begin{eqnarray}
    &&\{\tilde{H}^{(0)}_{\rm grav}[\bar{N}_1],\tilde{H}^{(1)}_\varphi[\delta N_2]\}_{\bar{A}}
    = \int {\rm d}^3 x \frac{\kappa}{3V_0} \left(\frac{\partial\tilde{H}^{(0)}_{\rm grav}[\bar{N}_1]}{\partial \bar{k}}\frac{\partial\tilde{H}^{(1)}_\varphi[\delta N_2]}{\partial\bar{p}}
    - 
    \frac{\partial\tilde{H}^{(0)}_{\rm grav}[\bar{N}_1]}{\partial \bar{p}}\frac{\partial\tilde{H}^{(1)}_\varphi[\delta N_2]}{\partial \bar{k}}\right)
    \nonumber\\
    &&\quad
    =
    \int {\rm d}^3 x \bar{N}_1 \delta N_2 
    \Bigg[
    \Bigg(- \frac{\partial {\cal K}^{(0)}}{\partial \bar{k}}
    \left(\frac{\partial{\cal P}_1}{\partial \bar{p}}+\frac{\partial{\cal P}_2}{\partial \bar{p}} - \frac{3 {\cal P}_1}{2\bar{p}} - \frac{3 {\cal P}_2}{2\bar{p}} \right)
    \\
    &&\qquad
    +
    \left( \frac{{\cal K}^{(0)}}{2\bar{p}} + \frac{\partial {\cal K}^{(0)}}{\partial \bar{p}}\right) 
    \left(
    \frac{\partial{\cal P}_1}{\partial \bar{k}}
    +
    \frac{\partial{\cal P}_2}{\partial \bar{k}}
    \right)
    \Bigg) \bar{\pi}
    \frac{\delta\pi}{2\bar{p}}
    \nonumber\\
    &&\qquad
    + \left(\frac{\partial {\cal K}^{(0)}}{\partial \bar{k}} \left(\frac{\partial{\cal P}_2}{\partial \bar{p}}
    - \frac{5 {\cal P}_2}{2\bar{p}} \right)
    - \left( \frac{{\cal K}^{(0)}}{2\bar{p}} + \frac{\partial {\cal K}^{(0)}}{\partial \bar{p}}\right) 
    \frac{\partial{\cal P}_2}{\partial \bar{k}} 
    \right) \bar{\pi}^2 \frac{\delta^k_d E^d_k}{4\bar{p}^2}
    \nonumber\\
    &&\qquad
    +
    \Bigg(
    -\frac{\partial {\cal K}^{(0)}}{\partial \bar{k}}
    \left(
    \frac{\partial{\Phi}_0}{\partial \bar{p}}
    -
    2{\Phi}_2
    -
    2\bar{p}\frac{\partial {\Phi}_2}{\partial \bar{p}}
    - 
    \bar{k}\frac{\partial {\Phi}_3}{\partial \bar{p}} 
    \right)
    \nonumber\\
    &&\qquad
    +
    \left( \frac{{\cal K}^{(0)}}{2\bar{p}} + \frac{\partial {\cal K}^{(0)}}{\partial \bar{p}}\right) 
    \left(
    \frac{\partial{\Phi}_0}{\partial \bar{k}}
    -
    2\bar{p}\frac{\partial {\Phi}_2}{\partial \bar{k}}
     -
    {\Phi}_3
    - 
    \bar{k}\frac{\partial {\Phi}_3}{\partial \bar{k}} 
    \right)
    \Bigg)\sqrt{\bar{p}}\delta \pi
    \nonumber\\
    &&\qquad
    +\left(
    -\frac{\partial {\cal K}^{(0)}}{\partial \bar{k}}
    \frac{\partial \Phi_2}{\partial \bar{p}}
    +
    \left( \frac{{\cal K}^{(0)}}{2\bar{p}} + \frac{\partial {\cal K}^{(0)}}{\partial \bar{p}}\right)
    \frac{\partial \Phi_2}{\partial \bar{k}}
    \right)
    \sqrt{\bar{p}}\bar{\pi}\delta^j_c \delta E^c_j
    \nonumber\\
    &&\qquad
    +\left(
    -\frac{\partial {\cal K}^{(0)}}{\partial \bar{k}}
    \frac{\partial \Phi_3}{\partial \bar{p}}
    +
    \left( \frac{{\cal K}^{(0)}}{2\bar{p}} + \frac{\partial {\cal K}^{(0)}}{\partial \bar{p}}\right)
    \frac{\partial \Phi_3}{\partial \bar{k}}
    \right)
    \sqrt{\bar{p}}\bar{\pi}\delta^c_j \delta K^j_c
    \Bigg]
    \nonumber\\
    &&\qquad
    + \int {\rm d}^3 x (\bar{N}_1 \partial_c\delta N_2)
     \Bigg[\frac{\partial {\cal K}^{(0)}}{\partial \bar{k}} 
     \frac{\partial \Theta_1}{\partial \bar{p}}
     -
     \left( \frac{{\cal K}^{(0)}}{2\bar{p}} + \frac{\partial {\cal K}^{(0)}}{\partial \bar{p}}\right)
     \frac{\partial \Theta_1}{\partial \bar{k}} \Bigg]
    \bar{\pi} \delta^{ck} \partial_d \delta E^d_k
    \nonumber\\
    &&\qquad
    + \int {\rm d}^3 x (\bar{N}_1 \partial_c\delta N_2)
     \frac{1}{2\bar{p}}
     \Bigg[\frac{\partial {\cal K}^{(0)}}{\partial \bar{k}} 
     \bigg(
     \frac{\partial \Theta_2}{\partial \bar{p}}
     -
     \frac{\Theta_2}{\bar{p}}
     \bigg)
     -
     \left( \frac{{\cal K}^{(0)}}{2\bar{p}} + \frac{\partial {\cal K}^{(0)}}{\partial \bar{p}}\right)
     \frac{\partial \Theta_2}{\partial \bar{k}} \Bigg]
    \bar{\pi} \delta^{ck} \partial_d \delta E^d_k
    \,,\nonumber
\end{eqnarray}
\begin{eqnarray}
    &&\{\tilde{H}^{(0)}_\varphi[\bar{N}_1],\tilde{H}^{(1)}_{\rm grav}[\delta N_2]\}_{\bar{A}}
    = \int {\rm d}^3 x \frac{\kappa}{3V_0} \left(\frac{\partial\tilde{H}^{(0)}_\varphi[\bar{N}_1]}{\partial \bar{k}}\frac{\partial\tilde{H}^{(1)}_{\rm grav}[\delta N_2]}{\partial\bar{p}} - \frac{\partial\tilde{H}^{(0)}_\varphi[\bar{N}_1]}{\partial \bar{p}}\frac{\partial\tilde{H}^{(1)}_{\rm grav}[\delta N_2]}{\partial \bar{k}}\right)
    \nonumber\\
    &&\quad =
    \int {\rm d}^3 x \bar{N}_1 \delta N_2
    \Bigg[
    - \frac{2}{3} \left(\frac{\partial{\cal P}_1}{\partial \bar{k}} \left(\frac{\partial {\cal K}^{(1)}_1}{\partial \bar{p}} + \frac{{\cal K}^{(1)}_1}{2 \bar{p}}\right) - \left( \frac{\partial{\cal P}_1}{\partial \bar{p}}
    - \frac{3 {\cal P}_1}{2\bar{p}}\right) \frac{\partial {\cal K}^{(1)}_1}{\partial \bar{k}} \right) \bar{\pi}^2 \frac{\delta^c_j\delta K_c^j}{2\bar{p}}
    \nonumber\\
    &&\qquad\qquad
    - \frac{1}{3} \left(\frac{\partial{\cal P}_1}{\partial \bar{k}} \left(\frac{\partial {\cal K}^{(1)}_2}{\partial \bar{p}} - \frac{{\cal K}^{(1)}_2}{2 \bar{p}}\right) - \left( \frac{\partial{\cal P}_1}{\partial \bar{p}}
    - \frac{3 {\cal P}_1}{2\bar{p}}\right) \frac{\partial {\cal K}^{(1)}_2}{\partial \bar{k}} \right) \bar{\pi}^2 \frac{\delta_c^j\delta E^c_j}{4 \bar{p}^2}
    \Bigg]
    \nonumber\\
    &&\qquad\qquad
    - \frac{2}{3} \left(\frac{\partial{\Phi}_0}{\partial \bar{k}} \left(\frac{\partial {\cal K}^{(1)}_1}{\partial \bar{p}} + \frac{{\cal K}^{(1)}_1}{2 \bar{p}}\right) -\frac{\partial{\Phi}_0}{\partial \bar{p}} \frac{\partial {\cal K}^{(1)}_1}{\partial \bar{k}} \right) \sqrt{\bar{p}}\bar{\pi}\delta^c_j\delta K_c^j
    \nonumber\\
    &&\qquad\qquad
    - \frac{1}{6} \left(\frac{\partial{\Phi}_0}{\partial \bar{k}} \left(\frac{\partial {\cal K}^{(1)}_2}{\partial \bar{p}} - \frac{{\cal K}^{(1)}_2}{2 \bar{p}}\right) - \frac{\partial{\Phi}_0}{\partial \bar{p}} \frac{\partial {\cal K}^{(1)}_2}{\partial \bar{k}} \right) \sqrt{\bar{p}}\bar{\pi}\delta_c^j\delta E^c_j
    \Bigg]
    \nonumber\\
    &&\qquad
    - \int {\rm d}^3 x (\bar{N}_1 \partial_c\delta N_2) \frac{1}{3}
    \Bigg[
    \left( \frac{\partial{\cal P}_1}{\partial \bar{k}} \left(\frac{\partial {\cal K}^{(1)}_3}{\partial \bar{p}} - \frac{{\cal K}^{(1)}_3}{2 \bar{p}}\right) - \left( \frac{\partial{\cal P}_1}{\partial \bar{p}}
    - \frac{3 {\cal P}_1}{2\bar{p}}\right) \frac{\partial {\cal K}^{(1)}_3}{\partial \bar{k}} \right)
    \frac{\bar{\pi}^2}{2\bar{p}^2}
    \nonumber\\
    &&\qquad
    +
    \left( \frac{\partial{\Phi}_0}{\partial \bar{k}} \left(\frac{\partial {\cal K}^{(1)}_3}{\partial \bar{p}} - \frac{{\cal K}^{(1)}_3}{2 \bar{p}}\right) -\frac{\partial{\Phi}_0}{\partial \bar{p}} \frac{\partial {\cal K}^{(1)}_3}{\partial \bar{k}} \right)\frac{\bar{\pi}}{\sqrt{\bar{p}}}
    \Bigg]\delta^{ck} \partial_d\delta E^d_k
    \,,
\end{eqnarray}
\begin{eqnarray}
    &&\{\tilde{H}^{(0)}_\varphi[\bar{N}_1],\tilde{H}^{(1)}_\varphi[\delta N_2]\}_{\bar{A}} 
    \nonumber\\
    &&=
    \kappa \int {\rm d}^3 x \bar{N}_1 \delta N_2
    \Bigg[
    \left(
    \frac{\partial{\cal P}_1}{\partial \bar{k}}
    \left(
    \frac{\partial{\cal P}_1}{\partial \bar{p}}
    +
    \frac{\partial{\cal P}_2}{\partial \bar{p}}
    - 
    \frac{3 {\cal P}_1}{2\bar{p}} 
    - 
    \frac{3 {\cal P}_2}{2\bar{p}} 
    \right)
    - \left( \frac{\partial{\cal P}_1}{\partial \bar{p}}
    - \frac{3 {\cal P}_1}{2\bar{p}}\right)
    \left(
    \frac{\partial{\cal P}_1}{\partial \bar{k}}
    +
    \frac{\partial{\cal P}_2}{\partial \bar{k}}
    \right)
    \right)
    \bar{\pi}^3 \frac{\delta\pi}{12\bar{p}^{3}}
    \nonumber\\
    &&
    + 
    \left(
    \left( 
    \frac{\partial{\cal P}_1}{\partial \bar{p}}
    - \frac{3 {\cal P}_1}{2\bar{p}}\right)
    \frac{\partial{\cal P}_2}{\partial \bar{k}}
    - \frac{\partial{\cal P}_1}{\partial \bar{k}} \left(\frac{\partial{\cal P}_2}{\partial \bar{p}}
    - \frac{5{\cal P}_2}{2\bar{p}} \right) 
    \right) 
    \bar{\pi}^4 \frac{\delta^j_c \delta E^c_j}{24\bar{p}^4}
    \nonumber\\
    &&
    +
    \left(
    \frac{\partial{\Phi}_0}{\partial \bar{k}}
    \left(
    \frac{\partial{\cal P}_1}{\partial \bar{p}}
    +
    \frac{\partial{\cal P}_2}{\partial \bar{p}}
    - 
    \frac{3 {\cal P}_1}{2\bar{p}} 
    - 
    \frac{3 {\cal P}_2}{2\bar{p}} 
    \right)
    - 
    \frac{\partial{\Phi}_0}{\partial \bar{p}}
    \left(
    \frac{\partial{\cal P}_1}{\partial \bar{k}}
    +
    \frac{\partial{\cal P}_2}{\partial \bar{k}}
    \right)
    \right)
    \bar{\pi}^2 \frac{\delta\pi}{6\bar{p}^{3/2}}
    \nonumber\\
    &&
    + 
    \left(
    \frac{\partial{\Phi}_0}{\partial \bar{p}}
    \frac{\partial{\cal P}_2}{\partial \bar{k}}
    - \frac{\partial{\Phi}_0}{\partial \bar{k}} \left(\frac{\partial{\cal P}_2}{\partial \bar{p}}
    - \frac{5{\cal P}_2}{2\bar{p}} \right)
    \right) \bar{\pi}^3 \frac{\delta^j_c \delta E^c_j}{12\bar{p}^{5/2}}
    \nonumber\\
    &&
    +
    \left(
    \frac{\partial{\cal P}_1}{\partial \bar{k}}
    \left(
    \frac{\partial{\Phi}_0}{\partial \bar{p}}
    -
    2{\Phi}_2
    -
    2\bar{p}\frac{\partial {\Phi}_2}{\partial \bar{p}}
    - 
    \bar{k}\frac{\partial {\Phi}_3}{\partial \bar{p}} 
    \right)
    - \left( \frac{\partial{\cal P}_1}{\partial \bar{p}}
    - \frac{3 {\cal P}_1}{2\bar{p}}\right)
    \left(
    \frac{\partial{\Phi}_0}{\partial \bar{k}}
    -
    2\bar{p}\frac{\partial {\Phi}_2}{\partial \bar{k}}
    -
    {\Phi}_3
    - 
    \bar{k}\frac{\partial {\Phi}_3}{\partial \bar{k}} 
    \right)
    \right)
    \bar{\pi}^3 \frac{\delta\pi}{6\bar{p}^{3/2}}
    \nonumber\\
    &&
    +
    \left(
    \frac{\partial{\Phi}_0}{\partial \bar{k}}
    \left(
    \frac{\partial{\Phi}_0}{\partial \bar{p}}
    -
    2{\Phi}_2
    -
    2\bar{p}\frac{\partial {\Phi}_2}{\partial \bar{p}}
    - 
    \bar{k}\frac{\partial {\Phi}_3}{\partial \bar{p}} 
    \right)
    - \frac{\partial{\Phi}_0}{\partial \bar{p}}
    \left(
    \frac{\partial{\Phi}_0}{\partial \bar{k}}
    -
    2\bar{p}\frac{\partial {\Phi}_2}{\partial \bar{k}}
    -
    {\Phi}_3
    - 
    \bar{k}\frac{\partial {\Phi}_3}{\partial \bar{k}} 
    \right)
    \right)
    \bar{\pi} \frac{\delta\pi}{3}
    \nonumber\\
    &&
    + 
    \left(
    \frac{\partial{\cal P}_1}{\partial \bar{k}}
    \frac{\partial{\Phi}_2}{\partial \bar{p}} 
    -\left( 
    \frac{\partial{\cal P}_1}{\partial \bar{p}}
    - \frac{3 {\cal P}_1}{2\bar{p}}\right)
    \frac{\partial{\Phi}_2}{\partial \bar{k}}
    \right) 
    \bar{\pi}^3 \frac{\delta^j_c \delta E^c_j}{6\bar{p}^{3/2}}
     \nonumber\\
    &&
    + 
    \left(
    \frac{\partial{\cal P}_1}{\partial \bar{k}}
    \frac{\partial{\Phi}_3}{\partial \bar{p}} 
    -\left( 
    \frac{\partial{\cal P}_1}{\partial \bar{p}}
    - \frac{3 {\cal P}_1}{2\bar{p}}\right)
    \frac{\partial{\Phi}_3}{\partial \bar{k}}
    \right) 
    \bar{\pi}^3 \frac{\delta_j^c \delta K_c^j}{6\bar{p}^{3/2}}
    \nonumber\\
    &&
    + 
    \left(
    \frac{\partial{\Phi}_0}{\partial \bar{k}}
    \frac{\partial{\Phi}_2}{\partial \bar{p}} 
    -
    \frac{\partial{\Phi}_0}{\partial \bar{p}}
    \frac{\partial{\Phi}_2}{\partial \bar{k}}
    \right) 
    \bar{\pi}^2 \frac{\delta^j_c \delta E^c_j}{3}
    \nonumber\\
    &&
    + 
    \left(
    \frac{\partial{\Phi}_0}{\partial \bar{k}}
    \frac{\partial{\Phi}_3}{\partial \bar{p}} 
    -
    \frac{\partial{\Phi}_0}{\partial \bar{p}}
    \frac{\partial{\Phi}_3}{\partial \bar{k}}
    \right) 
    \bar{\pi}^2 \frac{\delta^c_j \delta K^j_c}{3}
    \Bigg]
    \,,
\end{eqnarray}
\begin{eqnarray}
    \{\tilde{H}^{(2)}_{\rm grav}[\bar{N}_1],\tilde{H}^{(1)}_\varphi[\delta N_2]\}_\delta &=&
    \int{\rm d}^3x \bar{N}_1 \delta N_2
    \Bigg[ 
    \Bigg(
    {\cal P}_2{\cal K}^{(2)}_1\frac{\bar{\pi}^2}{2\bar{p}^2}
    -{2{\Phi}_2\cal K}^{(2)}_1 \bar{\pi} \sqrt{\bar{p}}
    + \Phi_3{\cal K}^{(2)}_3 \frac{\bar{\pi}}{\sqrt{\bar{p}}} 
    \Bigg)
    \delta^d_k \delta K_d^k
    \nonumber\\
    &&
    +
    \Bigg({\cal P}_2 {\cal K}^{(2)}_3\frac{\bar{\pi}^2}{4\bar{p}^3}
    -{\Phi}_2{\cal K}^{(2)}_3\frac{ \bar{\pi}}{\sqrt{\bar{p}}}
    - \Phi_3{\cal K}^{(2)}_4 \frac{\bar{\pi}}{4\bar{p}^{3/2}}\Bigg)
    \delta_k^d \delta E^d_k
    \Bigg]
    \\
    &&
    +\int{\rm d}^3x \bar{N}_1 \delta N_2  \left( {-2\sqrt{\bar{p}}\cal K}^{(2)}_1 \Theta_1
     -{\cal K}^{(2)}_1\frac{\Theta_2}{\sqrt{\bar{p}}}
     \right) \bar{\pi} \delta^{cd} \partial_k \delta K_d^k
     \nonumber\\
    &&
    +\int{\rm d}^3x \bar{N}_1 \delta N_2 \left( 
     -\frac{{\cal K}^{(2)}_3}{\sqrt{\bar{p}}} \Theta_1
     -\frac{{\cal K}^{(2)}_3}{2\bar{p}^{3/2}} \Theta_2
     + {\Phi}_3 \frac{{\cal K}^{(2)}_6}{2\bar{p}^{3/2}}
     \right) \bar{\pi} \delta^{ck} \partial_d \delta E_k^d
    \,,\nonumber
\end{eqnarray}
\begin{eqnarray}
    &&\{\tilde{H}^{(2)}_\varphi[\bar{N}_1],\tilde{H}^{(1)}_{\rm grav}[\delta N_2]\}_\delta
    = \int \bar{N}_1 \delta N_2 
    \Bigg[
    \left(
    -3\frac{{\Phi}_5 {\cal K}^{(1)}_2}{2 \sqrt{\bar{p}}}
    - \frac{3 {\cal P}_4 {\cal K}^{(1)}_1}{\bar{p}^2} \bar{\pi}
    -3\bar{k} {\Phi}_5 \frac{{\cal K}^{(1)}_1}{\sqrt{\bar{p}}}
    \right)\delta\pi
    \nonumber\\
    &&\qquad
    + \left(\frac{\partial {\cal P}_1}{\partial \bar{k}}
    \frac{{\cal K}^{(1)}_1}{3 \bar{p}^2}\bar{\pi}^2
    +\frac{2}{3} {\cal K}^{(1)}_1 \frac{\partial \Phi_0}{\partial \bar{k}}
    \frac{\bar{\pi}}{\sqrt{\bar{p}}}
    -3 {\cal K}^{(1)}_1 ({\Phi}_5 + 2\bar{k}{\Phi}_{10})
    \frac{\bar{\pi}}{\sqrt{\bar{p}}} 
    \right)\delta^d_k\delta K_d^k 
    \nonumber\\
    &&\qquad
    + \Bigg(
    \left(3 {\cal K}^{(1)}_2 
    \left(\Phi_5 + 2\bar{k}\Phi_{10}\right)
    - \frac{2}{3} {\cal K}^{(1)}_2 \frac{\partial \Phi_0}{\partial \bar{k}} \right)
    \frac{\bar{\pi}}{4 \bar{p}^{3/2}}
    \nonumber\\
    &&\qquad\qquad
    - \left(\frac{2}{3}
    \left(\bar{k}\frac{\partial {\cal P}_1}{\partial \bar{k}}
    -\bar{p}\frac{\partial {\cal P}_1}{\partial \bar{p}}
    \right) \frac{{\cal K}^{(1)}_1}{4\bar{p}^3}
    +
    \Bigg({\cal P}_1 - 6 {\cal P}_4
    \Bigg)\frac{ {\cal K}^{(1)}_1}{4\bar{p}^3}
    + \frac{\Phi {\cal K}^{(1)}_2}{8 \bar{p}^3} \right) \bar{\pi}^2
    \nonumber\\
    &&\qquad\qquad
    -\left(\frac{1}{3}\left(\bar{k}\frac{\partial {\cal P}_1}{\partial \bar{k}}
    -\bar{p}\frac{\partial {\cal P}_1}{\partial \bar{p}}\right)
    {\cal K}^{(1)}_1
    + \frac{3}{2} \bar{k}^2
    {\cal K}^{(1)}_1 {\Phi}_{10}\right) \frac{{\bar{\pi}}}{\bar{p}^{3/2}} 
    \Bigg)
    \delta_d^k \delta E^d_k
    \Bigg]
    \nonumber\\
    &&\qquad
    - \int {\rm d}^3x (\bar{N}_1 \partial_a\delta N_2)
    \bigg(
    \bar{p} \Phi_{11} {\cal K}^{(1)}_1
    +
    2\sqrt{\bar{p}} {\Theta}_9 {\cal K}^{(1)}_1 \bar{\pi} 
    \bigg) \partial^a \delta\varphi
    \nonumber\\
    &&\qquad
    + \int {\rm d}^3x (\bar{N}_1 \partial_c\delta N_2)
    \Bigg[
    {\cal K}^{(1)}_3 \left(
    \Phi  \frac{\bar{\pi}^2}{4 \bar{p}^3}
    + \left(\frac{1}{3}\frac{\partial \Phi_0}{\partial \bar{k}}
    -\frac{1}{2} (\Phi_{5} + 2\bar{k}\Phi_{10})\right) \frac{\bar{\pi}}{\bar{p}^{3/2}}
    \right)
    \nonumber\\
    &&\qquad\qquad
    -
    {\cal K}^{(1)}_1 \bigg(
    \Theta_3 
    \frac{2\bar{\pi} }{\sqrt{\bar{p}}}
    +
    \Theta_6
    \frac{\bar{\pi}^2}{\bar{p}^2}
    \bigg)
    -
    \bigg(
    \bar{p}\Phi_{11}
    +
    2\sqrt{\bar{p}} \Theta_9 \bar{\pi}
    \bigg)
    {\cal K}^{(1)}_1
    \Bigg]
    \delta^{ck}\partial_d\delta E^d_k
    \nonumber\\
    &&\qquad\qquad
    +\int {\rm d}^3x (\bar{N}_1 \partial_c\delta N_2)
    \Bigg[
    -2{\cal K}^{(1)}_3\Phi_{10}\frac{\bar{\pi}}{\sqrt{\bar{p}}}
    \delta^{ce} \delta^d_e \partial_k\delta K^k_d
    \Bigg]
    \,,
\end{eqnarray}
\begin{eqnarray}
    &&\{\tilde{H}^{(2)}_\varphi[\bar{N}_1],\tilde{H}^{(1)}_\varphi[\delta N_2]\}_\delta =
    \kappa \int{\rm d}^3x
    \bar{N}_1 \delta N_2\Bigg[ 
    \left(\frac{1}{3} \frac{\partial {\Phi}_0}{\partial \bar{k}}
    {\Phi}_2
    - \frac{3}{2} ({\Phi}_5 + 2\bar{k}{\Phi}_{10}) {\Phi}_2\right) \frac{\bar{\pi}^2}{\bar{p}}
    \nonumber\\
    &&\qquad\qquad\qquad
    + \left(\frac{\bar{p}}{6}
    \frac{\partial {\cal P}_1}{\partial \bar{k}} 
    {\Phi}_2
    -\frac{1}{12} \frac{\partial {\Phi}_0 }{\partial \bar{k}}{\cal P}_2
    + \frac{3{\cal P}_2}{8}
    ({\Phi}_5 + 2\bar{k}{\Phi}_{10})\right) \frac{\bar{\pi}^3}{\bar{p}^{7/2}}
    - \frac{{\cal P}_2}{24}
    \frac{\partial {\cal P}_1}{\partial \bar{k}}
    \frac{\bar{\pi}^4}{\bar{p}^5}
    \nonumber\\
    &&\qquad\qquad\qquad
    +
    \left(\frac{1}{12}
    \left(
    \bar{k}\frac{\partial {\cal P}_1}{\partial \bar{k}}
    -
    \bar{p}\frac{\partial {\cal P}_1}{\partial \bar{p}}
    \right)
    +
    \frac{1}{8}{\cal P}_1
    -
    \frac{3}{4} {\cal P}_4\right) {\Phi}_3 \frac{\bar{\pi}^3}{\bar{p}^{7/2}}
    \nonumber\\
    &&\qquad\qquad\qquad
    - \left(
    \frac{1}{6}
    \left(
    \bar{k}\frac{\partial {\cal P}_1}{\partial \bar{k}}
    -
    \bar{p}\frac{\partial {\cal P}_1}{\partial \bar{p}}
    \right)
    + \frac{3}{2}\bar{k}^2 {\Phi}_{10}\right) {\Phi}_3 \frac{\bar{\pi}^2}{\bar{p}^{2}}
    \Bigg]\delta^k_d\delta E^d_k
    \nonumber\\
    &&
    +\kappa \int{\rm d}^3x
    \bar{N}_1 \delta N_2
    \Bigg[
    -3 {\Phi}_{10} {\cal P}_2 \frac{\bar{\pi}^3}{2 \bar{p}^{5/2}}
    + 6 {\Phi}_{10} {\Phi}_2 \bar{\pi}^2
    -
   \frac{1}{6} \frac{\partial {\cal P}_1 }{\partial \bar{k}}
    \frac{\bar{\pi}^3}{\bar{p}^{5/2}}{\Phi}_3
    -
   \frac{1}{3} \frac{\partial {\Phi}_0 }{\partial \bar{k}}
    \frac{\bar{\pi}^2}{\bar{p}}{\Phi}_3
    \nonumber\\
    &&\qquad\qquad\qquad
    +
    \frac{3}{2}
    \left(
    {\Phi}_5 + 2\bar{k}{\Phi}_{10}
    \right)\frac{\bar{\pi}^2 }{\bar{p}}{\Phi}_3
    \Bigg]
    \delta^d_k\delta K^k_d
    \nonumber\\
    &&
    +\kappa \int{\rm d}^3x
    \bar{N}_1 \delta N_2
    \Bigg[
    + 3 \left({\Phi}_2 + \frac{\bar{k}}{2\bar{p}} {\Phi}_3\right) {\Phi}_5\bar{\pi}
    + \left(\frac{3}{2}{\cal P}_4 {\Phi}_3
    - \frac{3}{4} {\cal P}_2 {\Phi}_5 \right) \frac{\bar{\pi}^2}{\bar{p}^{5/2}}
    \Bigg]
    \delta \pi
     \nonumber\\
    &&
    + \int {\rm d}^3x (\bar{N}_1 \partial_c\delta N_2)
    3\bigg(
    \Theta_1 + \frac{\Theta_2}{2\bar{p}}
    \bigg)\Phi_5 \bar{\pi} \partial_c \delta \pi
    \nonumber\\
    &&
    + \int {\rm d}^3x (\bar{N}_1 \partial_c\delta N_2)
    \Bigg[\bigg(
    \Theta_1 + \frac{\Theta_2}{2\bar{p}}
    \bigg)
   \frac{\partial \Phi_0}{\partial \bar{k}} \frac{\bar{\pi}^2}{3\bar{p}}
   + \bigg(
    \Theta_1 + \frac{\Theta_2}{2\bar{p}}
    \bigg)
    \frac{\bar{\pi}^3 }{4\bar{p}^{5/2}}
   \nonumber\\
   &&\qquad\qquad
   + \left(2\Phi_{10}\bigg(
    \Theta_1 + \frac{\Theta_2}{2\bar{p}}
    \bigg)
    - \frac{\Theta_1}{2\bar{p}}(\Phi_5 + 2\bar{k} \Phi_{10})
   - \frac{\Theta_2}{4\bar{p}^2}(\Phi_5 + 2\bar{k} \Phi_{10})
    \right) \bar{\pi}^2
    \Bigg]
    \delta^{ck}\partial_d\delta E^d_k
    \nonumber\\
    &&
    - \int {\rm d}^3x (\bar{N}_1 \partial_c\delta N_2)
    \Bigg[
    \Theta_3
    \frac{\bar{\pi}^2 }{\bar{p}}
    +
    \Theta_6
    \frac{\bar{\pi}^3}{2\bar{p}^{5/2}}
    \Bigg]
    {\Phi}_3 \delta^{ck}\partial_d\delta E^d_k
    \nonumber\\
    &&
    - \int{\rm d}^3x (\bar{N}_1 \partial_a \delta N_2)
    \Bigg[
    ({\cal P}_1 + {\cal P}_2) \frac{\bar{\pi}}{\bar{2p}}
    +
    \sqrt{\bar{p}} {\Phi}_0
    -
    2\bar{p}^{3/2} {\Phi}_2
    -
    \bar{k} \sqrt{\bar{p}} {\Phi}_3
    \Bigg]
    {\cal P}_8 (\partial^a \delta\varphi)
    \nonumber\\
    &&
    -\int {\rm d}^3x (\bar{N}_1 \partial_a\delta N_2)
    \bigg[
    -
    \frac{\sqrt{\bar{p}}}{2} {\Phi}_3 \Phi_{11}
    \bar{\pi}
    -
    {\Theta}_9 {\Phi}_3
    \bar{\pi}^2
    \bigg] \partial^a \delta\varphi
    \\
    &&
    - \int{\rm d}^3x (\bar{N}_1 \partial_c \delta N_2)
    \Bigg[
    {\Phi}_0
    -
    2\bar{p} {\Phi}_2
    -
    \bar{k} {\Phi}_3
    + ({\cal P}_1 + {\cal P}_2) \frac{\bar{\pi}}{2\bar{p}^{3/2}}\Bigg]
    \bigg(
    \Phi_{11}\frac{\sqrt{\bar{p}}}{2}
    +
    \Theta_9 \bar{\pi}
    \bigg)
    \delta^{ck} \partial_d \delta E^d_k
    \,.
    \nonumber
\end{eqnarray}

\subsection{Scalar potential contributions}
\label{sec:Scalar potential contributions}

Here we assume that
${\cal P}_3 = \frac{1}{2}({\cal P}_1+{\cal P}_2)$
${\cal P}_{V4} = {\cal P}_{V3}$\,,
${\Theta}_{V4} = {\Theta}_{V3}$\,,
$\Phi_1=\Phi_0 -2\bar{p}\Phi_2-\bar{k}\Phi_3$\,
and
$\Theta_{1} = \Theta_{2}=\Theta_{4}=\Theta_{5} = \Theta_{7}=\Theta_{8} = \Phi_{V7} = 0$
\,.
%%%%%
%%%%%
%%%%%
%%%%%
\subsubsection{Derivatives}

\begin{eqnarray}
    \frac{\partial\tilde{H}^{(0)}_V[\bar{N}]}{\partial \bar{k}} &=&
    \bar{N} V_0 \bar{p}^{3/2}
    \left(
    \frac{\partial {\cal P}_{V1}}{\partial \bar{k}}
    +
    \frac{\partial {\Theta}_{V1}}{\partial \bar{k}}
    \bar{\pi}
    \right)
    \,,
    \nonumber\\
    \frac{\partial\tilde{H}^{(0)}_V[\bar{N}]}{\partial \bar{p}} &=&
    \bar{N} V_0 \bar{p}^{3/2} 
    \left[
    \left(\frac{3 {\cal P}_{V1}}{2\bar{p}} + \frac{\partial {\cal P}_{V1}}{\partial \bar{p}} \right)
    +
    \left(\frac{3 {\Theta}_{V1}}{2\bar{p}} + \frac{\partial {\Theta}_{V1}}{\partial \bar{p}} \right)\bar{\pi}
    \right]
    \,,
    \nonumber\\
    \frac{\partial\tilde{H}^{(0)}_V[\bar{N}]}{\partial \bar{\varphi}} &=&
    \bar{N} V_0 \bar{p}^{3/2} 
    \left(
    \frac{\partial {\cal P}_{V1}}{\partial \bar{\varphi}}
    +
    \frac{\partial {\Theta}_{V1}}{\partial \bar{\varphi}}
    \bar{\pi}
    \right)
    \,,
    \nonumber\\
    \frac{\partial\tilde{H}^{(0)}_V[\bar{N}]}{\partial \bar{\pi}} &=&
    \bar{N} V_0 \bar{p}^{3/2} 
    {\Theta}_{V1}
    \,,
\end{eqnarray}
\begin{eqnarray}
    \frac{\partial\tilde{H}^{(1)}_V[\delta N]}{\partial \bar{k}} &=&
    \int{\rm d}^3 x \delta N \bar{p}^{3/2} 
    \Bigg[
    \frac{\partial{\cal P}_{dV1}}{\partial\bar{k}}\delta\varphi
    +\frac{\partial{\Theta}_{dV1}}{\partial\bar{k}} \bar{\pi}\delta\varphi
    + \frac{\partial{\cal P}_{V2}}{\partial\bar{k}} \frac{\delta^j_c \delta E^c_j}{2 \bar{p}}
    + \frac{\partial{\Theta}_{V2}}{\partial\bar{k}} \bar{\pi}\frac{\delta^j_c \delta E^c_j}{2 \bar{p}}
    \nonumber\\
    &&\qquad\qquad\qquad
    + 
    \frac{\partial\Phi_{V1}}{\partial\bar{k}} \delta^d_k \delta K_d^k
    \Bigg] \,,
    \nonumber\\
    \frac{\partial\tilde{H}^{(1)}_V[\delta N]}{\partial \bar{p}} &=&
    \int{\rm d}^3 x \delta N \bar{p}^{3/2} \Bigg[
    \left( \frac{3 {\cal P}_{dV1}}{2\bar{p}} + \frac{\partial{\cal P}_{dV1}}{\partial\bar{p}}\right) \delta\varphi
    +
    \left( \frac{3 {\Theta}_{dV1}}{2\bar{p}} + \frac{\partial{\Theta}_{dV1}}{\partial\bar{p}}\right) \bar{\pi}\delta\varphi
    \nonumber\\
    &&\qquad
    +
    \left( \frac{{\cal P}_{V2}}{2\bar{p}} + \frac{\partial{\cal P}_{V2}}{\partial\bar{p}} \right) \frac{\delta^j_c \delta E^c_j}{2 \bar{p}}
    +
    \left( \frac{{\Theta}_{V2}}{2\bar{p}} + \frac{\partial{\Theta}_{V2}}{\partial\bar{p}} \right) \bar{\pi}\frac{\delta^j_c \delta E^c_j}{2 \bar{p}}
    \nonumber\\
    &&\qquad
    + \left( \frac{3\Phi_{V1}}{2\bar{p}} + \frac{\partial\Phi_{V1}}{\partial\bar{p}} \right) \delta^d_k \delta K_d^k
    \Bigg]\,,
    \nonumber\\
    \frac{\partial\tilde{H}^{(1)}_V[\delta N]}{\partial \bar{\varphi}} &=&
    \int{\rm d}^3x\;\delta N \bar{p}^{3/2} 
    \bigg[ 
    \frac{\partial {\cal P}_{dV1}}{\partial \bar{\varphi}} \delta\varphi
    +\frac{\partial {\Theta}_{dV1}}{\partial \bar{\varphi}} \bar{\pi} \delta\varphi
    + \frac{\partial {\cal P}_{V2}}{\partial \bar{\varphi}} \frac{\delta^j_c \delta E^c_j}{2 \bar{p}}
    + \frac{\partial{\Theta}_{V2}}{\partial\bar{\varphi}} \bar{\pi}\frac{\delta^j_c \delta E^c_j}{2 \bar{p}}
    \nonumber\\
    &&\qquad
    + 
    \frac{\partial \Phi_{V1}}{\partial \bar{\varphi}} \delta^d_k \delta K_d^k
    \bigg]
    \nonumber\\
    \frac{\partial\tilde{H}^{(1)}_V[\delta N]}{\partial \bar{\pi}}
    &=&
    \int{\rm d}^3 x \delta N
    \bigg[
    \bar{p}^{3/2}{\Theta}_{dV1} \delta\varphi
    + \sqrt{\bar{p}}{\Theta}_{V2} \delta^k_d \delta E_k^d
    \bigg]
    \,,
\end{eqnarray}
\begin{eqnarray}
    \frac{\delta\tilde{H}^{(1)}_V[\delta N]}{\delta \delta K_a^i} &=&
    \delta N \bar{p}^{3/2} 
    \Phi_{V1}
    \delta^a_i
    \,,
    \nonumber\\
    \frac{\delta\tilde{H}^{(1)}_V[\delta N]}{\delta \delta E^a_i} &=&
    \delta N \frac{\sqrt{\bar{p}}}{2}
    \bigg[
    {\cal P}_{V2}
    +
    \bar{\pi}{\Theta}_{V2}
    \bigg]
    \delta^i_a
    \,,
    \nonumber\\
    \frac{\delta\tilde{H}^{(1)}_V[\delta N]}{\delta \delta \varphi} &=&
    \delta N \bar{p}^{3/2}
    \left[
    {\cal P}_{dV1}
    +
    \bar{\pi}{\Theta}_{dV1}
    \right]
    \,,
    \nonumber\\
    \frac{\delta\tilde{H}^{(1)}_V[\delta N]}{\delta \delta \pi} &=& 
    0
    \,,
\end{eqnarray}

\begin{eqnarray}
    \frac{\partial\tilde{H}^{(2)}_V[\bar{N}]}{\partial \bar{k}} &=&
    \bar{N} V_0 \bar{p}^{3/2} \Bigg[ 
      \left(
    \frac{\partial{\cal P}_{V3}}{\partial \bar{k}}
    +
    \frac{\partial{\Theta}_{V3}}{\partial \bar{k}}
    \bar{\pi}
    \right) 
    \left( 
    \frac{(\delta^j_c\delta E^c_j)^2}{8\bar{p}^2}
    - \frac{\delta^k_c\delta^j_d\delta E^c_j\delta E^d_k}{4\bar{p}^2}\right)
    + 
    \left(
    \frac{\partial{\cal P}_{ddV}}{\partial \bar{k}}
    +
    \frac{\partial{\Theta}_{ddV}}{\partial \bar{k}}
    \bar{\pi}
    \right)
    \frac{1}{2} 
    \delta\varphi^2
    \nonumber\\
    &&
    + 
    \left(
    \frac{\partial{\cal P}_{dV2}}{\partial \bar{k}}
    +
    \frac{\partial{\Theta}_{dV2}}{\partial \bar{k}}
    \bar{\pi}
    \right)
    \delta\varphi\frac{\delta^j_c\delta E^c_j}{2\bar{p}}
    + \frac{\partial\Phi_{V3}}{\partial \bar{k}} \frac{\delta E^d_k \delta K^k_d}{2\bar{p}}
    + \frac{\partial\Phi_{V4}}{\partial \bar{k}} 
    \delta\varphi \delta^d_k \delta K^k_d
    \nonumber\\
    &&
    +
    \frac{\partial {\Phi}_{V6}}{\partial \bar{k}}
     \delta_k^c\delta_j^d\delta K_c^j\delta K_d^k
    \Bigg]
    \,,
    \nonumber\\
    \frac{\partial\tilde{H}^{(2)}_V[\bar{N}]}{\partial \bar{p}} &=&
    \bar{N} V_0 \bar{p}^{3/2} \Bigg[ 
    \left( - \frac{1}{2\bar{p}} {\cal P}_{V3} + \frac{\partial{\cal P}_{V3}}{\partial \bar{p}}\right) 
    \left( 
    \frac{(\delta^j_c\delta E^c_j)^2}{8\bar{p}^2}
    - \frac{\delta^k_c\delta^j_d\delta E^c_j\delta E^d_k}{4\bar{p}^2}\right)
    \nonumber\\
    &&
    +
     \left( - \frac{1}{2\bar{p}} {\Theta}_{V3} + \frac{\partial{\Theta}_{V3}}{\partial \bar{p}}\right) 
     \bar{\pi}
    \left( 
    \frac{(\delta^j_c\delta E^c_j)^2}{8\bar{p}^2}
    - \frac{\delta^k_c\delta^j_d\delta E^c_j\delta E^d_k}{4\bar{p}^2}\right)
    \nonumber\\
    &&
    + \left( \frac{3}{2\bar{p}} {\cal P}_{ddV} + \frac{\partial{\cal P}_{ddV}}{\partial \bar{p}}\right) \frac{1}{2} \delta\varphi^2
    + \left( \frac{3}{2\bar{p}} {\Theta}_{ddV} + \frac{\partial{\Theta}_{ddV}}{\partial \bar{p}}\right) \frac{1}{2} \bar{\pi} \delta\varphi^2
    \nonumber\\
    &&
    + \left( \frac{1}{2\bar{p}} {\cal P}_{dV2} + \frac{\partial{\cal P}_{dV2}}{\partial \bar{p}}\right) \delta\varphi\frac{\delta^j_c\delta E^c_j}{2\bar{p}}
    + \left( \frac{1}{2\bar{p}} {\Theta}_{dV2} + \frac{\partial{\Theta}_{dV2}}{\partial \bar{p}}\right)
    \bar{\pi}
    \delta\varphi\frac{\delta^j_c\delta E^c_j}{2\bar{p}}
    \nonumber\\
    &&
    + \left( \frac{1}{2\bar{p}} \Phi_{V3} + \frac{\partial\Phi_{V3}}{\partial \bar{p}}\right)  \frac{\delta E^d_k \delta K^k_d}{2\bar{p}}
    + 
    \left( \frac{3}{2\bar{p}} \Phi_{V4}
    +
    \frac{\partial\Phi_{V4}}{\partial \bar{p}}\right)
    \delta\varphi \delta^d_k \delta K^k_d
    \Bigg]
    \,,
    \nonumber\\
    \frac{\partial\tilde{H}^{(2)}_V[\bar{N}]}{\partial \bar{\varphi}} &=&
    \bar{N} V_0 \bar{p}^{3/2} \Bigg[
    \left(
    \frac{\partial {\cal P}_{V3}}{\partial\bar{\varphi}} 
    +
    \frac{\partial {\Theta}_{V3}}{\partial\bar{\varphi}}
    \bar{\pi}
    \right)
    \left( 
    \frac{(\delta^j_c\delta E^c_j)^2}{8\bar{p}^2}
    - \frac{\delta^k_c\delta^j_d\delta E^c_j\delta E^d_k}{4\bar{p}^2}\right)
    +  \frac{1}{2}
    \left(
    \frac{\partial {\cal P}_{ddV}}{\partial \bar{\varphi}}
    +
    \frac{\partial {\Theta}_{ddV}}{\partial \bar{\varphi}}
    \bar{\pi}
    \right)
    \delta\varphi^2
    \nonumber\\
    &&
    + 
    \left(
    \frac{\partial {\cal P}_{dV2} }{\partial \bar{\varphi}}
    +
    \frac{\partial {\Theta}_{dV2} }{\partial \bar{\varphi}}
    \bar{\pi}
    \right)
    \delta\varphi\frac{\delta^j_c\delta E^c_j}{2\bar{p}}
    + \frac{\partial \Phi_{V3}}{\partial \bar{\varphi}} \frac{\delta E^d_k \delta K^k_d}{2\bar{p}}
    + \frac{\partial \Phi_{V4}}{\partial \bar{\varphi}} \delta\varphi \delta^d_k \delta K^k_d
    \Bigg]
    \,,
    \nonumber\\
    \frac{\partial\tilde{H}^{(2)}_V[\bar{N}]}{\partial \bar{\pi}} &=&
    \bar{N} V_0 \bar{p}^{3/2} \Bigg[
    \Theta_{V3}
    \frac{(\delta^j_c\delta E^c_j)^2}{8\bar{p}^2}
    +
    \Theta_{V4}
    \frac{\delta^k_c \delta^j_d \delta E^c_j \delta E^d_k}{4\bar{p}^2}
    +
    \Theta_{ddV} \frac{1}{2} (\delta \varphi)^2
    +
    \Theta_{dV2} \delta\varphi
    \frac{\delta^j_c\delta E^c_j}{2\bar{p}}
    \Bigg]
    \,,
\end{eqnarray}

\begin{eqnarray}
    \frac{\delta\tilde{H}^{(2)}_V[\bar{N}]}{\delta \delta K_a^i} &=&
    \bar{N} \bar{p}^{3/2} 
    \left[ 
    \Phi_{V3}  \frac{\delta E^a_i}{2\bar{p}}
    + \Phi_{V4} \delta\varphi \delta^i_a 
    + 2 \Phi_{V5} \delta^c_j \delta K^j_c \delta^a_i
    + 2 \Phi_{V6} \delta K^j_c \delta^a_j \delta^c_i
    \right]
    \,,
    \nonumber\\
    \frac{\delta\tilde{H}^{(2)}_V[\bar{N}]}{\delta \delta E^a_i} &=&
    \bar{N} \bar{p}^{3/2} \Bigg[ 
    \left(
    {\cal P}_{V3}
    +
    {\Theta}_{V3} \bar{\pi}
    \right)
    \frac{\delta^j_c\delta E^c_j}{4\bar{p}^2} \delta^i_a
    -
    \left(
    {\cal P}_{V3}
    +
    {\Theta}_{V3} \bar{\pi}
    \right)
    \frac{\delta^k_a\delta^i_d \delta E^d_k}{2\bar{p}^2}
    \nonumber \\
    &&
    +
    \left(
    {\cal P}_{dV2}
    +
    {\Theta}_{dV2} \bar{\pi}
    \right)
    \frac{\delta^i_a}{2\bar{p}} \delta{\varphi}
    + \Phi_{V3} 
    \frac{\delta K^i_a}{2\bar{p}}
    \Bigg]
    \,,
    \nonumber\\
    \frac{\delta\tilde{H}^{(2)}_V[\bar{N}]}{\delta \delta \varphi} &=&
    \bar{N} \bar{p}^{3/2} \Bigg[ 
    \left(
    {\cal P}_{ddV}
    +
    {\Theta}_{ddV} \bar{\pi}
    \right)
    \delta\varphi
    +
    \left(
    {\cal P}_{dV2}
    +
    {\Theta}_{dV2} \bar{\pi}
    \right)
    \frac{\delta^j_c\delta E^c_j}{2\bar{p}}
    + 
    \Phi_{V4}  \delta^d_k \delta K^k_d
    \Bigg]
    \,,
    \nonumber\\
    \frac{\delta\tilde{H}^{(2)}_V[\bar{N}]}{\delta \delta \pi} &=& 0 \,,
\end{eqnarray}
%%%%%%%%%%%%%%%%%%%%%%%%%
%%%%%%%%%%%%%%%%%%%%%%%%%
%%%%%%%%%%%%%%%%%%%%%%%%%
%%%%%%%%%%%%%%%%%%%%%%%%%
\subsubsection{Brackets}

\begin{eqnarray}
    &&\{\tilde{H}^{(0)}_{\rm grav}[\bar{N}_1],\tilde{H}^{(1)}_V[\delta N_2]\}_{\bar{A}} =
    \frac{\kappa}{3V_0} \frac{\partial \tilde{H}^{(0)}_{\rm grav}[\bar{N}_1]}{\partial \bar{k}}
    \frac{\partial \tilde{H}^{(1)}_V[\delta N_2]}{\partial \bar{p}}
    - \frac{\kappa}{3V_0} \frac{\partial \tilde{H}^{(0)}_{\rm grav}[\bar{N}_1]}{\partial \bar{p}}
    \frac{\partial \tilde{H}^{(1)}_V[\delta N_2]}{\partial \bar{k}}
    \nonumber\\
    &&=
    \int{\rm d}^3 x\; \bar{N}_1 \delta N_2 \bar{p}^2
    \Bigg[
    \left( \left( \frac{{\cal K}^{(0)}}{2\bar{p}} + \frac{\partial {\cal K}^{(0)}}{\partial \bar{p}}\right) \frac{\partial{\cal P}_{dV1}}{\partial\bar{k}}
    - \frac{\partial {\cal K}^{(0)}}{\partial \bar{k}} \left( \frac{3 {\cal P}_{dV1}}{2\bar{p}} + \frac{\partial{\cal P}_{dV1}}{\partial\bar{p}}\right) \right) \delta\varphi
    \nonumber\\
    &&\qquad
    +
    \left( \left( \frac{{\cal K}^{(0)}}{2\bar{p}} + \frac{\partial {\cal K}^{(0)}}{\partial \bar{p}}\right) \frac{\partial{\Theta}_{dV1}}{\partial\bar{k}}
    - \frac{\partial {\cal K}^{(0)}}{\partial \bar{k}} \left( \frac{3 {\Theta}_{dV1}}{2\bar{p}} + \frac{\partial{\Theta}_{dV1}}{\partial\bar{p}}\right) \right) \bar{\pi} \delta\varphi
    \nonumber\\
    &&\qquad
    + \left( \left( \frac{{\cal K}^{(0)}}{2\bar{p}} + \frac{\partial {\cal K}^{(0)}}{\partial \bar{p}}\right) \frac{\partial{\cal P}_{V2}}{\partial\bar{k}}
    - \frac{\partial {\cal K}^{(0)}}{\partial \bar{k}} \left( \frac{3{\cal P}_{V2}}{2\bar{p}} + \frac{\partial{\cal P}_{V2}}{\partial\bar{p}} \right) \right) \frac{\delta^j_c \delta E^c_j}{2 \bar{p}}
    \nonumber\\
    &&\qquad
    \nonumber\\
    &&\qquad
    + \left( \left( \frac{{\cal K}^{(0)}}{2\bar{p}} + \frac{\partial {\cal K}^{(0)}}{\partial \bar{p}}\right) \frac{\partial\Phi_{V1}}{\partial\bar{k}}
    - \frac{\partial {\cal K}^{(0)}}{\partial \bar{k}} \left( \frac{3}{2\bar{p}} \Phi_{V1} + \frac{\partial\Phi_{V1}}{\partial\bar{p}} \right) \right) \delta^d_k \delta K_d^k
    \Bigg]\,,
\end{eqnarray}

\begin{eqnarray}
    &&\{\tilde{H}^{(0)}_V[\bar{N}_1],\tilde{H}^{(1)}_{\rm grav}[\delta N_2]\}_{\bar{A}} =
    \frac{\kappa}{3V_0} \frac{\partial \tilde{H}^{(0)}_V[\bar{N}_1]}{\partial \bar{k}}
    \frac{\partial \tilde{H}^{(1)}_{\rm grav}[\delta N_2]}{\partial \bar{p}}
    - \frac{\kappa}{3V_0} \frac{\partial \tilde{H}^{(0)}_V[\bar{N}_1]}{\partial \bar{p}}
    \frac{\partial \tilde{H}^{(1)}_{\rm grav}[\delta N_2]}{\partial \bar{k}}
    \nonumber\\
    &&=
    \int {\rm d}^3 x\; \bar{N}_1 \delta N_2 \bar{p}^2 \Bigg[
    \frac{2}{3}  \left( \left(\frac{3 {\cal P}_{V1}}{2\bar{p}} + \frac{\partial {\cal P}_{V1}}{\partial \bar{p}} \right) \frac{\partial {\cal K}^{(1)}_1}{\partial \bar{k}}
    - \frac{\partial {\cal P}_{V1}}{\partial \bar{k}} \left(\frac{\partial {\cal K}^{(1)}_1}{\partial \bar{p}} + \frac{{\cal K}^{(1)}_1}{2 \bar{p}}\right) \right) \delta^c_j\delta K_c^j
    \nonumber\\
    &&\qquad
    +
    \frac{2}{3}  \left( \left(\frac{3 {\Theta}_{V1}}{2\bar{p}} + \frac{\partial {\Theta}_{V1}}{\partial \bar{p}} \right) \frac{\partial {\cal K}^{(1)}_1}{\partial \bar{k}}
    - \frac{\partial {\Theta}_{V1}}{\partial \bar{k}} \left(\frac{\partial {\cal K}^{(1)}_1}{\partial \bar{p}} + \frac{{\cal K}^{(1)}_1}{2 \bar{p}}\right) \right) \bar{\pi} \delta^c_j\delta K_c^j
    \nonumber\\
    &&\qquad
    + \left( \frac{1}{3} \left(\frac{3 {\cal P}_{V1}}{2\bar{p}} + \frac{\partial {\cal P}_{V1}}{\partial \bar{p}} \right) \frac{\partial {\cal K}^{(1)}_2}{\partial \bar{k}}
    - \frac{1}{3} \frac{\partial {\cal P}_{V1}}{\partial \bar{k}} \left(\frac{\partial {\cal K}^{(1)}_2}{\partial \bar{p}} - \frac{{\cal K}^{(1)}_2}{2 \bar{p}}\right) \right) \frac{\delta_c^j\delta E^c_j}{2 \bar{p}}
    \nonumber\\
    &&\qquad
    + \left( \frac{1}{3} \left(\frac{3 {\Theta}_{V1}}{2\bar{p}} + \frac{\partial {\Theta}_{V1}}{\partial \bar{p}} \right) \frac{\partial {\cal K}^{(1)}_2}{\partial \bar{k}}
    - \frac{1}{3} \frac{\partial {\Theta}_{V1}}{\partial \bar{k}} \left(\frac{\partial {\cal K}^{(1)}_2}{\partial \bar{p}} - \frac{{\cal K}^{(1)}_2}{2 \bar{p}}\right) \right) \bar{\pi}\frac{\delta_c^j\delta E^c_j}{2 \bar{p}}
    \nonumber\\
    &&\qquad
    + \left( \frac{1}{3} \frac{\partial {\cal P}_{V1}}{\partial \bar{k}} \left(\frac{\partial {\cal K}^{(1)}_3}{\partial \bar{p}} - \frac{{\cal K}^{(1)}_3}{2 \bar{p}}\right)
    - \frac{1}{3} \left(\frac{3 {\cal P}_{V1}}{2\bar{p}} + \frac{\partial {\cal P}_{V1}}{\partial \bar{p}} \right) \frac{\partial {\cal K}^{(1)}_3}{\partial \bar{k}} \right)  \frac{\delta^{jk}\partial_j\partial_c\delta E^c_k}{\bar{p}}
    \\
    &&\qquad
    + \left( \frac{1}{3} \frac{\partial {\Theta}_{V1}}{\partial \bar{k}} \left(\frac{\partial {\cal K}^{(1)}_3}{\partial \bar{p}} - \frac{{\cal K}^{(1)}_3}{2 \bar{p}}\right)
    - \frac{1}{3} \left(\frac{3 {\Theta}_{V1}}{2\bar{p}} + \frac{\partial {\Theta}_{V1}}{\partial \bar{p}} \right) \frac{\partial {\cal K}^{(1)}_3}{\partial \bar{k}} \right) \bar{\pi}\frac{\delta^{jk}\partial_j\partial_c\delta E^c_k}{\bar{p}}
    \Bigg]\,,\nonumber
\end{eqnarray}

\begin{eqnarray}
    &&\{\tilde{H}^{(0)}_\varphi[\bar{N}_1],\tilde{H}^{(1)}_V[\delta N_2]\}_{\bar{A}}
    =
    \frac{\kappa}{3V_0} \frac{\partial \tilde{H}^{(0)}_\varphi[\bar{N}_1]}{\partial \bar{k}}
    \frac{\partial \tilde{H}^{(1)}_V[\delta N_2]}{\partial \bar{p}}
    - \frac{\kappa}{3V_0} \frac{\partial \tilde{H}^{(0)}_\varphi[\bar{N}_1]}{\partial \bar{p}}
    \frac{\partial \tilde{H}^{(1)}_V[\delta N_2]}{\partial \bar{k}}
    \nonumber\\
    &&\qquad\qquad\qquad\qquad\qquad\quad
    - \frac{1}{V_0} \frac{\partial \tilde{H}^{(0)}_\varphi[\bar{N}_1]}{\partial \bar{\pi}}
    \frac{\partial \tilde{H}^{(1)}_V[\delta N_2]}{\partial \bar{\varphi}}
    \nonumber\\
    &&=
    \int{\rm d}^3 x\; \bar{N}_1 \delta N_2
    \frac{\kappa}{3} \Bigg[
    \left(\frac{\partial{\cal P}_1}{\partial \bar{k}} \left( \frac{3 {\cal P}_{dV1}}{2\bar{p}} + \frac{\partial{\cal P}_{dV1}}{\partial\bar{p}}\right)
    + \left(\frac{3 {\cal P}_1}{2\bar{p}}-\frac{\partial{\cal P}_1}{\partial \bar{p}}\right) \frac{\partial{\cal P}_{dV1}}{\partial\bar{k}} \right) \frac{\bar{\pi}^2}{2}\delta\varphi
    \nonumber\\
    &&\qquad
    +
    \left( \frac{\partial{\cal P}_1}{\partial \bar{k}} \left( \frac{3 \Phi_{V1}}{2\bar{p}} + \frac{\partial\Phi_{V1}}{\partial\bar{p}} \right)
    + \left( \frac{\partial{\cal P}_1}{\partial \bar{p}}
    - \frac{3 {\cal P}_1}{2\bar{p}}\right) \frac{\partial\Phi_{V1}}{\partial\bar{k}} \right) \frac{\bar{\pi}^2}{2} \delta^d_k \delta K_d^k
    \nonumber\\
    &&\qquad
    +
    \left( \frac{\partial{\cal P}_1}{\partial \bar{k}} \left( \frac{{\cal P}_{V2}}{2\bar{p}} + \frac{\partial{\cal P}_{V2}}{\partial\bar{p}} \right)
    + \left( \frac{\partial{\cal P}_1}{\partial \bar{p}}
    - \frac{3 {\cal P}_1}{2\bar{p}}\right) \frac{\partial{\cal P}_{V2}}{\partial\bar{k}} \right) \frac{\bar{\pi}^2}{2} \frac{\delta^j_c \delta E^c_j}{2 \bar{p}}
\nonumber\\
    &&\qquad
    +
    \left( \frac{\partial{\cal P}_1}{\partial \bar{k}} \left( \frac{{\Theta}_{V2}}{2\bar{p}} + \frac{\partial{\Theta}_{V2}}{\partial\bar{p}} \right)
    + \left( \frac{\partial{\cal P}_1}{\partial \bar{p}}
    - \frac{3 {\cal P}_1}{2\bar{p}}\right) \frac{\partial{\Theta}_{V2}}{\partial\bar{k}} \right) \frac{\bar{\pi}^3}{2} \frac{\delta^j_c \delta E^c_j}{2 \bar{p}}
    \nonumber\\
    &&\qquad
    + \left(\frac{\partial{\cal P}_1}{\partial \bar{k}} \left( \frac{3 {\Theta}_{dV1}}{2\bar{p}} + \frac{\partial{\Theta}_{dV1}}{\partial\bar{p}}\right)
    + \left(\frac{3 {\cal P}_1}{2\bar{p}}-\frac{\partial{\cal P}_1}{\partial \bar{p}}\right) \frac{\partial{\Theta}_{dV1}}{\partial\bar{k}} \right) \frac{\bar{\pi}^3}{2}\delta\varphi
    \Bigg]
    \nonumber\\
    &&+
    \int{\rm d}^3 x\; \bar{N}_1 \delta N_2
    \frac{\kappa}{3} \Bigg[
    \left(\frac{\partial{\Phi}_0}{\partial \bar{k}} \left( \frac{3 {\cal P}_{dV1}}{2\bar{p}} + \frac{\partial{\cal P}_{dV1}}{\partial\bar{p}}\right)
+ \left(\frac{3 {\Phi}_0}{2\bar{p}}-\frac{\partial{\Phi}_0}{\partial \bar{p}}\right) \frac{\partial{\cal P}_{dV1}}{\partial\bar{k}} \right) \bar{p}^{3/2}\bar{\pi}\delta\varphi
    \nonumber\\
    &&\qquad
    +
    \left( \frac{\partial{\Phi}_0}{\partial \bar{k}} \left( \frac{3 \Phi_{V1}}{2\bar{p}} + \frac{\partial\Phi_{V1}}{\partial\bar{p}} \right)
    + \left( \frac{\partial{\Phi}_0}{\partial \bar{p}}
    - \frac{3 {\Phi}_0}{2\bar{p}}\right) \frac{\partial\Phi_{V1}}{\partial\bar{k}} \right) \bar{p}^{3/2}\bar{\pi} \delta^d_k \delta K_d^k
    \nonumber\\
    &&\qquad
    +
    \left( \frac{\partial{\Phi}_0}{\partial \bar{k}} \left( \frac{{\cal P}_{V2}}{2\bar{p}} + \frac{\partial{\cal P}_{V2}}{\partial\bar{p}} \right)
    + \left( \frac{\partial{\Phi}_0}{\partial \bar{p}}
    - \frac{3 {\Phi}_0}{2\bar{p}}\right) \frac{\partial{\cal P}_{V2}}{\partial\bar{k}} \right) \bar{p}^{3/2}\bar{\pi} \frac{\delta^j_c \delta E^c_j}{2 \bar{p}}
\nonumber\\
    &&\qquad
    +
    \left( \frac{\partial{\Phi}_0}{\partial \bar{k}} \left( \frac{{\Theta}_{V2}}{2\bar{p}} + \frac{\partial{\Theta}_{V2}}{\partial\bar{p}} \right)
    + \left( \frac{\partial{\Phi}_0}{\partial \bar{p}}
    - \frac{3 {\Phi}_0}{2\bar{p}}\right) \frac{\partial{\Theta}_{V2}}{\partial\bar{k}} \right) \bar{p}^{3/2}\bar{\pi}^2 \frac{\delta^j_c \delta E^c_j}{2 \bar{p}}
    \nonumber\\
    &&\qquad
    + \left(\frac{\partial{\Phi}_0}{\partial \bar{k}} \left( \frac{3 {\Theta}_{dV1}}{2\bar{p}} + \frac{\partial{\Theta}_{dV1}}{\partial\bar{p}}\right)
    + \left(\frac{3 {\Phi}_0}{2\bar{p}}-\frac{\partial{\Phi}_0}{\partial \bar{p}}\right) \frac{\partial{\Theta}_{dV1}}{\partial\bar{k}} \right) \bar{p}^{3/2}\bar{\pi}^2\delta\varphi
    \Bigg]
    \nonumber\\
    &&
    + \int{\rm d}^3x\; \bar{N}_1 \delta N_2
    \Bigg[ 
    - {\cal P}_1 \frac{\partial{\cal P}_{dV1}}{\partial \bar{\varphi}} \bar{\pi}\delta\varphi
    - {\cal P}_1 \frac{\partial{\Theta}_{dV1}}{\partial \bar{\varphi}} \bar{\pi}^2 \delta\varphi
    \nonumber\\
    &&\qquad
    - {\cal P}_1 \frac{\partial {\cal P}_{V2}}{\partial \bar{\varphi}} \bar{\pi}\frac{\delta^j_c \delta E^c_j}{2 \bar{p}}
    - {\cal P}_1 \frac{\partial \Phi_{V1}}{\partial \bar{\varphi}} \bar{\pi}\delta^d_k \delta K_d^k
    \Bigg]
    \nonumber\\
    &&
    + \int{\rm d}^3x\; \bar{N}_1 \delta N_2
    \bar{p}^{3/2}
    \Bigg[ 
    - {\Phi}_0 \frac{\partial{\cal P}_{dV1}}{\partial \bar{\varphi}} \delta\varphi
    - {\Phi}_0 \frac{\partial{\Theta}_{dV1}}{\partial \bar{\varphi}} \bar{\pi} \delta\varphi
    \nonumber\\
    &&\qquad
    - {\Phi}_0 \frac{\partial {\cal P}_{V2}}{\partial \bar{\varphi}} \frac{\delta^j_c \delta E^c_j}{2 \bar{p}}
    - {\Phi}_0 \frac{\partial \Phi_{V1}}{\partial \bar{\varphi}} \delta^d_k \delta K_d^k
    \Bigg]\,,
\end{eqnarray}

\begin{eqnarray}
        &&\{\tilde{H}^{(0)}_V[\bar{N}_1], \tilde{H}^{(1)}_\varphi[\delta N_2]\}_{\bar{A}} =
        \frac{\kappa}{3V_0} \frac{\partial \tilde{H}^{(0)}_V[\bar{N}_1]}{\partial \bar{k}}
        \frac{\partial \tilde{H}^{(1)}_\varphi[\delta N_2]}{\partial \bar{p}}
        - \frac{\kappa}{3V_0} \frac{\partial \tilde{H}^{(0)}_V[\bar{N}_1]}{\partial \bar{p}}
        \frac{\partial \tilde{H}^{(1)}_\varphi[\delta N_2]}{\partial \bar{k}}
        \nonumber\\
        &&\hspace{10em}\quad
        + \frac{1}{V_0} \frac{\partial \tilde{H}^{(0)}_V[\bar{N}_1]}{\partial \bar{\varphi}}
        \frac{\partial \tilde{H}^{(1)}_\varphi[\delta N_2]}{\partial \bar{\pi}}
        \nonumber\\
        &&= 
        \int{\rm d}^3 x\; \bar{N}_1 \delta N_2 \frac{\kappa}{3} \Bigg[\left( 
        \frac{\partial {\cal P}_{V1}}{\partial \bar{k}} \left( \frac{5 {\cal P}_2}{2\bar{p}} - \frac{\partial {\cal P}_2}{\partial \bar{p}} \right)
        - \frac{\partial {\cal P}_2}{\partial \bar{k}} \left( \frac{3 {\cal P}_{V1}}{2\bar{p}} + \frac{\partial {\cal P}_{V1}}{\partial \bar{p}} \right)
        \right) \frac{\bar{\pi}^2 \delta^j_c \delta E^c_j}{4 \bar{p}}
        \nonumber\\
        &&\qquad
        + \left( 
        \frac{\partial {\Theta}_{V1}}{\partial \bar{k}} \left( \frac{5 {\cal P}_2}{2\bar{p}} - \frac{\partial {\cal P}_2}{\partial \bar{p}} \right)
        - \frac{\partial {\cal P}_2}{\partial \bar{k}} \left( \frac{3 {\Theta}_{V1}}{2\bar{p}} + \frac{\partial {\Theta}_{V1}}{\partial \bar{p}} \right)
        \right) \frac{\bar{\pi}^3 \delta^j_c \delta E^c_j}{4 \bar{p}}
        \nonumber\\
        &&\qquad
        + \bigg[\frac{\partial {\cal P}_{V1}}{\partial \bar{k}} \left( \frac{\partial {\cal P}_1}{\partial \bar{p}} + \frac{\partial {\cal P}_2}{\partial \bar{p}} - \frac{3 {\cal P}_1}{2\bar{p}} - \frac{3 {\cal P}_2}{2\bar{p}} \right)
        - \left( \frac{\partial {\cal P}_1}{\partial \bar{k}} + \frac{\partial {\cal P}_2}{\partial \bar{k}} \right) \left( \frac{3 {\cal P}_{V1}}{2\bar{p}} + \frac{\partial {\cal P}_{V1}}{\partial \bar{p}} \right)
        \bigg] \frac{\bar{\pi} \delta\pi}{2 \bar{p}^{3/2}}
        \nonumber\\
        &&\qquad
        + \bigg[
        \frac{\partial {\Theta}_{V1}}{\partial \bar{k}} \left( \frac{\partial {\cal P}_1}{\partial \bar{p}} + \frac{\partial {\cal P}_2}{\partial \bar{p}} - \frac{3 {\cal P}_1}{2\bar{p}} - \frac{3 {\cal P}_2}{2\bar{p}} \right)
        - \left( \frac{\partial {\cal P}_1}{\partial \bar{k}} + \frac{\partial {\cal P}_2}{\partial \bar{k}} \right) \left( \frac{3 {\Theta}_{V1}}{2\bar{p}} + \frac{\partial {\Theta}_{V1}}{\partial \bar{p}} \right)
        \bigg] \frac{\bar{\pi}^2 \delta\pi}{2 \bar{p}^{3/2}}
        \nonumber\\
        &&\qquad
        + \left( 
        \frac{\partial {\cal P}_{V1}}{\partial \bar{k}} \frac{\partial \Phi_2}{\partial \bar{p}}
        - \frac{\partial \Phi_2}{\partial \bar{k}} \left( \frac{3 {\cal P}_{V1}}{2\bar{p}} + \frac{\partial {\cal P}_{V1}}{\partial \bar{p}} \right)
        \right) \bar{\pi} \delta^j_c \delta E^c_j
        \nonumber\\
        &&\qquad
        + \left( 
        \frac{\partial {\cal P}_{V1}}{\partial \bar{k}} \frac{\partial \Phi_3}{\partial \bar{p}}
        - \frac{\partial \Phi_3}{\partial \bar{k}} \left( \frac{3 {\cal P}_{V1}}{2\bar{p}} + \frac{\partial {\cal P}_{V1}}{\partial \bar{p}} \right)
        \right) \bar{\pi} \delta^c_j \delta K^j_c
        \nonumber\\
        &&\qquad
        + \left( 
        \frac{\partial {\Theta}_{V1}}{\partial \bar{k}} \frac{\partial \Phi_2}{\partial \bar{p}}
        - \frac{\partial \Phi_2}{\partial \bar{k}} \left( \frac{3 {\Theta}_{V1}}{2\bar{p}} + \frac{\partial {\Theta}_{V1}}{\partial \bar{p}} \right)
        \right) \bar{\pi}^2 \delta^j_c \delta E^c_j
        \nonumber\\
        &&\qquad
        + \left( 
        \frac{\partial {\Theta}_{V1}}{\partial \bar{k}} \frac{\partial \Phi_3}{\partial \bar{p}}
        - \frac{\partial \Phi_3}{\partial \bar{k}} \left( \frac{3 {\Theta}_{V1}}{2\bar{p}} + \frac{\partial {\Theta}_{V1}}{\partial \bar{p}} \right)
        \right) \bar{\pi}^2 \delta^c_j \delta K^j_c
        \nonumber\\
        &&\qquad
        + \bigg[
        \frac{\partial {\cal P}_{V1}}{\partial \bar{k}} \left(\frac{\partial \Theta_2}{\partial \bar{p}} - \frac{\Theta_2}{\bar{p}} \right)
        - \frac{\partial \Theta_2}{\partial \bar{k}} \left( \frac{3 {\cal P}_{V1}}{2\bar{p}} + \frac{\partial {\cal P}_{V1}}{\partial \bar{p}} \right)
        \bigg] \bar{\pi} \frac{\partial_c \partial^j \delta E^c_j}{2\bar{p}}
        \nonumber\\
        &&\qquad
        + \bigg[
        \frac{\partial {\Theta}_{V1}}{\partial \bar{k}} \left(\frac{\partial \Theta_2}{\partial \bar{p}} - \frac{\Theta_2}{\bar{p}} \right)
        - \frac{\partial \Theta_2}{\partial \bar{k}} \left( \frac{3 {\Theta}_{V1}}{2\bar{p}} + \frac{\partial {\Theta}_{V1}}{\partial \bar{p}} \right)
        \bigg] \bar{\pi}^2 \frac{\partial_c \partial^j \delta E^c_j}{2\bar{p}}\Bigg]
        \nonumber\\
        &&\quad
        + \int{\rm d}^3 x\; \bar{N}_1 \delta N_2 \bar{p}^{3/2} \Bigg[
        + \Phi_2 \frac{\partial {\cal P}_{V1}}{\partial \bar{\varphi}} \delta^j_c \delta E^c_j
        + \left(\Phi_2 \frac{\partial {\Theta}_{V1}}{\partial \bar{\varphi}} - \frac{{\cal P}_2}{2 \bar{p}} \frac{\partial {\cal P}_{V1}}{\partial \bar{\varphi}}\right) \bar{\pi} \delta^j_c \delta E^c_j
        \nonumber\\
        &&\qquad
        - {\cal P}_2 \frac{\partial {\Theta}_{V1}}{\partial \bar{\varphi}} \frac{\bar{\pi}^2 \delta^j_c \delta E^c_j}{2 \bar{p}}
        + \frac{{\cal P}_1 + {\cal P}_2}{2} \frac{\partial {\cal P}_{V1}}{\partial \bar{\varphi}} \delta\pi
        + \frac{{\cal P}_1 + {\cal P}_2}{2} \frac{\partial {\Theta}_{V1}}{\partial \bar{\varphi}} \bar{\pi} \delta\pi
        \\
        &&\qquad
        + \Phi_3 \frac{\partial {\cal P}_{V1}}{\partial \bar{\varphi}} \delta^c_j \delta K^j_c
        + \Phi_3 \frac{\partial {\Theta}_{V1}}{\partial \bar{\varphi}} \bar{\pi} \delta^c_j \delta K^j_c
        + \frac{\Theta_2}{2 \bar{p}} \frac{\partial {\cal P}_{V1}}{\partial \bar{\varphi}} \partial_c \partial^j \delta E^c_j
        + \frac{\Theta_2}{2 \bar{p}} \frac{\partial {\Theta}_{V1}}{\partial \bar{\varphi}} \bar{\pi} \partial_c \partial^j \delta E^c_j
        \Bigg]\,,\nonumber
    \end{eqnarray}

\begin{eqnarray}
    &&\{\tilde{H}^{(0)}_V[\bar{N}_1],\tilde{H}^{(1)}_V[\delta N_2]\}_{\bar{A}} =
    \frac{\kappa}{3V_0} \frac{\partial \tilde{H}^{(0)}_V[\bar{N}_1]}{\partial \bar{k}}
    \frac{\partial \tilde{H}^{(1)}_V[\delta N_2]}{\partial \bar{p}}
    - \frac{\kappa}{3V_0} \frac{\partial \tilde{H}^{(0)}_V[\bar{N}_1]}{\partial \bar{p}}
    \frac{\partial \tilde{H}^{(1)}_V[\delta N_2]}{\partial \bar{k}}
    \nonumber\\
    &&\qquad\qquad\qquad\qquad\qquad\quad
    + \frac{1}{V_0} \frac{\partial \tilde{H}^{(0)}_V[\bar{N}_1]}{\partial \bar{\varphi}}
    \frac{\partial \tilde{H}^{(1)}_V[\delta N_2]}{\partial \bar{\pi}}
    -
    \frac{1}{V_0} \frac{\partial \tilde{H}^{(0)}_V[\bar{N}_1]}{\partial \bar{\pi}}
    \frac{\partial \tilde{H}^{(1)}_V[\delta N_2]}{\partial \bar{\varphi}}
    \nonumber\\
    &&=
    \int{\rm d}^3 x\; \bar{N}_1 \delta N_2 \frac{\kappa}{3} \bar{p}^3
    \Bigg[
    \left(\frac{\partial {\cal P}_{V1}}{\partial \bar{k}} \left( \frac{3 {\cal P}_{dV1}}{2\bar{p}} + \frac{\partial{\cal P}_{dV1}}{\partial\bar{p}}\right)
    - \left(\frac{3 {\cal P}_{V1}}{2\bar{p}} + \frac{\partial {\cal P}_{V1}}{\partial \bar{p}} \right) \frac{\partial{\cal P}_{dV1}}{\partial\bar{k}} \right) \delta\varphi
    \nonumber\\
    &&\qquad
    +
    \left(\frac{\partial {\Theta}_{V1}}{\partial \bar{k}} \left( \frac{3 {\cal P}_{dV1}}{2\bar{p}} + \frac{\partial{\cal P}_{dV1}}{\partial\bar{p}}\right)
    - \left(\frac{3 {\Theta}_{V1}}{2\bar{p}} + \frac{\partial {\Theta}_{V1}}{\partial \bar{p}} \right) \frac{\partial{\cal P}_{dV1}}{\partial\bar{k}} \right) \bar{\pi}\delta\varphi
    \nonumber\\
    &&\qquad
    +
    \left(\frac{\partial {\cal P}_{V1}}{\partial \bar{k}} \left( \frac{3 {\Theta}_{dV1}}{2\bar{p}} + \frac{\partial{\Theta}_{dV1}}{\partial\bar{p}}\right)
    - \left(\frac{3 {\cal P}_{V1}}{2\bar{p}} + \frac{\partial {\cal P}_{V1}}{\partial \bar{p}} \right) \frac{\partial{\Theta}_{dV1}}{\partial\bar{k}} \right) \bar{\pi}\delta\varphi
    \nonumber\\
    &&\qquad
    +
    \left(\frac{\partial {\Theta}_{V1}}{\partial \bar{k}} \left( \frac{3 {\Theta}_{dV1}}{2\bar{p}} + \frac{\partial{\Theta}_{dV1}}{\partial\bar{p}}\right)
    - \left(\frac{3 {\Theta}_{V1}}{2\bar{p}} + \frac{\partial {\Theta}_{V1}}{\partial \bar{p}} \right) \frac{\partial{\Theta}_{dV1}}{\partial\bar{k}} \right) \bar{\pi}^2 \delta\varphi
    \nonumber\\
    &&\qquad
    + \left( \frac{\partial {\cal P}_{V1}}{\partial \bar{k}} \left( \frac{3 \Phi_{V1}}{2\bar{p}} + \frac{\partial\Phi_{V1}}{\partial\bar{p}} \right)
    - \left(\frac{3 {\cal P}_{V1}}{2\bar{p}} + \frac{\partial {\cal P}_{V1}}{\partial \bar{p}} \right) \frac{\partial\Phi_{V1}}{\partial\bar{k}} \right)  \delta^d_k \delta K_d^k
    \nonumber\\
    &&\qquad
    + \left( \frac{\partial {\Theta}_{V1}}{\partial \bar{k}} \left( \frac{3 \Phi_{V1}}{2\bar{p}} + \frac{\partial\Phi_{V1}}{\partial\bar{p}} \right)
    - \left(\frac{3 {\Theta}_{V1}}{2\bar{p}} + \frac{\partial {\Theta}_{V1}}{\partial \bar{p}} \right) \frac{\partial\Phi_{V1}}{\partial\bar{k}} \right)  \bar{\pi} \delta^d_k \delta K_d^k
     \nonumber\\
    &&\qquad
    + \left( \frac{\partial {\cal P}_{V1}}{\partial \bar{k}} \left( \frac{{\cal P}_{V2}}{2\bar{p}} + \frac{\partial{\cal P}_{V2}}{\partial\bar{p}} \right)
    - \left(\frac{3 {\cal P}_{V1}}{2\bar{p}} + \frac{\partial {\cal P}_{V1}}{\partial \bar{p}} \right) \frac{\partial{\cal P}_{V2}}{\partial\bar{k}} \right) \frac{\delta^j_c \delta E^c_j}{2 \bar{p}}
     \nonumber\\
    &&\qquad
    + \left( \frac{\partial {\Theta}_{V1}}{\partial \bar{k}} \left( \frac{{\cal P}_{V2}}{2\bar{p}} + \frac{\partial{\cal P}_{V2}}{\partial\bar{p}} \right)
    - \left(\frac{3 {\Theta}_{V1}}{2\bar{p}} + \frac{\partial {\Theta}_{V1}}{\partial \bar{p}} \right) \frac{\partial{\cal P}_{V2}}{\partial\bar{k}} \right) \bar{\pi}\frac{\delta^j_c \delta E^c_j}{2 \bar{p}}
     \nonumber\\
    &&\qquad
    + \left( \frac{\partial {\cal P}_{V1}}{\partial \bar{k}} \left( \frac{{\Theta}_{V2}}{2\bar{p}} + \frac{\partial{\Theta}_{V2}}{\partial\bar{p}} \right)
    - \left(\frac{3 {\cal P}_{V1}}{2\bar{p}} + \frac{\partial {\cal P}_{V1}}{\partial \bar{p}} \right) \frac{\partial{\Theta}_{V2}}{\partial\bar{k}} \right) 
    \bar{\pi}
    \frac{\delta^j_c \delta E^c_j}{2 \bar{p}}
     \nonumber\\
    &&\qquad
    + \left( \frac{\partial {\Theta}_{V1}}{\partial \bar{k}} \left( \frac{{\Theta}_{V2}}{2\bar{p}} + \frac{\partial{\Theta}_{V2}}{\partial\bar{p}} \right)
    - \left(\frac{3 {\Theta}_{V1}}{2\bar{p}} + \frac{\partial {\Theta}_{V1}}{\partial \bar{p}} \right) \frac{\partial{\Theta}_{V2}}{\partial\bar{k}} \right)
    \bar{\pi}^2
    \frac{\delta^j_c \delta E^c_j}{2 \bar{p}}
    \Bigg]
    \\
    &&\quad
    + 
    \int{\rm d}^3 x\; \bar{N}_1 \delta N_2 \bar{p}^3
    \Bigg[
    \left(\frac{\partial {\cal P}_{V1}}{\partial \bar{\varphi}}
    {\Theta}_{dV1}
    -
    \frac{\partial {\cal P}_{dV1}}{\partial \bar{\varphi}}  {\Theta}_{V1}
    + \left(\frac{\partial {\Theta}_{V1}}{\partial \bar{\varphi}}
    {\Theta}_{dV1}
    +\frac{\partial {\Theta}_{dV1}}{\partial \bar{\varphi}}  {\Theta}_{V1}\right) \bar{\pi}\right) \delta\varphi
    \nonumber \\
    &&\qquad
    + \left(\frac{\partial {\cal P}_{V1}}{\partial \bar{\varphi}}{\Theta}_{V2}
    + \frac{\partial {\cal P}_{V2}}{\partial \bar{\varphi}}  {\Theta}_{V1}
    + \left(\frac{\partial {\Theta}_{V1}}{\partial \bar{\varphi}}{\Theta}_{V2}+ \frac{\partial {\Theta}_{V2}}{\partial \bar{\varphi}} {\Theta}_{V1}\right)\bar{\pi}\right) \frac{\delta^j_c \delta E^c_j}{2\bar{p}}
    \nonumber \\
    &&\qquad
    + \frac{\partial \Phi_{V1}}{\partial \bar{\varphi}} 
     {\Theta}_{V1}\delta^d_k \delta K_d^k
    \Bigg]\nonumber
\end{eqnarray}

\begin{eqnarray}
    && \{\tilde{H}^{(2)}_{\rm grav}[\bar{N}_1],\tilde{H}^{(1)}_V[\delta N_2]\}_\delta
    = \int{\rm d}^3x \kappa \left[\frac{\delta \tilde{H}^{(2)}_{\rm grav}[\bar{N}_1]}{\delta \delta K_a^i} \frac{\delta \tilde{H}^{(1)}_V[\delta N_2]}{\delta \delta E^a_i}
    - \frac{\delta \tilde{H}^{(2)}_{\rm grav}[\bar{N}_1]}{\delta \delta E^a_i} \frac{\delta \tilde{H}^{(1)}_V[\delta N_2]}{\delta \delta K_a^i}\right]
    \nonumber\\
    &&
    = \int{\rm d}^3x \bar{N}_1 \delta N_2 \bar{p}^2 \Bigg[
    \left( \frac{\Phi_{V1}}{\bar{p}} {\cal K}^{(2)}_3
    - \frac{{\cal P}_{V2}}{\bar{p}} {\cal K}^{(2)}_1 \right)  \delta^d_k\delta K_d^k
    - \left(
    \frac{\Phi_{V1}}{2 \bar{p}} {\cal K}^{(2)}_4
    + \frac{{\cal P}_{V2}}{\bar{p}} {\cal K}^{(2)}_3 \right)
    \frac{\delta^k_d \delta E^d_k}{2 \bar{p}}
    \Bigg]
    \nonumber\\
    &&\qquad
    + \int{\rm d}^3x\; \bar{N}_1 \delta N_2 \bar{p} \left[ -\Phi_{V1} {\cal K}^{(2)}_6  \frac{\partial^i \partial_a \delta E^a_i}{2 \bar{p}} \right]\,,
\end{eqnarray}

\begin{eqnarray}
    && \{\tilde{H}^{(2)}_V[\bar{N}_1],\tilde{H}^{(1)}_{\rm grav}[\delta N_2]\}_\delta
    = \int{\rm d}^3x\; \kappa \left[\frac{\delta \tilde{H}^{(2)}_V[\bar{N}_1]}{\delta \delta K_a^i} \frac{\delta \tilde{H}^{(1)}_{\rm grav}[\delta N_2]}{\delta \delta E^a_i}
    - \frac{\delta \tilde{H}^{(2)}_V[\bar{N}_1]}{\delta \delta E^a_i} \frac{\delta \tilde{H}^{(1)}_{\rm grav}[\delta N_2]}{\delta \delta K_a^i}\right]
    \nonumber\\
    &&
    = \int{\rm d}^3x\; \bar{N}_1 \delta N_2 \bar{p} \Bigg[
    \left(
    3 {\cal P}_{dV2} {\cal K}^{(1)}_1
    + 3 {\Theta}_{dV2} \bar{\pi} 
    {\cal K}^{(1)}_1
    - 
    \frac{3 \Phi_{V4}}{2} {\cal K}^{(1)}_2 
    \right) 
    \delta\varphi
    \nonumber\\
    &&\qquad\qquad
    + \Phi_{V3} {\cal K}^{(1)}_1 \delta^d_k \delta K^k_d
    + \left({\cal P}_{V3} {\cal K}^{(1)}_1
    - \frac{\Phi_{V3}}{2} {\cal K}^{(1)}_2
    + {\Theta}_{V3}
    {\cal K}^{(1)}_1 \bar{\pi} \right) \frac{\delta_d^k \delta E^d_k}{2 \bar{p}}
    \Bigg]
    \nonumber\\
    &&\quad
    + \int{\rm d}^3x\; \bar{N}_1 \delta N_2 \bar{p} \left[ \Phi_{V3} {\cal K}^{(1)}_3 \frac{\partial^i\partial_a\delta E^a_i}{2\bar{p}}
    + \Phi_{V4} {\cal K}^{(1)}_3 
    \delta^{ab}\partial_a\partial_b\delta\varphi \right]
    \nonumber\\
    &&\quad
    - \int{\rm d}^3x\; \bar{N}_1 \delta N_2 \bar{p}
    \left[
    3 \Phi_{V5}
    +\Phi_{V6} 
    \right]  {\cal K}^{(1)}_2 \delta^c_j \delta K^j_c
    \nonumber\\
    &&\quad
    + \int{\rm d}^3x\; \bar{N}_1 \delta N_2 \bar{p}
    \left[
    2\Phi_{V5} {\cal K}^{(1)}_3 {\partial}^a {\partial}_a \left(\delta^c_j \delta K^j_c\right)
    + 2\Phi_{V6} {\cal K}^{(1)}_3 {\partial}^c {\partial}_j \delta K^j_c \right]  \,,
\end{eqnarray}

\begin{eqnarray}
    && \{\tilde{H}^{(2)}_\varphi[\bar{N}_1],\tilde{H}^{(1)}_V[\delta N_2]\}_\delta
    = \int{\rm d}^3x \left[\kappa \frac{\delta \tilde{H}^{(2)}_\varphi[\bar{N}_1]}{\delta \delta K_a^i} \frac{\delta \tilde{H}^{(1)}_V[\delta N_2]}{\delta \delta E^a_i}
    - \kappa \frac{\delta \tilde{H}^{(2)}_\varphi[\bar{N}_1]}{\delta \delta E^a_i} \frac{\delta \tilde{H}^{(1)}_V[\delta N_2]}{\delta \delta K_a^i}
    \right.
    \nonumber\\
    &&\left.\qquad\qquad\qquad\qquad\qquad\qquad
    + \frac{\delta \tilde{H}^{(2)}_\varphi[\bar{N}_1]}{\delta \delta \varphi} \frac{\delta \tilde{H}^{(1)}_V[\delta N_2]}{\delta \delta \pi}
    - \frac{\delta \tilde{H}^{(2)}_\varphi[\bar{N}_1]}{\delta \delta \pi} \frac{\delta \tilde{H}^{(1)}_V[\delta N_2]}{\delta \delta \varphi}\right]
    \nonumber\\
    &&
    - \int {\rm d}^3x \bar{N}_1\,\delta N_2\, \kappa \bar{p}^{3/2}\,\Phi_{V1} \Bigg[
    - \frac{3{\cal P}_4}{2\bar{p}} \frac{\bar{\pi} \delta\pi}{\bar{p}^{3/2}}
    + \left(\frac{2}{3}\left(\bar{k} \frac{\partial  {\cal P}_1}{\partial \bar{k}} - \bar{p} \frac{\partial  {\cal P}_1}{\partial \bar{p}}\right)+{\cal P}_1 \right) \frac{\bar{\pi}^2}{2\bar{p}^{3/2}} \frac{\delta^j_c\delta E^c_j}{2\bar{p}^2}  \nonumber \\ 
    && \quad - 3\left(\frac{2}{3}\left(\bar{k} \frac{\partial  {\cal P}_1}{\partial \bar{k}} - \bar{p} \frac{\partial  {\cal P}_1}{\partial \bar{p}}\right)+{\cal P}_1 -2{\cal P}_4\right) \frac{\bar{\pi}^2}{2\bar{p}^{3/2}} \frac{\delta^j_c\delta E^c_j}{4\bar{p}^2}
    + \frac{2}{3} \frac{\partial  {\cal P}_1}{\partial \bar{k}} \frac{\bar{\pi}^2}{2\bar{p}^{3/2}} \frac{\delta^a_i  \delta K_a^i }{2 \bar{p}}
    - 3\bar{k} \Phi_5 \frac{\delta\pi}{2\bar{p}}  \nonumber \\ 
    && \quad -\frac{2}{3} \left(\bar{k} \frac{\partial  {\cal P}_1}{\partial \bar{k}} - \bar{p} \frac{\partial  {\cal P}_1}{\partial \bar{p}}\right) \bar{\pi} \frac{\delta^j_c\delta E^c_j}{4\bar{p}^2}
    + 3 \left(
    \frac{2}{3} \left(\bar{k} \frac{\partial  {\cal P}_1}{\partial \bar{k}} - \bar{p} \frac{\partial  {\cal P}_1}{\partial \bar{p}}\right)
    + 2 \bar{k}^2 \Phi_{10}
    \right) \bar{\pi} \frac{\delta^j_c\delta E^c_j}{4\bar{p}^2} + \frac{2}{3} \frac{\partial \Phi_0}{\partial \bar{k}}\bar{\pi}
    \frac{\delta^a_i  \delta K_a^i }{2 \bar{p}}  \nonumber \\ 
    && \quad - 3 (\Phi_5 + 2\bar{k} \Phi_{10}) \frac{\bar{\pi} \delta^c_j \delta K_c^j}{2 \bar{p}}
    - \bigg(
    \Theta_3 \bar{\pi} + \Theta_6 \frac{\bar{\pi}^2}{2\bar{p}^{3/2}}
    \bigg)
    \frac{\delta^{ak}}{\bar{p}} \partial_a \partial_d
    \delta E^d_k
    - \bigg(\frac{\sqrt{\bar{p}}}{2} {\cal P}_9
    +
    \Phi_{11} \bar{\pi}
    \bigg)
    \delta^c_j 
    \partial_c \partial^j
    \delta \varphi
    \Bigg]
    \nonumber\\
    &&
    + \int {\rm d}^3x \bar{N}_1\,\delta N_2\, \kappa
    \Biggl[
    \frac{\Phi}{8\bar{p}^2}{\cal P}_{V2}\bar{\pi}^2
    +
    \frac{\Phi}{8\bar{p}^2}{\Theta}_{V2}\bar{\pi}^3
    +\frac{{\cal P}_{V2}}{6\bar{p}^{1/2}}\frac{\partial \Phi_0}{\partial \bar{k}} \bar{\pi}
    +\frac{{\Theta}_{V2}}{6\bar{p}^{1/2}}\frac{\partial \Phi_0}{\partial \bar{k}} \bar{\pi}^2
    \Biggr]
    \delta^j_c\delta E^c_j
    \nonumber\\
    &&
    - \int {\rm d}^3x\; \bar{N}_1 \delta N_2 \bar{p}^{3/2} \Bigg[
    {\cal P}_4 \left({\cal P}_{dV1} + {\Theta}_{dV1} \bar{\pi}\right) \frac{\delta \pi}{\bar{p}^{3/2}}
    - {\cal P}_4 \left({\cal P}_{dV1} + {\Theta}_{dV1} \bar{\pi}\right) \frac{\bar{\pi}}{\bar{p}^{3/2}} \frac{\delta^j_c\delta E^c_j}{2\bar{p}} 
    \nonumber \\
    &&\qquad
    - \bar{k} {\Phi}_5 \left({\cal P}_{dV1} + {\Theta}_{dV1} \bar{\pi}\right) \frac{\delta^j_c \delta E^c_j}{2\bar{p}}
    + {\Phi}_5 \left({\cal P}_{dV1} + {\Theta}_{dV1} \bar{\pi}\right) \delta^c_j \delta K^j_c
    \Bigg]\,,
\end{eqnarray}
\begin{eqnarray}
    && \{\tilde{H}^{(2)}_V[\bar{N}_1],\tilde{H}^{(1)}_\varphi[\delta N_2]\}_\delta
    = \int{\rm d}^3x\; \left[\kappa \frac{\delta \tilde{H}^{(2)}_V[\bar{N}_1]}{\delta \delta K_a^i} \frac{\delta \tilde{H}^{(1)}_\varphi[\delta N_2]}{\delta \delta E^a_i}
    - \kappa \frac{\delta \tilde{H}^{(2)}_V[\bar{N}_1]}{\delta \delta E^a_i} \frac{\delta \tilde{H}^{(1)}_\varphi[\delta N_2]}{\delta \delta K_a^i}
    \right.
    \nonumber\\
    &&\left.\qquad\qquad\qquad\qquad\qquad\qquad
    + \frac{\delta \tilde{H}^{(2)}_V[\bar{N}_1]}{\delta \delta \phi} \frac{\delta \tilde{H}^{(1)}_\varphi[\delta N_2]}{\delta \delta \pi}
    - \frac{\delta \tilde{H}^{(2)}_V[\bar{N}_1]}{\delta \delta \pi} \frac{\delta \tilde{H}^{(1)}_\varphi[\delta N_2]}{\delta \delta \phi}\right]
    \nonumber\\
    &&
    = \int{\rm d}^3x\; \bar{N}_1 \delta N_2 \frac{\bar{\pi}^2}{2} \kappa \left[ - \Phi_{V3} \frac{{\cal P}_2}{2 \bar{p}} \frac{\delta^k_d\delta E^d_k}{2\bar{p}}
    - 3 \Phi_{V4} \frac{{\cal P}_2}{2 \bar{p}} \delta\varphi \right]
    \nonumber\\
    &&\qquad
    + \int{\rm d}^3x\; \bar{N}_1 \delta N_2 \bar{\pi} \bar{p}^{3/2}\kappa \left[ - \Phi_{V3} \Phi_{2} \frac{\delta^k_d\delta E^d_k}{2\bar{p}}
    - 3 \Phi_{V4} \Phi_{2} \delta\varphi \right]
    \nonumber\\
    &&\qquad
    + \int{\rm d}^3x\; \bar{N}_1 \delta N_2 \bar{\pi} \Bigg[ 
    {\cal P}_{ddV}
    {\cal P}_3 \delta\varphi
    + {\cal P}_{dV2} {\cal P}_3 \frac{\delta^j_c\delta E^c_j}{2\bar{p}}
    + \Phi_{V4} {\cal P}_3 \delta^d_k \delta K^k_d
    \Bigg]
    \nonumber\\
    &&\qquad
    + \int{\rm d}^3x\; \bar{N}_1 \delta N_2 \bar{\pi}^2 \Bigg[ 
    {\Theta}_{ddV}
    {\cal P}_3 \delta\varphi
    + {\Theta}_{dV2} {\cal P}_3 \frac{\delta^j_c\delta E^c_j}{2\bar{p}}
    \Bigg]
    \nonumber\\
    &&\qquad
    - \int{\rm d}^3x\; \bar{N}_1 \delta N_2 
    \Bigg[
     \frac{{\cal P}_2}{2\bar{p}}
    (
    3 {\Phi}_{V5} 
    + {\Phi}_{V6} 
    )
    \bar{\pi}^2 \delta^c_j \delta K^j_c
    \Bigg]
    \nonumber\\
    &&\qquad
    \int{\rm d}^3x\; \bar{N}_1 \delta N_2 \bar{p}^{3/2} {\Phi}_2
    \Bigg[
    6 {\Phi}_{V5} + 2 {\Phi}_{V6}
    \Bigg] \bar{\pi} \delta^c_j \delta K^j_c
    \nonumber\\
    &&\qquad
    \int{\rm d}^3x\; \bar{N}_1 \delta N_2 \bar{p}^{3/2} {\Theta}_1
    \Bigg[
    6 {\Phi}_{V5} + 2 {\Phi}_{V6}
    \Bigg] \bar{\pi} \delta^c_j {\partial}^b {\partial}_b \delta K^j_c
    \nonumber\\
    &&\qquad
    \int{\rm d}^3x\; \bar{N}_1 \delta N_2 \bar{p}^{1/2} {\Theta}_2
    \Bigg[
    3 {\Phi}_{V5}
    +  {\Phi}_{V6} 
    \Bigg] \bar{\pi} {\partial}^c {\partial}_j \delta K^j_c\,,
\end{eqnarray}
\begin{eqnarray}
    && \{\tilde{H}^{(2)}_V[\bar{N}_1],\tilde{H}^{(1)}_V[\delta N_2]\}_\delta
    = \int{\rm d}^3x\; \kappa \left[\frac{\delta \tilde{H}^{(2)}_V[\bar{N}_1]}{\delta \delta K_a^i} \frac{\delta \tilde{H}^{(1)}_V[\delta N_2]}{\delta \delta E^a_i}
    - \frac{\delta \tilde{H}^{(2)}_V[\bar{N}_1]}{\delta \delta E^a_i} \frac{\delta \tilde{H}^{(1)}_V[\delta N_2]}{\delta \delta K_a^i}\right]
    \nonumber\\
    &&
    = \int{\rm d}^3x\; \bar{N}_1 \delta N_2 \bar{p}^2 \kappa \Bigg[ 
    \left( 
    \frac{3{\cal P}_{V2}}{2} \Phi_{V4}
    - \frac{3 {\cal P}_{dV2}}{2} \Phi_{V1}
    - \frac{3 {\Theta}_{dV2}}{2} \Phi_{V1} \bar{\pi}
    \right) \delta\varphi
    \nonumber\\
    &&\qquad
    - \frac{\Phi_{V1} \Phi_{V3}}{2} \delta^d_k\delta K^k_d
    + \left(\frac{{\cal P}_{V2}}{2} \Phi_{V3}
    - \frac{{\cal P}_{V3}}{2} \Phi_{V1}
    - \frac{{\Theta}_{V3}}{2} \Phi_{V1} \bar{\pi}
    \right) \frac{\delta^k_d \delta E^d_k}{2\bar{p}}
    \Bigg]
    \nonumber\\
    &&\qquad
    +\int{\rm d}^3x\; \bar{N}_1 \delta N_2 \bar{p}^2 \kappa
    \Bigg[
    3 {\Phi}_{V5} + {\Phi}_{V6}
    \Bigg]
    {\cal P}_{V2} \delta^c_j \delta K^j_c\,.
\end{eqnarray}

\end{appendix}

%\bibliographystyle{../preprint.bst}
%\bibliography{../Bib/QuantGra.bib}

\end{document}